\documentclass[aos]{aosart}
\makeatletter\def\journal@name{}\makeatother
\startlocaldefs
\RequirePackage{amsthm,amsmath,amsfonts,amssymb,mathtools}
\RequirePackage[authoryear]{natbib}
\RequirePackage{graphicx,booktabs,array}
\RequirePackage{enumitem}
\RequirePackage{xcolor}
\RequirePackage{algorithm}
\RequirePackage{algpseudocode}
\RequirePackage[colorlinks,citecolor=blue!50!black,urlcolor=blue!50!black,linkcolor=blue!50!black]{hyperref}

\numberwithin{equation}{section}
\theoremstyle{plain}
\newtheorem{theorem}{Theorem}[section]
\newtheorem{proposition}[theorem]{Proposition}
\newtheorem{lemma}[theorem]{Lemma}
\newtheorem{corollary}[theorem]{Corollary}
\theoremstyle{definition}

\newtheorem{assumption}[theorem]{Assumption}
\newtheorem{example}[theorem]{Example}
\newtheorem{remark}[theorem]{Remark}

\renewcommand{\topfraction}{0.85}

\renewcommand{\floatpagefraction}{0.75}
\newcommand{\rp}{\psi}

\newcommand{\Pp}{\mathbb{P}}
\newcommand{\E}{\mathbb{E}}
\newcommand{\Hh}{\mathcal{H}}
\newcommand{\R}{\mathbb{R}}
\newcommand{\MMD}{\operatorname{MMD}}
\newcommand{\TV}{d_{\mathrm{TV}}}

\newcommand{\dist}{\operatorname{dist}}
\newcommand{\supp}{\operatorname{supp}}

\newcommand{\cohNX}{2{,}184}
\newcommand{\cohNY}{3{,}551}

\newcommand{\rgSep}{3}

\newcommand{\regTV}{0.45}
\newcommand{\regAlphaLo}{0.20}
\newcommand{\regAlphaHi}{0.65}

\newcommand{\calibReps}{400}

\newcommand{\harderAloGtMax}{0.035}
\newcommand{\harderAloGtMaxNull}{0.065}

\newcommand{\harderBoundaryCells}{73}

\newcommand{\harderExtCells}{67}

\newcommand{\harderExtReps}{200}
\newcommand{\harderFlagK}{19}

\newcommand{\harderFlipEsPowOff}{1.00}
\newcommand{\harderFlipEsPowOn}{0.72}
\newcommand{\harderFlipEsReps}{60}

\newcommand{\harderNcap}{12{,}800}

\newcommand{\harderNmax}{800}

\newcommand{\harderNmin}{100}

\newcommand{\harderNtwo}{1600}

\newcommand{\harderNullExtCells}{15}

\newcommand{\harderNullExtMaxGbm}{9}
\newcommand{\harderNullExtMaxLogit}{12}
\newcommand{\harderNullExtMaxRamp}{12}

\newcommand{\harderNullExtReplicates}{3{,}000}

\newcommand{\harderReps}{200}

\newcommand{\harderRunAnalyticEight}{0.33}
\newcommand{\harderRunExtCap}{13}
\newcommand{\harderRunExtKernelCap}{11}

\newcommand{\harderRunSplitEight}{5.9}

\newcommand{\calibCPUpper}{0.007}

\newcommand{\rhcNtotal}{5{,}735}

\newcommand{\cstBetRatioEighty}{1.75}

\newcommand{\cstCzero}{2.73}

\newcommand{\regWeightsP}{0.60, 0.30, 0.10}
\newcommand{\regWeightsQ}{0.15, 0.35, 0.50}

\newcommand{\hvBdWorstRamp}{5}

\newcommand{\hvNullRamp}{2.8}

\newcommand{\hvBdWorstGbm}{4}

\newcommand{\hvNullGbm}{2.5}

\newcommand{\hvBdWorstLogit}{13}

\newcommand{\hvNullLogit}{3.0}

\newcommand{\loExSigma}{1}
\newcommand{\loExAtoms}{24}
\newcommand{\loExPzero}{0.4}
\newcommand{\loExCommon}{0.4}
\newcommand{\loExCommonLoc}{-100}
\newcommand{\loExSpread}{0.2}
\newcommand{\loExSpreadK}{20}
\newcommand{\loExSpreadLo}{30}
\newcommand{\loExSpreadNext}{40}
\newcommand{\loExSpreadHi}{220}
\newcommand{\loExQhalf}{0.3}
\newcommand{\loExQloc}{0.5}
\newcommand{\loExQfar}{0.3}
\newcommand{\loExFar}{300}
\newcommand{\loExTail}{50}
\newcommand{\loExTV}{0.6}
\newcommand{\loExKS}{0.4}
\newcommand{\loExCumTwo}{0.1}
\newcommand{\loExCumSpread}{0.3}
\newcommand{\loExFzero}{0.135}
\newcommand{\loExFhalf}{0.053}
\newcommand{\loExFspread}{0.019}

\newcommand{\allNdesigns}{19}

\newcommand{\estAtmRD}{0.035}

\newcommand{\estKernelRD}{0.022}

\newcommand{\dNineRampBdMax}{7}

\newcommand{\dNineGbmBdMax}{9}

\newcommand{\dNineLogitBdMax}{7}

\newcommand{\dNineAloGtMax}{0.015}

\newcommand{\abSvmNuMed}{0.80}

\newcommand{\abNdesigns}{19}
\newcommand{\abReps}{200}

\newcommand{\abPermB}{999}
\newcommand{\abAggL}{10}

\newcommand{\abSvmAboveRamp}{2}

\newcommand{\abWtlRecRampZeroCut}{5/8/6}

\newcommand{\abRecMedRamp}{0.31}

\newcommand{\abRecMedRampZero}{0.22}

\newcommand{\abRecMedRampZeroCut}{0.35}

\newcommand{\abRecMedLogitMatchedLabels}{0.35}

\newcommand{\abRecMedLogitMatchedProb}{0.15}

\newcommand{\abRampAboveBestMatchedPow}{3}
\newcommand{\abBestMatchedAboveRampPow}{5}

\newcommand{\abSingleBetAboveEb}{16}

\newcommand{\abBoundaryMax}{0.07}
\newcommand{\abBoundaryMaxK}{13}

\newcommand{\abNullMax}{0.08}

\newcommand{\abNullMaxP}{0.044}

\newcommand{\mgRows}{134}

\newcommand{\mgLapShiftEps}{1\times10^{-5}}
\newcommand{\mgLapShiftSlope}{1.24}
\newcommand{\mgGaussEpsHi}{1.1\times10^{-1}}
\newcommand{\mgGaussEpsLo}{2.9\times10^{-4}}
\newcommand{\mgGaussPathHi}{1.0\times10^{-1}}
\newcommand{\mgGaussPathLo}{1.05\times10^{-8}}

\newcommand{\mgFitPathLo}{0.68}
\newcommand{\mgFitPathHi}{0.87}

\newcommand{\mgPiEight}{1.23}

\newcommand{\exLapSlopeS}{0.51}
\newcommand{\exLapSlopeM}{0.47}
\newcommand{\exGaussSlopeS}{0.35}
\newcommand{\exGaussSlopeM}{0.38}
\newcommand{\exLapSlopeCommonS}{0.57}
\newcommand{\exLapSlopeCommonM}{0.46}
\newcommand{\exKSSlopeS}{0.50}
\newcommand{\exKSSlopeM}{0.46}
\newcommand{\exAnSlopeS}{0.45}
\newcommand{\exAnSlopeM}{0.36}
\newcommand{\exKSRatioS}{1.7}
\newcommand{\exKSRatioM}{1.3}
\newcommand{\exRepMin}{30}
\newcommand{\exRepMax}{41}
\newcommand{\exNmax}{51200}
\newcommand{\exNmaxGauss}{12800}
\newcommand{\exGridS}{0.4}
\newcommand{\exGridM}{0.12}

\newcommand{\exGridMinS}{0.002}
\newcommand{\exGridMinM}{0.0006}

\newcommand{\fbRdRhcFallbackLogWMax}{0.95}

\newcommand{\fbSyReps}{50}
\newcommand{\fbSyFallbackWins}{1}
\newcommand{\fbSyFallbackTies}{18}
\newcommand{\fbSyFallbackLosses}{0}
\newcommand{\fbSyPlainWins}{0}
\newcommand{\fbSyPlainTies}{7}
\newcommand{\fbSyPlainLosses}{12}

\newcommand{\fbLogInvDelta}{3.00}

\newcommand{\kRdRhcLaplaceOwnMed}{0.35}

\newcommand{\kRdRhcLaplaceFallbackMed}{0.32}

\newcommand{\kRdAgeLaplaceOwnMed}{0.40}

\newcommand{\kRdAgeLaplaceFallbackMed}{0.35}

\newcommand{\mnNeightyRampBest}{14}
\newcommand{\mnNeightyRampBestAlone}{13}
\newcommand{\mnNeightyPermBest}{13}

\newcommand{\mnNeightyRampOverPermLo}{0.40}
\newcommand{\mnNeightyRampOverPermHi}{1.97}
\newcommand{\mnNeightyRampOverPermMed}{1.75}

\newcommand{\mnAnaOverRampLo}{3.9}
\newcommand{\mnAnaOverRampHi}{29}
\newcommand{\mnAnaOverRampMed}{9}
\newcommand{\mnRepsPerSize}{200}
\newcommand{\mnNmin}{100}
\newcommand{\mnNcap}{12{,}800}

\newcommand{\mnMvDesigns}{15}
\newcommand{\mnNeightyLapBelowGaussMv}{4}
\newcommand{\mnNeightyLapAboveGaussMv}{7}

\newcommand{\mnTauFireRatioLo}{0.8}
\newcommand{\mnTauFireRatioHi}{1.5}

\newcommand{\mnRecRampMedian}{0.38}
\newcommand{\mnRecRampMax}{0.81}

\newcommand{\mnRecGaussMedian}{0.31}

\newcommand{\mnRecAnalyticMax}{0.35}

\newcommand{\mnRecRhcRamp}{0.77}

\newcommand{\mnRecNfourLap}{0.69}
\newcommand{\mnRecNfourGauss}{0.02}
\newcommand{\mnLapGaussWins}{4}
\newcommand{\mnLapGaussTies}{9}
\newcommand{\mnLapGaussLosses}{6}
\newcommand{\mnLapGaussWinsMv}{1}

\newcommand{\mnLapGaussLossesMv}{6}
\newcommand{\mnLapGaussWinsLine}{3}
\newcommand{\mnLapGaussLossMax}{0.025}

\newcommand{\mnOrderLinearTies}{17}
\newcommand{\mnOrderLinearLosses}{2}
\newcommand{\mnRampAboveBestMatched}{9}
\newcommand{\mnBestMatchedAboveRamp}{10}

\newcommand{\mnWtlRecLogitTuned}{6/4/9}

\newcommand{\mnRecReps}{200}

\newcommand{\mnBdMax}{8/200}

\newcommand{\mnBdMaxP}{0.787}

\newcommand{\mnBdTotal}{17}
\newcommand{\mnBdReps}{3800}
\newcommand{\mnAloGtMax}{0.010}
\newcommand{\mnNullCells}{30}
\newcommand{\mnNullMax}{11/200}
\newcommand{\mnNullPct}{3.0}

\newcommand{\mnNullLoPct}{0.5}
\newcommand{\mnNullHiPct}{5.5}

\newcommand{\mnAnBdCells}{77}

\newcommand{\mnRdRhcMin}{0.35}

\newcommand{\mnRdRhcGaussMin}{0.30}

\newcommand{\mnRdRhcNoBet}{8}
\newcommand{\mnRdRhcFullGaussMin}{0.30}

\newcommand{\mnPvRhcScale}{0.5}
\newcommand{\mnRdChildMin}{0.75}

\newcommand{\mnRdChildGaussMin}{0.75}

\newcommand{\mnRdChildNoBet}{8}

\newcommand{\mnPvChildLast}{0.30}

\newcommand{\mnPvChildScale}{0.7}
\newcommand{\mnRdAgeMin}{0.35}

\newcommand{\mnRdAgeGaussMin}{0.35}

\newcommand{\mnRdAgeNoBet}{5}

\newcommand{\mnPvAgeLast}{0.05}

\newcommand{\mnPvAgeScale}{0.5}
\newcommand{\mnRdBmiMin}{0.05}

\newcommand{\mnRdBmiGaussMin}{0.00}

\newcommand{\mnPvBmiDzero}{0.075}
\newcommand{\mnPvBmiTauzero}{0.109}
\newcommand{\mnPvBmiFire}{2{,}633}

\newcommand{\mnPvBmiScale}{0.5}

\newcommand{\lcCohAlpha}{0.35}

\newcommand{\lcPoolGap}{\ensuremath{1.9\times10^{-16}}}

\newcommand{\lcPoolSigma}{8.96}

\newcommand{\lcPoolFracX}{154}
\newcommand{\lcPoolKeptX}{1{,}497}

\newcommand{\lcPoolFracMassX}{0.054}

\newcommand{\lcPoolFracY}{100}
\newcommand{\lcPoolKeptY}{2{,}359}

\newcommand{\lcPoolFracMassY}{0.021}

\newcommand{\lcPoolSmoSteps}{9{,}023}

\newcommand{\lcPoolLamMin}{\ensuremath{2.0\times10^{-5}}}
\newcommand{\lcPoolLtwoBound}{\ensuremath{3.0\times10^{-6}}}
\newcommand{\lcPoolWitnessBound}{\ensuremath{1.4\times10^{-8}}}

\newcommand{\lcPoolFracSpreadX}{\ensuremath{1.8\times10^{-11}}}

\newcommand{\lcCohKeptX}{1{,}503}

\newcommand{\lcCohKeptY}{2{,}361}

\newcommand{\lcSensKeptXlo}{1{,}496}
\newcommand{\lcSensKeptXhi}{1{,}503}
\newcommand{\lcSensKeptYlo}{2{,}357}
\newcommand{\lcSensKeptYhi}{2{,}363}

\newcommand{\lcSensSmdMaxLo}{0.030}
\newcommand{\lcSensSmdMaxHi}{0.053}

\newcommand{\lcUnwRD}{0.074}

\newcommand{\lcTrimRD}{0.040}
\newcommand{\lcTrimRDlo}{0.011}
\newcommand{\lcTrimRDhi}{0.072}
\newcommand{\lcTrimSMD}{0.032}

\newcommand{\lcBoot}{2{,}000}

\newcommand{\lcRDmax}{0.056}

\newcommand{\lpPoolFracX}{607}

\newcommand{\lpPoolFracY}{469}

\newcommand{\rrRhcNcovPaper}{33}
\newcommand{\rrRhcNcovFull}{45}
\newcommand{\rrRhcNcovAdded}{12}
\newcommand{\rrAddedSmdAbove}{8}
\newcommand{\rrAddedSmdMax}{0.353}

\newcommand{\rrOrthoN}{7}
\newcommand{\rrTraumaN}{52}
\newcommand{\rrRhcPaperMin}{0.30}

\newcommand{\rrRhcFullMin}{0.35}

\newcommand{\rrCohPaperAlpha}{0.30}

\newcommand{\rrCohPaperSmdMax}{0.065}

\newcommand{\rrCohPaperRD}{0.028}

\newcommand{\rrCohPaperOmitSmdMax}{0.205}

\newcommand{\rrCohPaperOmitSmdAbove}{7}

\newcommand{\rrEvRhcAvgFinal}{0.35}

\newcommand{\rrClfPaperLogitHardMin}{0.30}

\newcommand{\rrClfFullLogitHardMin}{0.35}

\newcommand{\rrClfPaperGbmHardMin}{0.30}

\newcommand{\rrClfFullGbmHardMin}{0.35}

\newcommand{\rrClfPaperLogitSoftMin}{0.15}

\newcommand{\rrClfFullLogitSoftMin}{0.20}

\newcommand{\rrClfPaperGbmSoftMin}{0.25}

\newcommand{\rrClfFullGbmSoftMin}{0.25}

\newcommand{\rrClfPaperLogitTunedHardMin}{0.30}

\newcommand{\rrClfFullLogitTunedHardMin}{0.30}

\newcommand{\rrClfPaperGbmTunedHardMin}{0.35}

\newcommand{\rrClfFullGbmTunedHardMin}{0.35}

\newcommand{\rrClfPaperLogitTunedSoftMin}{0.15}

\newcommand{\rrClfFullLogitTunedSoftMin}{0.15}

\newcommand{\rrClfPaperGbmTunedSoftMin}{0.15}

\newcommand{\rrClfFullGbmTunedSoftMin}{0.20}

\newcommand{\rrPropPaperPlug}{0.38}

\newcommand{\rrPropFullPlug}{0.42}
\newcommand{\rrPropFullPlugLo}{0.41}
\newcommand{\rrPropFullPlugHi}{0.45}

\newcommand{\rrPropFullPlugCf}{0.43}

\newcommand{\rrClChildMin}{0.80}

\newcommand{\rrClChildUnits}{61}
\newcommand{\rrClAgeMin}{0.35}

\newcommand{\rrClBmiMin}{0.05}

\newcommand{\rcReps}{75}
\newcommand{\rcRepsMain}{200}

\newcommand{\rcGridCeiling}{0.875}
\newcommand{\rcLevelsetMC}{1{,}000{,}000}

\newcommand{\rcRatioMedHundred}{0.08}

\newcommand{\rcRatioMedTwoHundred}{0.30}

\newcommand{\rcRatioMedFourHundred}{0.53}

\newcommand{\rcOracleMedEightHundred}{0.65}

\newcommand{\rcRatioMedEightHundred}{0.62}
\newcommand{\rcEightyEightHundred}{4}

\newcommand{\rcRampTenthMedEightHundred}{0.25}
\newcommand{\rcOracleTenthMedEightHundred}{0.50}
\newcommand{\rcEightyDesignsEightHundred}{RHC-shaped $d=33$; overlap $d=2$ ($\alpha^\star=0.2$); oscillating $\nu=1$; oscillating $\nu=4$}
\newcommand{\rcRatioGrows}{17}
\newcommand{\rcRatioGrowsOf}{19}

\newcommand{\rcSmallRamp}{0.21}
\newcommand{\rcSmallPsi}{0.13}
\newcommand{\rcSmallLevelset}{0.34}

\newcommand{\dvZeroCutAboveLap}{9}
\newcommand{\dvZeroCutBelowLap}{7}
\newcommand{\dvZeroCutEqualLap}{3}
\newcommand{\dvKerAnLapBelow}{3}
\newcommand{\dvKerAnLapAbove}{0}
\newcommand{\dvKsCells}{24}
\newcommand{\dvKsAboveKernel}{0}
\newcommand{\dvGbmTunedAboveLapN}{7}
\newcommand{\dvGbmTunedAboveLapDesigns}{sparse shift, $d=50$; mixed types, $m=5n$; $t_3$ tails, $d=5$; curves, narrow bump; oscillating, $\nu=1$ ($\alpha^\star=0.5$); oscillating, $\nu=4$ ($\alpha^\star=0.5$); oscillating, $\nu=16$ ($\alpha^\star=0.5$)}

\newcommand{\mnRecRampMean}{0.36}
\newcommand{\mnRecGaussMean}{0.33}
\newcommand{\mnRecRampMedianNoNfour}{0.35}
\newcommand{\mnRecGaussMedianNoNfour}{0.34}
\newcommand{\mnPowBestClfAboveLap}{4}
\newcommand{\mnPowLapAboveBestClf}{3}
\newcommand{\fxAlphaStar}{0.683}
\newcommand{\fxLevelLap}{0.58}
\newcommand{\fxLevelGauss}{0.65}
\newcommand{\fxFracLap}{0.88}
\newcommand{\fxFracGauss}{0.27}
\newcommand{\fxLevelGaussAtLap}{0.58}
\newcommand{\fxFracGaussAtLap}{0.64}
\newcommand{\stCalibRatioTauP}{6.5}
\def\wtlfirst#1/#2/#3\relax{#1}
\def\wtlsecond#1/#2/#3\relax{#2}
\def\wtlthird#1/#2/#3\relax{#3}
\newcommand{\wtlwins}[1]{\expandafter\wtlfirst#1\relax}
\newcommand{\wtlties}[1]{\expandafter\wtlsecond#1\relax}
\newcommand{\wtllosses}[1]{\expandafter\wtlthird#1\relax}
\def\fourdigitsi#1#2#3#4{#1{,}#2#3#4}
\newcommand{\fourdigits}[1]{\expandafter\fourdigitsi#1}

\RequirePackage{placeins}
\endlocaldefs

\begin{document}

\begin{frontmatter}
\title{Certifying total-variation contamination\\ by betting on trimmed kernel witnesses}
\runtitle{Betting on trimmed kernel witnesses}

\begin{aug}
\author[A]{\fnms{Hristo}~\snm{Inouzhe}\ead[label=e1]{hristo.inouze@uam.es}}
\address[A]{Departamento de Matem\'aticas, Universidad Aut\'onoma de Madrid\printead[presep={,\ }]{e1}}
\end{aug}

\begin{abstract}
Two laws share a common component after each discards a fraction $\alpha$ of its mass, a trimming, exactly when
their total variation is at most $\alpha$, so the contamination level $\alpha^\star=\TV(P,Q)$ is the share of each population
without counterpart in the other. We bound it from below from two samples on any space with a bounded
characteristic kernel. The kernel mean embeddings of all such trimmings of a law form a convex region, the regions of two laws first touch at $\alpha^\star$, and in every direction a region
reaches as far as a trimmed mean. The difference of their closest points, the witness, selects the trimming. We fit a ramp of the witness on part of the data and bet on the rest, at each level of a grid. The bound is
valid at every sample size. The cut points of the tested level carry most of its finite-sample gain over an untrimmed witness, while refitting
the witness at that level makes it consistent, gives its rates and returns the trimming. When a function of the
kernel's Hilbert space separates the parts of the two laws without counterpart, the bound is
within order $(n\wedge m)^{-1/2}$ of $\alpha^\star$ on a grid of that mesh, which is minimax over such pairs. When
these parts touch, a margin condition guarantees a polynomial rate for Sobolev kernels and a logarithmic rate for the
Gaussian kernel. For the Laplace kernel on the line this guarantee is not sharp: the witness stays between its two cut
points, and for densities that cross once, with a likelihood ratio bounded away from one off the crossing, the error
on such a grid is again of that order up to logarithmic factors. No informative upper bound exists without
assumptions. On the Right Heart Catheterization data, with confidence $0.95$, each arm must lose more than a fraction
$\mnRdRhcMin$ of its patients, if both lose the same fraction, before their covariate distributions can coincide.
\space
\end{abstract}



\end{frontmatter}

\section{Introduction}
\label{sec:intro}

A two-sample test answers whether two populations are identical. In many applications they are known not to be: the
two arms of an observational study, a training sample and the population on which a model will be deployed, a
reference cohort and a new batch. The useful question is then how different they are, and the contamination model
\citep{huber1964robust} gives it a precise answer. The total-variation distance
\[
\TV(P,Q)=\sup_A\,|P(A)-Q(A)|
\]
is the largest difference between the probabilities that two laws $P$ and $Q$ give to one event. The two laws can be
written as mixtures $P=(1-\alpha)R+\alpha P_0$ and $Q=(1-\alpha)R+\alpha Q_0$ that share a component $R$ of weight
$1-\alpha$ exactly when $\TV(P,Q)\le\alpha$. The smallest such $\alpha$, $\alpha^\star=\TV(P,Q)$, which we call the
contamination level, is therefore the fraction of each population with no counterpart in the other. In an
observational study it is the smallest fraction whose removal from each arm lets the covariate distributions
coincide. We certify lower bounds on $\alpha^\star$ from two samples, of sizes $n$ and $m$, on any sample
space that carries a bounded characteristic kernel, with finite-sample guarantees and without a model for how the
populations differ, and we also return the trimming of each sample at that level.

Discarding part of a population is a trimming. An $\alpha$-trimming of $P$ is a law obtained from $P$ by removing at
most a fraction $\alpha$ of its mass, possibly only partially at each point, and renormalizing, that is, a law with
density at most $(1-\alpha)^{-1}$ relative to $P$ \citep{gordaliza1991best,alvarez2008trimmed}.
\citet{alvarez2012similarity} observed that two laws have a common $\alpha$-trimming exactly when
$\TV(P,Q)\le\alpha$, so $\alpha^\star$ is the smallest level at which $P$ and $Q$ can be trimmed into the same law.
Their test of $\TV(P,Q)\le\alpha$ trims both samples to minimize a Wasserstein distance, with a bootstrap guarantee
that is asymptotic, pointwise and univariate, and they ask whether ``a different choice of metric could lead to some
type of uniform bound'' \citep[p.~617]{alvarez2012similarity}. With a
kernel discrepancy the answer is positive at every sample size, which gives a lower confidence bound on $\alpha^\star$.
An upper bound needs assumptions: data can show that $\TV(P,Q)$ is large, not that it is small
(Theorem~\ref{thm:impossible}).

A bounded kernel $k$ on the sample space maps each point $x$ to a feature vector $\phi(x)$ in a Hilbert space $\Hh$,
the reproducing kernel Hilbert space (RKHS) of $k$, and each law $P$ to the mean of its feature vectors, the kernel
mean embedding $\mu_P=\E_P[\phi(X)]$. The distance $\|\mu_P-\mu_Q\|_\Hh$ is the maximum mean discrepancy (MMD) of
\citet{gretton2012kernel}, and for a characteristic kernel, such as the Gaussian or the Laplace kernel, the embedding
determines the law. Our approach is geometric: instead of an optimal trimming, we follow the set of everything the
trimmings can produce. The trimmed region $\mathcal C_\alpha(P)$, the set of the embeddings of all $\alpha$-trimmings
of $P$, is a convex compact subset of $\Hh$ that grows with $\alpha$ from the single point $\mu_P$, and the regions of
$P$ and $Q$ first meet exactly at $\alpha^\star$, since a point common to both is the embedding of a trimming of $P$
and of a trimming of $Q$, which must then be the same law (Figure~\ref{fig:regions}). The distance $D_\alpha(P,Q)$
between the regions is thus positive below $\alpha^\star$ and zero from $\alpha^\star$ on. For the empirical laws the
regions are polytopes, whose distance is one convex quadratic program, and to show that $\TV(P,Q)>\alpha$ is to show
that two compact convex sets are apart, which the direction that joins their closest points shows.

The difference of the two closest points is the witness, a function on the sample space that is large where the first
trimmed law has more mass than the second. Once the witness is fixed, choosing the trimming is a linear problem of
bathtub type, so the optimal trimmings keep at full weight the observations on one side of a cut of the witness and
discard those beyond it, only the observations at the cut sharing weight. The quadratic program that gives the
distance, solved on all the data at the certified level, therefore also gives the trimmed samples. They converge to
the best trimming at a fixed level, at the rate $(n\wedge m)^{-1/2}$ when the witness has a density at the cuts, and to
the common component of $P$ and $Q$ at the estimated level, the target of a cohort of comparable patients.

Any function $\rp$ with values in $[0,1]$ has $\E_P\rp-\E_Q\rp\le\TV(P,Q)$, so a test that shows
$\E_P\rp-\E_Q\rp>\alpha$ shows that $\TV(P,Q)>\alpha$, and the geometry says which $\rp$ to use. In a direction
$f\in\Hh$ a region reaches as far as a trimmed mean, the mean of $f(X)$ over the upper $(1-\alpha)$-fraction of its
values, so the separation of the two regions in that direction is a difference of two trimmed means. A trimmed mean is
an infimum of ordinary means, $\inf_t\{t+(1-\alpha)^{-1}\E_P[(f(X)-t)_+]\}$, one for each cut $t$
\citep{rockafellaruryasev2000}. At the two cuts at which the trimmings at the tested level cut, the separation in the
direction of the witness is at most a positive multiple of the excess $\E_P\rp-\E_Q\rp-\alpha$ of a ramp $\rp$ over the
level, where $\rp$ is $0$ below the lower cut, $1$ above the upper one and linear in between, so a positive separation
gives a positive excess. We fit the witness and its cuts on one part of the sample. As the ramp is bounded and fixed
before the rest of the data are seen, betting on it gives a test valid at every sample size with no assumption on the
laws, which rejects the faster the larger the excess. Our
lower confidence bound $\widehat\alpha$ runs this test on a grid of levels fixed in advance, read upward to the first
level not rejected, and its error is measured against the mesh of the grid, the largest gap between its levels.

Betting, like classification, is a problem of estimating a set. For every $[0,1]$-valued $\rp$,
$\E_P\rp-\E_Q\rp=\alpha^\star-\mathrm{err}(\rp)$, with $\mathrm{err}(\rp)$ the distance under $|P-Q|$ from $\rp$ to the
indicator of the set where $P$ has the larger density (Section~\ref{sec:linearmargin}), so betting on $\rp$ certifies
levels near $\alpha^\star$ only if $\rp$ is close to that indicator. The trimming enters this estimate twice. The cut
points of the tested level carry most of the finite-sample gain over a ramp of the untrimmed witness in our simulations
(Section~\ref{sec:numharder}), while refitting the witness at that level makes the bound consistent where a ramp of the
untrimmed witness need not be (Proposition~\ref{prop:loconsistency}, Section~\ref{sec:alphazero}), gives the rates
below and returns the trimming. Any such $\rp$ fitted on earlier data, a classifier for instance, can replace the
ramp, and tuned classifiers certify more where a half-space or a tree is close to that set. A threshold valid under
every pair in the null also turns the empirical distance $\widehat D_\alpha$ into an analytic estimator, which needs no
split but many more observations, our baseline.

The geometry also governs the rates. No valid lower bound has a rate uniform over all pairs
(Section~\ref{sec:upper}), and the rate of ours depends on how fast $D_\alpha$ grows as $\alpha$ decreases below
$\alpha^\star$. A margin $D_{\alpha^\star-\epsilon}(P,Q)\ge c_0\epsilon^\gamma$ guarantees an error of order
$(n\wedge m)^{-1/(2\gamma)}$ up to the mesh. The margin is linear, $\gamma=1$, when a function of the RKHS separates
the two contaminations, the parts of $P$ and $Q$ beyond their common component, and the bound computed on a grid of
mesh $(n\wedge m)^{-1/2}$ is then within order $(n\wedge m)^{-1/2}$ of $\alpha^\star$, which is minimax over such
pairs. When the contaminations touch, as for continuous densities that coincide on no open set, the path
$\alpha\mapsto D_\alpha$ leaves zero tangentially for every continuous kernel, and the margin guarantees only
$(n\wedge m)^{-2/(2s+3)}$ for a Sobolev kernel of order $s$ on the line and a logarithmic rate for the Gaussian kernel,
with no lower bound proved to match.

The margin is pessimistic: it bounds the excess of the population ramp, the ramp of the population witness between
its population cuts, through the gap between these cuts, which it replaces by its largest value. When
$D_\alpha$ is of exact order $(\alpha^\star-\alpha)^\gamma$, the population ramp keeps an excess of at least a fixed
fraction of $\alpha^\star-\alpha$ whatever $\gamma$ (Proposition~\ref{prop:excess}), where the margin sees an excess of
order $(\alpha^\star-\alpha)^\gamma$ only. For the Laplace kernel on the line we show
that the fitted ramp keeps it too: for densities that cross once, under a condition \eqref{eq:condR} on the likelihood
ratio that excludes contaminations local to a common core, the error on a grid of mesh $(n\wedge m)^{-1/2}$ is of order
$(n\wedge m)^{-1/2}$ up to logarithmic factors. A one-sided Kolmogorov--Smirnov bound attains this rate there, and a
bound that does not sort the data, and also returns the set, matches it. The Laplace witness hurts the trimming: it stays
between its cut points, so the observations at the cut tie, while our rate for the trimming at a fixed level assumes a
witness with a density at the cut, as with the Gaussian kernel. So the two kernels divide the work, the Laplace kernel
for the bound and the Gaussian kernel where the trimming needs a rate. In $\R^d$ we keep the Laplace kernel as the
default, because its witness still sees differences of high frequency that the Gaussian witness at the same bandwidth
misses, as on our one-dimensional oscillating densities, although it trails the Gaussian kernel slightly in our
multivariate designs; Section~\ref{sec:discussion} conjectures the rate attainable in $\R^d$.

On the Right Heart Catheterization study of \citet{connors1996rhc}, comparing $\rrRhcNcovFull$ baseline covariates of
the patients who received a catheter and of those who did not, our bound shows that $\TV(P,Q)>\mnRdRhcMin$.
Before the covariate distributions can coincide, one arm, or each if both lose the same fraction, must lose more than
that fraction. This is not what the rules that discard patients by their estimated propensity score control
\citep{crump2009overlap,lei2021overlap,damour2021overlap}: they aim at positivity for a weighted comparison, and a total
variation of this size is compatible with propensity scores bounded well away from $0$ and $1$. On accelerometer curves from the National Health and Nutrition Examination Survey the bound measures how
far apart age and body-mass groups are.

Our bound is an instance of the lower bounds on $\TV$ of \citet{naf2022hplb}, which evaluate a function fitted on
other data, calibrated by the betting test of \citet{shekhar2024nonparametric} with a trimmed null. Such bounds and the
classifier tests \citep{lopezpaz2017revisiting,kim2021classification} apply on any sample space. What the geometry adds
is a canonical function to bet on, fitted at the level tested, for which the bound is consistent and has rates, and
the trimmed samples that come with it. As the Bayes risk of telling $P$ from $Q$ with equal priors is
$\{1-\TV(P,Q)\}/2$, the bound is also an upper confidence bound on a Bayes error, usually estimated
\citep{berisha2016empirically,noshad2019learning,theisen2021evaluating}. \citet{horebarber2026blurred} bound a blurred
relaxation of $\TV$. Mixture proportion estimation \citep{blanchard2010semi,scott2015rate,ramaswamy2016mixture} targets a
one-sided relative, $1-\TV(P,Q)$ being the largest weight of a component common to $P$ and $Q$, with the same dichotomy
of rates (Section~\ref{sec:optimality}), and finite-sample lower bounds on a mixing proportion go back to the
proportion of true null hypotheses in multiple testing
\citep{genovese2004stochastic,meinshausen2006estimating,patra2016estimation}. None of them trims both samples, and
\citet{nafshi2024partial}, when a known fraction of one sample has been assigned to the other, trim the receiving sample
alone. In one sample, the mixture index of fit of \citet{rudas1994index} \citep[see also][]{liu2009tolerance} is the
smallest fraction of the data that must be trimmed for the rest to follow a model, and \citet{delbarrio2020ks} bound it
on the line with the Kolmogorov--Smirnov distance. Other related work is cited where it is used.

Sections~\ref{sec:regions} to~\ref{sec:trimming} treat the regions, the test and the analytic baseline, the
split-sample bound, its rates and the trimming, Section~\ref{sec:numerical} the simulations and the data, and the
supplement gives the proofs and the experimental details. All computations use \texttt{trimmedmmd}, a Python package
released at \url{https://github.com/HristoInouzhe/trimmed-mmd}.
\space
\section{Trimmed regions and the witness}
\label{sec:regions}

The lower bound and the trimming come from one object, the distance between the sets of kernel mean embeddings that
the trimmings of two laws produce. This section defines it and shows that the two sets first meet at the contamination
level, that in every direction each set reaches exactly as far as a trimmed mean, and that the direction of the
difference of their closest points selects the trimming. The last two facts, and the empirical regions as polytopes, are RKHS
forms of known properties of zonoid trimmed regions and reduced convex hulls, and the equivalence \eqref{eq:zeroeq}
follows at once from \eqref{eq:tvintersect} for a characteristic kernel. New are the uniform concentration
\eqref{eq:hausdorff} and the analysis of the path in Section~\ref{sec:rates}.

\subsection{Trimmings and total variation}

We observe $X_1,\dots,X_n$ i.i.d.\ $P$ and $Y_1,\dots,Y_m$ i.i.d.\ $Q$, two independent samples on a measurable space
$\mathcal X$, with empirical measures $P_n$ and $Q_m$; we write $u_+=\max(u,0)$, $a\vee b=\max(a,b)$ and
$a\wedge b=\min(a,b)$. For $0\le\alpha<1$ let $\beta=(1-\alpha)^{-1}$. Formally, the \emph{$\alpha$-trimmings} of $P$ are the laws
\begin{equation}
R_\alpha(P)\coloneqq\Big\{P'\ll P:\ \frac{dP'}{dP}\le\beta\ \ P\text{-almost surely}\Big\}.
\label{eq:trimset}
\end{equation}
This trimming is impartial, in that the data decide what is removed, with no region such as the tails prescribed
\citep{gordaliza1991best,cuesta1997trimmed}. The fact of \citet{alvarez2012similarity} recalled in the introduction reads
\begin{equation}
\TV(P,Q)\le\alpha
\iff
R_\alpha(P)\cap R_\alpha(Q)\neq\varnothing ,
\label{eq:tvintersect}
\end{equation}
or, equivalently, there are probability measures $R$, $P_0$ and $Q_0$ with
\begin{equation}
P=(1-\alpha)R+\alpha P_0,
\qquad
Q=(1-\alpha)R+\alpha Q_0 ,
\label{eq:contam}
\end{equation}
and from now on $\alpha^\star\coloneqq\TV(P,Q)$ always denotes the contamination level of the pair at hand. At
$\alpha^\star$ the common component is the \emph{common core} $R^\star=(P\wedge Q)/(1-\alpha^\star)$, where $P\wedge Q$ has density $\min(p,q)$ for
densities $p$ and $q$ of $P$ and $Q$ with respect to any common dominating measure. Its complement $1-\alpha^\star$ is the overlapping
coefficient of $P$ and $Q$ \citep{inman1989overlapping}. For instance, if $P$ puts mass $\tfrac12$ on each of two
points $x_1,x_2$ and $Q$ on each of $x_2,x_3$, then $\alpha^\star=\tfrac12$, and removing $x_1$ from $P$ and $x_3$
from $Q$ leaves the point mass at $x_2$ in both. The level is shared because both laws lose the same fraction. If
they lose fractions $\alpha_1$ and $\alpha_2$ and what remains coincides, then $\alpha_1\vee\alpha_2\ge\alpha^\star$
(Section~\ref{app:regions} of the supplement).

The kernel $k$ on $\mathcal X$ has reproducing kernel Hilbert space (RKHS) $\Hh$ and feature map
$\phi(x)=k(x,\cdot)$, so that $f(x)=\langle f,\phi(x)\rangle_\Hh$ for $f\in\Hh$, and as in the introduction
$\mu_P=\E_P[\phi(X)]$ and $\MMD(P,Q)=\|\mu_P-\mu_Q\|_\Hh$ \citep{gretton2012kernel,muandet2017kernel}. Two constants of the kernel
enter our bounds, $\kappa=\sup_xk(x,x)^{1/2}$, which bounds the norm of every feature vector, and the diameter
$\Delta_\phi=\sup_{x,x'}\|\phi(x)-\phi(x')\|_\Hh\le2\kappa$ of the feature vectors. The Laplace kernel
$k(x,y)=\exp(-\|x-y\|/\sigma)$ and the Gaussian kernel $k(x,y)=\exp(-\|x-y\|^2/(2\sigma^2))$ on $\R^d$, of bandwidth $\sigma$, both have
$\kappa=1$ and $\Delta_\phi=\sqrt2$, and Section~\ref{sec:numerical} says where each is used.

\begin{assumption}
\label{ass:regularity}
$\mathcal X$ is a measurable space, with no further structure. Moreover:
\begin{enumerate}[label=(A\arabic*),leftmargin=*]
\item $x\mapsto k(x,x')$ is measurable for every $x'$, and $k(x,x)\le\kappa^2$ for all $x$;
\item $\Hh$ is separable;
\item $k$ is characteristic, that is, $\MMD(P',Q')=0$ only if $P'=Q'$, for probability measures $P',Q'$ on
$\mathcal X$ \citep{sriperumbudur2010hilbert}.
\end{enumerate}
\end{assumption}

Assumption~\ref{ass:regularity} holds for the Gaussian, Laplace and inverse multiquadric kernels on $\R^d$, and for
any bounded continuous characteristic kernel on a separable metric space, so curves, or graphs with a characteristic
kernel, are covered on the same terms as vectors. Theorem~\ref{thm:plugin} uses only (A1) and (A2),
Theorem~\ref{thm:betting} needs only that the function bet on be measurable with values in $[0,1]$, and (A3) makes the
split-sample bound consistent (Proposition~\ref{prop:loconsistency}).

\subsection{Trimmed regions}

\begin{figure}[t]
\centering
\includegraphics[width=\textwidth]{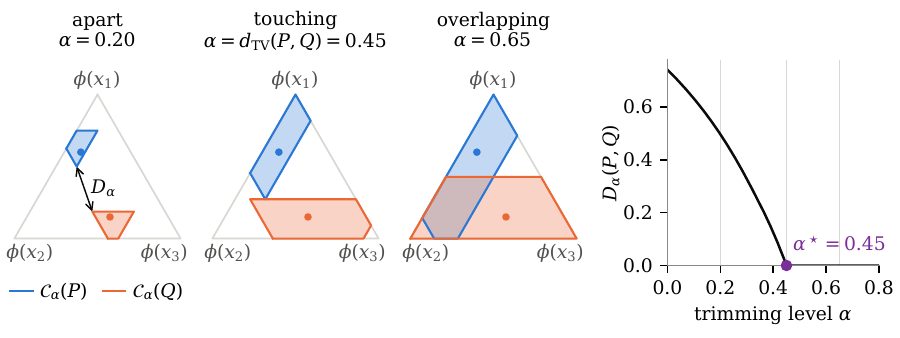}
\caption{Left: the regions $\mathcal C_\alpha(P)$ and $\mathcal C_\alpha(Q)$ of laws with weights $(\regWeightsP)$ and
$(\regWeightsQ)$ on three points whose feature vectors are unit vectors at $120$ degrees in the plane, drawn
isometrically (untrimmed embeddings as dots), at $\alpha=\regAlphaLo$, at $\alpha=\TV(P,Q)=\regTV$ and at
$\alpha=\regAlphaHi$. Right: the distance $D_\alpha(P,Q)$, computed exactly, with these levels marked.}
\label{fig:regions}
\end{figure}

The \emph{trimmed maximum mean discrepancy} is
\begin{equation}
D_\alpha(P,Q)\coloneqq\inf\big\{\MMD(P',Q'):\ P'\in R_\alpha(P),\ Q'\in R_\alpha(Q)\big\} ,
\label{eq:trimMMDpop}
\end{equation}
an analogue of optimal partial transport \citep{caffarelli2010free,figalli2010partial} with the MMD between the kept
parts in place of a transport cost, and with the Wasserstein distance the outlier-robust distance of
\citet{nietert2022robust}. \citet{mukherjee2021robust} penalize the change made to one law, kernel mean matching
\citep{huang2007correcting,gretton2008covariate} reweights one sample toward the other, and robust MMD estimators
\citep{lerasle2019monk,cherief2022finite} protect against a contamination, where we measure it. The \emph{trimmed
region} of $P$ is
\begin{equation}
\mathcal C_\alpha(P)\coloneqq\big\{\E_P[g(X)\phi(X)]:\ 0\le g\le\beta,\ \E_P[g(X)]=1\big\}\subset\Hh ,
\label{eq:region}
\end{equation}
the set of kernel mean embeddings of all $\alpha$-trimmings of $P$, since $g$ ranges over their densities. The regions
grow with $\alpha$ from the single point $\mu_P$, and Figure~\ref{fig:regions} draws them for two laws on three
points. For sets $A,B\subset\Hh$ we write $\dist(A,B)=\inf\{\|a-b\|_\Hh:\ a\in A,\ b\in B\}$, and
$\dist(a,B)=\dist(\{a\},B)$.

\begin{proposition}[Geometric representation]
\label{prop:georep}
Under Assumption~\ref{ass:regularity}, for every $\alpha\in[0,1)$,
\begin{equation}
D_\alpha(P,Q)=\dist\big(\mathcal C_\alpha(P),\mathcal C_\alpha(Q)\big),
\label{eq:setdist}
\end{equation}
and $\mathcal C_\alpha(P_n)=\{\sum_iw_i\phi(X_i):\ 0\le w_i\le\beta/n,\ \sum_iw_i=1\}$ is a polytope.
\end{proposition}

\begin{lemma}[Compactness and attainment]
\label{lem:attainment}
Under Assumption~\ref{ass:regularity}, $\mathcal C_\alpha(P)$ is convex, norm-bounded by $\kappa$ and
norm-compact. Hence the infimum in \eqref{eq:trimMMDpop} is attained at trimmings $P^\star\in R_\alpha(P)$ and
$Q^\star\in R_\alpha(Q)$, and if $D_\alpha(P,Q)=0$, then $\mathcal C_\alpha(P)\cap\mathcal C_\alpha(Q)$ is nonempty and
each of its elements is the embedding of a common trimming $R\in R_\alpha(P)\cap R_\alpha(Q)$.
\end{lemma}

With \eqref{eq:tvintersect} and (A3), these two results give
\begin{equation}
D_\alpha(P,Q)=0\iff\TV(P,Q)\le\alpha .
\label{eq:zeroeq}
\end{equation}
We call
$\alpha\mapsto D_\alpha(P,Q)$ the \emph{population path} of the trimmed distance, and its empirical counterpart
$\alpha\mapsto\widehat D_\alpha$ of Section~\ref{sec:statistic} the \emph{empirical path}. The population path is
positive below $\alpha^\star$ and zero from $\alpha^\star$ on, as in the right panel of Figure~\ref{fig:regions}. It
is nonincreasing, because the regions are nested, and Lipschitz with constant $4\kappa/(1-\bar\alpha)^2$ on
$[0,\bar\alpha]$ for every $\bar\alpha<1$ (Lemma~\ref{lem:continuity} of the supplement).

A closed convex set $A\subset\Hh$ is determined by its support function $h_A(f)=\sup_{a\in A}\langle a,f\rangle_\Hh$,
how far it reaches in the direction $f$. For the trimmed regions this function is a familiar risk measure, the
conditional value-at-risk of finance \citep{rockafellaruryasev2000}.

\begin{lemma}[Support function]
\label{lem:cvar}
Under Assumption~\ref{ass:regularity}, for every $f\in\Hh$,
\begin{equation}
h_{\mathcal C_\alpha(P)}(f)
=\sup_{0\le g\le\beta,\ \E_Pg=1}\E_P[g(X)f(X)]
=\inf_{t\in\R}\Big\{t+\beta\,\E_P\big[(f(X)-t)_+\big]\Big\},
\label{eq:cvar}
\end{equation}
the supremum is attained, and the second equality holds for every bounded measurable $f$.
\end{lemma}

The first equality is the reproducing property, and the right-hand side is the representation by
\citet{rockafellaruryasev2000} of the mean of $f(X)$ over the upper $(1-\alpha)$-fraction of its distribution, an atom
at the cut being split. The lemma is the RKHS instance of the defining property of zonoid trimmed regions
\citep{koshevoy1997zonoid,dyckerhoff2011weighted}.

Maximizing $\E_P[gf]$ over trimming densities is a bathtub problem, whose solution $g$ fills its cap $\beta$ on the
largest values of $f$ until the mass runs out and is zero below. In a fixed direction the optimal trimming therefore
keeps the part of the law on one side of a cut, the level set $\{f\ge t\}$ with $t$ the $\alpha$-quantile of $f(X)$,
and the minimizing $t$ in \eqref{eq:cvar} is that cut.

For each fixed $t$, moreover, the expression inside the infimum is an ordinary mean, of the bounded function
$t+\beta(f-t)_+$. A region is thus a family of trimmed means, one per direction, and a trimmed mean is an infimum of
ordinary means. Means of bounded functions, and of bounded vectors of $\Hh$, concentrate at the rate $n^{-1/2}$ with
constants that depend only on the bound $\kappa$, and this rate cannot be improved for the mean embedding or the MMD
\citep{tolstikhin2016minimax,tolstikhin2017minimax}. Every finite-sample validity statement below reduces to such
means, and the empirical regions are uniformly close to the population ones,
\begin{equation}
\sup_{0\le\alpha<1}(1-\alpha)\,d_H\big(\mathcal C_\alpha(P_n),\mathcal C_\alpha(P)\big)
\le\kappa\Big[\frac4{\sqrt n}+\sqrt{\frac{2\log(2/\delta)}n}\,\Big]
\label{eq:hausdorff}
\end{equation}
with probability at least $1-\delta$, for every $P$, where $d_H$ is the Hausdorff distance in $\Hh$
(Corollary~\ref{cor:uniform} of the supplement). No complexity of the sample space enters \eqref{eq:hausdorff}, because the
regions live in $\Hh$ and the kernel bound $\kappa$ is their only constant.

\subsection{The statistic, its witness and the trimming it selects}
\label{sec:statistic}

The statistic is $\widehat D_\alpha\coloneqq D_\alpha(P_n,Q_m)$, and $\alpha\mapsto\widehat D_\alpha$ is the empirical
path. By Proposition~\ref{prop:georep} it is the distance between two polytopes, the square root of
\begin{equation}
\widehat D_\alpha^{\,2}
=\min_{w,v}\big\{w^\top K_{XX}w+v^\top K_{YY}v-2w^\top K_{XY}v\big\}
\label{eq:qp}
\end{equation}
over the capped simplices $0\le w_i\le\beta/n$, $\sum_iw_i=1$ and $0\le v_j\le\beta/m$, $\sum_jv_j=1$, where $K_{XX}$,
$K_{YY}$ and $K_{XY}$ are the Gram matrices. The objective is $\|\sum_iw_i\phi(X_i)-\sum_jv_j\phi(Y_j)\|_\Hh^2$, so
\eqref{eq:qp} is a convex quadratic program, the problem of the nearest points of two reduced convex hulls, which is the
dual of the $\nu$-support vector machine \citep{bennett2000duality,crisp1999geometric,takeda2008nusvm}. The witness
below is the normal of its hyperplane, and Proposition~\ref{prop:bangbang} is the structure of its support vectors read
as a trimming. We solve it by accelerated projected gradient (Section~\ref{sec:computationdetails} of the supplement). For a minimizer $(w,v)$, the \emph{witness}, the name that \citet{gretton2012kernel} use at $\alpha=0$, is
the difference of the two closest points,
\begin{equation}
\widehat d=\sum_iw_i\phi(X_i)-\sum_jv_j\phi(Y_j),
\qquad
\widehat d(x)=\sum_iw_ik(X_i,x)-\sum_jv_jk(Y_j,x).
\label{eq:witness}
\end{equation}
It is a function on $\mathcal X$, large where the trimmed first sample has more mass than the trimmed second, and it
can be evaluated at points not used to compute it. Its direction $\widehat f=\widehat d/\widehat D_\alpha$, with $\widehat f=0$ when $\widehat D_\alpha=0$,
separates the two empirical regions. The population witness $d^\star$ is the element of minimal norm of
$\mathcal C_\alpha(P)-\mathcal C_\alpha(Q)$, with direction $f^\star=d^\star/D_\alpha(P,Q)$.

Let $\mathcal T_\alpha$ be the set of minimizers $(w,v)$ of \eqref{eq:qp}. Each of them is a pair of trimmed samples,
since the weighted empirical measures $P_n^w=\sum_iw_i\delta_{X_i}$ and $Q_m^v=\sum_jv_j\delta_{Y_j}$ are
$\alpha$-trimmings of $P_n$ and $Q_m$ whose embeddings are at distance $\widehat D_\alpha$. The witness determines
them.

\begin{proposition}[Structure of the empirical trimming]
\label{prop:bangbang}
The set $\mathcal T_\alpha$ is nonempty, convex and compact. Moreover:
\begin{enumerate}[label=(\roman*),leftmargin=*]
\item the difference $\widehat d=\sum_iw_i\phi(X_i)-\sum_jv_j\phi(Y_j)$ takes the same value at every
$(w,v)\in\mathcal T_\alpha$, namely the element of minimal norm of $\mathcal C_\alpha(P_n)-\mathcal C_\alpha(Q_m)$, and
$\|\widehat d\|_\Hh=\widehat D_\alpha$;
\item if $\widehat D_\alpha>0$, let $n_{1-\alpha}=\lfloor n(1-\alpha)\rfloor$ and let $\widehat t_n$ be the
$(n_{1-\alpha}+1)$-th smallest of the values $\widehat d(X_1),\dots,\widehat d(X_n)$, with $\widehat t_n=+\infty$ if
$n_{1-\alpha}=n$. Then every $(w,v)\in\mathcal T_\alpha$ has $w_i=\beta/n$ when $\widehat d(X_i)<\widehat t_n$ and $w_i=0$
when $\widehat d(X_i)>\widehat t_n$, the observations with $\widehat d(X_i)=\widehat t_n$ sharing the remaining mass.
Symmetrically, with $m_{1-\alpha}=\lfloor m(1-\alpha)\rfloor$, $v_j=\beta/m$ when $\widehat d(Y_j)$ is above the
$(m_{1-\alpha}+1)$-th largest of the values $\widehat d(Y_1),\dots,\widehat d(Y_m)$, $v_j=0$ when it is below, and the ties
share the remainder;
\item if $\widehat D_\alpha=0$, then $\mathcal T_\alpha$ is the set of all feasible $(w,v)$ with
$\sum_iw_i\phi(X_i)=\sum_jv_j\phi(Y_j)$.
\end{enumerate}
\end{proposition}

Part (ii) is the bathtub of Lemma~\ref{lem:cvar} in the direction of the witness. Once $\widehat d$ is fixed the
problem is linear: the weights on the $X$'s minimize $\sum_iw_i\widehat d(X_i)$ over the capped simplex, and those on
the $Y$'s maximize $\sum_jv_j\widehat d(Y_j)$. So every optimal trimming of the first sample keeps at full weight the
observations on which the witness is smallest, the most $Q$-like ones, discards those above the cut and splits weight
only among ties at the cut, and the trimming of the second sample is the level set above a cut of the same function.
The cuts reappear in the betting test of Section~\ref{sec:split}. Part (iii) does not arise for $\alpha<1$ with a
strictly positive definite kernel and pairwise distinct pooled observations $X_1,\dots,X_n,Y_1,\dots,Y_m$, since the
feature vectors are then linearly independent and the minimizer is unique. Many observations may still share the cut,
as support vectors share the margin of a $\nu$-support vector machine, and with the Laplace kernel on the line all the
$X$'s whose witness values have rank $\lceil n(1-\alpha)\rceil$ or higher share it (Lemma~\ref{lem:lapclip} of the
supplement). For kernels whose empirical regions can meet we return the minimizer closest to uniform
weights (Section~\ref{sec:computationdetails} of the supplement).

\subsection{No informative upper bound}
\label{sec:upper}

By \eqref{eq:zeroeq}, $\TV(P,Q)>\alpha$ exactly when the regions are apart, and Sections~\ref{sec:analytic}
and~\ref{sec:split} certify this by lower confidence bounds valid on any sample space. We certify only from below
because no upper confidence bound on $\TV(P,Q)$ can be valid for all pairs of laws and still informative, as
\citet[Theorem~1.1]{horebarber2026blurred} prove for $\TV$ and \citet[Theorem~9]{blanchard2010semi} for mixture
proportions, in a line that goes back to \citet{bahadursavage1956} and \citet{donoho1988onesided}. The version below
holds even for laws on $[0,1]$ with densities bounded by $2$, and its conclusion is at $P=Q$.

\begin{theorem}[No informative upper bound, even for bounded densities]
\label{thm:impossible}
Fix $n,m\ge1$ and $\delta\in(0,1)$, and let $\mathcal P_2$ be the set of pairs of probability measures on $[0,1]$
whose densities with respect to Lebesgue measure are bounded by $2$. Suppose that
$\widehat U=\widehat U(X_{1:n},Y_{1:m})$, measurable and possibly randomized, satisfies
$\Pp_{P,Q}(\TV(P,Q)\le\widehat U)\ge1-\delta$ for every $(P,Q)\in\mathcal P_2$. Then, with $P_{\mathrm u}$ the uniform
law on $[0,1]$, $\Pp_{P_{\mathrm u},P_{\mathrm u}}(\widehat U\ge1)\ge1-\delta$.
\end{theorem}

A fixed kernel is blind in the same way: densities that oscillate ever faster stay at total variation $\tfrac12$ from
the uniform law while their MMD to it tends to zero (Section~\ref{sec:upperimposs} of the supplement). An upper end
needs an assumption, such as a margin with known constants, since $D_\alpha(P,Q)\ge c_0(\alpha^\star-\alpha)^\gamma$
gives $\alpha^\star\le\alpha+(\check D_\alpha/c_0)^{1/\gamma}$ for any $\check D_\alpha\ge D_\alpha(P,Q)$, or a blurred
version of $\TV$, and the data do not identify $c_0$ and $\gamma$. Lower bounds escape the argument, because rejecting
$\TV(P,Q)\le\alpha$ needs only one set on which the samples differ, but their rates do not: a valid lower bound with a
rate uniform over $\mathcal P_2$ would give an informative upper bound, so some restriction such as the margin is needed
for any rate.
\space
\section{Testing a trimmed null, and an analytic baseline}
\label{sec:analytic}

For each level $\alpha\in[0,1)$ we test
\begin{equation}
H_{0,\alpha}:\ \TV(P,Q)\le\alpha
\qquad\text{against}\qquad
H_{1,\alpha}:\ \TV(P,Q)>\alpha ,
\label{eq:null}
\end{equation}
and by \eqref{eq:zeroeq} the null holds exactly when $D_\alpha(P,Q)=0$. The null is a total-variation neighborhood of
$P=Q$, a relevant hypothesis in the sense of \citet{dettekokotaue2020}, as in the trimming-based tests of
\citet{alvarez2012similarity} and \citet{delbarrio2020ks} and in robust tests between neighborhoods of laws
\citep{huber1965robust,huber1973minimax,schrab2025robust,liubriol2025robust}, which fix its size in advance, whereas
here it is the quantity to be estimated. Kernel tests of composite nulls \citep{key2025composite} calibrate over a
parametric family, and equivalence tests \citep{wellek2010equivalence} put such a neighborhood in the alternative
(Section~\ref{app:analytic} of the supplement). This section gives a threshold $\tau(\alpha)$, computed before the
data, that $\widehat D_\alpha$ exceeds with probability at most $\delta$ under every pair in the null. Inverted over
$\alpha$, it gives a baseline that needs no split but far more observations than our method.

\subsection{A threshold for the whole null}

Limit laws are of little help in calibrating $\widehat D_\alpha$, because at the boundary of the null the two regions
touch, and a minimum over trimmings is then only directionally differentiable in the laws, with a limit law that
depends on the unknown geometry of the contact \citep{fang2019inference}, and permutation is exact only under $P=Q$. We
bound $\widehat D_\alpha$ instead by a difference of sample means. Under $H_{0,\alpha}$ the laws share an
$\alpha$-trimming $R$ by \eqref{eq:tvintersect}, with densities $g_P=dR/dP$ and $g_Q=dR/dQ$ bounded by $\beta$, so the
weights $w_i=g_P(X_i)/n$ and $v_j=g_Q(Y_j)/m$ respect the caps of \eqref{eq:qp}, and the embeddings they give are
averages of independent vectors with the same mean $\mu_R$. Their totals fluctuate around one, and rescaling the weights,
or adding mass under the caps, gives a feasible pair at the price of a term proportional to the shortfalls
(Lemma~\ref{lem:plugin} of the supplement). So $\widehat D_\alpha$ is at most the norm of a centered difference of two
averages of vectors bounded by $\beta\kappa$, plus two shortfalls, and bounded-difference and Bernstein inequalities
give a threshold in closed form. Let $n_\star=nm/(n+m)$, so that $1/n_\star=1/n+1/m$, $c=\max(1/n,1/m)$, $L=\log(6/\delta)$ and $\bar\kappa^2=\min(\kappa^2,\Delta_\phi^2/2)$.

\begin{theorem}[Plug-in threshold]
\label{thm:plugin}
Let (A1) and (A2) of Assumption~\ref{ass:regularity} hold, and let
\begin{gather*}
\tau^{\mathrm P}(0)=\bar\kappa/\sqrt{n_\star}+\Delta_\phi\sqrt{\log(1/\delta)/(2n_\star)},\\
\tau^{\mathrm P}(\alpha)=\sqrt{\beta \bar\kappa^2/n_\star}+\beta \Delta_\phi\sqrt{L/(2n_\star)}+\Delta_\phi\Big[\sqrt{2(\beta-1)L/n_\star}+\frac{cL}3\Big]
\qquad(0<\alpha<1).
\end{gather*}
Then for every $\alpha\in[0,1)$, $n,m\ge1$, $\delta\in(0,1)$ and every pair with $\TV(P,Q)\le\alpha$,
$\Pp\big(\widehat D_\alpha>\tau^{\mathrm P}(\alpha)\big)\le\delta$. Moreover, $\alpha\mapsto\tau^{\mathrm P}(\alpha)$
is nondecreasing on $[0,1)$.
\end{theorem}

The threshold depends on $\kappa$, $\Delta_\phi$, $\alpha$, $n$, $m$ and $\delta$ alone and is, up to a term of order
$\Delta_\phi\log(1/\delta)/(n\wedge m)$, of order $\beta \Delta_\phi\{(1+\log(1/\delta))/(n\wedge m)\}^{1/2}$. It
jumps up at $0+$, where $L>\log(1/\delta)$ replaces $\log(1/\delta)$. The pair enters only through the existence of $R$. At
$\alpha=0$ the common trimming is $R=P=Q$, and Theorem~\ref{thm:plugin} is a finite-sample bound on the MMD of the
kind given by \citet[Corollary~9]{gretton2012kernel}. For the Gaussian kernel and $n=m$ it reads
$\tau^{\mathrm P}(0)=[1+\sqrt{\log(1/\delta)}\,]\sqrt{2/n}$.

We write $\tau=\tau^{\mathrm P}$ for the threshold of the test. A second valid threshold $\tau^{\mathrm S}(\alpha,\delta)$
comes from the two-sided bound of Theorem~\ref{thm:twosided} of the supplement, by the same reduction to means as
\eqref{eq:hausdorff}, but it is larger than $\tau^{\mathrm P}$ at the sample sizes we use. Since $\tau$ is
nondecreasing and the empirical path is nonincreasing, one event serves every level.

\begin{lemma}[One event for all levels]
\label{lem:oneevent}
Let $\tau:[0,1)\to[0,\infty)$ be nondecreasing, and suppose that $\Pp(\widehat D_\alpha>\tau(\alpha))\le\delta$ for
every $\alpha$ and every pair with $\TV(P,Q)\le\alpha$. Then for every pair with $\alpha^\star<1$, the event
$\{\widehat D_{\alpha^\star}\le\tau(\alpha^\star)\}$ has probability at least $1-\delta$, and on it
$\widehat D_\alpha\le\tau(\alpha)$ for every $\alpha\in[\alpha^\star,1)$.
\end{lemma}

The analytic test rejects $H_{0,\alpha}$ when
\begin{equation}
\widehat D_\alpha>\tau(\alpha) .
\label{eq:test}
\end{equation}
It has level $\delta$ for all $n,m\ge1$, uniformly over the null, by Lemma~\ref{lem:oneevent} one event carries the
guarantee for all levels at once, and its level does not use (A3). Its validity at every sample size, uniformly over the null, answers
the question of \citet{alvarez2012similarity} recalled in the introduction, as does, more simply, the betting test of
Section~\ref{sec:split}. The kernel must be fixed before the data are seen, for instance on a held-out pilot sample
(Section~\ref{sec:data}). One threshold for every pair in the null is a worst case: at the boundary of a
two-dimensional design, at $n=m=300$ and $\alpha=\alpha^\star$, $\tau(\alpha)$ is $\stCalibRatioTauP$ times the
empirical $1-\delta$ quantile of $\widehat D_\alpha$ (Section~\ref{app:experiments} of the supplement).

\subsection{The estimator}
\label{sec:alpha}

Inverting the test over $\alpha$, we define the analytic estimator, with a subscript that sets it apart from the
split-sample bound $\widehat\alpha$ of Section~\ref{sec:lower},
\begin{equation}
\widehat\alpha_{\mathrm{an}}\coloneqq\inf\big\{\alpha\in[0,1):\ \widehat D_\alpha\le\tau(\alpha)\big\} ,
\label{eq:alphaest}
\end{equation}
the first level at which the empirical path falls below the threshold. Since $\alpha\mapsto\widehat D_\alpha$ is
nonincreasing and $\tau$ nondecreasing, every level above $\widehat\alpha_{\mathrm{an}}$ is accepted and every level below it rejected,
so bisection brackets $\widehat\alpha_{\mathrm{an}}$. It is well defined, because
$\widehat D_\alpha\le \Delta_\phi$ while $\tau(\alpha)\to\infty$ as $\alpha\to1$ when $\Delta_\phi>0$. On the event of
Lemma~\ref{lem:oneevent}, of probability at least $1-\delta$ whenever $\alpha^\star<1$, every level from $\alpha^\star$
on is accepted, so $\widehat\alpha_{\mathrm{an}}\le\alpha^\star$ and every rejected level lies below $\alpha^\star$, at every sample
size, for every pair and however the levels tested were chosen. A grid of levels read upward, as in
Section~\ref{sec:data}, therefore gives a lower confidence bound too. It is consistent from below,
$\Pp(\widehat\alpha_{\mathrm{an}}<\alpha)\to0$ for $\alpha<\alpha^\star$, as $\widehat D_\alpha\ge
D_\alpha-\tau^{\mathrm S}(\alpha,\delta')$ with probability $1-\delta'$ (Theorem~\ref{thm:twosided} of the supplement)
and both thresholds tend to zero.

\subsection{How many observations}
\label{sec:samplesize}

Because $\tau$ does not depend on the data, the sample size at which the test fires can be computed before any data
are collected, given a guess at the effect. For the Gaussian kernel and $n=m$, Theorem~\ref{thm:plugin} gives
$\tau(0)\le C_0\sqrt{2/n}$, so the test rejects $H_{0,0}$ as soon as
\begin{equation}
n>2\Big(\frac{C_0}{\widehat D_0}\Big)^2,
\qquad
C_0=\bar\kappa+\Delta_\phi\sqrt{\log(1/\delta)/2}=1+\sqrt{\log(1/\delta)}\approx\cstCzero
\ \text{ at }\delta=0.05 .
\label{eq:samplesize}
\end{equation}
With a guess at $D_0=\MMD(P,Q)$ in place of $\widehat D_0$, \eqref{eq:samplesize} is a design rule, which scales like
$D_0^{-2}$.
\space
\section{The split-sample lower bound}
\label{sec:split}

The threshold of Section~\ref{sec:analytic} is a worst case over the whole null. Our method avoids it: we fit the
witness and its two cuts on part of the data and test a fixed function on the rest, refitting on each batch. The
batches and the number of bets are fixed in advance, so the test is valid at every sample size but not anytime-valid.
Validity rests on one inequality, $\E_P\rp-\E_Q\rp\le\TV(P,Q)$ for every $\rp$ with values in $[0,1]$, as in the
framework of \citet{naf2022hplb}, with the clipped witness in the role of their classifier.

\subsection{A ramp of the witness}
\label{sec:ramp}

Any unit vector $f\in\Hh$ gives a lower bound on the distance between two sets $A,B\subset\Hh$,
\[
\dist(A,B)\ge\mathrm{sep}(f;A,B)\coloneqq\inf_{a\in A}\langle f,a\rangle_\Hh-\sup_{b\in B}\langle f,b\rangle_\Hh ,
\]
with equality when $A,B$ are convex and weakly compact, $\dist(A,B)>0$, and $f$ is the direction of the minimal-norm
difference (Lemma~\ref{lem:separation} of the supplement). A direction chosen from data is therefore harmless once it
is independent of the data on which it is evaluated. For the trimmed regions both terms of $\mathrm{sep}$ are trimmed
means by Lemma~\ref{lem:cvar}, of $f(X)$ over the lower and of $f(Y)$ over the upper $(1-\alpha)$-fraction of its
distribution. A trimmed mean is cut at a quantile, which depends on the data and makes it hard to calibrate. With the cut fixed in advance it becomes an ordinary mean, as in the second reading of Lemma~\ref{lem:cvar}. For a bounded
measurable $f$ and reals $t,u$ put
\[
\theta(f;t,u)\coloneqq\E_P\big[t-\beta(t-f(X))_+\big]-\E_Q\big[u+\beta(f(Y)-u)_+\big].
\]

\begin{lemma}[Cut points turn the separation into a ramp]
\label{lem:ramp}
Let $f$ be bounded and measurable and $\alpha\in[0,1)$. Then:
\begin{enumerate}[label=(\roman*),leftmargin=*]
\item $\theta(f;t,u)\le\inf_{P'\in R_\alpha(P)}\E_{P'}f-\sup_{Q'\in R_\alpha(Q)}\E_{Q'}f$ for all reals $t,u$; the
supremum of the left side over $(t,u)$ equals the right side and is attained at a $(1-\alpha)$-quantile of $f(X)$
and an $\alpha$-quantile of $f(Y)$; and for $f$ in the unit ball of $\Hh$ the right side is
$\mathrm{sep}(f;\mathcal C_\alpha(P),\mathcal C_\alpha(Q))$;
\item if $t>u$, then with the clipped witness $\tilde f=(f\vee u)\wedge t$ and the ramp $\rp=(\tilde f-u)/(t-u)$, which
takes values in $[0,1]$, $\theta(\tilde f;t,u)=\beta(t-u)\big(\E_P\rp-\E_Q\rp-\alpha\big)\ge\theta(f;t,u)$;
\item every measurable $\rp$ with values in $[0,1]$, the ramp in particular, has $\E_P\rp-\E_Q\rp\le\TV(P,Q)$.
\end{enumerate}
\end{lemma}

In practice $f$ is the witness direction $\widehat f$. The \emph{clipped witness} $\tilde f$ is $f$ with its values
below $u$ raised to $u$ and above $t$ lowered to $t$, and the \emph{ramp} $\psi$ rescales it to $[0,1]$, a soft
classifier, as the witness is large where $P$ has more mass. By part (i), the trimmed separation in direction $f$ is the
largest value of $\theta(f;t,u)$ over the cut points, attained at the cuts of Proposition~\ref{prop:bangbang}, and once
these are fixed $\theta$ is a difference of two ordinary means. By part (ii), clipping never lowers $\theta$ and turns
it into a positive multiple of $\E_P\psi-\E_Q\psi-\alpha$, which by part (iii) never exceeds $\TV(P,Q)-\alpha$. So,
with the cut points fixed, a test that rejects $\E_P\psi-\E_Q\psi\le\alpha$ shows that $\TV(P,Q)>\alpha$, with no
further assumption on the laws.

\subsection{A betting test}
\label{sec:betting}

The test of $\E_P\psi-\E_Q\psi\le\alpha$ must use data that played no part in choosing $\psi$, and it should keep its
level at every sample size. We obtain both by splitting the samples. Before the data are examined, each sample is put in
a random order and cut into three batches, the first quarter (batch $0$), the second quarter (batch $1$) and the second
half (batch $2$). Batch $0$ is used only for fitting, and
batches $1$ and $2$ are bet on, each with a function fitted on the batches before it. Any number of batches at fractions
fixed in advance would do (Section~\ref{sec:bettingdetails} of the supplement).

Before betting on batch $b\in\{1,2\}$, we solve \eqref{eq:qp} at level $\alpha$ on the batches before it and take the
witness direction $\widehat f_b$. Its cut points are those at which a trimming at level $\alpha$ would cut: the
smallest $(1-\alpha)$-quantile $\widehat t_b$ of its values at the $X$'s of those batches, their
$\lceil n_{<b}(1-\alpha)\rceil$-th order statistic, and the largest $\alpha$-quantile $\widehat u_b$ of its values at
their $Y$'s, the $(\lfloor m_{<b}\alpha\rfloor+1)$-th, where $n_{<b},m_{<b}$ are the sizes of the batches before $b$.
Each value is computed without the point's own term in \eqref{eq:witness}, which pulls the value at a fitting point
toward its own sample, and removing it is a cheap leave-one-out correction of the in-sample bias of these quantiles.
If $\widehat t_b>\widehat u_b$, let $\psi_b$ be the ramp of $\widehat f_b$ between $\widehat u_b$ and $\widehat t_b$;
otherwise batch $b$ is skipped.

Within batch $b$ each $X$ is paired with a block of $Y$'s (or the reverse when the $Y$'s are fewer), and each pair
gives $Z=\psi_b(X)-\overline{\psi_b(Y)}$, the bar being the mean over the block. Given the earlier batches, $Z$ takes
values in $[-1,1]$ and has mean $\E_P\psi_b-\E_Q\psi_b$, at most $\alpha$ under $H_{0,\alpha}$ by
Lemma~\ref{lem:ramp}(iii). The pairs of both batches, numbered $k=1,\dots,N$, number about $\tfrac34\min(n,m)$, fixed
before any data are seen. A gambler starts with wealth $K_0=1$ and at pair $k$ stakes a fraction $\lambda_k$ of
the current wealth on $Z_k$ exceeding $\alpha$,
\[
K_k=K_{k-1}\{1+\lambda_k(Z_k-\alpha)\}=\prod_{l\le k}\{1+\lambda_l(Z_l-\alpha)\}.
\]
When the fitted ramp has $\E_P\rp_b-\E_Q\rp_b>\alpha$, as it does below $\alpha^\star$ for large samples
(Proposition~\ref{prop:loconsistency}), the logarithm of the wealth grows linearly in $k$ with well-chosen fractions.
The test rejects
$H_{0,\alpha}$ as soon as the wealth reaches $1/\delta$. Each fraction $\lambda_k\in[0,\lambda_{\max}/(1+\alpha)]$,
$\lambda_{\max}\in(0,1]$, may depend on earlier batches and pairs only, and the bound keeps the wealth nonnegative.
Such fitting rules and fractions are called predictable (Section~\ref{sec:bettingdetails} of the supplement). Ours stake more when the earlier pairs vary little, through a
running estimate of their variance started from the variance of the ramp on the fitting batches
(Equation~\eqref{eq:fractions} of the supplement). Algorithm~\ref{alg:split} lists the method with
every default.

\begin{theorem}[Validity of the betting test]
\label{thm:betting}
Let $X_1,\dots,X_n$ be i.i.d.\ $P$ and $Y_1,\dots,Y_m$ i.i.d.\ $Q$, the two samples independent, cut into batches as
above. Fix $\alpha\in[0,1)$ and $\delta\in(0,1)$. For every predictable fitting rule and fractions, all sample sizes and
every pair with $\TV(P,Q)\le\alpha$,
\[
\Pp\Big(\max_{k\le N}K_k\ge1/\delta\ \Big|\ \text{batch }0\Big)\le\delta .
\]
Hence rejecting $H_{0,\alpha}$ when $\max_{k\le N}K_k\ge1/\delta$ has level at most $\delta$ for every sample size,
uniformly over $H_{0,\alpha}$, with no assumption on the laws, the kernel or the witness, and a rejection shows that
$\TV(P,Q)>\alpha$.
\end{theorem}

Given the earlier batches and pairs, $\E Z_k\le\alpha$ by Lemma~\ref{lem:ramp}(iii), so $K_k$ is a nonnegative
supermartingale started at one, and the maximal inequality of \citet{ville1939} bounds the probability that it ever
reaches $1/\delta$. The test is the two-sample test by betting of \citet{shekhar2024nonparametric}, with the ramp as
the predictable function and the null $\E_P\rp-\E_Q\rp\le\alpha$ in place of $P=Q$. Any function with
values in $[0,1]$ fitted on earlier batches may replace the ramp, such as the predicted labels or probabilities of a
classifier; \citet{podkopaev2023sequential} bet on classifiers in this way in their predictive tests, and
Section~\ref{sec:numharder} compares such classifiers with the ramp. Betting tests of a tolerance on a difference of means audit fairness
\citep{chugg2023auditing}, and \citet{pandeva2024deep} bet with learned functions on nulls defined by operators. Unlike
these and \citet{shekhar2024nonparametric}, our test fixes $N$ and the batches in advance, and $\lambda_k$ depends on
$N$. Our fractions, adapted from \citet[Remark~1]{waudbysmith2024estimating} with $\lambda_{\max}=\tfrac12$, are
explicit and enter the rates of Section~\ref{sec:rates}. Against a small effect, at $\delta=0.05$ and power $0.8$, the
betting test needs, heuristically, about $\cstBetRatioEighty$ times as many bet-on pairs as a test calibrated by the
central limit theorem, and the quarter of each sample used only for fitting comes on top
(Section~\ref{sec:bettingdetails} of the supplement). Online Newton steps or mixtures over fractions may recover part
of this loss.

\subsection{The lower confidence bound}
\label{sec:lower}

Running the test of Theorem~\ref{thm:betting} on a grid of levels gives our lower confidence bound on $\alpha^\star$, the
split-sample bound $\widehat\alpha$.

\begin{corollary}[Lower confidence bound for the contamination level]
\label{cor:lower}
Fix a grid $\alpha_1<\dots<\alpha_J$ in advance, test each $H_{0,\alpha_j}$ by Theorem~\ref{thm:betting} on the same
batches, with any fitting rule at each level, and read the grid upward, stopping at the first failure:
\begin{equation}
\widehat\alpha\coloneqq\max\big\{\alpha_j:\ H_{0,\alpha_i}\text{ is rejected for every }i\le j\big\},
\label{eq:downclosed}
\end{equation}
with $\widehat\alpha=0$ if $H_{0,\alpha_1}$ is not rejected. Then
$\Pp(\widehat\alpha>\TV(P,Q))\le\delta$ for all sample sizes and every pair $(P,Q)$, with no
multiplicity correction. Moreover, the probability that $H_{0,\widehat\alpha}$ is rejected and
$\TV(P,Q)\le\widehat\alpha$ is at most $\delta$, so a rejected $\widehat\alpha$ shows
$\TV(P,Q)>\widehat\alpha$.
\end{corollary}

When the grid starts at $\alpha_1=0$, the value $\widehat\alpha=0$ means either that $H_{0,0}$ was rejected, which
shows $\TV(P,Q)>0$, or that nothing was rejected, and we report which. Only the first true null on the grid can make
$\widehat\alpha$ too large, which is why no correction is needed, whereas bisection could test several true nulls, as
the refitted ramp may accept a level and reject a larger one. The grid may be chosen from batch $0$
(Remark~\ref{rem:batch0grid} of the supplement). The bound is of the kind of the lower bound of
\citet[Theorem~7]{blanchard2010semi} on a mixture proportion, with one function fitted on earlier batches in place of
their uniform bound over a class, so no complexity term enters. With a witness fitted at each level it is also
consistent, up to the grid.

\begin{proposition}[Consistency of the lower bound]
\label{prop:loconsistency}
Let Assumption~\ref{ass:regularity} hold, and compute $\widehat\alpha$ with a kernel fixed before the data are seen,
an exact minimizer of \eqref{eq:qp} at each level, cut points within one rank of the quantiles of
Section~\ref{sec:betting} (Equation~\eqref{eq:cutrank} of the supplement), and the fractions of
Section~\ref{sec:betting} with their running variance estimate floored at a constant $v^2_{\min}\in(0,4]$ that does
not depend on $n,m$. Let $\alpha_{j^\star}=\max\{\alpha_j:\alpha_j<\alpha^\star\}$,
or $0$ if no level is below $\alpha^\star$. Then $\Pp(\widehat\alpha>\alpha_{j^\star})\le\delta$ for all
$n,m$, and $\Pp(\widehat\alpha<\alpha_{j^\star})\to0$ as $n,m\to\infty$; if $\alpha^\star$ is not a grid
level, also $\Pp(\widehat\alpha>\alpha_{j^\star})\to0$. On grids fixed in advance whose mesh tends to
zero, with $\log J=o(n\wedge m)$, $\widehat\alpha\to\alpha^\star$ in probability.
\end{proposition}

Below $\alpha^\star$ the excess of the population ramp over $\alpha$ is at least $D_\alpha/(\beta \Delta_\phi)$
(Lemma~\ref{lem:popramp} of the supplement), once the fitting batches are large the fitted ramp keeps half of this
excess, with no density condition, and betting has power tending to one against a fixed excess. Fitting the witness at
the level tested matters here. A single witness fitted at $\alpha=0$, the one a test of $P=Q$ would use, is valid and cheaper but need not be consistent:
Section~\ref{sec:alphazero} of the supplement gives laws with $\TV(P,Q)=\loExTV$ on which every ramp of its limit has
$\E_P\rp-\E_Q\rp\le\loExKS$.

The proposition, like the rates of Section~\ref{sec:rates}, assumes a kernel fixed in advance and exact minimizers,
and its constants grow like $1/v^2_{\min}$ through the bound $\lambda_k\le\{2\log(1/\delta)/(Nv^2_{\min})\}^{1/2}$, an
artifact of the proof, as the fractions keep the quadratic term of the wealth of order $\log(1/\delta)$ once the
running variance tracks the second moment. Our runs use the floor $10^{-8}$, a median-distance
bandwidth computed on the fitting batches, and accelerated projected gradient stopped at a relative tolerance
(Algorithm~\ref{alg:split} of the supplement). Validity is unaffected, but for these runs Proposition~\ref{prop:loconsistency} and
Theorems~\ref{thm:loratemain} and~\ref{thm:laplacerate} describe the trend only.
\space
\section{Rates and optimality}
\label{sec:rates}

The error $\alpha^\star-\widehat\alpha$ is parametric on a grid of matching mesh when a function of the RKHS
separates the two contaminations, and no valid lower bound does better over such pairs (Section~\ref{sec:bothrates}). When they touch, a margin argument guarantees slower rates that depend on
the kernel, polynomial for Sobolev kernels and logarithmic for the Gaussian kernel (Sections~\ref{sec:linearmargin}
and~\ref{sec:touching}). The excess of the bound's own ramp recovers the parametric rate up to logarithmic factors on
the line with the Laplace kernel, for densities that cross once and under a condition on their likelihood ratio
(Theorem~\ref{thm:laplacerate}).

\subsection{A margin condition}
\label{sec:rate}

A rate needs a condition on the pair, and we use one, on how fast the regions separate below $\alpha^\star$.

\begin{assumption}[Polynomial margin]
\label{ass:margin}
There are constants $c_0>0$ and $\gamma\ge1$ such that $D_{\alpha^\star-\epsilon}(P,Q)\ge c_0\,\epsilon^\gamma$ for
all $\epsilon\in[0,\alpha^\star]$.
\end{assumption}

Validity and consistency hold for every pair, but by Section~\ref{sec:upper} no rate holds uniformly over all pairs,
and Assumption~\ref{ass:margin} restricts the pair only through its path. It suffices for $\epsilon\le\epsilon_0$, as the path is
nonincreasing, with $c_0(\epsilon_0/\alpha^\star)^\gamma$ in place of $c_0$. A margin of some form always exists when
$0<\alpha^\star<1$ (Lemma~\ref{lem:marginexists} of the supplement), Proposition~\ref{prop:newrates} of the supplement
gives the rate under any margin, and the assumption adds the polynomial speed. Exponents below $1$ are excluded, because
the rescaled path $(1-\alpha)D_\alpha(P,Q)$ is convex (Lemma~\ref{lem:pathconvex} of the supplement), so the path
leaves zero at most linearly, with a kink ($\gamma=1$) or tangentially ($\gamma>1$). The linear case holds when a function of the RKHS
separates the two contaminations (Section~\ref{sec:linearmargin}) and, for laws with continuous densities and a
continuous kernel, only if they are apart. In the following example the separating function takes three values.

\begin{example}[Separated contamination]
\label{ex:separated}
Let (A1) and (A2) of Assumption~\ref{ass:regularity} hold. Let $R$, $P_0$ and $Q_0$ be probability measures
concentrated on pairwise disjoint measurable sets $S_R$, $S_P$ and $S_Q$, let $\alpha^\star\in(0,1)$, and put
$P=(1-\alpha^\star)R+\alpha^\star P_0$ and $Q=(1-\alpha^\star)R+\alpha^\star Q_0$. Then $\TV(P,Q)=\alpha^\star$. If
moreover some $f\in\Hh$ takes the value $0$ on $S_R$, $1$ on $S_P$ and $-1$ on $S_Q$, then for every
$\alpha\in[0,\alpha^\star]$,
\[
D_\alpha(P,Q)\ \ge\ \frac{2(\alpha^\star-\alpha)}{(1-\alpha)\|f\|_\Hh}\ \ge\ \frac{2(\alpha^\star-\alpha)}{\|f\|_\Hh},
\]
so Assumption~\ref{ass:margin} holds with $\gamma=1$ and $c_0=2/\|f\|_\Hh$.
\end{example}

Such an $f$ exists for finite sets and a strictly positive definite kernel, as in
Section~\ref{sec:regions}, and for Mat\'ern kernels when the sets are bounded and apart (Section~\ref{sec:exproof} of
the supplement).

\subsection{Linear and slower margins}
\label{sec:linearmargin}

Whether the margin is linear depends on whether the contaminations touch. Write $P=P\wedge Q+\alpha^\star P_0$ and $Q=P\wedge Q+\alpha^\star Q_0$, where the contaminations $P_0$ and $Q_0$ are
the probability measures with densities $(p-q)_+/\alpha^\star$ and $(q-p)_+/\alpha^\star$.

\begin{proposition}[Touching contaminations]
\label{prop:newslope}
Let (A1) and (A2) of Assumption~\ref{ass:regularity} hold and let $0<\alpha^\star<1$, let $\mathcal X$ be a topological
space with its Borel sets, and let $\phi$ be continuous, as it is for a continuous kernel. Then
\[
\lim_{\epsilon\downarrow0}\frac{(1-\alpha^\star+\epsilon)\,D_{\alpha^\star-\epsilon}(P,Q)}{\epsilon}
\ \le\ \inf\big\{\|\phi(x)-\phi(y)\|_\Hh:\ x\in\supp P_0,\ y\in\supp Q_0\big\} .
\]
If the supports of $P_0$ and $Q_0$ meet, the limit is $0$ and Assumption~\ref{ass:margin} fails with $\gamma=1$. They
meet when $\mathcal X$ is a connected open subset of $\R^d$ and $P\neq Q$ have continuous Lebesgue densities $p,q$
such that $\{p=q\}$ has empty interior, as for any two distinct finite mixtures of nondegenerate Gaussian laws.
\end{proposition}

The limit exists by Lemma~\ref{lem:pathconvex}(ii) of the supplement. At the level $\alpha^\star-\epsilon$ one may keep
$P\wedge Q$ and add mass $\epsilon$ of $P_0$ near a point $x$ to the first law and of $Q_0$ near $y$ to the second,
whose embeddings then differ by about $\epsilon\{\phi(x)-\phi(y)\}$. With continuous densities and kernel, a linear
margin thus needs an open set on which $p=q$ that separates $\{p>q\}$ from $\{q>p\}$, and for a Mat\'ern kernel and
$P_0$ with compact support disjoint supports of $P_0$ and $Q_0$ also suffice (Proposition~\ref{prop:newgammaone} of the
supplement). Overlapping smooth laws, such as the overlap and shift designs of Section~\ref{sec:numerical}, fall
outside the linear case.

When the contaminations touch, the path leaves zero tangentially, at a speed set by how well the RKHS approximates
the indicator of $A=\{p>q\}$, with $p,q$ densities with respect to $P+Q$. For measurable $f$ with $0\le f\le1$ let
$\mathrm{err}(f)=\int|\mathbf 1_A-f|\,d|P-Q|$, where $|P-Q|$ is the total variation measure of $P-Q$, so that
$\mathrm{err}(f)=\alpha^\star-(\E_Pf-\E_Qf)$, and $\mathrm{err}(f)/2$ is the excess risk of the randomized classifier
that assigns $x$ to $P$ with probability $f(x)$, in the balanced problem whose Bayes classifier is $\mathbf 1_A$.

\begin{proposition}[Margin from an approximate witness]
\label{prop:newapprox}
Let (A1) and (A2) hold and let $0<\alpha^\star<1$. For every $f\in\Hh$ with $0\le f\le1$ and $f\neq0$, and every
$\epsilon\in[0,\alpha^\star]$,
\[
D_{\alpha^\star-\epsilon}(P,Q)\ \ge\ \frac{\epsilon-\mathrm{err}(f)}{(1-\alpha^\star+\epsilon)\,\|f\|_\Hh} .
\]
Hence $D_{\alpha^\star-\epsilon}(P,Q)\ge\epsilon/\{2\mathcal W(\epsilon/2)\}$, where
$\mathcal W(t)=\inf\{\|f\|_\Hh:\ f\in\Hh,\ 0\le f\le1,\ \mathrm{err}(f)\le t\}$.
\end{proposition}

In particular, if some $f\in\Hh$ with $0\le f\le1$ equals $1$ $P_0$-almost surely and $0$ $Q_0$-almost surely,
then $\mathrm{err}(f)=0$, and Assumption~\ref{ass:margin} holds with $\gamma=1$ and $c_0=1/\|f\|_\Hh$, since
$1-\alpha^\star+\epsilon\le1$. This is what we mean by a function of the RKHS separating the two contaminations;
Example~\ref{ex:separated} gives the constant $2/\|f\|_\Hh$ for an $f\in\Hh$ equal to $0$ on the common component, $1$
on one contamination and $-1$ on the other.

A trimming at level $\alpha$, before it is renormalized, is a part $P'\le P$ of mass $1-\alpha$, so it lowers the
integral of a $[0,1]$-valued function under $P$ by at most $\alpha$ and does not raise it under $Q$, and the
reproducing property turns the remaining gap into a lower bound on $(1-\alpha)D_\alpha$. The function $\mathcal W$ is
the approximation error behind the rates of kernel classifiers \citep{smale2003estimating,steinwart2008svm}: the closer
$\alpha$ is to $\alpha^\star$, the sharper $f$ must be across $\{p=q\}$, and the smoother the kernel, the more this
costs in $\Hh$. For a single transversal crossing on the line, Sobolev and Gaussian kernels make this quantitative.

\begin{proposition}[Sobolev kernels]
\label{prop:newsobolev}
Let $\mathcal X=\R$ and $k(x,y)=k_0(x-y)$, with $k_0$ continuous and integrable and
$c(1+\xi^2)^{-s}\le\widehat{k_0}(\xi)\le C(1+\xi^2)^{-s}$ for some $s>1/2$, as for the Laplace kernel, with
$s=1$, and the Mat\'ern kernel of smoothness $s-1/2$. Let $P,Q$ have densities $p,q$ that are continuous near a
point $x_0$, with $p\ge q$ almost everywhere on $(x_0,\infty)$ and $p\le q$ almost everywhere on $(-\infty,x_0)$, $p-q$
differentiable at $x_0$ with
$(p-q)'(x_0)>0$, $p(x_0)>0$, and $|P-Q|(\{x:|x-x_0|>t\})\le C'e^{-t/C'}$ for some $C'$ and all $t>0$. Then there are
$0<c_1\le c_2$ such that for all small $\epsilon>0$,
\[
c_1\,\epsilon^{(2s+3)/4}\ \le\ D_{\alpha^\star-\epsilon}(P,Q)\ \le\ c_2\,\epsilon^{(2s+3)/4} .
\]
\end{proposition}

Assumption~\ref{ass:margin} therefore holds with $\gamma=(2s+3)/4$ and with no smaller exponent, $\gamma=5/4$ for the
Laplace kernel. For the lower bound, an indicator smoothed at width $w$ has $\mathrm{err}(f)$ of order $w^2$, as
$|p-q|$ vanishes linearly at $x_0$, and $\|f\|_\Hh$ of order $w^{1/2-s}$, and $w\asymp\epsilon^{1/2}$ balances the two, while
the upper bound comes from explicit trimmings whose difference has vanishing moments near $x_0$. For $s>d/2$ the lower
bound extends to $\R^d$ when $|P-Q|$ has compact support and $\{p>q\}$ has a smooth compact boundary, with the exponent $1+(2s-1)/\{2(r+1)\}$ when
$p-q$ vanishes at least to order $r$ there (Theorem~\ref{thm:newsobolev} and Corollary~\ref{cor:newsobolev} of the
supplement), which is sharp on the line, at least for $s<3/2$, when it vanishes exactly to order $r$
(Proposition~\ref{prop:newupperone} of the supplement).

\begin{proposition}[Gaussian kernel]
\label{prop:newgauss}
Let $P,Q$ satisfy the hypotheses of Proposition~\ref{prop:newsobolev} on the laws, put $\ell_0=(p-q)'(x_0)$, the slope of $p-q$ at $x_0$, and let
$k(x,y)=\exp\{-(x-y)^2/(2\sigma^2)\}$ on $\R$. For every $c>\pi^2\sigma^2\ell_0/8$ and every $j\ge1$ there are
$\epsilon_0>0$ and $C_j<\infty$ such that, for $0<\epsilon\le\epsilon_0$,
\[
\exp(-c/\epsilon)\ \le\ D_{\alpha^\star-\epsilon}(P,Q)\ \le\ C_j\,\epsilon^j .
\]
\end{proposition}

With the Gaussian kernel the path is thus flatter at $\alpha^\star$ than every power, yet a margin of exponential type holds.
The approximant is an indicator smoothed by a band-limited density, which lies in the Gaussian RKHS
\citep{steinwart2006explicit} with a norm that grows like $\exp(c/\epsilon)$.

\subsection{The rate, and when it is optimal}
\label{sec:bothrates}
\label{sec:optimality}

Under the margin the error of the bound is at most of order $(n\wedge m)^{-1/(2\gamma)}$, up to the mesh of its
grid.

\begin{theorem}[Rate of the split-sample lower bound]
\label{thm:loratemain}
Let Assumptions~\ref{ass:regularity} and~\ref{ass:margin} hold, let $\alpha^\star<1$, and compute
$\widehat\alpha$ at level $\delta$ as in Proposition~\ref{prop:loconsistency}, with the batches of
Section~\ref{sec:betting}, $\lambda_{\max}=\tfrac12$ and the fractions \eqref{eq:fractions} of the supplement with initial
variance $v_0^2\le4$, on a grid of $J$ levels
fixed in advance whose mesh below $\alpha^\star$ is $\Delta_{\mathcal G}$. If $n\wedge m\ge10$, then for every $\eta\in(0,1)$, with
probability at least $1-\delta-\eta$,
\[
\alpha^\star-\Delta_{\mathcal G}-O\bigg(\bigg(\frac{\kappa}{c_0(1-\alpha^\star)v^2_{\min}}
\sqrt{\frac{1+\log(1/\delta)+\log(J/\eta)}{n\wedge m}}\bigg)^{1/\gamma}\bigg)
\ \le\ \widehat\alpha\ \le\ \alpha^\star ,
\]
with an absolute constant in the $O$.
\end{theorem}

This is parts (ii) and (iii) of Theorem~\ref{thm:lorate} of the supplement, which gives the constants. Below
$\alpha^\star$, on an event of high probability, the excess $\E_P\rp-\E_Q\rp-\alpha$ of the fitted ramps stays above a
multiple of $c_0(\alpha^\star-\alpha)^\gamma$ as soon as this exceeds a multiple of $(n\wedge m)^{-1/2}$, and betting
then rejects. Equating the two gives the rate, and the split adds the mesh and the factor $1/v^2_{\min}$. Part (i) of
that theorem removes $\log J$ when consecutive levels are far apart compared with the rate, as on the grid of
Corollary~\ref{cor:minimax} below. The constants are far from sharp.

When the regions separate linearly, no lower confidence bound improves the rate $(n\wedge m)^{-1/2}$, by the two-point
method of \citet{lecam1973convergence}: pairs whose total variations differ by $h$ of order $n^{-1/2}$ cannot be told
apart, and a bound valid at the pair with the smaller value must often fall short at the other.

\begin{proposition}[Two-point lower bound]
\label{prop:lower2pt}
Let $R$, $P_0$ and $Q_0$ be probability measures on $\mathcal X$ concentrated on pairwise disjoint measurable sets.
For $a\in(0,1)$ and $h\in[0,1-a)$ put $P_a=(1-a)R+aP_0$, $Q_a=(1-a)R+aQ_0$ and
\[
P_{a,h}=(1-a-h)R+(a+h)P_0,\qquad Q_{a,h}=(1-a-h)R+(a+h)Q_0 .
\]
Then $\TV(P_a,Q_a)=a$ and $\TV(P_{a,h},Q_a)=\TV(P_a,Q_{a,h})=a+h$. Let $\tilde\alpha$ be any measurable function of
the two samples and of an independent randomization such that $\Pp_{P_a,Q_a}(\tilde\alpha>a)\le\delta$. Then for every
$h\in(0,1-a)$,
\begin{align*}
\Pp_{P_{a,h},Q_a}\big(\TV(P_{a,h},Q_a)-\tilde\alpha\ge h\big)&\ge1-\delta-h\sqrt{n/\{2a(1-a)\}},\\
\Pp_{P_a,Q_{a,h}}\big(\TV(P_a,Q_{a,h})-\tilde\alpha\ge h\big)&\ge1-\delta-h\sqrt{m/\{2a(1-a)\}}.
\end{align*}
\end{proposition}

With $h=\tfrac12(1-\delta)\{2a(1-a)/(n\wedge m)\}^{1/2}$, when this $h$ is below $1-a$, a lower confidence bound
valid at $(P_a,Q_a)$ falls short by at least $h$, with probability at least $(1-\delta)/2$, at the pair that perturbs
the law of the smaller sample, and the proof bounds a $\chi^2$-divergence (Section~\ref{sec:optproofs} of the supplement).
The perturbed pairs satisfy Assumption~\ref{ass:margin} with $\gamma=1$ and $c_0=1/\|f\|_\Hh$, for $f$ as in
Example~\ref{ex:separated} (Lemma~\ref{lem:twopointmargin} of the supplement), the class on which
Theorem~\ref{thm:loratemain} gives the parametric rate.

\begin{corollary}[Minimax rate]
\label{cor:minimax}
Let Assumption~\ref{ass:regularity} hold, suppose that $\mathcal X$ contains three points $x_1,x_2,x_3$ and that some
$f\in\Hh$ has $f(x_1)=1$, $f(x_2)=0$ and $f(x_3)=-1$, as for a strictly positive definite kernel. Let
$0<a_0<a_1<1$ and $0<c_0\le1/\|f\|_\Hh$, let $\mathcal M$ be the set of pairs $(P,Q)$ with
$a_0\le\TV(P,Q)\le a_1$ that satisfy Assumption~\ref{ass:margin} with $\gamma=1$ and this $c_0$, and fix
$\delta\in(0,1)$.
\begin{enumerate}[label=(\roman*),leftmargin=*]
\item Let $c_{\mathrm{lo}}=\tfrac12(1-\delta)\{2a_0(1-a_0)\}^{1/2}$. If $n\wedge m\ge c_{\mathrm{lo}}^2/(a_1-a_0)^2$, every
$\tilde\alpha$ with $\sup_{\mathcal M}\Pp_{P,Q}(\tilde\alpha>\TV(P,Q))\le\delta$ has
\[
\sup_{(P,Q)\in\mathcal M}\Pp_{P,Q}\big(\TV(P,Q)-\tilde\alpha\ge c_{\mathrm{lo}}(n\wedge m)^{-1/2}\big)\ge\frac{1-\delta}2 .
\]
\item For every $\eta\in(0,1)$ there are $C$ and $n_1$, depending only on $\kappa$, $\Delta_\phi$, $c_0$, $a_1$,
$v^2_{\min}$, $\delta$ and $\eta$, such that for $n\wedge m\ge\max(n_1,10)$ the split-sample bound of
Theorem~\ref{thm:loratemain} at level $\delta$, on the grid $\{j(n\wedge m)^{-1/2}:j=1,2,\dots\}\cap(0,1)$, satisfies
$\sup_{\mathcal M}\Pp_{P,Q}(\widehat\alpha>\TV(P,Q))\le\delta$ and
$\sup_{\mathcal M}\Pp_{P,Q}(\TV(P,Q)-\widehat\alpha>C(n\wedge m)^{-1/2})\le\eta$.
\end{enumerate}
\end{corollary}

Over $\mathcal M$, on this grid, $(n\wedge m)^{-1/2}$ is thus the minimax rate for lower confidence bounds valid at
level $\delta$, and the split-sample bound attains it, and with $\eta<(1-\delta)/2$ the two parts concern the same
quantile of the error. The class contains every pair with $\alpha^\star\in[a_0,a_1]$ whose rescaled path leaves zero
with slope at least $c_0$ (Lemma~\ref{lem:pathconvex}(ii) of the supplement), and for a continuous kernel its
contaminations are apart (Proposition~\ref{prop:newslope}). Part (i) asks validity only on $\mathcal M$, so it also
covers procedures that use the margin. Nothing is said about $\gamma>1$, and minimax rates over a margin class are
open. Minimax rates for the $L_1$ distance between discrete laws \citep{valiant2011power,jiao2018minimax} and for
kernel two-sample tests \citep{li2024optimality,schrab2023aggregated} concern other problems.

Mixture proportion estimation shows the same dichotomy. Its rate can be arbitrarily slow without a condition on the
laws \citep{blanchard2010semi}, and is $n^{-1/2}$ when the support of the component is not contained in that of the
rest \citep{scott2015rate}, a condition slightly stronger than the irreducibility that identifies the components
\citep{blanchard2016classification}. \citet{ramaswamy2016mixture} estimate the proportion as the point where an RKHS
distance between a reweighted embedding and a convex hull of embeddings, a convex function, leaves zero, linearly under
a separability condition. The linear margin is the two-sample counterpart of separability, and touching contaminations
that of a proportion with no anchor set.

\begin{figure}[t]
\centering
\includegraphics[width=\textwidth]{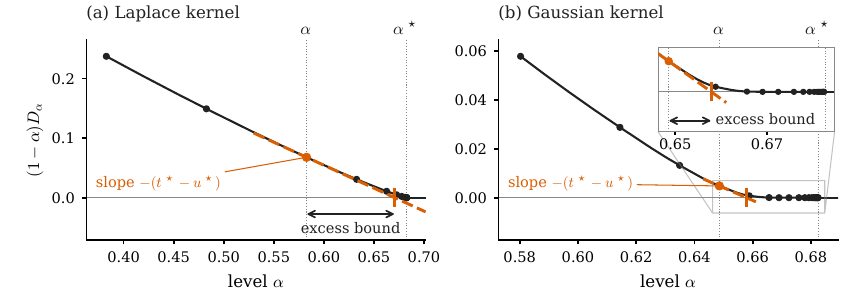}
\caption{The rescaled path $(1-\alpha)D_\alpha$ for $P=N(1,1)$ and $Q=N(-1,1)$, whose total variation is
$\alpha^\star=2\Phi(1)-1\approx\fxAlphaStar$, with the Laplace kernel (a) and the Gaussian kernel (b), both with
$\sigma=1$ (dots: computed levels; line: monotone interpolation). At $\alpha=\fxLevelLap$ in (a) and
$\alpha=\fxLevelGauss$ in (b) the dashed line of slope $-(t^\star-u^\star)$ supports the path; the excess of the
population ramp over $\alpha$ is at least the distance (arrow) from $\alpha$ to the zero of the line (tick), by
Proposition~\ref{prop:excess}, and the inset of (b) enlarges the region near $\alpha^\star$. That distance is a fraction
$\fxFracLap$ of $\alpha^\star-\alpha$ for the Laplace kernel and $\fxFracGauss$ for the Gaussian kernel at this level,
closer to $\alpha^\star$. The Gaussian fraction, $\fxFracGaussAtLap$ at $\alpha=\fxLevelGaussAtLap$, falls toward zero
as $\alpha\uparrow\alpha^\star$, where its path is flatter than every power.}
\label{fig:excess}
\end{figure}

\subsection{Touching contaminations, and why the margin is pessimistic}
\label{sec:touching}

Under the hypotheses of Proposition~\ref{prop:newsobolev}, Theorem~\ref{thm:loratemain} guarantees at least the rate
$(n\wedge m)^{-2/(2s+3)}$, $(n\wedge m)^{-2/5}$ for the Laplace kernel, which Theorem~\ref{thm:laplacerate} improves to
$(n\wedge m)^{-1/2}$, up to powers of $\log(n\vee m)$, under one more condition, for samples of comparable size. Under those of Proposition~\ref{prop:newgauss} the margin
guarantees at least the logarithmic rate $c/\log(n\wedge m)$ for every $c>\pi^2\sigma^2\ell_0/4$, on grids with at
most $(n\wedge m)^b$ levels for a fixed $b$ (Proposition~\ref{prop:newrates} of the supplement), in both cases up to the
mesh. For the split-sample bound we prove no converse. For the analytic baseline run with the larger threshold
$\tau^{\mathrm S}$ the margin rate cannot be improved for Sobolev kernels, and with the Gaussian kernel its error is
not $O((n\wedge m)^{-a})$ for any $a>0$ (Proposition~\ref{prop:newconverse} of the supplement); for
$\tau^{\mathrm P}$ no converse is known.

The margin is pessimistic. Theorem~\ref{thm:loratemain} bounds the excess $\E_P\rp-\E_Q\rp-\alpha$ of the population
ramp from below by $D_\alpha/(\beta\Delta_\phi)$ (Lemma~\ref{lem:popramp} of the supplement), replacing the gap
$t^\star-u^\star$ between its cut points by its largest value $\Delta_\phi$. The gap is a slope of the rescaled
path $(1-\alpha)D_\alpha$, the distance between the parts of mass $1-\alpha$ that the trimmings keep, which is convex in
$\alpha$ (Lemma~\ref{lem:pathconvex} of the supplement). The line through the point of the rescaled path at $\alpha$
with slope $-(t^\star-u^\star)$ supports it, and the excess of the ramp is at least the distance from $\alpha$ to the
zero of this line (Proposition~\ref{prop:excess}, Figure~\ref{fig:excess}). When the path lies between two multiples of one power of
$\alpha^\star-\alpha$, this distance is a fixed fraction of $\alpha^\star-\alpha$.

\begin{proposition}[Excess of the population ramp]
\label{prop:excess}
Let Assumption~\ref{ass:regularity} hold, let $0<\alpha^\star<1$, let $\alpha<\alpha^\star$, let $f^\star$ be the
direction of the population witness, $t^\star$ the smallest $(1-\alpha)$-quantile of $f^\star(X)$ and $u^\star$ the
largest $\alpha$-quantile of $f^\star(Y)$, so that $t^\star-u^\star\ge D_\alpha>0$ (Lemma~\ref{lem:popramp} of the
supplement), and let $\rp^\star$ be the ramp of $f^\star$ between $u^\star$ and $t^\star$. Then:
\begin{enumerate}[label=(\roman*),leftmargin=*]
\item $\E_P\rp^\star-\E_Q\rp^\star-\alpha\ge(1-\alpha)D_\alpha/(t^\star-u^\star)$, and
$(1-\alpha')D_{\alpha'}\ge(1-\alpha)D_\alpha-(t^\star-u^\star)(\alpha'-\alpha)$ for every $\alpha'\in[0,1)$;
\item if $c_1\epsilon^\gamma\le D_{\alpha^\star-\epsilon}\le c_2\epsilon^\gamma$ for $0<\epsilon\le\epsilon_0\le\alpha^\star$, then
at $\alpha=\alpha^\star-\epsilon$ with $\epsilon\le\epsilon_0/2$ the excess in (i) is at least
$(1-\alpha^\star)c_1\epsilon/(2^\gamma c_2)$, and this holds under the hypotheses of Proposition~\ref{prop:newsobolev};
\item for the Laplace kernel $e^{-|x-y|/\sigma}$ on $\R$, $u^\star\le f^\star\le t^\star$ on $\R$ and the first
inequality of (i) is an equality.
\end{enumerate}
\end{proposition}

On the line, by part (iii), the Laplace witness is already clipped at its cut points, and so is the witness fitted on
the data, whose ramp maximizes the ratio of empirical excess to RKHS norm over functions of oscillation at most one
(Lemma~\ref{lem:lapclip} of the supplement). Comparing this ratio with the population path at two levels shows that the
fitted ramp keeps a fixed fraction of the excess of the population ramp (Proposition~\ref{prop:lapfitted}).

\begin{theorem}[Near-parametric rate with the Laplace kernel]
\label{thm:laplacerate}
Let $P,Q$ satisfy the hypotheses of Proposition~\ref{prop:newsobolev} on the laws, let $n\wedge m\ge10$, and compute
$\widehat\alpha$ as in Theorem~\ref{thm:loratemain}, on a grid of $J$ levels in $(0,1)$ with mesh $\Delta_{\mathcal G}$
below $\alpha^\star$, with the kernel $e^{-|x-y|/\sigma}$, $\sigma>0$ fixed in advance, an exact minimizer of
\eqref{eq:qp} at each level, and, in place of the one-rank tolerance \eqref{eq:cutrank}, cut points that are genuine
empirical quantiles of the witness values, computed with or without each point's own term in \eqref{eq:witness}. Let
$\eta\in(0,1)$, $L_\delta=\log(1/\delta)$, $L_{n,m}=\log\{(n\vee m)/\eta\}$ and
$B_{n,m}=v_{\min}^{-2}\{L_\delta^{1/2}+(\log(2J/\eta))^{1/2}\}(n\wedge m)^{-1/2}$.
\begin{enumerate}[label=(\roman*),leftmargin=*]
\item If for some $r_\ast\in(0,1]$ and $c_\ast>0$
\begin{equation}
|p(x)-q(x)|\ \ge\ c_\ast\min(r_\ast,|x-x_0|)\,\{p(x)+q(x)\}\qquad\text{for almost every }x,
\label{eq:condR}
\end{equation}
then with probability at least $1-\delta-\eta$,
\[
\alpha^\star-\Delta_{\mathcal G}-O\big(L_{n,m}^{3/4}(n\wedge m)^{-1/2}+B_{n,m}\big)\ \le\ \widehat\alpha\
\le\ \alpha^\star .
\]
\item If instead $(P+Q)(\{x:|x-x_0|>t\})\le C'e^{-t/C'}$ for all $t>0$, the same holds with
$L_{n,m}^{2/3}(n\wedge m)^{-4/9}$ in place of $L_{n,m}^{3/4}(n\wedge m)^{-1/2}$.
\end{enumerate}
The constant in the $O$ depends only on $\sigma$ and on the pair, as stated in Theorem~\ref{thm:laplaceconst} of the
supplement.
\end{theorem}

Under these hypotheses $\TV(P,Q)=F_Q(x_0)-F_P(x_0)$ is a one-sided Kolmogorov distance, which the one-sided
Kolmogorov--Smirnov statistic bounds from below at the rate $(n\wedge m)^{-1/2}$, with no condition on the likelihood
ratio and no logarithmic factor. The interest of the theorem is that a kernel bound that does not sort the
data, and returns the set, attains this rate up to logarithmic factors, which the margin does not see; it is no reason
to prefer the method on the line. Condition \eqref{eq:condR} asks that $p/q$ stay away from one off every neighborhood of $x_0$, so it excludes
contaminations local to a common core, with $p=q>0$ on an interval, where only part (ii) applies. No multivariate theorem
singles out the Laplace kernel in $\R^d$, where its choice rests on this one-dimensional theory and on its power against
differences of high frequency, seen on a one-dimensional oscillating density (Section~\ref{sec:numharder}). Section~\ref{sec:laplaceproof} of the supplement discusses
the exact minimizers the theorem needs, its constants, whose uniformity over a class is open, and a simulation.
\space
\section{The trimming}
\label{sec:trimming}

Once we know that at least a fraction $\widehat\alpha$ of each population has no counterpart in the other, the next
question is which observations these are. By Proposition~\ref{prop:bangbang} the minimizer of \eqref{eq:qp} answers it
with a pair of trimmed samples, which keep at full weight the observations on one side of a cut of the witness and split
weight among those at the cut. They converge, at a fixed level to the best trimming at that level and at the estimated
level to the common core. The rate we prove at a fixed level assumes a witness with a density at the cut, condition
\eqref{eq:qdens} below, as for the Gaussian kernel (Remark~\ref{rem:qdens} of the supplement). With the Laplace kernel
on the line the population witness equals its cut on a set of $P$-mass at least $\alpha$ (Section~\ref{sec:excessproofs}
of the supplement), so \eqref{eq:qdens} fails at every level, Theorem~\ref{cor:trimrate} does not apply, and the
trimming is decided among tied observations; Section~\ref{sec:trimest}, in MMD and without a rate, covers every kernel
of Assumption~\ref{ass:regularity}.

\subsection{A parametric rate at a fixed level}

Fix $\alpha$ with $D=D_\alpha(P,Q)>0$, which under (A3) means $\alpha<\alpha^\star$, let $d^\star$ be the population
witness of Section~\ref{sec:statistic}, with direction $f^\star=d^\star/D$, and let $t_\alpha$ be a $(1-\alpha)$-quantile
of $d^\star(X)$ under $P$. The population version of Proposition~\ref{prop:bangbang} holds: every optimal trimming
density of $P$ has the threshold form (Lemma~\ref{lem:popthreshold} of the supplement)
\begin{equation}
g^\star=\beta\,\mathbf 1\{d^\star<t_\alpha\}+\vartheta\,\mathbf 1\{d^\star=t_\alpha\},
\qquad 0\le\vartheta\le\beta\ \text{a function with}\ \E_P[g^\star]=1 .
\label{eq:popdensity}
\end{equation}
The observations that an optimal trimming keeps at full weight form the \emph{kept sets}
\[
A^\star\coloneqq\{x:\ d^\star(x)<t_\alpha\}
\qquad\text{and}\qquad
\widehat A_n\coloneqq\{x:\ \widehat d(x)<\widehat t_n\},
\]
population and empirical, whose symmetric difference $\widehat A_n\triangle A^\star$ measures the error, and we let
$\widehat g_n$ be the density \eqref{eq:popdensity} with $(\widehat d,\widehat t_n)$ in place of $(d^\star,t_\alpha)$
and any tie function $0\le\widehat\vartheta\le\beta$, a rule that need not integrate to one under $P$. Level sets are
stable where the witness has a density at the cut, which we ask at both cuts:
\begin{gather}
\tag{Q}\label{eq:qdens}
\parbox[t]{0.84\textwidth}{\raggedright the law of $d^\star(X)$ under $P$ has, on $[t_\alpha-r_Q,t_\alpha+r_Q]$ for some
$r_Q>0$, a density with values in $[c_f,C_f]$, where $0<c_f\le C_f<\infty$;}\\[4pt]
\tag{Q$'$}\label{eq:qdensY}
\parbox[t]{0.84\textwidth}{\raggedright the law of $d^\star(Y)$ under $Q$ has, on $[u_\alpha-r_Q,u_\alpha+r_Q]$, a
density at least $c'_f>0$, where $u_\alpha$ is an $\alpha$-quantile of $d^\star(Y)$.}
\end{gather}
Condition \eqref{eq:qdens} is the classical quantile-density condition for the cut of the $X$ sample, and
\eqref{eq:qdensY} its lower half for the cut of the $Y$ sample. Laws with finite support, such as those of
Example~\ref{ex:separated} with finitely supported $R$, $P_0$, $Q_0$, violate \eqref{eq:qdens}, so the margin does not imply the density conditions. The rate comes from a localization argument of the kind behind rates for M-estimators
\citep[Theorem~3.2.5]{vandervaart1996weak}: the separation of Section~\ref{sec:ramp} falls quadratically away from its
maximizer $f^\star$, while its fluctuation grows linearly in the distance to $f^\star$, times $(n\wedge m)^{-1/2}$.

\begin{theorem}[A parametric rate for the witness and the kept set]
\label{cor:trimrate}
Let (A1) and (A2) of Assumption~\ref{ass:regularity} hold, fix $\alpha\in(0,1)$ with $D=D_\alpha(P,Q)>0$, and suppose
\eqref{eq:qdens} and \eqref{eq:qdensY}. Put $V_\alpha=\kappa\beta+[\beta/\min(c_f,c'_f)]^{1/2}$ and
$W_\alpha=C_f(\kappa V_\alpha+c_f^{-1})$. Then, with $(P,Q,k,\alpha)$ fixed and $n\wedge m\to\infty$,
\begin{gather*}
\big\|\widehat f-f^\star\big\|_\Hh=O_\Pp\Big(\frac{V_\alpha}{D\sqrt{n\wedge m}}\Big),
\qquad
\|\widehat d-d^\star\|_\Hh=O_\Pp\Big(\frac{V_\alpha}{\sqrt{n\wedge m}}\Big),\\
P(\widehat A_n\triangle A^\star)=O_\Pp\Big(\frac{W_\alpha}{\sqrt{n\wedge m}}\Big),
\qquad
\frac1\beta\int|\widehat g_n-g^\star|\,dP=O_\Pp\Big(\frac{W_\alpha}{\sqrt{n\wedge m}}\Big),
\end{gather*}
with constants in the $O_\Pp$ that depend only on the probability level.
\end{theorem}

The rate is that of a sample quantile, free of the dimension of $\mathcal X$, because the witness converges in $\Hh$,
as a mean of feature vectors, and the kept set inherits its rate through the density at the cut. As
$\alpha\uparrow\alpha^\star$ the bound on the direction grows like $1/D$, and the sample size from which the bounds hold,
not explicit (Section~\ref{app:trimming} of the supplement states what it depends on), grows as $D$ vanishes. The kept set then has in general no limit, because the common
core keeps only a fraction of $P$ on the set $\{0<q<p\}$, and an indicator of a set cannot converge to such a
fractional weight. With the Wasserstein distance, optimal trimmings are consistent and computable
\citep{alvarez2011incomplete,agullo2018trimming}, and rates are known only toward a known target
\citep{delbarrio2013rates}.

\subsection{The trimming at the estimated level}
\label{sec:trimest}

In practice the trimming is read at a data-driven level, the lower bound $\widehat\alpha$, where
Theorem~\ref{cor:trimrate} does not apply: its level is fixed in advance, and its bounds degrade as $D_\alpha(P,Q)$
vanishes near $\alpha^\star$. There the trimmed sample still converges as a measure. The common core $R^\star$ is the unique common
trimming at $\alpha^\star$, and
\[
\Omega(\epsilon)=\sup\big\{\|a-\mu_{R^\star}\|_\Hh:\ a\in\mathcal C_{\alpha^\star}(P),\
\dist\big(a,\mathcal C_{\alpha^\star}(Q)\big)\le\epsilon\big\}
\]
is a modulus of the contact of the two regions at $\alpha^\star$.

\begin{proposition}[The trimmed sample at the data-driven level]
\label{prop:alphahat}
Let Assumption~\ref{ass:regularity} hold and let $0<\alpha^\star<1$. Then $\Omega(\epsilon)\to0$ as
$\epsilon\downarrow0$. Fix $\delta'\in(0,1)$ and put $r_n=\kappa(1-\alpha^\star)^{-1}[4n^{-1/2}+(2\log(2/\delta')/n)^{1/2}]$,
and $r_m$ likewise. There is an event $E$ of probability at least $1-2\delta'$, not depending on the level, on which for
every level $\tilde\alpha\le\alpha^\star$, possibly chosen from the data, and every $(w,v)\in\mathcal T_{\tilde\alpha}$,
\[
\MMD\big(P_n^w,R^\star\big)\le r_n+\Omega\big(\widehat D_{\tilde\alpha}+r_n+r_m\big).
\]
The same holds for the second sample, with $P$ and $Q$ exchanged.
\end{proposition}

On $E$ each empirical region is within $r_n$, respectively $r_m$, of its population counterpart, at all levels up to
$\alpha^\star$ at once, by \eqref{eq:hausdorff}. The bound covers the trimming read at $\widehat\alpha$ or at any other
valid lower bound. It carries no rate unless $\Omega$ is known, and it places the reweighted sample near the common core
without saying which observations are kept.

\begin{corollary}[The trimmed sample at the lower bound]
\label{cor:alphalo}
Let Assumption~\ref{ass:regularity} hold, let $0<\alpha^\star<1$, and compute $\widehat\alpha$ at level
$\delta$ as in Proposition~\ref{prop:loconsistency}, on grids fixed in advance whose mesh tends to zero, with
$\log J=o(n\wedge m)$. Then for every $\epsilon>0$ and $\delta'\in(0,1)$, for all $n,m$ large enough, with probability
at least $1-\delta-\delta'$ every $(w,v)\in\mathcal T_{\widehat\alpha}$ has
$\MMD(P_n^w,R^\star)\le\epsilon$ and $\MMD(Q_m^v,R^\star)\le\epsilon$.
\end{corollary}

By Lemma~\ref{lem:pathconvex}(ii) of the supplement the path falls to zero at most linearly, so a consistent
$\widehat\alpha$ has a small trimmed distance whatever the laws and no margin is needed. The proof uses
$\widehat\alpha$ only through its validity and its consistency, so the level may come from a bound computed with
another kernel than the one that trims, as in Section~\ref{sec:appcohort}.
\space
\section{Numerical results}
\label{sec:numerical}

The theory says that the split-sample bound is valid at every sample size, but not how large the sample must be for
it to be useful. We measure this on simulated data whose total variation is known, and on two public data sets,
where the trimming at the certified level also gives a cohort.

\subsection{Simulations}
\label{sec:numharder}

Every design follows the contamination model \eqref{eq:contam}, with a common component of weight $1-a$ shared by $P$
and $Q$ and two design components $P_0$ and $Q_0$ of weight $a$, so that
$\TV(P,Q)=a\,\TV(P_0,Q_0)$ is known in closed form; $\alpha^\star=0.2$ unless a design says otherwise. The $\allNdesigns$ designs, listed in Table~\ref{tab:hardern80} and described in
Table~\ref{tab:harderdesigns} of the supplement, vary one feature at a time: where the contamination sits, the
dimension, how the signal is spread, its size, the type of data, and how smooth the difference of the densities is. In
the overlap and shift designs $P_0$ and $Q_0$ are $N(\pm\mu,I)$ on a $N(0,I)$ core, with $\mu$ along one
coordinate, along two or spread over all of them, so that $\{p>q\}$ is a half-space; the correlation flip changes the
sign of one correlation, the $t_3$ design the scale of one coordinate of a $t_3$ law, and the curves add or subtract a
bump to Brownian paths. The oscillating densities are those of Section~\ref{sec:upper}, which a fixed kernel sees less
and less as their frequency $\nu$ grows. One design is shaped on the RHC data of Section~\ref{sec:data}: it keeps the $\rrRhcNcovPaper$
covariates stored as numbers, takes a smoothed resample of the patients as the common component, and places $P_0$ and $Q_0$
in boxes at $\pm\rgSep$ pooled standard deviations in four coordinates.

Four methods certify a level of $\TV$. The split-sample bound $\widehat\alpha$ is Algorithm~\ref{alg:split} of the
supplement with the Laplace kernel, whose bandwidth is the median distance among the points each witness is fitted on.
The analytic baseline $\widehat\alpha_{\mathrm{an}}$ uses the threshold of Theorem~\ref{thm:plugin} and a Gaussian
kernel with its bandwidth fixed on a separate pilot draw of $1{,}000$ points from each law ($400$ and $2{,}000$ for
mixed types), not counted in $n$ or $m$. The other two run the same betting test on the labels of a
classifier, gradient boosting (GBM) or logistic regression \citep{pedregosa2011sklearn}, trained on the first half of
each sample as in \citet{naf2022hplb}. For reference we add two tests of $P=Q$, the permutation test of the MMD, with
the Gaussian kernel at the pilot bandwidth, and MMDAgg \citep{schrab2023aggregated}; they detect a difference but
certify nothing about its size. All methods see the same replicates, $\mnRepsPerSize$ per design and size, with $n=m$
doubling from $\harderNmin$ to $\harderNmax$ (mixed types: $n$ up to $400$, with $m=5n$). The four that certify are also
run at $\fourdigits{\harderNtwo}$ in the two designs of Table~\ref{tab:harderdesigns} that list it and, in every
design, at sizes doubling from $\fourdigits{\harderNtwo}$ to $\harderNcap$ until their power against $H_{0,0}$ reaches
$0.8$. The two tests of $P=Q$ are also run at $\fourdigits{\harderNtwo}$ on the narrow bump. The four that certify
test $H_{0,\alpha}$ at the levels $\alpha=j\alpha^\star/8$, $j=0,\dots,10$, with $\delta=0.05$.

We first check validity: a test may reject a true null, and the bound exceed $\alpha^\star$, with probability at most
$\delta$. At the level $\alpha=\alpha^\star$ the null $H_{0,\alpha^\star}$ holds, as $\TV(P,Q)$ is
exactly $\alpha^\star$: there, at $n=m=800$ (mixed types: $n=400$, $m=5n$), the betting test on the Laplace ramp
rejected $\mnBdTotal$ times in $\mnBdReps$ replicates in all, and at most $\mnBdMax$ in any one design. When $P=Q$, a
rejection of $H_{0,0}$ falsely claims a difference: on four designs in which both samples come from the same law,
over $\mnNullCells$ combinations of design and sample size, the same test rejected in $\mnNullPct\%$ of the replicates,
and at most $\mnNullMax$ in one combination. The analytic test never rejected in either case. For no method and no
combination does a binomial test at level $0.01$ find more rejections than $\delta$ allows, and in every design the
split-sample bound exceeded $\alpha^\star$ in at most a fraction $\mnAloGtMax$ of the replicates, well within the
$\delta$ that Corollary~\ref{cor:lower} allows (Section~\ref{app:experiments} of the supplement).

\begin{table}[t]
\centering
\footnotesize
\setlength{\tabcolsep}{5pt}
\begin{tabular}{lccccccc}
\toprule
design & $\alpha^\star$ & $\widehat\alpha$ & $\widehat\alpha_{\mathrm{an}}$ & GBM & logistic & perm. & MMDAgg \\
 &  &  &  & + betting & + betting & MMD &  \\
\midrule
RHC-shaped, $d=33$ & $0.2$ & $\mathbf{{\le}100}$ & $450$ & $250$ & $230$ & $\mathit{{\le}100}$ & $\mathit{{\le}100}$ \\
overlap, $d=2$ & $0.1$ & $\mathbf{670}$ & $2{,}600$ & $5{,}100$ & $1{,}100$ & $\mathit{340}$ & $400$ \\
overlap, $d=2$ & $0.2$ & $\mathbf{160}$ & $620$ & $720$ & $280$ & $\mathit{{\le}100}$ & $\mathit{{\le}100}$ \\
overlap, $d=10$ & $0.1$ & $\mathbf{890}$ & $11{,}000$ & $4{,}900$ & $1{,}300$ & $\mathit{510}$ & $670$ \\
overlap, $d=10$ & $0.2$ & $\mathbf{180}$ & $1{,}500$ & $690$ & $320$ & $\mathit{{\le}100}$ & $130$ \\
dense shift, $d=20$ & $0.2$ & $\mathbf{210}$ & $2{,}800$ & $650$ & $360$ & $\mathit{120}$ & $150$ \\
dense shift, $d=50$ & $0.2$ & $\mathbf{300}$ & $5{,}800$ & $730$ & $500$ & $\mathit{160}$ & $190$ \\
dense shift, $d=100$ & $0.2$ & $\mathbf{380}$ & $11{,}000$ & $1{,}000$ & $690$ & $\mathit{210}$ & $240$ \\
sparse shift, $d=50$ & $0.2$ & $\mathbf{300}$ & $5{,}800$ & $660$ & $530$ & $\mathit{160}$ & $180$ \\
correlation flip, $d=50$ & $0.2$ & ${>}12{,}800$ & ${>}12{,}800$ & $\mathbf{5{,}200}$ & ${>}12{,}800$ & ${>}800$ & ${>}800$ \\
small contamination, $d=2$ & $0.05$ & $\mathbf{900}$ & $5{,}100$ & $5{,}500$ & $3{,}500$ & $600$ & $\mathit{500}$ \\
mixed types, $m=5n$ & $0.2$ & $\mathbf{140}$ & $660$ & $390$ & $170$ & $\mathit{{\le}100}$ & $\mathit{{\le}100}$ \\
$t_3$ tails, $d=5$ & $0.2$ & $\mathbf{320}$ & $2{,}800$ & $1{,}000$ & ${>}12{,}800$ & $\mathit{220}$ & $300$ \\
curves, wide bump & $0.2$ & $\mathbf{270}$ & $1{,}300$ & $690$ & $390$ & $\mathit{180}$ & $240$ \\
curves, narrow bump & $0.2$ & $1{,}200$ & $11{,}000$ & $1{,}100$ & $\mathbf{420}$ & $1{,}200$ & $\mathit{1{,}000}$ \\
oscillating, $\nu=1$ & $0.5$ & $\mathbf{{\le}100}$ & $\mathbf{{\le}100}$ & $\mathbf{{\le}100}$ & $\mathbf{{\le}100}$ & $\mathit{{\le}100}$ & $\mathit{{\le}100}$ \\
oscillating, $\nu=4$ & $0.5$ & $230$ & $2{,}600$ & $\mathbf{150}$ & $10{,}000$ & $580$ & $\mathit{{\le}100}$ \\
oscillating, $\nu=16$ & $0.5$ & $2{,}900$ & ${>}12{,}800$ & $\mathbf{290}$ & ${>}12{,}800$ & ${>}800$ & ${>}800$ \\
oscillating, $\nu=64$ & $0.5$ & ${>}12{,}800$ & ${>}12{,}800$ & $\mathbf{720}$ & ${>}12{,}800$ & ${>}800$ & ${>}800$ \\
\bottomrule
\end{tabular}
\space
\caption{Sample size $n=m$ (mixed types: $n$, with $m=5n$) at which each method reaches power $0.8$ against
$H_{0,0}$ at $\delta=0.05$, from $\mnRepsPerSize$ replicates per size, interpolated between the two sizes that bracket
$0.8$ and rounded to two significant digits; ${\le}\mnNmin$, ${>}\mnNcap$: reached at the smallest size, not reached
at the largest. Bold: the smallest among the four tests that certify a level of $\TV$; italic: the smaller of the last
two, which test $P=Q$ only. $\widehat\alpha$: the betting test of Theorem~\ref{thm:betting} on the Laplace ramp;
$\widehat\alpha_{\mathrm{an}}$: the analytic test \eqref{eq:test} with the Gaussian kernel; GBM, logistic: the betting test on the labels of
gradient boosting and logistic regression trained once on half of the data, a handicap removed in Figure~\ref{fig:recovery}; perm.\ MMD: permutation test
with the Gaussian kernel at the pilot bandwidth of the analytic test and $\abPermB$ relabelings; MMDAgg: the test of
\citet{schrab2023aggregated}; the last two were run up to $n=800$ ($\fourdigits{\harderNtwo}$ on the narrow bump), so
${>}800$ means not reached there. The analytic and permutation tests take their bandwidth from a separate pilot draw,
not counted in $n$ (Section~\ref{sec:numharder}).}
\label{tab:hardern80}
\end{table}

The next question is how many observations each method needs to detect the contamination, that is, to reject
$H_{0,0}$ with power $0.8$ (Table~\ref{tab:hardern80}). Against classifiers trained once on half of the data, the
split-sample bound needs the fewest observations among the methods that certify a level of $\TV$ in
$\mnNeightyRampBest$ of the $\allNdesigns$ designs, alone in $\mnNeightyRampBestAlone$. Gradient boosting needs fewer
where the kernel at its bandwidth barely sees the difference, on the correlation flip, whose population MMD is about
zero, and on the oscillating densities with $\nu\ge4$, and both classifiers need fewer on the narrow bump. Refitted on
the batches of the ramp, the best of the classifiers, default or tuned, has more power at $n=m=800$ than the
split-sample bound in $\mnPowBestClfAboveLap$ designs and less in $\mnPowLapAboveBestClf$
(Section~\ref{sec:baselines} of the supplement). The analytic baseline needs $\mnAnaOverRampLo$ to $\mnAnaOverRampHi$ times
as many observations as the split-sample bound (median $\mnAnaOverRampMed$, over the designs where both sizes are
interpolated), the price of a threshold that holds for every pair in the null. It fires close to where the design rule
\eqref{eq:samplesize} predicts, at $\mnTauFireRatioLo$ to $\mnTauFireRatioHi$ times the size at which $\tau(0)$ meets
the population MMD.

Detecting a difference without certifying its size takes fewer observations. The permutation test of the MMD is the
fastest of all methods, or tied for fastest, in $\mnNeightyPermBest$ designs, and where both sizes are interpolated the
split-sample bound needs $\mnNeightyRampOverPermLo$ to $\mnNeightyRampOverPermHi$ times as many observations (median
$\mnNeightyRampOverPermMed$). The split-sample bound spends a quarter of each sample on fitting alone and pays for
betting, while the permutation test also takes its bandwidth from the pilot draw; the ratio falls below one only on the
oscillating density with $\nu=4$.

\begin{figure}[t]
\centering
\includegraphics[width=\textwidth]{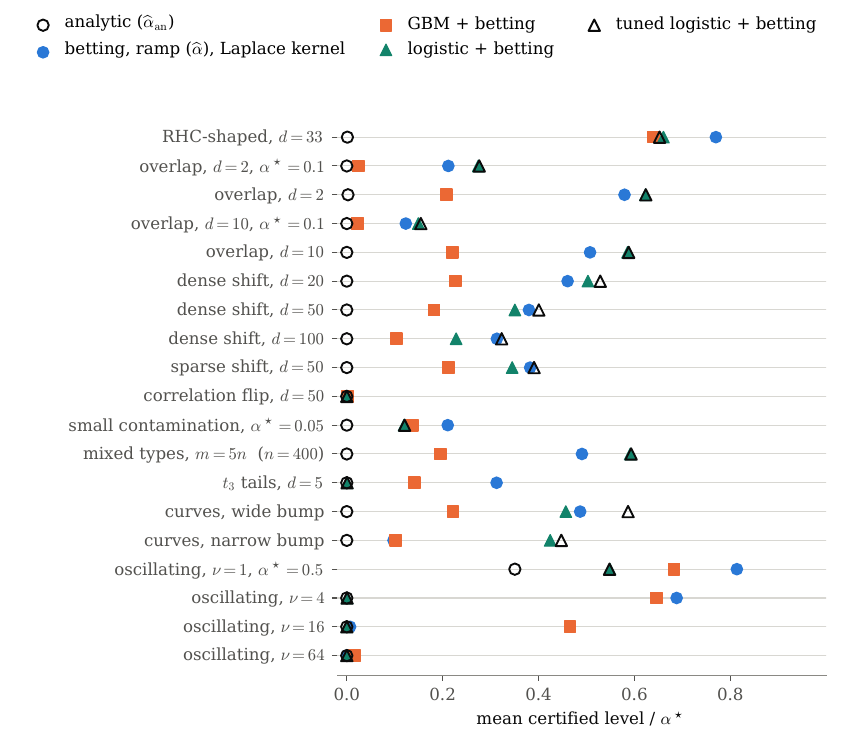}
\caption{How much of $\alpha^\star$ each lower bound recovers at $n=m=800$ (mixed types: $n=400$, $m=5n$): the mean
over $\mnRecReps$ replicates of the largest level $j\alpha^\star/8$ it certifies divided by $\alpha^\star$, with $0$
when it certifies none, on a grid that uses the known $\alpha^\star$; $\alpha^\star=0.2$ unless shown, $\delta=0.05$;
classifiers: labels bet on the batches of the ramp.}
\label{fig:recovery}
\end{figure}

Detection certifies only the first level of the grid, and the bound is meant to certify as much of $\alpha^\star$ as
it can. We measure this by the recovery, the mean over replicates of the largest certified level divided by
$\alpha^\star$. The grid $j\alpha^\star/8$ uses the known $\alpha^\star$ only to express the levels as fractions of
it, and its largest level below $\alpha^\star$ is $7\alpha^\star/8$, so a valid bound recovers at most about
$\rcGridCeiling$. At $n=m=800$ the split-sample bound recovers a median $\mnRecRampMedian$ and a mean
$\mnRecRampMean$ over the designs, at most $\mnRecRampMax$, and $\mnRecRhcRamp$ on the RHC-shaped design
(Figure~\ref{fig:recovery}), while the
analytic baseline recovers a visible fraction only on the oscillating density with $\nu=1$, $\mnRecAnalyticMax$.

No bound recovers all of $\alpha^\star$ at this size. Near $\alpha^\star$ the excess of every function over $\alpha$ is
small, and a bet needs many pairs to detect a small excess: even the oracle that runs the same test on the better of
$\mathbf 1\{p>q\}$ and a level set of $p/q$ recovers a median $\rcOracleMedEightHundred$. Relative to this oracle,
with $\rcReps$ replicates per size below $800$, the recovery of the split-sample bound has median $\rcRatioMedHundred$, $\rcRatioMedTwoHundred$, $\rcRatioMedFourHundred$
and $\rcRatioMedEightHundred$ at $n=m=100$, $200$, $400$ and $800$, and the ratio is larger at $800$ than at $100$ in
$\rcRatioGrows$ of the $\rcRatioGrowsOf$ designs; in the other two the bound certifies almost nothing at any size
(Figure~\ref{fig:recoverycurves} of the supplement shows the curves).

Tuned classifiers can recover more than the split-sample bound where their model fits $\{p>q\}$. Refitted on the
batches of the ramp, unlike in Table~\ref{tab:hardern80}, so that all methods bet on the same observations, tuned
logistic regression recovers more than the bound in $\wtllosses{\mnWtlRecLogitTuned}$ designs and less in
$\wtlwins{\mnWtlRecLogitTuned}$, differences within two paired standard errors counting as ties, mostly more where
$\{p>q\}$ is a half-space, and tuned gradient boosting recovers more, at two decimals, on several designs, among them
the $t_3$ tails and the oscillating densities with $\nu\le16$; with default settings, the bound's mean recovery is
higher than that of the better of the two classifiers in $\mnRampAboveBestMatched$ of the $\allNdesigns$ designs and
lower in the others (Section~\ref{sec:baselines} of the supplement).
We keep the ramp as the default because it needs no tuning beyond the median-distance bandwidth, the rates of Section~\ref{sec:rates} are proved for it, and the program
that fits it also returns the trimming of Section~\ref{sec:trimming}. Validity does not depend on this choice: an
analyst who trusts a classifier on their data can bet on its labels instead (Section~\ref{sec:betting}).

In these designs the gain over an untrimmed witness comes mostly from where the witness is cut. The Gaussian witness
fitted once at $\alpha=0$, with no quadratic program, recovers a median $\abRecMedRampZero$ when cut at $\alpha=0$ and
$\abRecMedRampZeroCut$ when cut at the quantiles of the tested level, and the witness refitted at each level
$\abRecMedRamp$. At two paired standard errors refitting recovers more than the witness fitted at $\alpha=0$ and cut
at the tested level in $\wtlwins{\abWtlRecRampZeroCut}$ designs, as much in $\wtlties{\abWtlRecRampZeroCut}$ and less
in $\wtllosses{\abWtlRecRampZeroCut}$, and, at two decimals, that witness recovers more than the Laplace split-sample
bound in $\dvZeroCutAboveLap$ designs and less in $\dvZeroCutBelowLap$. Refitting gains on the $t_3$ tails, on both bumps and on
the oscillating density with $\nu=1$, and it is what consistency requires in general: a witness fitted at $\alpha=0$
can stop short of $\alpha^\star$ (Sections~\ref{sec:lower} and~\ref{sec:alphazero}).

The theory that favors the Laplace kernel holds on the line. Over the $\allNdesigns$ designs the Laplace kernel
recovers more than the Gaussian kernel in $\mnLapGaussWins$, $\mnLapGaussWinsLine$ of them oscillating densities on the
line, as much in $\mnLapGaussTies$, and less in $\mnLapGaussLosses$, where it is behind by at most
$\mnLapGaussLossMax$. In the $\mnMvDesigns$ multivariate designs, where our theory says nothing about the kernel, it is
slightly behind the Gaussian kernel: it needs fewer observations to detect in $\mnNeightyLapBelowGaussMv$ of them and
more in $\mnNeightyLapAboveGaussMv$, and recovers less in $\mnLapGaussLossesMv$ and more in $\mnLapGaussWinsMv$. Its
higher median recovery, $\mnRecRampMedian$ against $\mnRecGaussMedian$, comes from one design: on the oscillating
density with $\nu=4$ the recovery rises from $\mnRecNfourGauss$ to $\mnRecNfourLap$, which alone lifts the median.
Without it the medians are $\mnRecRampMedianNoNfour$ and $\mnRecGaussMedianNoNfour$, and the means over all designs
$\mnRecRampMean$ and $\mnRecGaussMean$. We keep the Laplace kernel as the default because its witness still sees
differences of high frequency that the Gaussian witness at the same bandwidth misses, as on this one-dimensional
density.

\subsection{Real data}
\label{sec:data}

We apply the split-sample bound and the analytic baseline to two public data sets, which give four contrasts. The
Right Heart Catheterization (RHC) study \citep{connors1996rhc} is an observational study of critically ill adults in
which the treatment is a catheter placed in the right side of the heart; of its $\rhcNtotal$ patients, $\cohNX$
received a catheter. We use its $\rrRhcNcovFull$ numeric and binary baseline covariates, the binary ones coded $0/1$,
standardized (Section~\ref{sec:rdbw} of the supplement lists them). The National Health and Nutrition Examination
Survey (NHANES) \citep{nhanes2012,john2019mims}, cycles 2011--2014 pooled, provides wrist accelerometry, from which we
take curves: for each participant, the mean over the valid days of $\log(1+\mathrm{MIMS})$ in each hour of the day,
with MIMS the monitor-independent movement summary, a curve on $24$ points. As an illustration on functional data, we
compare children aged 6--11 with older adults aged 60--79, younger adults aged 20--39 with the same older adults, and
body mass index below $25$ with $30$ or above within ages 20--59, which holds age roughly fixed; we call these the
child, the adult age and the body-mass contrasts. NHANES is a multistage survey, and we ignore its weights and design,
so these bounds concern the laws of the sampling process rather than the population of the United States. The
independence the betting test needs then holds only approximately, since participants share households and sampling
units; assigning whole sampling units to the batches gives the same or higher minima, though pairs within a batch can
still share a unit (Section~\ref{sec:rdbw}).

Every reported split-sample bound uses the Laplace kernel, with the median distance on batch $0$ as bandwidth for both
witnesses, a rule that, as in the simulations, uses only fitting data. The analytic baseline uses a Gaussian kernel with its
bandwidth fixed on a $30\%$ pilot sample of each group, and is computed on the other $70\%$ only. Each contrast is
reported at $\delta=0.05$, with no correction for the three NHANES contrasts.

\begin{table}[t]
\centering
\small
\setlength{\tabcolsep}{2.3pt}
\begin{tabular}{lrrcccccccc}
\toprule
 & & & \multicolumn{5}{c}{analytic ($\widehat\alpha_{\mathrm{an}}$), pilot bandwidth} & & \multicolumn{2}{c}{split-sample} \\
\cmidrule(lr){4-8}\cmidrule(lr){10-11}
contrast & $n$ & $m$ & $\widehat D_0$ & $\tau(0)$ & $n_{\mathrm{fire}}$ & $\alpha_{\mathrm{last}}$ & bisection & & $\widehat\alpha$ (min) & $\widehat\alpha$ (avg.\ e-value) \\
\midrule
RHC arms & 2{,}184 & 3{,}551 & 0.122 & 0.089 & 1{,}006 & 0.00 & 0 &  & 0.35 & 0.35 \\
ages 6--11 vs.\ 60--79 & 1{,}969 & 2{,}300 & 0.457 & 0.100 & 72 & 0.30 & 0.3375 &  & 0.75 & 0.80 \\
ages 20--39 vs.\ 60--79 & 2{,}519 & 2{,}300 & 0.209 & 0.094 & 342 & 0.05 & 0.0625 &  & 0.35 & 0.40 \\
BMI ${<}25$ vs.\ ${\ge}30$ & 1{,}627 & 2{,}008 & 0.075 & 0.109 & 2{,}633 & -- & -- &  & 0.05 & 0.05 \\
\bottomrule
\end{tabular}
\space
\caption{The two bounds on the four contrasts (the last three from NHANES), $\delta=0.05$, grid mesh $0.05$; $n,m$:
group sizes, of which the analytic baseline uses $70\%$ after a $30\%$ pilot. Analytic baseline: $\widehat D_0$ and
$\tau(0)$, its statistic and threshold at $\alpha=0$; $n_{\mathrm{fire}}$, the balanced size of \eqref{eq:samplesize};
$\alpha_{\mathrm{last}}$, the largest rejected grid level, and bisection, the largest level rejected when bisecting to
width $0.0125$ (Section~\ref{sec:alpha}), both at most the infimum \eqref{eq:alphaest}; `--': none. Split-sample
bound, Laplace kernel: the smallest bound over eight random splits and the level certified by the average of their
final wealths, each $\delta$-level for this kernel. The three NHANES statements hold jointly with probability at least
$1-3\delta$.}
\label{tab:realdata}
\end{table}

One split of the observations into the batches of Section~\ref{sec:betting} makes the bound depend on the split, so we
compute it on eight random splits and summarize them in two ways. Their minimum is a $\delta$-level bound, as it
exceeds $\alpha^\star$ only if every split does. The average rests on e-values, nonnegative statistics whose mean under
the null is at most one, as is the wealth of a betting test at any fixed time: at each level the wealths of the eight
tests at their horizon are e-values for $H_{0,\alpha}$, so by Markov's inequality their average reaches $1/\delta$
with probability at most $\delta$, whatever the dependence between the splits \citep{vovk2021evalues}. Either summary
is valid if fixed in advance, while reporting the larger of the two is guaranteed only at level $2\delta$. We report
the minimum. The average, shown in Table~\ref{tab:realdata} for comparison, certifies as much, and more on the two age
contrasts.

The minima say that more than $\mnRdChildMin$ of the distribution of daily profiles of children, and likewise of older
adults, has no counterpart in the other group, more than $\mnRdAgeMin$ for younger against older adults, and more than
$\mnRdBmiMin$ for the body-mass groups (Table~\ref{tab:realdata}). On RHC every split certifies $\mnRdRhcMin$, so more
than $\mnRdRhcMin$ of the covariate distribution of each arm has no counterpart in the other. On RHC and on the child
and adult age contrasts, $\mnRdRhcNoBet$, $\mnRdChildNoBet$ and $\mnRdAgeNoBet$ of the eight splits stop at a level
where the fitted cut points satisfy $\widehat t_b\le\widehat u_b$ in both batches, so that there is no ramp to bet on
and the bound is limited by the fitted witness rather than by the number of pairs; the others stop at a failed bet.
The minima depend on the kernel on two of the contrasts: with the Gaussian kernel they are the same on the age
contrasts, but $\mnRdRhcFullGaussMin$ on RHC, with the same $\rrRhcNcovFull$ covariates, and $\mnRdBmiGaussMin$ on body
mass (Section~\ref{sec:kernelfallback}). The analytic baseline rejects a positive grid level only on the age contrasts, $\alpha^\star>\mnPvChildLast$ and
$\alpha^\star>\mnPvAgeLast$; on body mass $\widehat D_0=\mnPvBmiDzero$ lies below $\tau(0)=\mnPvBmiTauzero$, and the
design rule asks for $\mnPvBmiFire$ observations per group. Bet on the batches of the ramp, the labels of the
classifiers of Section~\ref{sec:numharder} certify $\rrClfFullGbmHardMin$ on RHC, as the ramp does, except tuned
logistic regression, $\rrClfFullLogitTunedHardMin$, and their predicted probabilities certify between
$\rrClfFullLogitTunedSoftMin$ and $\rrClfFullGbmSoftMin$ (minima over the eight splits; Section~\ref{sec:rdbw}).

\subsection{From the contamination level to a cohort}
\label{sec:appcohort}

At the level $\alpha^\star$ itself, trimming both arms would give the common core $R^\star$, with density proportional
to $\min(p,q)$ for the covariate densities $p,q$ of the two arms. With $e$ the propensity score, the probability $e(x)$ that a patient with covariates $x$ receives the catheter
\citep{rosenbaum1983central}, $\pi$ the fraction of
catheterized patients and $\rho$ the pooled density, $\min(p,q)=\rho\min\{e/\pi,(1-e)/(1-\pi)\}$, an overlap population
in the sense of \citet{li2018balancing}, and for arms of equal size that of the matching weights of
\citet{ligreene2013matching}. As $p/q$ increases with $e$, $\alpha^\star$ is also the total variation between the laws
of $e(X)$ in the two arms. Replacing $e$ by a logistic fit and $\rho$ by the pooled sample in
$\alpha^\star=1-\int\min(p,q)$ gives $\rrPropFullPlug$ (bootstrap interval $(\rrPropFullPlugLo,\rrPropFullPlugHi)$),
an estimate that rests on the propensity model and certifies nothing (Section~\ref{sec:rdbw}). Trimming near
$\alpha^\star$ targets the patients for whom both treatments are plausible, as propensity trimming does with a model
\citep{yang2018asymptotic}, while the rules of Section~\ref{sec:intro} aim at positivity.

At the level $\alpha=\lcCohAlpha$ of the lower bound on RHC we trim both arms with the Gaussian kernel at the pooled
median distance, which Corollary~\ref{cor:alphalo} allows for any valid and consistent level, here one from the Laplace bound. The program \eqref{eq:qp} has a unique minimizer, as
the Gram matrix of the patients is positive definite. It gives positive weight to $\lcPoolKeptX$ of the $\cohNX$ catheterized patients and
$\lcPoolKeptY$ of the $\cohNY$ others, and fractional weight only to the $\lcPoolFracX$ and $\lcPoolFracY$ on which the
witness equals its cut (Proposition~\ref{prop:bangbang}(ii)), which carry fractions $\lcPoolFracMassX$ and
$\lcPoolFracMassY$ of the trimmed samples (Section~\ref{sec:rdbw}). With the Laplace kernel, whose witness on the line
is flat on the set it trims (Proposition~\ref{prop:excess}(iii)), the program at the same level leaves $\lpPoolFracX$ and $\lpPoolFracY$ patients with fractional
weight.

The cohort is a $\lcCohAlpha$-trimming of each arm. Proposition~\ref{prop:alphahat} and Corollary~\ref{cor:alphalo}
describe such trimmings for a kernel fixed in advance, while here the bandwidth is the pooled median. The cohort leaves the
largest standardized difference of the $\rrRhcNcovFull$ covariates at $\lcTrimSMD$, below the usual threshold $0.1$.
A trimming balances only what it sees. In a separate run on the $\rrRhcNcovPaper$ covariates stored as numbers, the
Laplace trimming at the level $\rrCohPaperAlpha$ of the bound on them leaves $\rrCohPaperOmitSmdAbove$ of the other $\rrRhcNcovAdded$ above $0.1$, against
$\rrAddedSmdAbove$ before trimming. Matching and overlap weights balance more tightly but need a propensity model,
whose covariates the overlap weights balance exactly by construction (Table~\ref{tab:cohort}). The $30$-day mortality
difference, $\lcUnwRD$ in the raw arms, is $\lcTrimRD$ under the trimming, with interval $(\lcTrimRDlo,\lcTrimRDhi)$,
and up to $\lcRDmax$ under the other weightings, with overlapping intervals. These are illustrations, not causal
estimates: they would estimate effects only under no unmeasured confounding, and the intervals ignore the estimation of
the weights. As $\lcCohAlpha$ is a lower bound on $\alpha^\star$, the trimming approaches the common core only as the
level approaches $\alpha^\star$ (Corollary~\ref{cor:alphalo}), and the level should be reported with any effect
estimated on the cohort.
\space
\section{Final comments}
\label{sec:discussion}

We have bounded $\alpha^\star=\TV(P,Q)$ from below by following the embeddings of all trimmings of the two laws
rather than one optimal trimming. They form convex regions that meet exactly at $\alpha^\star$ and fluctuate like
sample means, which gives guarantees at every sample size on any sample space with a kernel as in
Assumption~\ref{ass:regularity}, and the witness of their closest points chooses both the trimming and the function
bet on.

The common fraction is a choice. If the two laws may lose different fractions, the larger of the two is still at
least $\alpha^\star$ (Section~\ref{sec:regions}), so equal fractions minimize the loss of the arm that loses more.
Fractions $\alpha_1$ of $P$ and $\alpha_2$ of $Q$ leave a common remainder
exactly when $\int\min\{p/(1-\alpha_1),q/(1-\alpha_2)\}\ge1$, that is, when
$E_\zeta(P\,\|\,Q)=\sup_A\{P(A)-\zeta Q(A)\}\le\alpha_1$ with $\zeta=(1-\alpha_1)/(1-\alpha_2)$, the hockey-stick
divergence of information theory and differential privacy \citep{polyanskiy2010channel}, whose values over $\zeta$
trace the trade-off curve of testing $P$ against $Q$ \citep{dong2022gaussian}. The regions, their witness and the
betting test carry over, with the weights and cuts of each law at its own fraction and bets on
$\psi(X)-\zeta\psi(Y)$, whose best choice is the indicator of $\{p>\zeta q\}$. A bound for each $\zeta$ would say how much of one arm must go to
keep more of the other. The rates of Section~\ref{sec:rates} are proved for equal fractions, and we expect them to carry
over along each ray of fixed $\zeta$, where the rescaled distance is still convex and the paired statistics bounded.

The dimension can enter the rate $(n\wedge m)^{-1/(2\gamma)}$ of Theorem~\ref{thm:loratemain} only through the margin
exponent $\gamma$. Contaminations separated by a function of the RKHS give $\gamma=1$ in every
dimension. When they touch, the margin proved for Sobolev kernels in Section~\ref{sec:linearmargin} has
$\gamma=1+(2s-1)/\{2(r+1)\}$ when $p-q$ vanishes to order $r$ where the contaminations meet, and a Sobolev kernel on
$\R^d$ is bounded only for $s>d/2$, so the guaranteed rate is slower than $(n\wedge m)^{-(r+1)/(2r+1+d)}$: for a
transversal crossing, $(n\wedge m)^{-2/(d+3)}$ at best, and $(n\wedge m)^{-2/(d+4)}$ with the Laplace kernel. Only a
contact that is flat compared with $d$ keeps the dimension out of this guarantee. Whether the dimension enters the rate
itself is open; on the line, for the Laplace kernel, densities that cross once and samples of comparable size, the
margin is pessimistic (Theorem~\ref{thm:laplacerate}).

What rate is possible in $\R^d$? We expect that of classification. Betting on a function $\hat f$ fitted on half of the data certifies
at most $\E_P\hat f-\E_Q\hat f=\alpha^\star-\mathrm{err}(\hat f)$, except with probability $\delta$, so a plug-in
classifier falls short of $\alpha^\star$ by at least $\mathrm{err}(\hat f)$, unless the bound is zero, and, up to logarithmic factors, by at most
$\mathrm{err}(\hat f)$ plus the mesh of the grid and a betting error of order $(n\wedge m)^{-1/2}$. On
the class of \citet[Section~3]{audibert2007fast}, with $p/(p+q)$ H\"older of order $\beta_{\mathrm H}$, margin exponent
$\kappa_{\mathrm M}$, $(P+Q)/2$ under their strong density assumption and $\kappa_{\mathrm M}\beta_{\mathrm H}\le d$, their
bounds make this shortfall of order
$\max\{(n\wedge m)^{-1/2},(n\wedge m)^{-\beta_{\mathrm H}(1+\kappa_{\mathrm M})/(2\beta_{\mathrm H}+d)}\}$, in which the
regularity $\beta_{\mathrm H}$ again offsets the dimension. We conjecture
that every lower bound valid at level $\delta$ uniformly over this class, even one that reports only a value, has a
worst-case error at least of this order up to a logarithmic factor, so that locating the difference costs nothing in
rate. The natural route to this lower bound is that of nonsmooth functionals, fuzzy hypotheses with matched moments,
where polynomial approximation gains
exactly a logarithmic factor over plug-in \citep{lepski1999estimation,cai2011testing,han2020estimation}.

Whether our kernel bounds attain this rate, whether the near-parametric rate of Theorem~\ref{thm:laplacerate}(i) holds without
condition \eqref{eq:condR}, for other kernels and in $\R^d$, and how to choose the kernel from the data with a
guarantee, are open, and so is whether averaging the e-values of fits with the batches' roles exchanged, which lets
every observation be bet on without losing validity \citep{vovk2021evalues}, gains power over a single split.
\space
\label{mainend}

\clearpage
\phantomsection\pdfbookmark[1]{Supplement}{supplement}
\begin{center}
\normalfont\normalsize\bfseries\mathversion{bold}\MakeTextUppercase{Supplement}
\end{center}
\setcounter{subsection}{0}\setcounter{equation}{0}\setcounter{theorem}{0}
\setcounter{section}{0}
\setcounter{table}{0}
\setcounter{figure}{0}
\setcounter{algorithm}{0}
\expandafter\def\csname theHsection\endcsname{S\arabic{section}}
\expandafter\def\csname theHtable\endcsname{S\arabic{table}}
\expandafter\def\csname theHfigure\endcsname{S\arabic{figure}}
\expandafter\def\csname theHalgorithm\endcsname{S\arabic{algorithm}}
\renewcommand{\thesection}{S\arabic{section}}
\renewcommand{\thetable}{S\arabic{table}}
\renewcommand{\thefigure}{S\arabic{figure}}
\renewcommand{\thealgorithm}{S\arabic{algorithm}}
\setcounter{topnumber}{4}
\setcounter{totalnumber}{6}
\renewcommand{\floatpagefraction}{0.5}
\renewcommand{\topfraction}{0.85}
\makeatletter
\let\unified@toclevel@section\toclevel@section
\let\unified@toclevel@subsection\toclevel@subsection
\let\unified@toclevel@subsubsection\toclevel@subsubsection
\def\toclevel@section{2}\def\toclevel@subsection{3}\def\toclevel@subsubsection{4}
\makeatother

Sections~\ref{app:regions} to~\ref{app:trimming} prove the results of Sections~\ref{sec:regions} to~\ref{sec:trimming}
of the main text in their order, with three exceptions: Proposition~\ref{prop:bangbang} and the uniform bound
\eqref{eq:hausdorff} are proved with the trimming in Section~\ref{app:trimming}; the lemma on extreme trimmed means,
Lemma~\ref{lem:trimmedmean}, which Sections~\ref{app:regions}, \ref{app:analytic} and~\ref{app:trimming} also use, is
proved with the betting test in Section~\ref{app:split}; and Section~\ref{app:rates} proves the rate of
Theorem~\ref{thm:loratemain} first, Proposition~\ref{prop:newsobolev} after Proposition~\ref{prop:newgauss}, and Proposition~\ref{prop:lower2pt} and Corollary~\ref{cor:minimax} last.
Section~\ref{app:regions} establishes the geometry of the trimmed regions and the concentration inequality,
Lemma~\ref{lem:hinge}, behind most deviation bounds, and ends with the proof of Theorem~\ref{thm:impossible} on upper
bounds. Section~\ref{app:analytic} turns the inequality into a two-sided bound on the statistic before proving
Theorem~\ref{thm:plugin}. Section~\ref{app:split} gives the betting test in full, with Algorithm~\ref{alg:split}, its
validity and consistency, and an example in which a witness fitted at $\alpha=0$ is not consistent; validity rests on
Lemma~\ref{lem:ramp} and Theorem~\ref{thm:betting} alone. Section~\ref{app:rates}, the longest, proves the rates and
their limits, with the constants of Theorem~\ref{thm:loratemain} in Theorem~\ref{thm:lorate} and the proof of
Theorem~\ref{thm:laplacerate} in Section~\ref{sec:laplaceproof}, and Section~\ref{app:experiments} describes the
experiments.

A few ideas carry most of the proofs. The trimmed regions are convex, weakly compact sets of kernel mean embeddings, so
the trimmed distance is the largest gap between them along a unit direction, attained in the direction of the
difference of their closest points, the witness; each closest point minimizes, or maximizes, the integral of the
witness over the trimmings of its law. Optimizing the integral of a fixed function over trimmings is a bathtub problem,
solved by putting the largest weight allowed on the largest values until the mass runs out, that is, by a cut at a
quantile. This turns choices among trimmings into level sets of the witness and writes the extreme trimmed means as
$t+\beta\,\E(f-t)_+$ and $t-\beta\,\E(t-f)_+$ at the cut $t$, means of hinge functions. In the deviation bounds,
sample fluctuations therefore enter through means of bounded functions, which concentrate at the rate
$(n\wedge m)^{-1/2}$ by symmetrization and McDiarmid's inequality, with constants that depend on the kernel through
$\kappa$ and $\Delta_\phi$ and on the level only through $\beta$, so that one event serves all levels. The bets are
valid because under the null each has conditional mean at most $\alpha$ given the past, whatever the fitted ramp, so
the wealth is a nonnegative supermartingale, which by Ville's inequality reaches $1/\delta$ with probability at most
$\delta$. The rate of the trimming at a fixed level, Theorem~\ref{cor:trimrate}, rests on a quadratic margin of the
separation on the unit sphere around the direction of the witness (Lemma~\ref{lem:sepmargin}), which a peeling over
shells around that direction turns into a rate.

The rates of the bound rest on the path. A mixture of the parts of mass $1-\alpha_0$ and $1-\alpha_1$ kept by two
trimmings is a part kept at the level in between, so the rescaled path $\bar D_\alpha=(1-\alpha)D_\alpha(P,Q)$ is
convex; as it vanishes at $\alpha^\star$, it leaves zero at most linearly, and the gap between the cut points of the
population witness is the slope of a line supporting it from below. Margins come from approximating the indicator of
$\{p>q\}$ by functions of the RKHS, whose norms Fourier analysis bounds for Sobolev and Gaussian kernels. On the line
the Laplace kernel is the Green function of $1-\sigma^2\partial_x^2$, so its witness cannot overshoot its cut points.
The fitted ramp then maximizes a ratio of excess to norm, and comparing this ratio with the population path at two
levels shows that the fitted ramp keeps a fixed fraction of the excess of the population ramp, which gives
Theorem~\ref{thm:laplacerate}; its empirical-process steps use their own bounds on intervals of the line
(Lemmas~\ref{lem:lapintervals} to~\ref{lem:lapdev}). The acknowledgments of the main text say where the Lean~4
formalization of the results is.
\space

\section{Proofs for Section~\ref{sec:regions}}
\label{app:regions}

The notation is that of the main text, and symbols are recalled where they reappear far from their definition. We
write $B_\Hh$ for the closed unit ball of $\Hh$ and $\mathcal G_\alpha(P)=\{g:0\le g\le\beta,\ \E_P[g]=1\}$ for the set
of densities of the $\alpha$-trimmings of $P$ with respect to $P$, so that
$\mathcal C_\alpha(P)=\{\E_P[g\phi]:g\in\mathcal G_\alpha(P)\}$, and for independent Rademacher signs $\varepsilon_i$
we put $e_k=\E|\varepsilon_1+\dots+\varepsilon_k|/k\le k^{-1/2}$. Several proofs use the separation
$\mathrm{sep}(f;A,B)=\inf_{a\in A}\langle f,a\rangle_\Hh-\sup_{b\in B}\langle f,b\rangle_\Hh$ of
Section~\ref{sec:ramp} through the elementary fact, proved in Lemma~\ref{lem:separation}, that
$\dist(A,B)=\max_{f\in B_\Hh}\mathrm{sep}(f;A,B)$ for convex weakly compact $A$ and $B$, which is \eqref{eq:distmax}.
Proposition~\ref{prop:bangbang} and the uniform bound \eqref{eq:hausdorff} of Section~\ref{sec:regions} are proved
later, in Section~\ref{app:trimming}, next to the tools they need.

Two technical points are settled once and for all. By (A1) and (A2) of Assumption~\ref{ass:regularity}, $\phi$ is
weakly measurable with $\|\phi\|_\Hh\le\kappa$, hence Bochner integrable by Pettis' theorem, so $\E_P[g\phi]$ is well
defined for bounded measurable $g$. And the suprema over directions and cuts that appear below are measurable. Indeed,
$|f(x)-f'(x)|\le\kappa\|f-f'\|_\Hh$ and $|(u-t)_+-(u-t')_+|\le|t-t'|$ make $(f,t)\mapsto(f(x)-t)_+$ jointly
continuous, and dominated convergence makes $f\mapsto\E_P[(f-t)_+]$ continuous, so every such supremum may be taken over
a countable dense subset without changing its value.

\subsection{The results of Section~\ref{sec:regions}}

We first prove \eqref{eq:tvintersect}. A law $R$ is a common $\alpha$-trimming exactly when $(1-\alpha)R$ lies below
both $P$ and $Q$, so the question is one of mass. Let $P\wedge Q$ be the largest measure below both $P$ and $Q$; it has
density $\min(p,q)$ with respect to any measure dominating both, and mass $1-\TV(P,Q)$. If $\TV(P,Q)\le\alpha<1$, the
law $R=(P\wedge Q)/(1-\TV(P,Q))$ has densities at most $1/(1-\TV(P,Q))\le\beta$ with respect to $P$ and to $Q$, so it
lies in $R_\alpha(P)\cap R_\alpha(Q)$. Conversely, if $R$ lies in both sets, then $(1-\alpha)R\le P$ and
$(1-\alpha)R\le Q$, hence $(1-\alpha)R\le P\wedge Q$ by maximality of $P\wedge Q$, and comparing the masses of the two
sides gives $1-\alpha\le1-\TV(P,Q)$.

The same comparison of masses shows that laws which lose different fractions $\alpha_1$ and $\alpha_2$ and agree on what
remains must have $\alpha_1\vee\alpha_2\ge\alpha^\star$. Their common remainder is a probability measure with density
at most $\min\{p/(1-\alpha_1),q/(1-\alpha_2)\}\le\min(p,q)/(1-\alpha_1\vee\alpha_2)$, whose integral is
$(1-\alpha^\star)/(1-\alpha_1\vee\alpha_2)$; as it must be at least one, $\alpha_1\vee\alpha_2\ge\alpha^\star$.

\begin{proof}[Proof of Proposition~\ref{prop:georep}]
The map $g\mapsto g\cdot P$ is a bijection from $\mathcal G_\alpha(P)$ onto $R_\alpha(P)$, and Bochner
integrability gives $\mu_{g\cdot P}=\E_P[g\phi]$, so $P'\mapsto\mu_{P'}$ maps $R_\alpha(P)$ onto
$\mathcal C_\alpha(P)$. The infimum in \eqref{eq:trimMMDpop} is therefore the infimum of $\|u-v\|_\Hh$ over
$u\in\mathcal C_\alpha(P)$ and $v\in\mathcal C_\alpha(Q)$, which is \eqref{eq:setdist}.

For $P=P_n$ the only care needed is with sample points that no measurable set separates, so we group the sample points
into such classes. A density with respect to $P_n$ is measurable, hence constant on each class, and so defines a capped
weight vector $w_i=g(X_i)/n$ with the same embedding. Conversely, averaging a capped weight vector within each class
preserves the cap, the normalization and $\sum_iw_i\phi(X_i)$, because $k(\cdot,x')$ is measurable for every $x'$ by
(A1), so that $\phi$ is constant on a class; the averaged vector is a density. The set is a polytope, as the
image of the capped simplex, itself a polytope, under the linear map $w\mapsto\sum_iw_i\phi(X_i)$.
\end{proof}

\begin{proof}[Proof of Lemma~\ref{lem:attainment}]
Everything except norm-compactness follows from the weak compactness of the regions.

Convexity and the bound are immediate: $\mathcal G_\alpha(P)$ is convex and $g\mapsto\E_P[g\phi]$ is linear, and
$\|\E_P[g\phi]\|_\Hh\le\E_P[g\|\phi\|_\Hh]\le\kappa$. For compactness, regard $\mathcal G_\alpha(P)$ as a subset of
$L^\infty(P)=(L^1(P))^*$. It is bounded, and it is weak$^*$-closed, since each constraint has the form
$\int gh\,dP\ \{\ge,\le,=\}\ c$ for a fixed $h\in L^1(P)$, with $h=\mathbf 1_A$ for the pointwise bounds and $h\equiv1$
for the normalization; so it is weak$^*$-compact by the Banach--Alaoglu theorem. The map $T:g\mapsto\E_P[g\phi]$ is
weak$^*$-to-weak continuous, because $\langle Tg,e\rangle_\Hh=\int ge\,dP$ for $e\in\Hh$, and $e$ is a bounded
function, $|e(x)|\le\kappa\|e\|_\Hh$. So $\mathcal C_\alpha(P)$ is weakly compact, hence weakly
closed and norm-closed. It is even norm-compact, because the embedding is a Hilbert--Schmidt operator. The adjoint of
$T:L^2(P)\to\Hh$, $Tg=\E_P[g\phi]$, sends $h\in\Hh$ to the function $h$ in $L^2(P)$. For an orthonormal basis $(h_j)$
of $\Hh$, Parseval's identity gives $\sum_j\|T^*h_j\|^2_{L^2(P)}=\E_P\sum_jh_j(X)^2=\E_P\|\phi(X)\|_\Hh^2\le\kappa^2$,
so $T^*$, and hence $T$, is Hilbert--Schmidt and therefore compact. The set $\mathcal G_\alpha(P)$ is bounded in
$L^2(P)$, so $\mathcal C_\alpha(P)=T(\mathcal G_\alpha(P))$ is relatively compact, and it is norm-closed.

The norm is weakly lower semicontinuous on the weakly compact product of the two regions, so the distance is attained,
at points $\mu_{P^\star},\mu_{Q^\star}$ for trimmings $P^\star,Q^\star$ by Proposition~\ref{prop:georep}. If
$D_\alpha(P,Q)=0$, the two points coincide, so the intersection is nonempty, and every element of it is
$\mu_{P^\star}=\mu_{Q^\star}$ for some trimmings $P^\star,Q^\star$, which are equal by (A3).
\end{proof}

The equivalence \eqref{eq:zeroeq} follows. If $\TV(P,Q)\le\alpha$, \eqref{eq:tvintersect} gives a common trimming,
whose embedding lies in both regions, so $D_\alpha(P,Q)=0$; this direction uses neither attainment nor (A3).
Conversely, if $D_\alpha(P,Q)=0$, Lemma~\ref{lem:attainment} gives a common trimming, and \eqref{eq:tvintersect} gives
$\TV(P,Q)\le\alpha$.

\begin{proof}[Proof of Lemma~\ref{lem:cvar}]
The first equality holds because $\langle\E_P[g\phi],f\rangle_\Hh=\E_P[gf]$. The second is a bathtub problem, in
which every cut gives an upper bound and the bound at one cut is attained. For any $t$ and $g\in\mathcal G_\alpha(P)$, since
$0\le g\le\beta$ and $\E_Pg=1$,
\[
\E_P[gf]=t+\E_P[g(f-t)]\le t+\beta\,\E_P[(f-t)_+] ,
\]
so the supremum is at most the infimum. For the converse we fill the values of $f$ above a quantile with the largest
weight allowed. Let $t^\star$ be an $\alpha$-quantile of $f(X)$, so that
$\Pp_P(f>t^\star)\le1-\alpha\le\Pp_P(f\ge t^\star)$, and put $g^\star=\beta\mathbf 1\{f>t^\star\}+c\mathbf
1\{f=t^\star\}$ with $c=(1-\beta\Pp_P(f>t^\star))/\Pp_P(f=t^\star)$ when $\Pp_P(f=t^\star)>0$, and $c=0$ otherwise. Multiplied by $\beta$, the quantile inequalities read
$\beta\Pp_P(f>t^\star)\le1\le\beta\Pp_P(f\ge t^\star)$, hence $c\in[0,\beta]$ and $\E_P[g^\star]=1$ (when
$\Pp_P(f=t^\star)=0$ they force $\beta\Pp_P(f>t^\star)=1$). Since $(f-t^\star)_+=(f-t^\star)\mathbf 1\{f>t^\star\}$,
we get $\E_P[g^\star f]=t^\star+\beta\E_P[(f-t^\star)_+]$, so $g^\star$ attains the supremum and $t^\star$ the
infimum. Only boundedness and measurability of $f$ were used.
\end{proof}

\subsection{The cut, continuity in the level and a hinge class}

The cut $t$ in \eqref{eq:cvar}, the threshold of the next lemma, can be confined to $[-\kappa,\kappa]$.

\begin{lemma}[Compact range of the threshold]
\label{lem:tcompact}
Let Assumption~\ref{ass:regularity} hold, $\alpha\in[0,1)$, and $R$ any probability measure. For every $f\in B_\Hh$
the infimum in \eqref{eq:cvar}, with $R$ in place of $P$, is attained at some $t^\star\in[-\kappa,\kappa]$.
\end{lemma}

\begin{proof}
The function $A(t)=t+\beta\E_R[(f(X)-t)_+]$ is convex, with right derivative $1-\beta\Pp_R(f(X)>t)$ by dominated
convergence. Since $|f|\le\kappa$, $A(t)=t$ for $t>\kappa$, while for $t<-\kappa$ the right derivative is
$1-\beta=-\alpha\beta\le0$. A continuous convex function that is nonincreasing on $(-\infty,-\kappa]$ and increasing
on $[\kappa,\infty)$ attains its infimum in $[-\kappa,\kappa]$.
\end{proof}

\begin{lemma}[Local Lipschitz continuity in $\alpha$]
\label{lem:continuity}
Under Assumption~\ref{ass:regularity}, for every $\bar\alpha\in[0,1)$ and $\alpha,\alpha'\in[0,\bar\alpha]$,
\[
\left|D_\alpha(P,Q)-D_{\alpha'}(P,Q)\right|\le\frac{4\kappa}{(1-\bar\alpha)^2}\,|\alpha-\alpha'| .
\]
In particular $\alpha\mapsto D_\alpha(P,Q)$ is continuous on $[0,1)$.
\end{lemma}

\begin{proof}
The level enters the support functions only through $\beta$, as an infimum of functions affine in $\beta$, so it
suffices to bound their slopes. Write $\beta'=(1-\alpha')^{-1}$. For $f\in B_\Hh$, Lemmas~\ref{lem:cvar} and~\ref{lem:tcompact} give
\[
h_{\mathcal C_\alpha(P)}(f)=\inf_{t\in[-\kappa,\kappa]}\big\{t+\beta\,\E_P[(f-t)_+]\big\},
\]
an infimum of affine functions of $\beta$ with slopes $\E_P[(f-t)_+]\in[0,2\kappa]$. An infimum of functions that are
$2\kappa$-Lipschitz is $2\kappa$-Lipschitz, so $|h_{\mathcal C_\alpha(P)}(f)-h_{\mathcal C_{\alpha'}(P)}(f)|\le
2\kappa|\beta-\beta'|$ uniformly in $f$, and likewise for $Q$. By \eqref{eq:distmax} in Section~\ref{app:split},
\[
D_\alpha(P,Q)=\max_{f\in B_\Hh}\big\{-h_{\mathcal C_\alpha(P)}(-f)-h_{\mathcal C_\alpha(Q)}(f)\big\},
\]
and a difference of two maxima is at most the largest difference of the maximized functions, so
$|D_\alpha(P,Q)-D_{\alpha'}(P,Q)|\le4\kappa|\beta-\beta'|$. Finally
$\tfrac{d}{d\alpha}(1-\alpha)^{-1}\le(1-\bar\alpha)^{-2}$ on $[0,\bar\alpha]$, which gives the claim. Only (A1) and (A2)
were used, for arbitrary probability measures, so the lemma applies to $P_n$ and $Q_m$.
\end{proof}

The deviation bounds of Theorem~\ref{thm:twosided}, Corollary~\ref{cor:uniform} and Lemma~\ref{lem:localmod} are
instances of one lemma, in which $n_\star=nm/(n+m)$, as in Section~\ref{sec:analytic}.

\begin{lemma}[A two-sample hinge class]
\label{lem:hinge}
Let (A1) and (A2) of Assumption~\ref{ass:regularity} hold, let $X_1,\dots,X_n$ be i.i.d.\ $P$ and $Y_1,\dots,Y_m$
i.i.d.\ $Q$, the two samples independent, and let $r,\rho_1,\rho_2\ge0$. For each $x\in\mathcal X$ let
$\varphi^1_x,\varphi^2_x:\R\to\R$ be $1$-Lipschitz, with $(x,z)\mapsto\varphi^i_x(z)$ measurable and $\sup_x|\varphi^i_x(0)|<\infty$.
For $(v,a,b)$ in $\mathcal S=\{(v,a,b)\in\Hh\times\R^2:\|v\|_\Hh\le r,\ |a|\le\rho_1,\ |b|\le\rho_2\}$ put
$h^1_{v,a}(x)=\varphi^1_x(a-v(x))$ and $h^2_{v,b}(y)=\varphi^2_y(v(y)-b)$, and let
\[
\Gamma=\sup_{\mathcal S}\big[(P-P_n)h^1_{v,a}+(Q-Q_m)h^2_{v,b}\big],
\qquad
\Gamma^-=\sup_{\mathcal S}\big[(P_n-P)h^1_{v,a}+(Q_m-Q)h^2_{v,b}\big].
\]
Then $\Gamma$ and $\Gamma^-$ are measurable and their means are at most $2[r\kappa/\sqrt{n_\star}+\rho_1e_n+\rho_2e_m]$. If moreover
$|h^1_{v,a}(x)-h^1_{v,a}(x')|\le\Delta_1$ and $|h^2_{v,b}(y)-h^2_{v,b}(y')|\le\Delta_2$ for all points and all
$(v,a,b)\in\mathcal S$, then for every $\chi>0$ each of $\Gamma$ and $\Gamma^-$ exceeds its mean by more than
$\sqrt{\chi(\Delta_1^2/n+\Delta_2^2/m)/2}$ with probability at most $e^{-\chi}$. For one sample the same holds with the
terms of the second sample removed: $n_\star=n$, and no $\rho_2e_m$ and no $\Delta_2^2/m$.
\end{lemma}

\begin{proof}
The mean is bounded by symmetrization, contraction and a bound on the norm of a Rademacher sum in $\Hh$, and the
deviation by McDiarmid's inequality. Measurability is settled first: for fixed samples the bracket is continuous in
$(v,a,b)$, and the population terms are continuous by dominated convergence with the envelope
$\sup_x|\varphi^i_x(0)|+\rho_i+\kappa r$; as $\Hh\times\R^2$ is separable, $\Gamma$ and $\Gamma^-$ are measurable.

For the mean, $\Gamma$ is a supremum of sums of $n+m$ independent centered terms, and symmetrization holds in the signed
form, with no absolute value inside the supremum. We introduce an independent copy of the samples, apply Jensen's
inequality, multiply the differences by independent signs $\varepsilon_i,\varepsilon'_j$, and split the supremum, to get
\[
\E \Gamma\le2\,\E\sup_{\mathcal S}\Big[\frac1n\sum_i\varepsilon_i\varphi^1_{X_i}\big(a-v(X_i)\big)+\frac1m\sum_j\varepsilon'_j
\varphi^2_{Y_j}\big(v(Y_j)-b\big)\Big].
\]
Conditionally on the samples, we remove the maps $\varphi^i$ by the contraction principle for signed Rademacher
averages with a different $1$-Lipschitz map in each coordinate \citep[Lemma~26.9]{shalevshwartzbendavid2014}, applied
to the coordinates $(a-v(X_i))/n$ and $(v(Y_j)-b)/m$ and the maps $z\mapsto\varphi^1_{X_i}(nz)/n$ and
$z\mapsto\varphi^2_{Y_j}(mz)/m$. Its constant is $1$, not the $2$ of the version with absolute values
\citep[Theorem~4.12]{ledouxtalagrand1991}, and it needs no normalization $\varphi(0)=0$. It replaces the bracket by the
linear expression
\[
\langle v,S_\varepsilon\rangle_\Hh+\frac an\sum_i\varepsilon_i-\frac bm\sum_j\varepsilon'_j,
\qquad
S_\varepsilon=\frac1m\sum_j\varepsilon'_j\phi(Y_j)-\frac1n\sum_i\varepsilon_i\phi(X_i),
\]
whose supremum over $\mathcal S$ is $r\|S_\varepsilon\|_\Hh+\rho_1|n^{-1}\sum_i\varepsilon_i|+\rho_2|m^{-1}\sum_j\varepsilon'_j|$.
The means of the two sign averages are $e_n$ and $e_m$ by definition, and
\[
\E\|S_\varepsilon\|_\Hh\le(\E\|S_\varepsilon\|_\Hh^2)^{1/2}\le\kappa/\sqrt{n_\star}
\]
by Jensen's inequality, then independence of the signs and $k(x,x)\le\kappa^2$, since $1/n+1/m=1/n_\star$. This gives the
bound on the mean. The argument for $\Gamma^-$ is the same, with the sample and its copy exchanged.

For the deviation, replacing one $X_i$ moves $\Gamma$ and $\Gamma^-$ by at most $\Delta_1/n$, and one $Y_j$ by at most
$\Delta_2/m$, so McDiarmid's inequality \citep{mcdiarmid1989} gives the tail bound (if $\Delta_1=\Delta_2=0$ every $h$ is
constant and $\Gamma=\Gamma^-=0$).
\end{proof}

\subsection{No informative upper bound}
\label{sec:upperimposs}

The triviality of distribution-free upper bounds on $\TV$ persists on the much
smaller class of pairs whose densities are bounded by $2$, which contains every $\tfrac12$-trimming of the uniform law
and in which no atom can be planted: a procedure valid over the whole class must return the trivial bound, with
probability at least $1-\delta$, even when the two laws are identical.

\begin{proof}[Proof of Theorem~\ref{thm:impossible}]
We hide a pair of laws at total variation one in a fine random partition. Fix an even integer $M\ge2(n+m)$, cut $[0,1]$ into $M$ cells of equal length, and let $S$ be a uniformly random set of
$M/2$ cells. Let $P_S$ and $Q_S$ be the uniform laws on the union of the cells in $S$ and in its complement, with
densities $2\cdot\mathbf 1_S$ and $2\cdot\mathbf 1_{S^c}$. Then $(P_S,Q_S)\in\mathcal P_2$ and $\TV(P_S,Q_S)=1$, so the
coverage requirement gives $\Pp_{P_S,Q_S}(\widehat U\ge1)\ge1-\delta$ for each $S$, and hence under the mixture $\mu_M$
of the laws of the data over the random $S$.

Data drawn from $\mu_M$ are nearly indistinguishable from two uniform samples, because they rarely put two
observations in one cell. Let $E$ be the event that the $n+m$ observations fall in $n+m$ distinct cells. Given $E$, $\mu_M$ and the product
$P_{\mathrm u}^{\otimes(n+m)}$ of uniform laws coincide. Under $\mu_M$, an ordered tuple of distinct cells receives
$X_1,\dots,X_n,Y_1,\dots,Y_m$ with probability $(2/M)^{n+m}$ times the probability that its first $n$ cells lie in $S$
and its last $m$ in $S^c$, which depends only on $n$ and $m$, so under both laws the tuple is uniform given $E$, with
independent uniform positions within the cells. Laws that agree given an event differ in total variation by at most
the sum of the masses of its complement. Under $\mu_M$ an $X$ and a $Y$ never share a cell, and two $X$'s, or two
$Y$'s, do so with probability at most $2/M$; under the product law two observations do so with probability $1/M$. Hence
$\TV(\mu_M,P_{\mathrm u}^{\otimes(n+m)})\le\{n(n-1)+m(m-1)\}/M+(n+m)(n+m-1)/(2M)\le2(n+m)^2/M$, also with an independent
randomizer. Finally,
$\Pp_{P_{\mathrm u},P_{\mathrm u}}(\widehat U\ge1)\ge1-\delta-2(n+m)^2/M$ for every even $M\ge2(n+m)$, and letting
$M\to\infty$ gives the claim.
\end{proof}

Lower bounds escape Theorem~\ref{thm:impossible}, but their rates do not. If $\tilde\alpha$ were a lower confidence
bound valid at level $\delta$ for every pair in $\mathcal P_2$ and also had
$\Pp_{P,Q}(\TV(P,Q)-\tilde\alpha>r)\le\delta'$ for every pair in $\mathcal P_2$, with $r<1$ and $\delta+\delta'<1$,
then $\tilde\alpha+r$ would be an upper bound valid at level $\delta'$, so by Theorem~\ref{thm:impossible} it would
reach $1$ at the uniform pair with probability at least $1-\delta'$, and there $\tilde\alpha\ge1-r>0=\TV$ with
probability larger than $\delta$, against its validity.

For a kernel the obstruction holds even without sampling. For any kernel satisfying (A1) and (A2), let $Q_\nu$ have
density $1+\varsigma(\nu\,\cdot)$ on $[0,1]$, for a measurable $1$-periodic mean-zero $\varsigma$ with $|\varsigma|=1$ and
an integer $\nu\ge1$. Then $\TV(P_{\mathrm u},Q_\nu)=\tfrac12$ for every $\nu$, while $\mu_{Q_\nu}-\mu_{P_{\mathrm u}}=\int_0^1\phi(x)\,
\varsigma(\nu x)\,dx\to0$ in $\Hh$ by the Riemann--Lebesgue lemma for Bochner integrals, since the bounded, strongly
measurable $\phi$ lies in $L^1([0,1];\Hh)$, where step functions are dense, and $\int_a^b\varsigma(\nu x)\,dx\to0$. As
$\mu_P\in\mathcal C_\alpha(P)$ and $\mu_Q\in\mathcal C_\alpha(Q)$, $D_\alpha(P,Q)\le\MMD(P,Q)$ for every $\alpha$, so the
trimmed path is uniformly small while the total variation stays at $\tfrac12$.
\space
\section{Proofs for Section~\ref{sec:analytic}}
\label{app:analytic}

The proofs of this section use the constants $\beta$, $\kappa$, $\Delta_\phi$, $\bar\kappa^2$, $n_\star$, $c$ and $L$ of
Section~\ref{sec:analytic} and the separation $\mathrm{sep}$ of \eqref{eq:sep}. We first prove a two-sided bound on the
statistic, the tool behind the proofs of Sections~\ref{app:split} to~\ref{app:trimming} and behind the threshold
$\tau^{\mathrm S}$, which is valid but larger than $\tau^{\mathrm P}$ at our sample sizes, and then the results of
Section~\ref{sec:analytic}. We write $D_\alpha=D_\alpha(P,Q)$ and add
the subscript $n,m$ and the argument $\delta$ to the thresholds of Section~\ref{sec:analytic} where they vary, as in
$\tau_{n,m}(\alpha,\delta)$ and $\tau^{\mathrm S}_{n,m}(\alpha,\delta)$.

The null hypotheses of Section~\ref{sec:analytic} are total-variation neighborhoods of $P=Q$. Equivalence tests \citep{wellek2010equivalence}
put a neighborhood of fixed size in the alternative instead, with a trimmed Mallows distance in
\citet{munkczado1998} and the MMD in \citet{liugandy2026equivalence}, so that a rejection shows that two laws are
close. In \eqref{eq:null} a rejection shows that they are far apart, and $\alpha$ is varied to locate $\alpha^\star$.

\subsection{A two-sided bound on the statistic}

\begin{theorem}[Concentration of the trimmed discrepancy]
\label{thm:twosided}
Under (A1) and (A2) of Assumption~\ref{ass:regularity}, for every pair $(P,Q)$, all $n,m\ge1$ and $\delta\in(0,1)$
there are events $E^+$ and $E^-$, not depending on $\alpha$, each of probability at least $1-\delta$, such that for every
$\alpha\in[0,1)$,
\[
\widehat D_\alpha\le D_\alpha+\tau^{\mathrm S}_{n,m}(\alpha,\delta)\ \text{ on }E^+,
\qquad
\widehat D_\alpha\ge D_\alpha-\tau^{\mathrm S}_{n,m}(\alpha,\delta)\ \text{ on }E^-,
\]
where
\begin{equation}
\tau^{\mathrm S}_{n,m}(\alpha,\delta)=\beta\Big[\,2\kappa/\sqrt{n_\star}+2\kappa(e_n+e_m)+\Delta_\phi\sqrt{\log(1/\delta)/(2n_\star)}\,\Big]
\label{eq:tauA}
\end{equation}
and $e_k=\E|\varepsilon_1+\dots+\varepsilon_k|/k\le k^{-1/2}$ for independent Rademacher signs $\varepsilon_i$.
\end{theorem}

The statistic is the largest separation over directions, a difference of two extreme trimmed means, each the best of
a family of ordinary means of hinge functions. So one supremum of centered means over the two samples controls it at
every level, a change of one observation moves that supremum by at most $\Delta_\phi/n$, and the level enters only
through $\beta$, the largest weight a trimming can put on one observation. The suprema $\Gamma$ and $\Gamma^-$ of the proof do not involve $\alpha$; computed on the fitting batches,
they control the fitted ramps in the proofs of Proposition~\ref{prop:loconsistency} and Theorem~\ref{thm:lorate}.

\begin{proof}
We first bound the difference of the two paths, at every level at once, by $\beta$ times a supremum that does not
involve $\alpha$, and then bound that supremum by Lemma~\ref{lem:hinge}. Fix $\alpha$ and write
$\mathcal A=\mathcal C_\alpha(P)$, $\mathcal B=\mathcal C_\alpha(Q)$, $\mathcal A_n=\mathcal C_\alpha(P_n)$ and
$\mathcal B_m=\mathcal C_\alpha(Q_m)$. Recall from \eqref{eq:sep} that the separation of two sets in the direction $f$ is
$\inf_a\langle f,a\rangle_\Hh-\sup_b\langle f,b\rangle_\Hh$, so that in terms of support functions
$\mathrm{sep}(f)=\mathrm{sep}(f;\mathcal A,\mathcal B)=-h_{\mathcal A}(-f)-h_{\mathcal B}(f)$ and
$\mathrm{sep}_n(f)=\mathrm{sep}(f;\mathcal A_n,\mathcal B_m)=-h_{\mathcal A_n}(-f)-h_{\mathcal B_m}(f)$.
By \eqref{eq:distmax}, $\widehat D_\alpha=\max_{f\in B_\Hh}\mathrm{sep}_n(f)$, and $\mathrm{sep}(f)\le D_\alpha$ for every
$f\in B_\Hh$ by \eqref{eq:sep}. It is therefore enough to bound $\mathrm{sep}_n(f)-\mathrm{sep}(f)$ uniformly in $f$.

Fix $f\in B_\Hh$. Each support function is an infimum over cuts (Lemma~\ref{lem:cvar}), and evaluating the population
infimum at the cut that is optimal for the sample compares the two. By Lemmas~\ref{lem:cvar} and~\ref{lem:tcompact},
applied to $P_n$ and the direction $-f$, there is $t_n\in[-\kappa,\kappa]$ with
\[
h_{\mathcal A_n}(-f)=t_n+\beta P_n(-f-t_n)_+,
\qquad\text{while}\qquad
h_{\mathcal A}(-f)\le t_n+\beta P(-f-t_n)_+ .
\]
Hence $h_{\mathcal A}(-f)-h_{\mathcal A_n}(-f)\le\beta(P-P_n)(-f-t_n)_+$, and in the same way
$h_{\mathcal B}(f)-h_{\mathcal B_m}(f)\le\beta(Q-Q_m)(f-t'_m)_+$ for the cut $t'_m\in[-\kappa,\kappa]$ optimal for $Q_m$
and $f$. Adding the two bounds and using $\mathrm{sep}(f)\le D_\alpha$,
\[
\mathrm{sep}_n(f)-D_\alpha\le\beta\Gamma,
\qquad
\Gamma=\sup_{f\in B_\Hh,\ t,t'\in[-\kappa,\kappa]}\Big[(P-P_n)(-f-t)_++(Q-Q_m)(f-t')_+\Big],
\]
for every $f\in B_\Hh$. Taking the maximum over $f$ gives $\widehat D_\alpha-D_\alpha\le\beta\Gamma$ for every $\alpha$,
and $\Gamma$ does not involve $\alpha$. The reverse inequality $D_\alpha-\widehat D_\alpha\le\beta\Gamma^-$, with
$\Gamma^-=\sup[(P_n-P)(-f-t)_++(Q_m-Q)(f-t')_+]$ over the same set, follows in the same way, with the roles of the
population and the sample exchanged: one evaluates at a maximizer of $\mathrm{sep}$, which exists by \eqref{eq:distmax}
and Lemma~\ref{lem:attainment}, and at the population minimizers of Lemma~\ref{lem:tcompact}.

We now bound $\Gamma$ and $\Gamma^-$. The supremum $\Gamma$ is that of Lemma~\ref{lem:hinge} with
$\varphi^1_x=\varphi^2_y=(\cdot)_+$, $v=f$, $a=-t$, $b=t'$, $r=1$ and $\rho_1=\rho_2=\kappa$, and $\Gamma^-$ is its
counterpart. Replacing $X_i$ by $X_i'$ changes $(-f(X_i)-t)_+$ by at most
$|f(X_i)-f(X_i')|\le\|\phi(X_i)-\phi(X_i')\|_\Hh\le\Delta_\phi$, and likewise for the $Y$'s, so
$\Delta_1=\Delta_2=\Delta_\phi$. The lemma gives $\E\Gamma,\E\Gamma^-\le2\kappa/\sqrt{n_\star}+2\kappa(e_n+e_m)$ and, at $e^{-\chi}=\delta$, as $\Delta_\phi^2(1/n+1/m)/2=\Delta_\phi^2/(2n_\star)$, it gives
$\Pp\big(\Gamma>\E\Gamma+\Delta_\phi\sqrt{\log(1/\delta)/(2n_\star)}\big)\le\delta$, and the same for $\Gamma^-$. The
bracket in \eqref{eq:tauA} is the sum of these two bounds, so the events $E^+=\{\beta\Gamma\le\tau^{\mathrm S}_{n,m}(\alpha,\delta)\}$
and $E^-=\{\beta\Gamma^-\le\tau^{\mathrm S}_{n,m}(\alpha,\delta)\}$ have probability at least $1-\delta$. They do not
depend on $\alpha$, because both sides of each inequality carry the factor $\beta$.
\end{proof}

\subsection{A feasible point near the common embedding}

The reduction behind Theorem~\ref{thm:plugin}, explained in Section~\ref{sec:analytic}, is the following lemma.

\begin{lemma}[A feasible point near the common embedding]
\label{lem:plugin}
Let $\alpha\in[0,1)$, and let $R$ be a probability measure with $R\ll P$ and $R\ll Q$ whose densities have versions
$g_P=dR/dP$ and $g_Q=dR/dQ$ with values in $[0,(1-\alpha)^{-1}]$. Put $\mu_R=\E_R[\phi(X)]$,
$\rho_R=\sup_x\|\phi(x)-\mu_R\|_\Hh$,
\begin{gather*}
U_X=\frac1n\sum_{i=1}^ng_P(X_i)\big(\phi(X_i)-\mu_R\big),\qquad
U_Y=\frac1m\sum_{j=1}^mg_Q(Y_j)\big(\phi(Y_j)-\mu_R\big),\\
S_X=\frac1n\sum_{i=1}^ng_P(X_i),\qquad
S_Y=\frac1m\sum_{j=1}^mg_Q(Y_j).
\end{gather*}
Then, for every realization of the samples,
$\widehat D_\alpha\le\max\{\|U_X-U_Y\|_\Hh,\|U_X\|_\Hh,\|U_Y\|_\Hh\}+\rho_R\big[(1-S_X)_++(1-S_Y)_+\big]$.
\end{lemma}

\begin{proof}
Any feasible pair of weights bounds the statistic from above. The weights $g_P(X_i)/n$ and $g_Q(Y_j)/m$ respect the
caps but their totals need not be one, so we repair the totals and pay in proportion to the shortfall. By
Proposition~\ref{prop:georep} the capped simplex $\{0\le w_i\le\beta/n,\ \sum_iw_i=1\}$ parametrizes
$\mathcal C_\alpha(P_n)$. Let $w^0_i=g_P(X_i)/n\in[0,\beta/n]$, so that $\sum_iw^0_i\phi(X_i)=U_X+S_X\mu_R$.

If $S_X\ge1$, put $w=w^0/S_X$, which respects the caps, and then $\sum_iw_i\phi(X_i)-\mu_R=U_X/S_X$. If $S_X<1$, the
room under the caps is $\sum_i(\beta/n-w^0_i)=\beta-S_X\ge1-S_X$, so there is $u$ with $0\le u_i\le\beta/n-w^0_i$ and
$\sum_iu_i=1-S_X$; put $w=w^0+u$, and then $\sum_iw_i\phi(X_i)-\mu_R=U_X+\sum_iu_i(\phi(X_i)-\mu_R)$. In both cases
$\sum_iw_i\phi(X_i)-\mu_R=\lambda_XU_X+r_X$ with $\lambda_X\in(0,1]$ and $\|r_X\|_\Hh\le\rho_R(1-S_X)_+$, the bound on
$r_X$ because $u$ has total mass $1-S_X$ and each $\phi(X_i)-\mu_R$ has norm at most $\rho_R$.

Doing the same on the $Y$ side gives a feasible pair $(w,v)$, so
\[
\widehat D_\alpha\le\Big\|\sum_iw_i\phi(X_i)-\sum_jv_j\phi(Y_j)\Big\|_\Hh\le\|\lambda_XU_X-\lambda_YU_Y\|_\Hh+\|r_X\|_\Hh
+\|r_Y\|_\Hh .
\]
The unknown factors $\lambda_X$ and $\lambda_Y$ are removed as follows. The map
$(\lambda,\lambda')\mapsto\|\lambda U_X-\lambda'U_Y\|_\Hh$ is convex on $[0,1]^2$, so it is at most its largest value at a
vertex, $\max\{0,\|U_X\|_\Hh,\|U_Y\|_\Hh,\|U_X-U_Y\|_\Hh\}$.
\end{proof}

\subsection{Proof of Theorem~\ref{thm:plugin}}

The proof bounds the three norms and the two shortfalls of Lemma~\ref{lem:plugin} separately, and uses four facts
about the summands. Write $\xi=g_P(X)(\phi(X)-\mu_R)$ and $W=1-g_P(X)$, so that $U_X$ and $1-S_X$ are averages of $n$
independent copies of $\xi$ and $W$, and recall that $\bar\kappa^2=\min(\kappa^2,\Delta_\phi^2/2)$. Then
\begin{gather*}
\E\xi=0,\qquad\E\|\xi\|_\Hh^2\le\beta\bar\kappa^2, \tag{i}\\
\|\xi(x)\|_\Hh\le\beta\rho_R,\qquad\|\xi(x)-\xi(x')\|_\Hh\le\beta\Delta_\phi, \tag{ii}\\
\E W=0,\qquad 1-\beta\le W\le1,\qquad\operatorname{Var}W\le\beta-1, \tag{iii}\\
\rho_R\le\Delta_\phi . \tag{iv}
\end{gather*}
For (i), $\E\|\xi\|_\Hh^2=\E_P[g_P^2\|\phi-\mu_R\|_\Hh^2]\le\beta\,\E_R\|\phi-\mu_R\|_\Hh^2$ since $g_P\le\beta$, and the last
variance is at most $\kappa^2$, as $\E_R\|\phi-\mu_R\|_\Hh^2=\E_Rk(X,X)-\|\mu_R\|_\Hh^2$, and at most $\Delta_\phi^2/2$, as
it equals $\tfrac12\E_{R\otimes R}\|\phi(X)-\phi(X')\|_\Hh^2$. For (ii), the map
$(p,q)\mapsto\|p(\phi(x)-\mu_R)-q(\phi(x')-\mu_R)\|_\Hh$ is convex on $[0,\beta]^2$, and its values at the vertices are at
most these bounds, using $\rho_R\le\Delta_\phi$. In (iii),
$\operatorname{Var}W=\E_Pg_P^2-1\le\beta-1$ as $g_P^2\le\beta g_P$. For (iv), $\mu_R$ lies in the closed convex hull of
$\phi(\mathcal X)$. The same holds for $g_Q$, $Q$ and the $Y$ side.

\begin{proof}
Under the null, \eqref{eq:tvintersect} gives a common trimming $R$, so Lemma~\ref{lem:plugin} applies, and it bounds
the statistic by three norms and two shortfalls. We bound six tails in closed form, each by $\delta/6$, which explains
$L=\log(6/\delta)$ in the threshold.

The level $\alpha=0$ is simpler and is treated first. There $0\le g_P\le1$ and $\E_Pg_P=1$ force $g_P=1$
$P$-almost surely, so $S_X=S_Y=1$ and $\widehat D_0\le\|U_X-U_Y\|_\Hh$ almost surely. As $U_X-U_Y$ is a sum of $n+m$
independent centered vectors, (i) gives $\E\|U_X-U_Y\|_\Hh\le(\E\|U_X-U_Y\|_\Hh^2)^{1/2}\le\bar\kappa/\sqrt{n_\star}$.
Changing one $X_i$ changes $\|U_X-U_Y\|_\Hh$ by at most $\Delta_\phi/n$ by (ii), and one $Y_j$ by at most
$\Delta_\phi/m$, so McDiarmid's inequality gives
$\Pp(\|U_X-U_Y\|_\Hh>\bar\kappa/\sqrt{n_\star}+\Delta_\phi\sqrt{\log(1/\delta)/(2n_\star)})\le\delta$, which is the bound
at $\alpha=0$.

Now let $\alpha>0$, and split the threshold as $\tau^{\mathrm P}(\alpha)=T_1+\Delta_\phi T_2$, with
\[
T_1=\sqrt{\beta\bar\kappa^2/n_\star}+\beta\Delta_\phi\sqrt{L/(2n_\star)},
\qquad
T_2=\sqrt{2(\beta-1)L/n_\star}+cL/3 .
\]
By Lemma~\ref{lem:plugin} and (iv), the event $\{\widehat D_\alpha>T_1+\Delta_\phi T_2\}$ is contained in
$\{\max(\|U_X-U_Y\|_\Hh,\|U_X\|_\Hh,\|U_Y\|_\Hh)>T_1\}\cup\{(1-S_X)_++(1-S_Y)_+>T_2\}$. The term $T_1$ handles the norms. For each of $\|U_X-U_Y\|_\Hh$, $\|U_X\|_\Hh$ and $\|U_Y\|_\Hh$ the mean is at most
$\sqrt{\beta\bar\kappa^2/n_\star}$ by (i), the sum of squared bounded differences is at most $\beta^2\Delta_\phi^2/n_\star$ by
(ii), and McDiarmid's inequality bounds the probability of exceeding $T_1$ by $e^{-L}=\delta/6$. The term $T_2$ handles
the shortfalls. The event $\{(1-S_X)_++(1-S_Y)_+>T_2\}$ is contained in
$\{1-S_X>T_2\}\cup\{1-S_Y>T_2\}\cup\{(1-S_X)+(1-S_Y)>T_2\}$, and each of these is the upper tail of a sum of
independent centered terms bounded above by $c$, with variance sum at most $V=(\beta-1)/n_\star$ by (iii). Bernstein's
inequality in sub-gamma form, $\Pp(S\ge\sqrt{2VL}+cL/3)\le e^{-L}$, which needs only the upper bound $c$ on the summands
\citep[Section~2.4 and Theorem~2.10]{boucheron2013concentration}, bounds each by $e^{-L}=\delta/6$. The six tails add up
to at most $\delta$.

The threshold is monotone: on $(0,1)$, $T_1$ and $T_2$ increase with $\beta$, hence with
$\alpha$, and as $\alpha\downarrow0$ the threshold $\tau^{\mathrm P}(\alpha)=T_1+\Delta_\phi T_2$ decreases to
$\bar\kappa/\sqrt{n_\star}+\Delta_\phi\sqrt{L/(2n_\star)}+\Delta_\phi cL/3$, which is at least
$\tau^{\mathrm P}_{n,m}(0,\delta)$ because $L>\log(1/\delta)$.
\end{proof}

\subsection{Proof of Lemma~\ref{lem:oneevent}}

\begin{proof}
Above $\alpha^\star$ the empirical path can only fall and the threshold can only rise, so the single level
$\alpha^\star$ carries the guarantee. The probability bound is the hypothesis at $\alpha=\alpha^\star$, which is a true
null. The regions are nested in
$\alpha$, so $\alpha\mapsto\widehat D_\alpha$ is nonincreasing, and for $\alpha\ge\alpha^\star$ on the event,
\[
\widehat D_\alpha\ \le\ \widehat D_{\alpha^\star}\ \le\ \tau(\alpha^\star)\ \le\ \tau(\alpha),
\]
by monotonicity of the path, by the event, and since $\tau$ is nondecreasing.
\end{proof}
\space
\section{Proofs for Section~\ref{sec:split}}
\label{app:split}

As in Section~\ref{app:regions}, $B_\Hh$ is the unit ball of $\Hh$ and $\mathcal G_\alpha(P)$ the set of
trimming densities; the betting notation is that of Section~\ref{sec:betting}. We begin with two facts about separation in a fixed direction, which the other sections also use.

\subsection{Separation in a fixed direction}

The distance between two convex sets is the largest gap between them along a unit direction.

\begin{lemma}[Directional separation]
\label{lem:separation}
Let $A,B\subset\Hh$ be nonempty. For every $f\in B_\Hh$,
\begin{equation}
\dist(A,B)\ \ge\ \inf_{a\in A}\langle f,a\rangle_\Hh-\sup_{b\in B}\langle f,b\rangle_\Hh\ =:\ \mathrm{sep}(f;A,B).
\label{eq:sep}
\end{equation}
If $A,B$ are also convex and weakly compact and $\dist(A,B)>0$, the bound is attained: taking
$f^\star=(a^\star-b^\star)/\|a^\star-b^\star\|_\Hh$ for nearest points $a^\star\in A$, $b^\star\in B$ gives
$\mathrm{sep}(f^\star;A,B)=\dist(A,B)$.
\end{lemma}

\begin{proof}
For $a\in A$ and $b\in B$, the Cauchy--Schwarz inequality gives
$\mathrm{sep}(f;A,B)\le\langle f,a-b\rangle_\Hh\le\|a-b\|_\Hh$, and the infimum over $a,b$ yields \eqref{eq:sep}. For
the equality, nearest points exist, as the norm is weakly lower semicontinuous on the weakly compact $A\times B$, and
we check that $f^\star$ separates the sets exactly at them. First-order optimality of $a^\star$ for
$\min_{a\in A}\|a-b^\star\|_\Hh^2$ gives $\langle a-a^\star,a^\star-b^\star\rangle_\Hh\ge0$ for all $a\in A$, that is,
$\inf_{a\in A}\langle f^\star,a\rangle_\Hh=\langle f^\star,a^\star\rangle_\Hh$; symmetrically
$\sup_{b\in B}\langle f^\star,b\rangle_\Hh=\langle f^\star,b^\star\rangle_\Hh$. Subtracting,
$\mathrm{sep}(f^\star;A,B)=\langle f^\star,a^\star-b^\star\rangle_\Hh=\|a^\star-b^\star\|_\Hh=\dist(A,B)$.
\end{proof}

Since $\mathrm{sep}(0;A,B)=0$, for convex weakly compact $A,B$ the lemma gives
\begin{equation}
\dist(A,B)=\max_{f\in B_\Hh}\mathrm{sep}(f;A,B),
\label{eq:distmax}
\end{equation}
the maximum being attained at $f^\star$ when $\dist(A,B)>0$ and at $f=0$ otherwise. By Lemma~\ref{lem:attainment} and
Proposition~\ref{prop:georep} this applies to the population and to the empirical regions, and by Lemma~\ref{lem:cvar},
$\mathrm{sep}(f;\mathcal C_\alpha(P),\mathcal C_\alpha(Q))=-h_{\mathcal C_\alpha(P)}(-f)-h_{\mathcal C_\alpha(Q)}(f)$. For
empirical regions both terms are trimmed means of the values of $f$, which gives the structure of the empirical
trimming (Proposition~\ref{prop:bangbang}, proved in Section~\ref{app:trimming}).

\begin{lemma}[Support functions are extreme trimmed means]
\label{lem:trimmedmean}
Let $v_1,\dots,v_n$ be reals with order statistics $v_{(1)}\le\dots\le v_{(n)}$, let $\alpha\in[0,1)$ and
$c_\alpha=[n(1-\alpha)]^{-1}=\beta/n$. Then
\begin{gather*}
\max_{\substack{0\le w_i\le c_\alpha\\ \sum_iw_i=1}}\ \sum_iw_iv_i
\ =\ \sum_{i>n-n_{1-\alpha}}c_\alpha v_{(i)}+\big(1-n_{1-\alpha} c_\alpha\big)\,v_{(n-n_{1-\alpha})},\\
n_{1-\alpha}=\lfloor1/c_\alpha\rfloor=\lfloor n(1-\alpha)\rfloor,
\end{gather*}
where the final term is $0$ when $n_{1-\alpha}=n$ (at $\alpha=0$); the minimum is the same expression applied to
the smallest values. Moreover, every maximizer satisfies $w_i=c_\alpha$ when $v_i>v_{(n-n_{1-\alpha})}$ and $w_i=0$ when
$v_i<v_{(n-n_{1-\alpha})}$, the values equal to $v_{(n-n_{1-\alpha})}$ sharing the remaining mass $1-c_\alpha\#\{i:v_i>v_{(n-n_{1-\alpha})}\}$ in any way. So the maximizer is unique for distinct values.
\end{lemma}

\begin{proof}
Moving weight from a smaller value to a larger one below its cap does not decrease the objective, so an optimum loads
the cap $c_\alpha$ onto the largest values until the mass runs out, which gives the value. For the structure of the
maximizers, let $n_{1-\alpha}<n$ (otherwise the feasible set is a point) and suppose that a maximizer $w$ has
$w_i<c_\alpha$ for some $v_i>v_{(n-n_{1-\alpha})}$. At most $n_{1-\alpha}$ values are at least $v_i$, and they carry less than
$n_{1-\alpha} c_\alpha\le1$, so some $j$ with $v_j<v_i$ has $w_j>0$; moving weight from $j$ to $i$ increases the objective,
a contradiction. The case $v_i<v_{(n-n_{1-\alpha})}$ with $w_i>0$ is symmetric. The objective depends on the mass at the
value $v_{(n-n_{1-\alpha})}$ only through its total, and the minimum follows by applying all this to $-v_1,\dots,-v_n$.
\end{proof}

\subsection{Proof of Lemma~\ref{lem:ramp}}

\begin{proof}
Part (i) is the variational form of trimmed means of Lemma~\ref{lem:cvar}, part (ii) an exact computation, and
part (iii) the layer-cake formula. For the inequality in (i), let $P'\in R_\alpha(P)$ and $g=dP'/dP\in[0,\beta]$, so
that $\E_Pg=1$. Then $\E_{P'}f=t+\E_P[g(f-t)]\ge t-\E_P[g(t-f)_+]\ge t-\beta\E_P(t-f)_+$, and symmetrically $\E_{Q'}f\le u+\beta\E_Q(f-u)_+$ for $Q'\in R_\alpha(Q)$; subtracting and optimizing over $P'$ and $Q'$
gives the inequality. For the equality, Lemma~\ref{lem:cvar} applied to the bounded function $-f$, with its variable
written $-t$, gives $\inf_{P'\in R_\alpha(P)}\E_{P'}f=-\sup_{g\in\mathcal G_\alpha(P)}\E_P[-gf]=\sup_{t\in\R}\{t-\beta\E_P(t-f)_+\}$,
attained at $t=-t'$ for an $\alpha$-quantile $t'$ of $-f(X)$, that is, at a $(1-\alpha)$-quantile of $f(X)$. Applied to
$f$ under $Q$, the lemma gives $\sup_{Q'\in R_\alpha(Q)}\E_{Q'}f=\inf_{u\in\R}\{u+\beta\E_Q(f-u)_+\}$, attained at an
$\alpha$-quantile of $f(Y)$. For $f\in B_\Hh$, the reproducing property and Proposition~\ref{prop:georep} turn the two
extreme trimmed means into $\inf_{a\in\mathcal C_\alpha(P)}\langle f,a\rangle_\Hh$ and
$\sup_{b\in\mathcal C_\alpha(Q)}\langle f,b\rangle_\Hh$, whose difference is the separation \eqref{eq:sep}.

For (ii), clipping makes both hinges linear in the ramp: since $\tilde f\le t$ and $\tilde f\ge u$,
\[
t-\beta(t-\tilde f)_+=t-\beta(t-u)(1-\rp),
\qquad
u+\beta(\tilde f-u)_+=u+\beta(t-u)\rp ,
\]
and taking expectations and using $1-\beta=-\alpha\beta$ gives the identity. For the inequality, pointwise
$(t-\tilde f)_+=(t-f)_+-(u-f)_+$ and $(\tilde f-u)_+=(f-u)_+-(f-t)_+$, so
$\theta(\tilde f;t,u)-\theta(f;t,u)=\beta\E_P(u-f)_++\beta\E_Q(f-t)_+\ge0$.

For (iii), if $\rp$ takes values in $[0,1]$, the layer-cake formula and the definition of total variation give
$\E_P\rp-\E_Q\rp=\int_0^1\{P(\rp>r)-Q(\rp>r)\}\,dr\le\TV(P,Q)$.
\end{proof}

\subsection{The betting test in full}
\label{sec:bettingdetails}

We describe the test of Section~\ref{sec:betting} for any batch fractions $0=\pi_0<\pi_1<\dots<\pi_{B+1}=1$ fixed in advance;
Section~\ref{sec:betting} uses $\pi=(0,\tfrac14,\tfrac12,1)$, and the single split $(0,\tfrac12,1)$ is the case $B=1$. Each
sample is put in a random order drawn independently of the data and cut into batches $0,\dots,B$ at these fractions of
its size, rounded to the nearest integers. Batch $0$ is used
only for fitting, and batch $b\ge1$ is bet on with a function $\rp_b$ with values in $[0,1]$ fitted on batches
$0,\dots,b-1$, or not at all.

Before betting on batch $b$ we solve \eqref{eq:qp} on batches $0,\dots,b-1$ and form the direction
$\widehat f_b$ of the witness \eqref{eq:witness}. Its cut points are $\widehat t_b$, a genuine empirical
$(1-\alpha)$-quantile of its values at the $X$'s of those batches, and $\widehat u_b$, a genuine empirical
$\alpha$-quantile of its values at their $Y$'s, each value computed with the point's own term removed from
\eqref{eq:witness}. If $\widehat
t_b>\widehat u_b$, $\rp_b$ is the ramp of $\widehat f_b$ between $\widehat u_b$ and $\widehat t_b$ of
Lemma~\ref{lem:ramp}(ii); otherwise there is no bet on batch $b$.

Pairs are formed within each batch. Let batch $b$ hold $n_{[b]}$ observations $X$ and $m_{[b]}$ observations $Y$. If $n_{[b]}\le m_{[b]}$,
its $Y$'s are cut, in their fixed order, into $n_{[b]}$ consecutive blocks whose sizes differ by at most one, and the
$X$'s are paired one to one with the blocks, an $X$ and its block $\mathcal J$ giving
$Z=\rp_b(X)-|\mathcal J|^{-1}\sum_{j\in\mathcal J}\rp_b(Y_j)$. If $n_{[b]}>m_{[b]}$, the roles are exchanged: each $Y$ is
paired with a block $\mathcal J$ of $X$'s, giving $Z=|\mathcal J|^{-1}\sum_{i\in\mathcal J}\rp_b(X_i)-\rp_b(Y)$. Batch $b$
thus gives $N_b=\min(n_{[b]},m_{[b]})$ pairs, and the pairs of all batches are numbered $k=1,\dots,N$ in the order of the
batches, with $N=\sum_bN_b$ fixed by the sample sizes. Each $Z_k$ takes values in $[-1,1]$.

The wealth after $k$ pairs is $K_k=\prod_{l\le k}\{1+\lambda_l(Z_l-\alpha)\}$, with
$K_0=1$, and both the function bet on and the fractions must be predictable. A fitting rule is predictable if each $\rp_b$ is a jointly measurable function of batches $0,\dots,b-1$ and of the
point, with values in $[0,1]$; the ramp above is one, and so are the predicted labels of a classifier trained on batches
$0,\dots,b-1$. Fractions are predictable if, for pair $k$ in batch $b$, $\lambda_k\in[0,\lambda_{\max}/(1+\alpha)]$ for a
fixed $\lambda_{\max}\in(0,1]$, $\lambda_k$ is a jointly measurable function of batches $0,\dots,b-1$ and of the earlier
$Z$'s, and $\lambda_k=0$ on a batch without a ramp. Theorem~\ref{thm:betting} holds, with the same statement, for every
choice of batch fractions $\pi$; we prove it in this generality.

\begin{proof}[Proof of Theorem~\ref{thm:betting}]
Under the null every bet has a nonpositive expected gain, so the wealth is a nonnegative supermartingale, and such a
process rarely reaches $1/\delta$. To set up the filtration, for pair $k$ in batch $b$ let $\mathcal F_{k-1}$ be generated
by batches $0,\dots,b-1$ and the earlier pairs of batch $b$. These $\sigma$-fields increase with $k$, $\rp_b$ and
$\lambda_k$ are $\mathcal F_{k-1}$-measurable, and the observations of pair $k$ are independent of $\mathcal F_{k-1}$.
So, given $\mathcal F_{k-1}$, $Z_k$ takes values in $[-1,1]$ and has mean
$\Delta_k=\E_P\rp_b-\E_Q\rp_b\le\TV(P,Q)\le\alpha$, by Lemma~\ref{lem:ramp}(iii) and the null. Consequently
$\E[K_k\mid\mathcal F_{k-1}]=K_{k-1}\{1+\lambda_k(\Delta_k-\alpha)\}\le K_{k-1}$, while
$\lambda_k(Z_k-\alpha)\ge-\lambda_k(1+\alpha)\ge-\lambda_{\max}$ gives $K_k\ge(1-\lambda_{\max})K_{k-1}\ge0$. So $(K_k)$ is
a nonnegative supermartingale with $K_0=1$.

Optional stopping at $T=\min\{k\le N:K_k\ge1/\delta\}\wedge N$, a bounded time as $N$ is fixed in advance, gives
$\E[K_T\mid\text{batch }0]\le1$. As $\{\max_{k\le N}K_k\ge1/\delta\}\subseteq\{K_T\ge1/\delta\}$, Markov's inequality
gives the bound, which is Ville's maximal inequality \citep{ville1939}. Integrating over batch $0$ ends the proof.
\end{proof}

The fractions we use are
\begin{equation}
\lambda_k=\min\Big\{\frac{\lambda_{\max}}{1+\alpha},\ \sqrt{\frac{2\log(1/\delta)}{N\widehat v^2_{k-1}}}\Big\},
\qquad \lambda_{\max}=\tfrac12,
\label{eq:fractions}
\end{equation}
adapted from \citet[Remark~1]{waudbysmith2024estimating}, where
$\widehat v^2_{k-1}=(w_0v_0^2+\sum_{i<k}(Z_i-\alpha-\widehat\mu_{i-1})^2)/(w_0+k-1)$,
$\widehat\mu_{i-1}=\sum_{l<i}(Z_l-\alpha)/(w_0+i-1)$, $w_0=20$, $v_0^2\le4$ is an estimate of the variance of one
$Z_k$ from the batches before that of $Z_k$ (Algorithm~\ref{alg:split} gives ours), the sums and the counts $k-1$ and
$i-1$ run over the pairs actually bet on before $Z_k$, in its batch and in the earlier ones, so that they carry over
from batch $1$ to batch $2$ while $v_0^2$ is that of the batch of $Z_k$, and $\widehat v^2_{k-1}$ is floored at a constant $v^2_{\min}>0$. Our runs use
$v^2_{\min}=10^{-8}$, a guard against division by zero, for which the constants of Proposition~\ref{prop:loconsistency}
and Theorems~\ref{thm:lorate} and~\ref{thm:laplacerate}, of order $1/v^2_{\min}$, are not usable and those results
describe the trend (Section~\ref{sec:lower} of the main text). Any other predictable choice in
$[0,\lambda_{\max}/(1+\alpha)]$ is equally valid, and this one comes close to the best constant fraction for large $N$.

\begin{algorithm}[!tb]
\caption{The split-sample lower bound $\widehat\alpha$, with the defaults of our runs}
\label{alg:split}
\begin{algorithmic}[1]
\Require samples $X_{1:n}$, $Y_{1:m}$; level $\delta$; a grid $\alpha_1<\dots<\alpha_J$, fixed in advance or chosen from
batch $0$ (Remark~\ref{rem:batch0grid})
\State Put each sample in a random order drawn independently of the data and cut it into batch $0$ (first quarter),
batch $1$ (second quarter) and batch $2$ (second half)
\State Pair within each batch $b$: if $n_{[b]}\le m_{[b]}$, cut its $Y$'s in order into $n_{[b]}$ consecutive blocks of
sizes differing by at most one and pair the $i$-th $X$ with the $i$-th block, and the reverse otherwise; $N$ is the
total number of pairs
\For{$j=1,\dots,J$, with $\alpha=\alpha_j$}
\For{$b=1,2$, fitting on batches $0,\dots,b-1$, of sizes $n_{<b}$, $m_{<b}$}
\State Kernel $e^{-\|x-y\|/\sigma}$, with $\sigma$ the median pairwise distance of the fitting points (simulations) or
of batch $0$ (real data, both batches; RHC standardized by the moments of batch $0$), on at most $2{,}000$ points
\State Minimizer $(w,v)$ of \eqref{eq:qp} at $\alpha$ by accelerated projected gradient with restarts and an exact
projection onto the capped simplex, warm-started from the previous level, stopped when the objective changes by at most
a relative $10^{-9}$ over ten iterations or after $1000$ iterations ($10^{-10}$ and $2000$ on real data)
\State Witness values at the fitting points without their own term, $\tilde h_i=\widehat f(X_i)-w_ik(X_i,X_i)/\widehat D_\alpha$
and $\tilde h'_j=\widehat f(Y_j)+v_jk(Y_j,Y_j)/\widehat D_\alpha$
\State Cut points $\widehat t_b=\tilde h_{(\lceil n_{<b}(1-\alpha)\rceil)}$ and $\widehat u_b=\tilde h'_{(\lfloor m_{<b}\alpha\rfloor+1)}$,
order statistics
\If{$\widehat t_b\le\widehat u_b$} skip batch $b$: $\lambda_k=0$ on its pairs
\Else{} $\rp_b$ is the ramp of $\widehat f_b$ between $\widehat u_b$ and $\widehat t_b$, each pair gives
$Z=\rp_b(X)-\overline{\rp_b(Y)}$ (or the mirror), and $v_0^2=\max\{s_X^2+s_Y^2n_{[b]}/m_{[b]},10^{-4}\}$ (mirrored
if $n_{[b]}>m_{[b]}$), with $s_X^2,s_Y^2$ the variances of $\rp_b$ at the fitting $X$'s and $Y$'s
\EndIf
\EndFor
\State Bet: $K_0=1$, $K_k=K_{k-1}\{1+\lambda_k(Z_k-\alpha)\}$, with \eqref{eq:fractions}, $\lambda_{\max}=\tfrac12$,
$w_0=20$ and $v^2_{\min}=10^{-8}$
\If{$\max_{k\le N}K_k<1/\delta$} stop: $H_{0,\alpha_j}$ is not rejected \EndIf
\EndFor
\State \Return $\widehat\alpha$ of \eqref{eq:downclosed}; whether the reading stopped at a failed bet or with both batches
skipped; the trimming at $\widehat\alpha$, the minimizer of \eqref{eq:qp} on all the data closest to uniform weights
(Section~\ref{sec:computationdetails})
\end{algorithmic}
\end{algorithm}

Both bandwidth rules are predictable, each witness using a bandwidth computed from the batches before the one bet on,
but the consistency and rate results, which fix the kernel in advance, cover neither of them, nor random features
scaled by such a bandwidth. Section~\ref{app:experiments} gives the grids of our experiments; in practice we
recommend a grid of mesh $0.05$ fixed in advance, or a short grid around a guess at $\alpha^\star$ from batch $0$
(Remark~\ref{rem:batch0grid}).

The constant $\cstBetRatioEighty$ of Section~\ref{sec:betting} comes from the best constant fraction: $\log
K_N=\lambda S_1-\tfrac12\lambda^2S_2+O(\lambda^3N)$, with $S_1=\sum_k(Z_k-\alpha)$ and $S_2=\sum_k(Z_k-\alpha)^2$, is at
most about $S_1^2/(2S_2)$, so the test rejects roughly when $S_1/S_2^{1/2}\ge\sqrt{2\log(1/\delta)}$, an approximately
$N(h,1)$ ratio if $\E Z_k-\alpha=hsN^{-1/2}$, with $s^2=\operatorname{Var}Z_k$. Power $p$ then needs $h\approx\sqrt{2\log(1/\delta)}+z_p$, against
$z_{1-\delta}+z_p$ by the central limit theorem, and the squared ratio at $\delta=0.05$, $p=0.8$ is the ratio of sizes.

\subsection{Proof of Corollary~\ref{cor:lower}}

\begin{proof}
A bound that is too large requires the rejection of one particular true null, the first grid level at or
above $\TV(P,Q)$, so no union over the levels is needed. Let $E$ be the event that $H_{0,\widehat\alpha}$ is rejected and $\TV(P,Q)\le\widehat\alpha$; it
contains $\{\widehat\alpha>\TV(P,Q)\}$, since then $\widehat\alpha>0$ is a rejected grid level. On
$E$, $\widehat\alpha=\alpha_j$ with $\alpha_j\ge\TV(P,Q)$ and $H_{0,\alpha_i}$ rejected for every $i\le j$. The
smallest index $i_0$ with $\alpha_{i_0}\ge\TV(P,Q)$ is at most $j$ and does not depend on the data, because the grid is
fixed in advance, so $E$ lies in the event that the true null $H_{0,\alpha_{i_0}}$ is rejected, of probability at most
$\delta$ by Theorem~\ref{thm:betting}. No multiplicity correction is needed, and if no grid level is at least
$\TV(P,Q)$, $E$ is empty.
\end{proof}

\begin{remark}[A grid chosen from batch $0$]
\label{rem:batch0grid}
Corollary~\ref{cor:lower} holds, with the same proof, when the grid $\alpha_1<\dots<\alpha_J$ is a measurable function
of batch $0$ instead of being fixed in advance. Given batch $0$ the grid is fixed, so the smallest index $i_0$ with
$\alpha_{i_0}\ge\TV(P,Q)$ is a function of batch $0$, and the conditional bound of Theorem~\ref{thm:betting} gives
$\Pp(H_{0,\alpha_{i_0}}\text{ rejected}\mid\text{batch }0)\le\delta$; integrating over batch $0$ gives the claim. For
instance, a bisection on batch $0$ alone, with any statistic, may place a short grid around a guess at $\alpha^\star$,
which the betting batches then read upward.
\end{remark}

\subsection{Proof of Proposition~\ref{prop:loconsistency}}

We allow any batch fractions fixed in advance. For $b\ge1$ let $n_{<b},m_{<b}$ be the sizes of batches $0,\dots,b-1$,
$r_b=1/n_{<b}+1/m_{<b}\le r_1$, and $\Gamma_b,\Gamma_b^-$ the suprema $\Gamma,\Gamma^-$ of the proof of
Theorem~\ref{thm:twosided} on these batches, free of $\alpha$. At a level $\alpha$ the witness fitted on them takes
values $h_i,h'_j$ there, and $\tilde h_i=h_i-w_ik(X_i,X_i)/\widehat D$ and
$\tilde h'_j=h'_j+v_jk(Y_j,Y_j)/\widehat D$ with the own term removed. With $\iota^-(n,p)=\max(1,\lceil np\rceil-1)$ and
$\iota^+(n,p)=\min(n,\lfloor np\rfloor+2)$, the cut points satisfy
\begin{equation}
\tilde h_{(\iota^-(n_{<b},1-\alpha))}\le\widehat t_b\le\tilde h_{(\iota^+(n_{<b},1-\alpha))},\qquad
\tilde h'_{(\iota^-(m_{<b},\alpha))}\le\widehat u_b\le\tilde h'_{(\iota^+(m_{<b},\alpha))},
\label{eq:cutrank}
\end{equation}
as do the empirical quantiles in the range of the values and their linear interpolations.

\begin{lemma}[The population ramp]
\label{lem:popramp}
Under (A1) and (A2), if $D_\alpha>0$, the ramp $\rp^\star$ of the normalized difference $f^\star$ of nearest points of
$\mathcal C_\alpha(P)$ and $\mathcal C_\alpha(Q)$, between the largest $\alpha$-quantile $u^\star$ of $f^\star(Y)$ and the
smallest $(1-\alpha)$-quantile $t^\star$ of $f^\star(X)$, has $D_\alpha\le t^\star-u^\star\le \Delta_\phi$ and
$\E_P\rp^\star-\E_Q\rp^\star-\alpha\ge D_\alpha/(\beta \Delta_\phi)$.
\end{lemma}

\begin{proof}
The right derivatives $1-\beta P(f^\star\le t)$ and $1-\beta Q(f^\star>u)$ make $t^\star$ maximize
$t-\beta\E_P(t-f^\star)_+$ and $u^\star$ minimize $u+\beta\E_Q(f^\star-u)_+$. So by Lemmas~\ref{lem:ramp}(i)
and~\ref{lem:separation} the cut points attain the separation, and since $\theta(f;t,u)\le t-u$,
$\Delta_\phi\ge\sup f^\star-\inf f^\star\ge t^\star-u^\star\ge\theta(f^\star;t^\star,u^\star)=D_\alpha$, the first
inequality because $f^\star$ has unit norm. Lemma~\ref{lem:ramp}(ii) then gives
$\beta(t^\star-u^\star)(\E_P\rp^\star-\E_Q\rp^\star-\alpha)\ge D_\alpha$, and $t^\star-u^\star\le\Delta_\phi$ gives the
excess.
\end{proof}

\begin{proof}[Proof of Proposition~\ref{prop:loconsistency}]
Below $\alpha^\star$ the population ramp has a fixed positive excess over $\alpha$ (Lemma~\ref{lem:popramp}). We show
that the fitted ramp keeps part of this excess on an event that does not depend on the level, that betting turns a fixed excess
into a rejection, and then apply both to the grid.

We begin with the fitted ramp at a fixed level $\alpha$ and batch $b$, writing $n,m$ for $n_{<b},m_{<b}$, and assume
$\beta\Gamma_b^-<D_\alpha$. By the deterministic reduction in the proof of Theorem~\ref{thm:twosided},
$\widehat D\ge D_\alpha-\beta\Gamma_b^->0$. Let $\theta_n$ be the criterion $\theta$ of Section~\ref{sec:ramp} with
$P_n,Q_m$ in place of $P,Q$. In the fitted direction $\widehat f$ it separates into a concave function of the first cut
and a convex function of the second,
\[
\theta_n(\widehat f;t,u)=T(t)-U(u),\quad T(t)=t-\frac\beta n\sum_i(t-h_i)_+,\quad U(u)=u+\frac\beta m\sum_j(h'_j-u)_+,
\]
and $\max T-\min U=\widehat D$ by Lemmas~\ref{lem:separation} and~\ref{lem:ramp}(i) applied to $P_n,Q_m$. We first show
that the cut points lose little against the best cuts. The concave $T$ has right derivative $1-\beta F_n$, with $F_n(x)=\#\{i:h_i\le x\}/n$, so it
is $\beta$-Lipschitz and maximal at $h_{(\lceil n(1-\alpha)\rceil)}$. Between $h_{(\iota^-)}$ and $h_{(\iota^+)}$ we have
$\iota^-/n\le F_n\le(\iota^+-1)/n$, within $1/n$ of $1-\alpha$, so the slope of $T$ there is at most $\beta/n$; as the
values span at most $\Delta_\phi$, $T\ge\max T-\beta\Delta_\phi/n$ on that interval. The own terms shift the values only
a little: the caps $\beta/n$ and $k(x,x)\le\kappa^2$ give $0\le h_i-\tilde h_i\le\beta\kappa^2/(n\widehat D)$, the order
statistics move by at most as much, and concavity, with the $\beta$-Lipschitz bound, gives
\[
T(\widehat t_b)\ge\max T-\frac{\beta\Delta_\phi}n-\frac{\beta^2\kappa^2}{n\widehat D} .
\]
The convex $U$, minimal at $h'_{(\lfloor m\alpha\rfloor+1)}$, is treated alike, with $m$ in place of $n$. Hence
$\theta_n(\widehat f;\widehat t_b,\widehat u_b)\ge\widehat D-\beta\Delta_\phi r_b-\beta^2\kappa^2r_b/\widehat D$.

Next we pass from the empirical criterion to the population one. For all $(f,t,u)$,
$\theta-\theta_n=-\beta[(P-P_n)(t-f)_++(Q-Q_m)(f-u)_+]\ge-\beta\Gamma_b$, the centered terms being constant in $t$
and $u$ outside $[-\kappa,\kappa]$. As $x\mapsto x-c/x$ increases and $\widehat D\ge D_\alpha-\beta\Gamma_b^-$, we may
replace $\widehat D$ by $D_\alpha-\beta\Gamma_b^-$. Finally $\theta(f;t,u)\le t-u$, and $\widehat t_b\le\max_ih_i$,
$\widehat u_b\ge\min_jh'_j$, values of a unit vector, are at most $\Delta_\phi$ apart. Together,
\begin{equation}
\Delta_\phi\ge\widehat t_b-\widehat u_b\ge\theta(\widehat f;\widehat t_b,\widehat u_b)\ge\Theta_b=D_\alpha-\beta(\Gamma_b
+\Gamma_b^-)-\beta \Delta_\phi r_b-\frac{\beta^2\kappa^2r_b}{D_\alpha-\beta\Gamma_b^-},
\label{eq:fittedramp}
\end{equation}
and if $\Theta_b>0$, Lemma~\ref{lem:ramp}(ii) under $P,Q$, given batches $0,\dots,b-1$, yields a ramp $\rp_b$ with
$\E_P\rp_b-\E_Q\rp_b-\alpha\ge\Theta_b/(\beta \Delta_\phi)$. The terms of $\Theta_b$ after $D_\alpha$ are the costs of the
fluctuation, of the rank tolerance and of the own terms.

The second step concerns the bets: a fixed excess is turned into a rejection, and a strict null is rarely rejected.
Recall that $\Delta_k$ is the conditional mean of $Z_k$ in the proof of Theorem~\ref{thm:betting}, and put
$L_\delta=\log(1/\delta)$ and
\[
\varepsilon_N(\varrho)=\exp\Big\{-\frac{v^2_{\min}}{4L_\delta}\Big(\varrho\sqrt{NL_\delta/2}-L_\delta
-\frac{16L_\delta}{v^2_{\min}}\Big)_+^2\Big\}.
\]
We show that for $N\ge2L_\delta/\lambda_{\max}^2$,
\begin{equation}
\Pp\big(\mathcal E\cap\{\max_{k\le N}K_k<1/\delta\}\big)\le\varepsilon_N(\varrho),\qquad
\Pp\big(\max_{k\le N}K_k\ge1/\delta\big)\le e^{-\zeta v_{\min}(2NL_\delta)^{1/2}},
\label{eq:betbounds}
\end{equation}
the first for any event $\mathcal E$ on which every batch has a ramp and $\Delta_k\ge\alpha+\varrho>\alpha$ for all $k$,
the second if $\TV(P,Q)\le\alpha-\zeta<\alpha$.

Two bounds on the fractions are used. Each square in \eqref{eq:fractions} is at most $4$, as $Z_i-\alpha$ and
$\widehat\mu_{i-1}$ lie in $[-1-\alpha,1-\alpha]$, so for $v^2_{\min}\le4$ the running variance lies in
$[v^2_{\min},4]$ and
\[
\Big(\frac{L_\delta}{2N}\Big)^{1/2}\ \le\ \lambda_k\ \le\ \overline\lambda_N
=\min\Big\{\frac{\lambda_{\max}}{1+\alpha},\Big(\frac{2L_\delta}{Nv^2_{\min}}\Big)^{1/2}\Big\},
\]
the lower bound holding on batches with a ramp, where it uses $N\ge2L_\delta/\lambda_{\max}^2$. The centered gains
$M_k=\sum_{l\le k}\lambda_l(Z_l-\Delta_l)$ have increments in intervals of length $2\lambda_l$, so by Hoeffding's
lemma \citep{hoeffding1963}
\[
W_k(s)=\exp\Big(sM_k-\frac{s^2}2\sum_{l\le k}\lambda_l^2\Big)
\]
is a nonnegative supermartingale for every real $s$.

On $\mathcal E$ the wealth grows. The inequality $\log(1+y)\ge y-2y^2$ for $y\ge-\frac12$, the lower bound on
$\lambda_k$ and $\Delta_k-\alpha\ge\varrho$ give
\[
\log K_N\ \ge\ \varrho\Big(\frac{NL_\delta}2\Big)^{1/2}+M_N-\frac{4L_\delta(1+\alpha)^2}{v^2_{\min}},
\]
the last term from $2\sum_k\lambda_k^2(Z_k-\alpha)^2\le2N\overline\lambda_N^{\,2}(1+\alpha)^2$. As $(1+\alpha)^2\le4$,
failing to reject on $\mathcal E$ forces $M_N\le-x_N$, with
$x_N=(\varrho\sqrt{NL_\delta/2}-L_\delta-16L_\delta/v^2_{\min})_+$. Put $V=2L_\delta/v^2_{\min}\ge\sum_l\lambda_l^2$.
Markov's inequality for $W_N(-x_N/V)$, which is at least $e^{x_N^2/(2V)}$ when $M_N\le-x_N$, bounds the probability
by $e^{-x_N^2/(2V)}=\varepsilon_N(\varrho)$, without conditioning on $\mathcal E$.

Under the strict null the wealth shrinks. Now $\log(1+y)\le y$ and $\Delta_l\le\alpha-\zeta$ give
$\log K_k\le M_k-\zeta\sum_{l\le k}\lambda_l$. With $s=2\zeta/\overline\lambda_N$ and $\lambda_l\le\overline\lambda_N$,
the quadratic term of $W_k(s)$ is at most $\frac{s^2}2\overline\lambda_N\sum_{l\le k}\lambda_l=s\zeta\sum_{l\le k}
\lambda_l$, so $W_k(s)\ge K_k^{\,s}\ge\delta^{-s}$ wherever $K_k\ge1/\delta$. The maximal inequality in the proof of
Theorem~\ref{thm:betting} bounds the probability of this by $\delta^{2\zeta/\overline\lambda_N}$, which is at most the
second bound of \eqref{eq:betbounds} by the bound on $\overline\lambda_N$.

We now apply the two steps to the grid, starting with the upper side. If $\widehat\alpha>\alpha_{j^\star}$, then by
\eqref{eq:downclosed} the true null at the first grid level $\alpha_{i_0}\ge\alpha^\star$ was rejected, with probability
at most $\delta$ by Theorem~\ref{thm:betting}, and tending to zero if $\alpha^\star$ is off the grid, by
\eqref{eq:betbounds} with $\zeta=\alpha_{i_0}-\alpha^\star$.

On the lower side every grid level below $\alpha^\star$ must be rejected. Let $\alpha_{j^\star}$ be, as in the
statement, the largest grid level below $\alpha^\star$, $\underline D=D_{\alpha_{j^\star}}>0$ by \eqref{eq:zeroeq}, $\bar\beta=(1-\alpha_{j^\star})^{-1}$ and
$\beta_j=(1-\alpha_j)^{-1}$, so that $D_{\alpha_j}\ge \underline D$ and $\beta_j\le\bar\beta$ for $j\le j^\star$. Let $\mathcal E_{\mathrm{fit}}$ be the
event that $\Gamma_b,\Gamma_b^-\le \underline D/(8\bar\beta)$ for all $b$, and let the samples be large enough
that $\bar\beta\Delta_\phi r_1\le \underline D/8$ and $\bar\beta^2\kappa^2r_1\le7\underline D^2/64$. On
$\mathcal E_{\mathrm{fit}}$, at every level $\alpha_j$ with $j\le j^\star$, the four terms of $\Theta_b$ are at least
$\underline D$, $-\underline D/4$, $-\underline D/8$ and $-\underline D/8$, the last because the denominator is at
least $7\underline D/8$; so every batch has a ramp with excess at least $\varrho_*=\underline D/(2\bar\beta\Delta_\phi)$.
By Lemma~\ref{lem:hinge}, as in the proof of Theorem~\ref{thm:twosided},
$\E\Gamma_b,\E\Gamma_b^-\le\mu_1=2(1+\sqrt2)\kappa\sqrt{r_1}$, and a deviation $y$ above the mean has probability
at most $e^{-2y^2/(\Delta_\phi^2r_1)}$. So by \eqref{eq:downclosed} and \eqref{eq:betbounds},
\[
\Pp(\widehat\alpha<\alpha_{j^\star})\le\Pp(\mathcal E_{\mathrm{fit}}^c)+J\varepsilon_N(\varrho_*)\le
2Be^{-2(\underline D/(8\bar\beta)-\mu_1)_+^2/(\Delta_\phi^2r_1)}+J\varepsilon_N(\varrho_*)\to0 .
\]

Finer grids need one more step. For $\epsilon\in(0,\alpha^\star)$, the lower side with
$\underline D=D_{\alpha^\star-\epsilon/2}$ and $\bar\beta=(1-\alpha^\star+\epsilon/2)^{-1}$ rejects all grid levels in
$[0,\alpha^\star-\epsilon/2]$ with probability tending to one, as $\log J=o(N)$. Then
$\widehat\alpha\ge\alpha^\star-\epsilon/2-2h_{\mathcal G}\ge\alpha^\star-\epsilon$ once the mesh
$h_{\mathcal G}=\sup_{a\in[0,1)}\dist(a,\mathcal G\cup\{0\})$ of the grid $\mathcal G$, a two-sided mesh over $[0,1)$,
not the $\Delta_{\mathcal G}$ of Theorem~\ref{thm:lorate}, is at most $\epsilon/4$, while
$\widehat\alpha>\alpha^\star+\epsilon$ needs the rejection of the first grid level above $\alpha^\star+\epsilon/2$, of
probability at most $e^{-\epsilon v_{\min}(NL_\delta/2)^{1/2}}$ by \eqref{eq:betbounds}.
\end{proof}

The constants grow like $1/v^2_{\min}$: with a floor that shrinks with $N$, $\overline\lambda_N$ tends to
$\lambda_{\max}/(1+\alpha)$, and the second bound of \eqref{eq:betbounds} only to
$\delta^{2\zeta(1+\alpha)/\lambda_{\max}}$. The proof also allows an inexact minimizer of \eqref{eq:qp}: if its
objective is $d^2$ and its Frank--Wolfe gap is at most $\rho d^2$ with $\rho<1$, the gap equals $2d^2-2ds$ for the
empirical separation $s$ of its witness, so $s\ge(1-\rho/2)d$, and the argument goes through with
$(1-\rho/2)\underline D$ in place of $\underline D$.

\subsection{A witness fitted at $\alpha=0$}
\label{sec:alphazero}

Consistency needs the witness fitted at the level tested. A single witness fitted at $\alpha=0$, with direction
$f_0=(\mu_P-\mu_Q)/\MMD(P,Q)$ as in a test of $P=Q$, can keep $\widehat\alpha$ away from $\alpha^\star$. We show that
its ramps certify at most a one-sided Kolmogorov distance of the values of $f_0$, and then give laws for which this
distance is smaller than $\TV(P,Q)$. A ramp $\rp$ of $f\in B_\Hh$ with cuts $t>u$ has
$\{\rp>r\}=\{f>u+r(t-u)\}$ for $r\in[0,1)$, so by the layer-cake formula
\[
\E_P\rp-\E_Q\rp\ \le\ \mathrm{KS}^+(f)=\sup_x\{P(f(X)>x)-Q(f(Y)>x)\}.
\]
Witnesses $\widehat f_{0,b}$ fitted at $\alpha=0$ have uniform weights and tend to $f_0$ uniformly, almost surely.
Multiplying the fractions by $\mathbf 1\{\mathrm{KS}^+(\widehat f_{0,b})\le\alpha\}$, which is predictable, gives a
test that has $\Delta_k\le\alpha$ wherever its fraction is positive. By the proof of Theorem~\ref{thm:betting}, this
test has level $\delta$ even when $H_{0,\alpha}$ is false, and it has the same wealth as the original test when
$\max_b\mathrm{KS}^+(\widehat f_{0,b})\le\alpha$. Hence the original test rejects $H_{0,\alpha}$ with probability at
most $\delta+\Pp(\max_b\mathrm{KS}^+(\widehat f_{0,b})>\alpha)$.

For the laws, take the Gaussian kernel with $\sigma=\loExSigma$ on $\R$, let
$P$ put mass $\loExPzero$ at $0$, $\loExCommon$ at $\loExCommonLoc$ and $\loExSpread$ on
$\loExSpreadLo,\loExSpreadNext,\dots,\loExSpreadHi$ in proportion to $1,2,\dots,\loExSpreadK$, and $Q$ put $\loExQhalf$
at $\loExQloc$, $\loExQfar$ at $\loExFar$ and $\loExCommon$ at $\loExCommonLoc$, so $\TV(P,Q)=\loExTV$. Up to
$O(e^{-\loExTail})$, $\mu_P-\mu_Q$ is $\loExPzero-\loExQhalf e^{-1/8}=\loExFzero$ at $0$, $\loExPzero
e^{-1/8}-\loExQhalf=\loExFhalf$ at $\loExQloc$, increasing and at most $\loExFspread$ on the spread points, $0$ at
$\loExCommonLoc$ and $-\loExQfar$ at $\loExFar$. So the $\loExAtoms$ values of $f_0$ are distinct, its up-sets have
$P-Q$ equal to $\loExPzero$, $\loExCumTwo$, at most $\loExCumSpread$, $\loExCumSpread$ and $0$, and
$\mathrm{KS}^+(\widehat f_{0,b})=\mathrm{KS}^+(f_0)=\loExKS$ once $\widehat f_{0,b}$ is uniformly within half the smallest
gap of these values, so every level $\alpha\ge\loExKS$ is rejected with probability at most $\delta+o(1)$, the false
nulls below $\TV(P,Q)$ included.
\space
\section{Proofs for Section~\ref{sec:rates}}
\label{app:rates}

This section proves the results of Section~\ref{sec:rates}: the margin condition and its examples, the rate of the
split-sample bound under a margin, the margins of touching contaminations, the excess of the population ramp, the
parametric rate, up to logarithmic factors, with the Laplace kernel, and the two-point lower bound and the minimax
corollary. It also states and proves the converse for the analytic baseline.

\subsection{Proof of Example~\ref{ex:separated}}
\label{sec:exproof}

\begin{proof}
Every $\alpha$-trimming keeps enough mass where $f=\pm1$ for $f$ to separate the trimmed means. Since
$P-Q=\alpha^\star(P_0-Q_0)$ and $P_0$, $Q_0$ are concentrated on the disjoint sets $S_P$, $S_Q$,
$\TV(P,Q)=\alpha^\star\TV(P_0,Q_0)=\alpha^\star$. Fix $\alpha\in[0,\alpha^\star]$ and an $\alpha$-trimming $P'=gP$, with
$0\le g\le\beta$. The law $P$ is concentrated on $S_R\cup S_P$ and gives $S_R$ the mass $1-\alpha^\star$, and $f$ is $0$
on $S_R$ and $1$ on $S_P$, so
\[
\E_{P'}f=\int_{S_P}g\,dP=1-\int_{S_R}g\,dP\ge1-\beta(1-\alpha^\star)=\frac{\alpha^\star-\alpha}{1-\alpha}.
\]
In the same way $\E_{Q'}f\le-(\alpha^\star-\alpha)/(1-\alpha)$ for every $Q'\in R_\alpha(Q)$. The function $f$ is not $0$,
since it equals $1$ on $S_P$, and by the reproducing property and the Cauchy--Schwarz inequality
\[
\MMD(P',Q')=\|\mu_{P'}-\mu_{Q'}\|_\Hh\ge\frac{\langle\mu_{P'}-\mu_{Q'},f\rangle_\Hh}{\|f\|_\Hh}
=\frac{\E_{P'}f-\E_{Q'}f}{\|f\|_\Hh}\ge\frac{2(\alpha^\star-\alpha)}{(1-\alpha)\|f\|_\Hh}.
\]
The infimum over the two trimmings gives the first inequality of the example, and $1-\alpha\le1$ the second. At
$\alpha=\alpha^\star-\epsilon$ it reads $D_{\alpha^\star-\epsilon}(P,Q)\ge2\epsilon/\|f\|_\Hh$, which is
Assumption~\ref{ass:margin} with $c_0=2/\|f\|_\Hh$ and $\gamma=1$.
\end{proof}

For finite sets and a strictly positive definite kernel the Gram matrix of their points is invertible, so a combination
$\sum_ic_ik(\cdot,z_i)$ takes any prescribed values on them. For a Mat\'ern kernel on $\R^d$ the RKHS is a Sobolev space
with an equivalent norm \citep[Chapter~10]{wendland2005scattered}, which contains a smooth compactly supported function equal to
$1$ on the $\varepsilon/3$-neighborhood of $S_P$, to $-1$ on that of $S_Q$ and to $0$ outside their
$2\varepsilon/3$-neighborhoods, when $S_P$ and $S_Q$ are bounded and $\varepsilon>0$ is the smallest distance between
two of the three sets.

\subsection{A general margin and the convexity of the path}
\label{sec:generalmargin}

Assumption~\ref{ass:margin} asks for a polynomial lower bound on the population path below $\alpha^\star$. A lower
bound with no speed always exists, and Proposition~\ref{prop:newrates}(i), applied to $(1-\alpha^\star)\omega$, gives
the split-sample bound a rate for it.

\begin{lemma}[A margin always exists]
\label{lem:marginexists}
Let Assumption~\ref{ass:regularity} hold, let $0<\alpha^\star<1$, and for $\epsilon\in[0,\alpha^\star]$ put
\[
m(\epsilon)\coloneqq\min_{\epsilon'\in[\epsilon,\alpha^\star]}D_{\alpha^\star-\epsilon'}(P,Q),
\qquad
\omega(\epsilon)\coloneqq\frac{\epsilon}{\alpha^\star}\,m(\epsilon).
\]
Then $\omega$ is continuous and strictly increasing on $[0,\alpha^\star]$, $\omega(0)=0$, and
$D_{\alpha^\star-\epsilon}(P,Q)\ge\omega(\epsilon)$ for every $\epsilon\in[0,\alpha^\star]$.
\end{lemma}

\begin{proof}
The map $\epsilon'\mapsto D_{\alpha^\star-\epsilon'}$ is continuous on $[0,\alpha^\star]$ by Lemma~\ref{lem:continuity},
zero at $\epsilon'=0$ and positive elsewhere by \eqref{eq:zeroeq}. Its running minimum $m$ is therefore attained,
continuous and nondecreasing, with $m(0)=0$ and $m(\epsilon)>0$ for $\epsilon>0$. Hence $\omega$ is continuous with
$\omega(0)=0$, and for $0<\epsilon_1<\epsilon_2$, $\omega(\epsilon_2)\ge(\epsilon_2/\alpha^\star)m(\epsilon_1)>\omega(\epsilon_1)>0$.
Finally $\omega(\epsilon)\le m(\epsilon)\le D_{\alpha^\star-\epsilon}$, as $\epsilon\le\alpha^\star$.
\end{proof}

The next lemma, stated in words in Section~\ref{sec:rate}, shows which margins are possible at all, through the
convexity of the rescaled path. For $\alpha\in[0,1)$ let
\[
\bar{\mathcal C}_\alpha(P)\coloneqq(1-\alpha)\,\mathcal C_\alpha(P)=\big\{\E_P[h(X)\phi(X)]:\ 0\le h\le1,\ \E_P[h]=1-\alpha\big\},
\]
the embeddings of the parts $P'\le P$ of mass $1-\alpha$, that is, of the $\alpha$-trimmings before they are
renormalized, with $h=(1-\alpha)g$ for $g$ as in \eqref{eq:region}, and likewise $\bar{\mathcal C}_\alpha(Q)$. By
Proposition~\ref{prop:georep}, which holds for any probability measures and so for $P_n$ and $Q_m$, the rescaled path
is $\bar D_\alpha=(1-\alpha)D_\alpha(P,Q)=\dist(\bar{\mathcal C}_\alpha(P),\bar{\mathcal C}_\alpha(Q))$.

\begin{lemma}[Convexity of the path]
\label{lem:pathconvex}
Let Assumption~\ref{ass:regularity} hold, let $P,Q$ be any probability measures on $\mathcal X$, empirical measures
included, and put $\bar D_{\alpha}\coloneqq(1-\alpha)D_\alpha(P,Q)$ for $\alpha\in[0,1)$. Then $\bar D$ is convex and
nonincreasing. If moreover $0<\alpha^\star<1$, then:
\begin{enumerate}[label=(\roman*),leftmargin=*]
\item $\bar D$ is strictly decreasing on $[0,\alpha^\star]$, and for all $0\le\alpha_1<\alpha_2<\alpha^\star$,
\[
\alpha^\star\ \ge\ \alpha_2+\frac{(\alpha_2-\alpha_1)\,\bar D_{\alpha_2}}{\bar D_{\alpha_1}-\bar D_{\alpha_2}} ;
\]
\item $\epsilon\mapsto \bar D_{\alpha^\star-\epsilon}/\epsilon$ is nondecreasing on $(0,\alpha^\star]$, so it has a limit
$\mathfrak s\ge0$ as $\epsilon\downarrow0$, and $D_{\alpha^\star-\epsilon}(P,Q)\le D_0(P,Q)\,\epsilon/\{\alpha^\star(1-\alpha^\star)\}$
for all $\epsilon\in[0,\alpha^\star]$. Assumption~\ref{ass:margin} holds with $\gamma=1$ if and only if $\mathfrak s>0$, and then
it holds with $c_0=\mathfrak s$.
\end{enumerate}
\end{lemma}

\begin{proof}[Proof of Lemma~\ref{lem:pathconvex}]
A mixture of two trimmings at two levels is a trimming at the level in between, and the norm is convex; everything
else follows from convexity and the location of the zeros of the path.

We show first that $\bar D$ is convex. Let $\alpha_0,\alpha_1\in[0,1)$, $\lambda\in[0,1]$ and $\alpha_\lambda=(1-\lambda)\alpha_0+\lambda\alpha_1$.
If $h_0$ and $h_1$ are admissible weights at the levels $\alpha_0$ and $\alpha_1$, then $(1-\lambda)h_0+\lambda h_1$
lies between $0$ and $1$ and has mean $1-\alpha_\lambda$ under $P$, so
$(1-\lambda)\bar{\mathcal C}_{\alpha_0}(P)+\lambda \bar{\mathcal C}_{\alpha_1}(P)\subset \bar{\mathcal C}_{\alpha_\lambda}(P)$, and likewise for $Q$. For
$a_i\in \bar{\mathcal C}_{\alpha_i}(P)$ and $b_i\in \bar{\mathcal C}_{\alpha_i}(Q)$, the points $(1-\lambda)a_0+\lambda a_1$ and
$(1-\lambda)b_0+\lambda b_1$ therefore lie in the sets at level $\alpha_\lambda$, and the triangle inequality gives
\[
\bar D_{\alpha_\lambda}\le\big\|(1-\lambda)(a_0-b_0)+\lambda(a_1-b_1)\big\|_\Hh\le(1-\lambda)\|a_0-b_0\|_\Hh+\lambda\|a_1-b_1\|_\Hh .
\]
Taking the infimum over $a_0,a_1,b_0,b_1$ gives $\bar D_{\alpha_\lambda}\le(1-\lambda)\bar D_{\alpha_0}+\lambda \bar D_{\alpha_1}$. The path
$D_\alpha$ is nonnegative and nonincreasing, and so is $1-\alpha$, so their product $\bar D$ is nonincreasing.

From here on $0<\alpha^\star<1$, so that $\bar D_{\alpha^\star}=0$ and $\bar D>0$ on $[0,\alpha^\star)$ by
\eqref{eq:zeroeq}. The path is strictly
decreasing on $[0,\alpha^\star]$: for $\alpha_1<\alpha_2\le\alpha^\star$ write
$\alpha_2=(1-\lambda)\alpha_1+\lambda\alpha^\star$ with $\lambda=(\alpha_2-\alpha_1)/(\alpha^\star-\alpha_1)\in(0,1]$;
convexity gives $\bar D_{\alpha_2}\le(1-\lambda)\bar D_{\alpha_1}<\bar D_{\alpha_1}$, since $\bar D_{\alpha_1}>0$.

For the secant bound of (i), let $\alpha_1<\alpha_2<\alpha^\star$, so that $\bar D_{\alpha_1}>\bar D_{\alpha_2}>0$, and let
$\alpha\in(\alpha_2,1)$. Since $\alpha_2$ lies between $\alpha_1$ and $\alpha$, convexity gives
$(\alpha-\alpha_1)\bar D_{\alpha_2}\le(\alpha-\alpha_2)\bar D_{\alpha_1}+(\alpha_2-\alpha_1)\bar D_{\alpha}$, that is,
\[
\bar D_{\alpha}\ \ge\ \bar D_{\alpha_2}-(\alpha-\alpha_2)\,\frac{\bar D_{\alpha_1}-\bar D_{\alpha_2}}{\alpha_2-\alpha_1} .
\]
The right side, the secant through the two points, is positive for $\alpha<\alpha_s=\alpha_2+(\alpha_2-\alpha_1)\bar D_{\alpha_2}
/\{\bar D_{\alpha_1}-\bar D_{\alpha_2}\}$. So $\bar D>0$ on $[\alpha_2,\alpha_s)\cap[0,1)$, and since $\bar D_{\alpha^\star}=0$ with
$\alpha^\star<1$, this forces $\alpha^\star\ge\alpha_s$.

The slope at $\alpha^\star$ in (ii) comes from the same mixture argument with one endpoint at $\alpha^\star$. Put $r(\epsilon)=\bar D_{\alpha^\star-\epsilon}/\epsilon$ for
$\epsilon\in(0,\alpha^\star]$. For $0<\epsilon_1<\epsilon_2\le\alpha^\star$ the level $\alpha^\star-\epsilon_1$ is the
mixture of $\alpha^\star-\epsilon_2$ and $\alpha^\star$ with weights $\epsilon_1/\epsilon_2$ and $1-\epsilon_1/\epsilon_2$,
so convexity and $\bar D_{\alpha^\star}=0$ give $\bar D_{\alpha^\star-\epsilon_1}\le(\epsilon_1/\epsilon_2)\,\bar D_{\alpha^\star-\epsilon_2}$,
that is, $r(\epsilon_1)\le r(\epsilon_2)$. Hence $r$ is nondecreasing and nonnegative, its limit $\mathfrak s\ge0$ at $0$ exists, and
$\mathfrak s\le r(\epsilon)\le r(\alpha^\star)=D_0(P,Q)/\alpha^\star$ for every $\epsilon$. The upper bound gives
$D_{\alpha^\star-\epsilon}=\bar D_{\alpha^\star-\epsilon}/(1-\alpha^\star+\epsilon)\le D_0\,\epsilon/\{\alpha^\star(1-\alpha^\star)\}$
for $\epsilon\in(0,\alpha^\star]$, and at $\epsilon=0$ both sides vanish.

For the equivalence in (ii), if $\mathfrak s>0$, then $\bar D_{\alpha^\star-\epsilon}\ge \mathfrak s\epsilon$, and since $1-\alpha^\star+\epsilon\le1$,
$D_{\alpha^\star-\epsilon}\ge \bar D_{\alpha^\star-\epsilon}\ge \mathfrak s\epsilon$ for $\epsilon\in(0,\alpha^\star]$; at $\epsilon=0$ the
bound is $0\ge0$. So Assumption~\ref{ass:margin} holds with $\gamma=1$ and $c_0=\mathfrak s$. Conversely, if it holds with
$\gamma=1$ and some $c_0>0$, then $r(\epsilon)=(1-\alpha^\star+\epsilon)D_{\alpha^\star-\epsilon}/\epsilon\ge(1-\alpha^\star)c_0$
for every $\epsilon$, so $\mathfrak s\ge(1-\alpha^\star)c_0>0$. The bound $c_0\epsilon^\gamma\le D_0\,\epsilon/\{\alpha^\star(1-\alpha^\star)\}$
for all small $\epsilon>0$ rules out every exponent $\gamma<1$, which is why Assumption~\ref{ass:margin} takes
$\gamma\ge1$.
\end{proof}

The convexity and the monotonicity of $\bar D$, and $\bar D_{\alpha^\star}=0$, need only (A1) and (A2), since a
common trimming has one embedding; so does part (ii), which uses nothing else, while part (i) also needs
$\bar D>0$ below $\alpha^\star$, that is (A3).

\subsection{The rate of the lower bound}
\label{sec:lorate}

\begin{theorem}[Rate of the lower bound under the margin condition]
\label{thm:lorate}
Let Assumptions~\ref{ass:regularity} and~\ref{ass:margin} hold, let $\alpha^\star<1$, and compute
$\widehat\alpha$ as in Proposition~\ref{prop:loconsistency}: on a grid $\alpha_1<\dots<\alpha_J$ fixed in
advance, with a witness fitted at each level by a minimizer of \eqref{eq:qp} on the batches before the one bet on, cut
points satisfying \eqref{eq:cutrank}, and the fractions \eqref{eq:fractions} with $\lambda_{\max}=\tfrac12$,
$v_0^2\le4$ and the running variance floored at a constant $v^2_{\min}\in(0,4]$. Let $B$ be the number of
batches bet on, $N$ the number of pairs and $r_1=1/n_{<1}+1/m_{<1}$, where $n_{<1},m_{<1}$ are the sizes of batch $0$,
and suppose $r_1\le1$. Put $\beta_\star=(1-\alpha^\star)^{-1}$, $L_\delta=\log(1/\delta)$ and, for $\eta\in(0,1)$ and
$J'\ge1$,
\begin{gather}
T_{\mathrm{fit}}(\eta)=8\Big[2(1+\sqrt2)\kappa+\Delta_\phi\sqrt{\tfrac12\log(4B/\eta)}\,\Big]\sqrt{r_1},\notag\\
T_{\mathrm{bet}}(\eta,J')=2\Delta_\phi\sqrt{\frac2N}\,\Big[\Big(1+\frac{16}{v^2_{\min}}\Big)\sqrt{L_\delta}
+\frac2{v_{\min}}\sqrt{\log(2J'/\eta)}\,\Big].
\label{eq:loconst}
\end{gather}
For $\epsilon\in(0,\alpha^\star]$ let $a_\epsilon=\max(\{0\}\cup\{\alpha_j:\alpha_j\le\alpha^\star-\epsilon\})$, let
$\Delta_{\min}$ be the smallest gap between consecutive grid levels ($\Delta_{\min}=\infty$ if $J=1$), and let
$J_\epsilon=\min\{J,2(1+\epsilon/\Delta_{\min})\}$.
\begin{enumerate}[label=(\roman*),leftmargin=*]
\item If $c_0\epsilon^\gamma\ge\beta_\star\max\{T_{\mathrm{fit}}(\eta),T_{\mathrm{bet}}(\eta,J_\epsilon)\}$, then
$\Pp(\widehat\alpha<a_\epsilon)\le\eta$.
\item Let $t_{n,m}=\beta_\star\max\{T_{\mathrm{fit}}(\eta),T_{\mathrm{bet}}(\eta,J)\}$ and let
$\Delta_{\mathcal G}=\sup_{a\in[0,\alpha^\star]}\big(a-\max\{g\in\{0,\alpha_1,\dots,\alpha_J\}:g\le a\}\big)$ be the mesh of the grid
below $\alpha^\star$. Then, with probability at least $1-\delta-\eta$,
\[
\alpha^\star-\Delta_{\mathcal G}-\Big(\frac{t_{n,m}}{c_0}\Big)^{1/\gamma}\ \le\ \widehat\alpha\ \le\ \alpha^\star .
\]
\item With the batches of Section~\ref{sec:betting}, their boundaries rounded to the nearest integers, and
$n\wedge m\ge10$, one has $r_1\le\min\{1,12/(n\wedge m)\}$ and $N\ge(n\wedge m)/2$, so that, with an absolute
constant in the $O$,
\[
\Big(\frac{t_{n,m}}{c_0}\Big)^{1/\gamma}
=O\bigg(\bigg(\frac{\kappa}{c_0(1-\alpha^\star)v^2_{\min}}
\sqrt{\frac{1+\log(1/\delta)+\log(J/\eta)}{n\wedge m}}\bigg)^{1/\gamma}\bigg).
\]
\end{enumerate}
\end{theorem}

The error is of order $(n\wedge m)^{-1/(2\gamma)}$, plus the mesh, with the factor $1/v^2_{\min}$ of the betting
constants and a factor $\log J$, which part (i) removes when the grid is not finer than the rate: if $\Delta_{\min}\ge\epsilon'_{n,m}=(\beta_\star\max\{T_{\mathrm{fit}}(\eta),T_{\mathrm{bet}}(\eta,4)\}/c_0)^{1/\gamma}$,
then $J_{\epsilon'_{n,m}}\le4$, and with probability at least $1-\delta-\eta$,
$\alpha^\star-\Delta_{\mathcal G}-\epsilon'_{n,m}\le\widehat\alpha\le\alpha^\star$, with no $\log J$. The proof needs no
density of the witness at the cut points, unlike Theorem~\ref{cor:trimrate}, which uses \eqref{eq:qdens}, and it does
not use (A3), whose role the margin takes over.

\begin{proof}[Proof of Theorem~\ref{thm:lorate}]
We run the lower side of the proof of Proposition~\ref{prop:loconsistency} at the level $\alpha^\star-\epsilon$, with the
margin in place of a fixed positive distance, and keep the constants. One event, which does not depend on the level,
makes every fitted ramp good, and only the bets need a union over the levels.

Fix $\epsilon$ as in (i), and put $D_*=c_0\epsilon^\gamma$, $\tau_*=D_*/(8\beta_\star)$ and
$\varrho_*=D_*/(2\beta_\star\Delta_\phi)$: $D_*$ is the margin at distance $\epsilon$ from $\alpha^\star$, $\tau_*$ the
fluctuation allowed to each fitting batch, and $\varrho_*$ the excess that the bets will see. The two thresholds of
(i) are these choices solved for $D_*$: $T_{\mathrm{fit}}$ makes the fitting suprema smaller than $\tau_*$, and
$T_{\mathrm{bet}}$ makes an excess $\varrho_*$ win the bets, each with probability $\eta/2$ to spare. Every grid level $\alpha_j\le\alpha^\star-\epsilon$ has
$\beta_j=(1-\alpha_j)^{-1}\le\beta_\star$, and with $x_j=(\alpha^\star-\alpha_j)/\epsilon\ge1$,
Assumption~\ref{ass:margin} at $\epsilon x_j\in[\epsilon,\alpha^\star]$ gives
$D_{\alpha_j}\ge c_0\epsilon^\gamma x_j^\gamma=D_*x_j^\gamma\ge D_*$. Moreover $D_*\le D_{\alpha^\star-\epsilon}\le
D_0=\MMD(P,Q)=\|\E[\phi(X)-\phi(Y)]\|_\Hh\le\Delta_\phi$, as the path is nonincreasing, so $\Delta_\phi>0$. By
\eqref{eq:downclosed}, if $H_{0,\alpha_j}$ is rejected at every grid level $\alpha_j\le\alpha^\star-\epsilon$, then
$\widehat\alpha\ge a_\epsilon$, which also holds trivially when no grid level is that small. Hence
\[
\{\widehat\alpha<a_\epsilon\}\subseteq\bigcup_{j:\,\alpha_j\le\alpha^\star-\epsilon}
\{H_{0,\alpha_j}\text{ is not rejected}\}.
\]

The fits are controlled by the suprema $\Gamma_b,\Gamma_b^-$ of the proof of Proposition~\ref{prop:loconsistency},
those of the proof of Theorem~\ref{thm:twosided} computed on batches $0,\dots,b-1$, which do not depend on $\alpha$. Let
$\mathcal E_{\mathrm{fit}}=\bigcap_{b=1}^B\{\Gamma_b\le\tau_*,\ \Gamma_b^-\le\tau_*\}$. We show that on
$\mathcal E_{\mathrm{fit}}$ every batch has a ramp at every level $\alpha_j\le\alpha^\star-\epsilon$, with a large
excess. Fix such a level $\alpha=\alpha_j$ and a batch $b$, and write $\beta=\beta_j$. The hypothesis of (i) gives
$D_*\ge\beta_\star T_{\mathrm{fit}}(\eta)\ge16(1+\sqrt2)\beta_\star\kappa\sqrt{r_1}$, and since $\Delta_\phi\le2\kappa$
and $r_b\le r_1\le1$,
\[
\beta\Delta_\phi r_b\le2\beta_\star\kappa\sqrt{r_1}\le\frac{D_*}8,
\qquad
\beta^2\kappa^2r_b\le\beta_\star^2\kappa^2r_1\le\frac{D_*^2}{256(1+\sqrt2)^2}\le\frac{7D_*^2}{64};
\]
on $\mathcal E_{\mathrm{fit}}$ moreover $\beta(\Gamma_b+\Gamma_b^-)\le2\beta_\star\tau_*=D_*/4$ and
$D_\alpha-\beta\Gamma_b^-\ge D_*-D_*/8=7D_*/8>0$. The last inequality is the hypothesis
$\beta\Gamma_b^-<D_\alpha$ under which the proof of Proposition~\ref{prop:loconsistency} establishes
\eqref{eq:fittedramp}, with no condition on the law of the witness near the cut points:
\[
\Delta_\phi\ge\widehat t_b-\widehat u_b\ge\theta(\widehat f;\widehat t_b,\widehat u_b)\ge\Theta_b
=D_\alpha-\beta(\Gamma_b+\Gamma_b^-)-\beta\Delta_\phi r_b-\frac{\beta^2\kappa^2r_b}{D_\alpha-\beta\Gamma_b^-},
\]
and the four bounds above give $\Theta_b\ge D_\alpha-\frac{D_*}4-\frac{D_*}8-\frac{7D_*^2/64}{7D_*/8}=D_\alpha-\frac{D_*}2
\ge\frac{D_\alpha}2>0$. So $\widehat t_b>\widehat u_b$, batch $b$ has a ramp $\rp_b$, and Lemma~\ref{lem:ramp}(ii),
applied under $P$ and $Q$ given batches $0,\dots,b-1$ to the direction $\widehat f$ and the cut points $\widehat
t_b,\widehat u_b$, gives $\beta(\widehat t_b-\widehat u_b)(\E_P\rp_b-\E_Q\rp_b-\alpha)\ge\theta(\widehat f;\widehat
t_b,\widehat u_b)\ge\Theta_b$, hence
\begin{equation}
\E_P\rp_b-\E_Q\rp_b-\alpha_j\ \ge\ \frac{\Theta_b}{\beta\Delta_\phi}\ \ge\ \frac{D_{\alpha_j}}{2\beta_\star\Delta_\phi}
\ \eqqcolon\ \varrho_j\ \ge\ \varrho_*x_j^\gamma .
\label{eq:excessj}
\end{equation}
The fitting event is likely. By Lemma~\ref{lem:hinge} with $\varphi^1_x=\varphi^2_y=(\cdot)_+$, $r=1$ and
$\rho_1=\rho_2=\kappa$, as in the proof of Theorem~\ref{thm:twosided}, $\E\Gamma_b$ and $\E\Gamma_b^-$ are at most
$2\kappa\sqrt{r_b}+2\kappa(e_{n_{<b}}+e_{m_{<b}})\le2(1+\sqrt2)\kappa\sqrt{r_b}\le\mu_1=2(1+\sqrt2)\kappa\sqrt{r_1}$, by
$e_k\le k^{-1/2}$ and $n^{-1/2}+m^{-1/2}\le\{2(1/n+1/m)\}^{1/2}$, and, the bounded differences being $\Delta_\phi$, each
exceeds its mean by more than $y$ with probability at most $e^{-2y^2/(\Delta_\phi^2r_b)}\le e^{-2y^2/(\Delta_\phi^2r_1)}$.
Divided by $8\beta_\star$, the hypothesis $D_*\ge\beta_\star T_{\mathrm{fit}}(\eta)$ reads
$\tau_*-\mu_1\ge\Delta_\phi\{r_1\log(4B/\eta)/2\}^{1/2}$, so each of the $2B$ suprema exceeds $\tau_*$ with probability at
most $\eta/(4B)$, and $\Pp(\mathcal E_{\mathrm{fit}}^c)\le\eta/2$.

We turn to the bets, where only a union over the levels is needed. Fix a grid level $\alpha_j\le\alpha^\star-\epsilon$. On $\mathcal E_{\mathrm{fit}}$ every batch has a
ramp, and by \eqref{eq:excessj} every pair has conditional mean $\Delta_k\ge\alpha_j+\varrho_j$ in the notation of the
proof of Theorem~\ref{thm:betting}. The sample is large enough for \eqref{eq:betbounds}: as $v^2_{\min}\le4$,
$\Delta_\phi\ge D_*\ge\beta_\star T_{\mathrm{bet}}(\eta,J_\epsilon)\ge10\Delta_\phi(2L_\delta/N)^{1/2}$, so $N\ge200L_\delta\ge
2L_\delta/\lambda_{\max}^2$. The first bound of \eqref{eq:betbounds}, with $\mathcal E=\mathcal E_{\mathrm{fit}}$ and
$\varrho=\varrho_j$, gives
\[
\Pp\big(\mathcal E_{\mathrm{fit}}\cap\{H_{0,\alpha_j}\text{ is not rejected}\}\big)\le\varepsilon_N(\varrho_j)
=\exp\Big\{-\frac{v^2_{\min}}{4L_\delta}\big(\varrho_jS-C\big)_+^2\Big\},
\]
with $S=(NL_\delta/2)^{1/2}$ and $C=L_\delta(1+16/v^2_{\min})$. The hypothesis $D_*\ge\beta_\star T_{\mathrm{bet}}(\eta,J_\epsilon)$, divided by $2\beta_\star\Delta_\phi$ and multiplied
by $S$, reads $\varrho_*S\ge C+2v_{\min}^{-1}\{L_\delta\log(2J_\epsilon/\eta)\}^{1/2}$, so
$a=\varrho_*S-C\ge2v_{\min}^{-1}\{L_\delta\log(2J_\epsilon/\eta)\}^{1/2}>0$. Since $x_j\ge1$, $\gamma\ge1$ and
$\varrho_*S\ge a$, \eqref{eq:excessj} gives $\varrho_jS-C\ge\varrho_*x_jS-C=a+\varrho_*S(x_j-1)\ge ax_j$, hence
$\varepsilon_N(\varrho_j)\le e^{-Ax_j^2}$ with $A=v^2_{\min}a^2/(4L_\delta)\ge\log(2J_\epsilon/\eta)>\log2$.

It remains to add over the levels, and the polynomial growth of the excess away from $\alpha^\star$ keeps the sum
within a factor $J_\epsilon$ of $e^{-A}$. If $J_\epsilon=J$, the at most $J$ terms $e^{-Ax_j^2}\le e^{-A}$ add up to at most
$Je^{-A}\le\eta/2$. Otherwise $J_\epsilon=2(1+\epsilon/\Delta_{\min})$; the $x_j$ are at least $\Delta_{\min}/\epsilon$ apart, so at most
$1+\epsilon/\Delta_{\min}$ of them lie in each interval $[1+k,2+k)$, $k=0,1,\dots$, and as $(1+k)^2\ge1+3k$ and
$e^{-3A}\le\tfrac12$,
\[
\sum_{j:\,\alpha_j\le\alpha^\star-\epsilon}e^{-Ax_j^2}\le\Big(1+\frac\epsilon{\Delta_{\min}}\Big)\sum_{k\ge0}e^{-A(1+3k)}
\le2\Big(1+\frac\epsilon{\Delta_{\min}}\Big)e^{-A}=J_\epsilon e^{-A}\le\frac\eta2 .
\]
With the inclusion above, $\Pp(\widehat\alpha<a_\epsilon)\le\Pp(\mathcal E_{\mathrm{fit}}^c)
+\sum_j\Pp(\mathcal E_{\mathrm{fit}}\cap\{H_{0,\alpha_j}\text{ is not rejected}\})\le\eta$, which is (i).

For (ii), put $\epsilon_{n,m}=(t_{n,m}/c_0)^{1/\gamma}$. If $\epsilon_{n,m}>\alpha^\star$ the lower inequality holds
because $\widehat\alpha\ge0$. Otherwise $\epsilon=\epsilon_{n,m}$ satisfies the hypothesis of (i), because
$J_\epsilon\le J$ and $T_{\mathrm{bet}}$ increases in $J'$, so with probability at least $1-\eta$,
$\widehat\alpha\ge a_{\epsilon_{n,m}}\ge\alpha^\star-\epsilon_{n,m}-\Delta_{\mathcal G}$, the last step by the definition of
$\Delta_{\mathcal G}$. Corollary~\ref{cor:lower} gives $\widehat\alpha\le\alpha^\star$ with probability at least $1-\delta$,
and a union bound gives (ii).

For (iii), boundaries at the nearest integers to $n/4$ and $n/2$ give batch $0$ at least $(n-2)/4$ observations $X$,
batch $1$ at least $n/4-1$ and batch $2$ at least $n/2-\tfrac12$, and likewise for $Y$. Hence $r_1\le8/(n\wedge m-2)$,
which is at most $1$ for $n\wedge m\ge10$ and at most $12/(n\wedge m)$ for $n\wedge m\ge6$, and
$N\ge\tfrac34(n\wedge m)-\tfrac32\ge(n\wedge m)/2$ for $n\wedge m\ge6$. Inserting these bounds, $B=2$,
$\Delta_\phi\le2\kappa$, and $1+16/v^2_{\min}\le20/v^2_{\min}$ and $2/v_{\min}\le4/v^2_{\min}$, which
hold as $v^2_{\min}\le4$, into \eqref{eq:loconst} bounds $t_{n,m}$ by an absolute constant times
$\beta_\star\kappa v_{\min}^{-2}\{(1+L_\delta+\log(J/\eta))/(n\wedge m)\}^{1/2}$.
\end{proof}

\subsection{Touching contaminations}
\label{sec:newproofs}

This subsection and the six that follow prove Propositions~\ref{prop:newslope}--\ref{prop:newgauss} of the main
text, in a more general form, state and prove the rates under a general margin (Proposition~\ref{prop:newrates}), and
state and prove the converse Proposition~\ref{prop:newconverse}. Throughout, $P\wedge Q$ has mass $1-\alpha^\star$,
$P_0=(P-P\wedge Q)/\alpha^\star$ and $Q_0=(Q-P\wedge Q)/\alpha^\star$, so that $P=P\wedge Q+\alpha^\star P_0$,
$Q=P\wedge Q+\alpha^\star Q_0$, $P_0$ is concentrated on $A=\{p>q\}$, $Q_0$ on its complement, and
$|P-Q|=\alpha^\star(P_0+Q_0)$. The rescaled path $\bar D$ and the sets $\bar{\mathcal C}_\alpha$ are those of
Section~\ref{sec:generalmargin}, and for $\alpha\le\alpha^\star$, $\bar D_{\alpha}\le D_\alpha(P,Q)\le
\bar D_{\alpha}/(1-\alpha^\star)$.

On $\R^d$ we use translation-invariant kernels $k(x,y)=k_0(x-y)$ with $k_0$ continuous, integrable and positive
definite, and the Fourier transform $\widehat g(\xi)=\int g(x)e^{-i\xi\cdot x}\,dx$. Then $\widehat{k_0}\ge0$ is
integrable, $k_0(z)=(2\pi)^{-d}\int\widehat{k_0}(\xi)e^{i\xi\cdot z}\,d\xi$, and by
\citet[Theorem~10.12]{wendland2005scattered} the RKHS consists of the continuous $f\in L^2(\R^d)$ with
$\int|\widehat f|^2/\widehat{k_0}<\infty$, with
\begin{equation}
\|f\|_\Hh^2=(2\pi)^{-d}\int\frac{|\widehat f(\xi)|^2}{\widehat{k_0}(\xi)}\,d\xi .
\label{eq:newnative}
\end{equation}
For a finite signed measure $\nu$ with Fourier transform $\widehat\nu(\xi)=\int e^{-i\xi\cdot x}\,d\nu(x)$, Fubini's
theorem and the inversion formula give
\begin{equation}
\Big\|\int\phi\,d\nu\Big\|_\Hh^2=\iint k_0(x-y)\,d\nu(x)\,d\nu(y)=(2\pi)^{-d}\int|\widehat\nu(\xi)|^2\,\widehat{k_0}(\xi)\,d\xi .
\label{eq:newbochner}
\end{equation}

The first result needs no topology on $\mathcal X$. By (A1), (A2) and Pettis' theorem, $\phi$ is Borel measurable
into the separable space $\Hh$, so its image laws $\phi_\#P_0$ and $\phi_\#Q_0$ are Borel probability measures on a
separable metric space, whose supports are closed, nonempty and of full mass.

\begin{proposition}[Slope bound]
\label{prop:newslopegeneral}
Let (A1) and (A2) hold and let $0<\alpha^\star<1$. Then
$\lim_{\epsilon\downarrow0}\bar D_{\alpha^\star-\epsilon}/\epsilon\le\dist\big(\supp\phi_\#P_0,\supp\phi_\#Q_0\big)$.
\end{proposition}

\begin{proof}
We trim at the level $\alpha^\star-\epsilon$ by adding to the common part a little of each contamination near two
chosen feature vectors. Fix $z\in\supp\phi_\#P_0$, $z'\in\supp\phi_\#Q_0$ and $\varrho>0$, and let
$U=\phi^{-1}(B(z,\varrho))$ and $V=\phi^{-1}(B(z',\varrho))$, so that $m_U=P_0(U)>0$ and $m_V=Q_0(V)>0$. Write
$P_0^U=P_0(\cdot\cap U)/m_U$ and $Q_0^V=Q_0(\cdot\cap V)/m_V$. For $0<\epsilon\le\alpha^\star\min(m_U,m_V)$ the measure
$P_\epsilon=P\wedge Q+\epsilon P_0^U$ satisfies $P_\epsilon\le P\wedge Q+\alpha^\star P_0=P$, because $\epsilon/m_U\le\alpha^\star$, and has
mass $1-(\alpha^\star-\epsilon)$, so $\int\phi\,dP_\epsilon\in \bar{\mathcal C}_{\alpha^\star-\epsilon}(P)$. Likewise,
$Q_\epsilon=P\wedge Q+\epsilon Q_0^V$ gives a point of $\bar{\mathcal C}_{\alpha^\star-\epsilon}(Q)$. The common part cancels, and
\begin{align*}
\bar D_{\alpha^\star-\epsilon}&\le\Big\|\int\phi\,dP_\epsilon-\int\phi\,dQ_\epsilon\Big\|_\Hh
=\epsilon\,\Big\|\iint\{\phi(x)-\phi(y)\}\,dP_0^U(x)\,dQ_0^V(y)\Big\|_\Hh\\
&\le\epsilon\,(\|z-z'\|_\Hh+2\varrho),
\end{align*}
since $\|\phi(x)-z\|_\Hh<\varrho$ on $U$ and $\|\phi(y)-z'\|_\Hh<\varrho$ on $V$. Dividing by $\epsilon$ and letting
$\epsilon\downarrow0$, and then $\varrho\downarrow0$, bounds the limit, which exists by Lemma~\ref{lem:pathconvex}(ii) (that part
of its proof uses only the convexity of $\bar D$ and $\bar D_{\alpha^\star}=0$, which hold under (A1) and (A2)), by $\|z-z'\|_\Hh$, and the infimum over $z$ and $z'$ gives the claim.
\end{proof}

\begin{proof}[Proof of Proposition~\ref{prop:newslope}]
The quantity bounded in the proposition is $\lim \bar D_{\alpha^\star-\epsilon}/\epsilon$. If $\phi$ is continuous and
$x\in\supp P_0$, then for every $\varrho>0$ the set $\phi^{-1}(B(\phi(x),\varrho))$ is an open neighborhood of $x$ and has
positive $P_0$-mass, so $\phi(x)\in\supp\phi_\#P_0$. Likewise $\phi(y)\in\supp\phi_\#Q_0$ for $y\in\supp Q_0$, and
Proposition~\ref{prop:newslopegeneral} gives the bound. If the supports meet, take $y=x$: the limit is $0$, and by
Lemma~\ref{lem:pathconvex}(ii) Assumption~\ref{ass:margin} does not hold with $\gamma=1$.

For the last claim let $\mathcal X$ be an open connected subset of $\R^d$ and let $p,q$ be continuous. The density
$(p-q)_+/\alpha^\star$ of $P_0$ is continuous, so $\supp P_0$ is the closure $F_P$ in $\mathcal X$ of the open set
$\{p>q\}$: a neighborhood of a point of $F_P$ meets $\{p>q\}$ in a nonempty open set, which has positive Lebesgue
measure and on which the density is positive, while a point outside $F_P$ has a neighborhood of $P_0$-mass $0$.
Likewise $\supp Q_0$ is the closure $F_Q$ of $\{q>p\}$. Both sets are nonempty, as $\int(p-q)_+=\int(q-p)_+=\alpha^\star>0$.
If $F_P\cap F_Q$ were empty, the open set $\mathcal X\setminus(F_P\cup F_Q)$ would lie in $\{p=q\}$, because
$\{p\neq q\}\subseteq F_P\cup F_Q$, and so would be empty. Then $\mathcal X$ would be the union of the disjoint nonempty
sets $F_P$ and $F_Q$, closed in $\mathcal X$, against its connectedness. For two distinct finite Gaussian mixtures,
$p-q$ is real-analytic on $\R^d$ and not identically zero, so its zero set has empty interior, and $\alpha^\star<1$
because both densities are positive.
\end{proof}

When $\operatorname{supp}P_0$ is compact, the bound is an equivalence for continuous kernels whose RKHS contains the
smooth functions with compact support.

\begin{proposition}[Disjoint contaminations give a linear margin]
\label{prop:newgammaone}
Let $\mathcal X=\R^d$ and $k(x,y)=k_0(x-y)$ with $\widehat{k_0}(\xi)\ge c(1+|\xi|^2)^{-s}$ for some $s>d/2$ and $c>0$,
as for the Mat\'ern kernels, let $0<\alpha^\star<1$ and let $\supp P_0$ be compact. If $\supp P_0\cap\supp Q_0=\varnothing$,
there is $f\in\Hh$ with $0\le f\le1$, $f=1$ on $\supp P_0$ and $f=0$ on $\supp Q_0$, and every such $f$ gives
$D_{\alpha^\star-\epsilon}(P,Q)\ge \bar D_{\alpha^\star-\epsilon}\ge\epsilon/\|f\|_\Hh$ for $\epsilon\in[0,\alpha^\star]$, that is,
Assumption~\ref{ass:margin} with $\gamma=1$ and $c_0=1/\|f\|_\Hh$. Together with Proposition~\ref{prop:newslope}, the
rescaled path then has a positive slope at $\alpha^\star$ if and only if $\supp P_0\cap\supp Q_0=\varnothing$.
\end{proposition}

\begin{proof}
A compact set and a disjoint closed set are at a positive distance $d_0$. Let $U$ be the open
$d_0/2$-neighborhood of $\supp P_0$, which is bounded, and $f=\mathbf 1_U*b_{d_0/3}$ with $b\in C_c^\infty$ a
probability density supported in the closed unit ball and $b_w=w^{-d}b(\cdot/w)$. Then $f$ is smooth with compact
support and $0\le f\le1$. Moreover $f=1$ on $\supp P_0$, since the closed ball of radius $d_0/3$ about such a point lies in $U$,
and $f=0$ on $\supp Q_0$, since the closed ball of radius $d_0/3$ about such a point stays at distance at least
$2d_0/3$ from $\supp P_0$ and so misses $U$. By \eqref{eq:newnative},
$\|f\|_\Hh^2\le c^{-1}(2\pi)^{-d}\int|\widehat f|^2(1+|\xi|^2)^s<\infty$, so $f\in\Hh$. For any such $f$,
$\E_Pf-\E_Qf=\alpha^\star(\E_{P_0}f-\E_{Q_0}f)=\alpha^\star$, so $\mathrm{err}(f)=0$, and Proposition~\ref{prop:newapproxgen}(ii)
below gives the bound. The equivalence follows from Lemma~\ref{lem:pathconvex}(ii) and
Proposition~\ref{prop:newslope}, as $\phi$ is continuous.
\end{proof}

This sharpens Example~\ref{ex:separated}, which also separates the core.

\subsection{Margins from approximate witnesses}

\begin{lemma}[Trimmed means of a bounded function]
\label{lem:newtrimmean}
Let $\alpha\in[0,1)$, let $P$ be a probability measure and $f$ a bounded measurable function. Every
$P'\in R_\alpha(P)$ satisfies
\[
\frac{\E_Pf-\alpha\sup f}{1-\alpha}\ \le\ \E_{P'}f\ \le\ \frac{\E_Pf-\alpha\inf f}{1-\alpha} .
\]
\end{lemma}

\begin{proof}
Write $P'=gP$ with $0\le g\le\beta$ and $\E_Pg=1$. Then $\E_{P'}f=\beta\E_Pf-\E_P[(\beta-g)f]$, where $\beta-g\ge0$ and
$\E_P(\beta-g)=\beta-1=\alpha\beta$, so $\alpha\beta\inf f\le\E_P[(\beta-g)f]\le\alpha\beta\sup f$.
\end{proof}

These are the crude forms of Lemma~\ref{lem:cvar}: a trimming can at worst remove the mass $\alpha$ at the largest, or
the smallest, value of $f$. We write $\operatorname{osc}(f)=\sup f-\inf f$, $\mathcal F=\{f\in\Hh:0\le f\le1\}$, and, for
$\epsilon\in[0,\alpha^\star]$,
\[
\omega_\Hh(\epsilon)=\sup\Big\{\frac{(\epsilon-\mathrm{err}(f))_+}{\|f\|_\Hh}:\ f\in\mathcal F,\ f\neq0\Big\},
\qquad
\mathcal K(\lambda)=\inf_{f\in\mathcal F}\big\{\mathrm{err}(f)+\lambda\|f\|_\Hh\big\} ,
\]
with $\sup\varnothing=0$.

\begin{proposition}[Approximation margin]
\label{prop:newapproxgen}
Let (A1) and (A2) hold.
\begin{enumerate}[label=(\roman*),leftmargin=*]
\item For every $f\in\Hh$, $f\neq0$, and $\alpha\in[0,1)$,
$(1-\alpha)\|f\|_\Hh D_\alpha(P,Q)\ge\E_Pf-\E_Qf-\alpha\operatorname{osc}(f)$.
\item Let $0<\alpha^\star<1$. For every $f\in\mathcal F$, $\E_Pf-\E_Qf=\alpha^\star-\mathrm{err}(f)$, and if $f\neq0$, then
$\bar D_{\alpha^\star-\epsilon}\ge\{\epsilon-\mathrm{err}(f)\}/\|f\|_\Hh$ for every $\epsilon\in[0,\alpha^\star]$.
\item $\omega_\Hh$ is nondecreasing, $D_{\alpha^\star-\epsilon}(P,Q)\ge \bar D_{\alpha^\star-\epsilon}\ge\omega_\Hh(\epsilon)\ge
\epsilon/\{2\mathcal W(\epsilon/2)\}$, read as $0$ when $\mathcal W(\epsilon/2)=\infty$, and for $\lambda\ge0$ and
$\epsilon\in(0,\alpha^\star]$, $\omega_\Hh(\epsilon)>\lambda$ if and only if $\mathcal K(\lambda)<\epsilon$.
\end{enumerate}
\end{proposition}

\begin{proof}
Part (i) bounds each trimmed mean by Lemma~\ref{lem:newtrimmean}, and part (ii) evaluates the bound for
$[0,1]$-valued $f$. For (i), $f$ is bounded, by $\kappa\|f\|_\Hh$. For $P'\in R_\alpha(P)$ and $Q'\in R_\alpha(Q)$,
Lemma~\ref{lem:newtrimmean} gives $\E_{P'}f-\E_{Q'}f\ge\{\E_Pf-\E_Qf-\alpha\operatorname{osc}(f)\}/(1-\alpha)$, and by the
reproducing property and the Cauchy--Schwarz inequality $\MMD(P',Q')\ge(\E_{P'}f-\E_{Q'}f)/\|f\|_\Hh$, and the infimum over
$P'$ and $Q'$ gives (i).

For (ii), $\int\mathbf 1_A\,d(P-Q)=\int(p-q)_+=\alpha^\star$, so $\E_Pf-\E_Qf=\alpha^\star-\int(\mathbf 1_A-f)(p-q)\,d(P+Q)$. On $A$
both $1-f$ and $p-q$ are nonnegative, and off $A$ both $-f$ and $p-q$ are nonpositive, so the integrand is
$|\mathbf 1_A-f|\,|p-q|$ and the integral is $\mathrm{err}(f)$. As $\operatorname{osc}(f)\le1$, part (i) at
$\alpha=\alpha^\star-\epsilon$ gives $\|f\|_\Hh \bar D_{\alpha^\star-\epsilon}\ge\alpha^\star-\mathrm{err}(f)-(\alpha^\star-\epsilon)$.

For (iii), each map $\epsilon\mapsto(\epsilon-\mathrm{err}(f))_+/\|f\|_\Hh$ is nondecreasing, and so is their supremum,
which is at most $\bar D_{\alpha^\star-\epsilon}$ by (ii), and $\bar D_{\alpha}\le D_\alpha$. If $\mathcal W(\epsilon/2)<\infty$, every
$f\in\mathcal F$ with $\mathrm{err}(f)\le\epsilon/2$ is nonzero, because $\mathrm{err}(0)=\alpha^\star>\epsilon/2$, and has
$(\epsilon-\mathrm{err}(f))/\|f\|_\Hh\ge\epsilon/(2\|f\|_\Hh)$. The supremum over these $f$ is
$\epsilon/\{2\mathcal W(\epsilon/2)\}$. If $\mathcal K(\lambda)<\epsilon$, some $f\in\mathcal F$ has
$\mathrm{err}(f)+\lambda\|f\|_\Hh<\epsilon$, and it is nonzero because $\mathrm{err}(0)=\alpha^\star\ge\epsilon$, so
$\omega_\Hh(\epsilon)\ge(\epsilon-\mathrm{err}(f))/\|f\|_\Hh>\lambda$. Conversely, $\omega_\Hh(\epsilon)>\lambda$ gives a
nonzero $f\in\mathcal F$ with $\epsilon-\mathrm{err}(f)>\lambda\|f\|_\Hh$, that is, $\mathcal K(\lambda)<\epsilon$.
\end{proof}

\begin{proof}[Proof of Proposition~\ref{prop:newapprox}]
This is Proposition~\ref{prop:newapproxgen}(ii) and (iii), with $\bar D_{\alpha^\star-\epsilon}=(1-\alpha^\star+\epsilon)D_{\alpha^\star-\epsilon}$.
\end{proof}

\subsection{Rates under a general margin}

The rate of Theorem~\ref{thm:loratemain} extends to any margin, polynomial or not. For a nondecreasing $\omega$ we
write $\omega^{\leftarrow}(t)=\sup\{\epsilon\in[0,\alpha^\star]:\omega(\epsilon)\le t\}$.

\begin{proposition}[Rates under a general margin]
\label{prop:newrates}
Let Assumption~\ref{ass:regularity} hold, let $\alpha^\star<1$, compute $\widehat\alpha$ as in
Theorem~\ref{thm:loratemain}, and let $\omega:[0,\alpha^\star]\to[0,\infty)$ be nondecreasing with
$(1-\alpha^\star+\epsilon)\,D_{\alpha^\star-\epsilon}(P,Q)\ge\omega(\epsilon)$ for every $\epsilon\in[0,\alpha^\star]$.
\begin{enumerate}[label=(\roman*),leftmargin=*]
\item Parts (i) and (ii) of Theorem~\ref{thm:lorate} hold with $c_0\epsilon^\gamma$ replaced by
$\omega(\epsilon)$ and $(t_{n,m}/c_0)^{1/\gamma}$ by $\omega^{\leftarrow}(t_{n,m})$.
\item If moreover $\omega(\epsilon)\ge\exp(-c'/\epsilon)$ for $0<\epsilon\le\epsilon_0$, for some $c'>0$ and
$\epsilon_0\in(0,\alpha^\star]$, and the grid has at most $(n\wedge m)^b$ levels for a fixed $b>0$, then for every
$\eta\in(0,1)$ and every $c>2c'$ there is $n_0$, depending on $c,c',\epsilon_0,b,\eta,\delta,\kappa,\alpha^\star$ and
$v^2_{\min}$, such that
$\Pp\{\alpha^\star-\Delta_{\mathcal G}-c/\log(n\wedge m)\le\widehat\alpha\le\alpha^\star\}\ge1-\delta-\eta$
whenever $n\wedge m\ge n_0$.
\item Under the hypotheses of Proposition~\ref{prop:newgauss}, with a grid as in (ii) and no $\omega$ assumed, the
conclusion of (ii) holds for every $c>\pi^2\sigma^2\ell_0/4$.
\end{enumerate}
\end{proposition}

\begin{proof}[Proof of Proposition~\ref{prop:newrates}(i)]
In the proof of Theorem~\ref{thm:lorate}, the margin enters only through $D_*=c_0\epsilon^\gamma$ and two facts: $D_*\le
D_{\alpha^\star-\epsilon}$, and $D_{\alpha_j}\ge D_*x_j$ for the grid levels $\alpha_j\le\alpha^\star-\epsilon$, with
$x_j=(\alpha^\star-\alpha_j)/\epsilon\ge1$. That proof writes $D_{\alpha_j}\ge D_*x_j^\gamma$ and then uses only
$x_j^\gamma\ge x_j$, in the step $\varrho_jS-C\ge\varrho_*x_jS-C$. Put $D_*=\omega(\epsilon)$. Then
$D_*\le \bar D_{\alpha^\star-\epsilon}\le D_{\alpha^\star-\epsilon}$. By Lemma~\ref{lem:pathconvex}(ii), $\epsilon'\mapsto
\bar D_{\alpha^\star-\epsilon'}/\epsilon'$ is nondecreasing on $(0,\alpha^\star]$, and $\epsilon x_j=\alpha^\star-\alpha_j\in[\epsilon,\alpha^\star]$,
so
\[
D_{\alpha_j}\ \ge\ \bar D_{\alpha_j}\ =\ \bar D_{\alpha^\star-\epsilon x_j}\ \ge\ x_j\,\bar D_{\alpha^\star-\epsilon}\ \ge\ x_jD_* .
\]
The rest of the proof of part (i) of Theorem~\ref{thm:lorate} uses $D_*\le\Delta_\phi$, which holds since
$D_{\alpha^\star-\epsilon}\le D_0\le\Delta_\phi$, and the hypothesis on $D_*$, and is unchanged. For part (ii), every
$\epsilon\in(\omega^{\leftarrow}(t_{n,m}),\alpha^\star]$ has $\omega(\epsilon)>t_{n,m}$ and $J_\epsilon\le J$, so part (i) gives
$\Pp(\widehat\alpha<\alpha^\star-\epsilon-\Delta_{\mathcal G})\le\Pp(\widehat\alpha<a_\epsilon)\le\eta$. Along a sequence
$\epsilon_k\downarrow\omega^{\leftarrow}(t_{n,m})$ these events increase to
$\{\widehat\alpha<\alpha^\star-\omega^{\leftarrow}(t_{n,m})-\Delta_{\mathcal G}\}$, whose probability is therefore at most $\eta$, and
Corollary~\ref{cor:lower} and a union bound finish the proof as in Theorem~\ref{thm:lorate}.
\end{proof}

Applied to a polynomial margin of $D$, part (i) loses a factor $(1-\alpha^\star)^{1/\gamma}$ against
Theorem~\ref{thm:lorate}, since a margin of $D$ gives one of $\bar D$ only after multiplication by $1-\alpha^\star$. Its use is
for margins that are not polynomial, such as the exponential margins of part (ii). Part (iii) is proved after
Theorem~\ref{thm:newgauss}.

\begin{proof}[Proof of Proposition~\ref{prop:newrates}(ii)]
Write $\underline n=n\wedge m$ and let $\underline n\ge10$. By Theorem~\ref{thm:lorate}(iii), the quantity $t_{n,m}$
of its part (ii) is at most an absolute constant times
$\beta_\star\kappa v_{\min}^{-2}\{(1+L_\delta+\log(J/\eta))/\underline n\}^{1/2}$, and with $J\le\underline n^b$
this is at most $C_t(\log\underline n/\underline n)^{1/2}$, with $C_t$ depending only on $\kappa$, $\alpha^\star$,
$v_{\min}$, $\delta$, $\eta$ and $b$. Let $0<t<\min\{1,\omega(\epsilon_0)\}$. As $\omega$ is nondecreasing, every
$\epsilon\in[0,\alpha^\star]$ with $\omega(\epsilon)\le t$ has $\epsilon<\epsilon_0$, and then, if $\epsilon>0$,
$\exp(-c'/\epsilon)\le t$, that is, $\epsilon\le c'/\log(1/t)$; so $\omega^{\leftarrow}(t)\le c'/\log(1/t)$. For
$t=t_{n,m}$, $\log(1/t_{n,m})\ge\frac12\log\underline n-\frac12\log\log\underline n-\log C_t$, which exceeds
$(c'/c)\log\underline n$ for all large $\underline n$, because $c'/c<\frac12$; then also $t_{n,m}<\min\{1,\omega(\epsilon_0)\}$,
and $\omega^{\leftarrow}(t_{n,m})\le c/\log\underline n$. Part (i), in the form of part (ii) of
Theorem~\ref{thm:lorate}, gives the claim.
\end{proof}

\subsection{Sobolev kernels}

Let $\|f\|_{H^s}^2=(2\pi)^{-d}\int|\widehat f(\xi)|^2(1+|\xi|^2)^s\,d\xi$. If
$\widehat{k_0}(\xi)\ge c(1+|\xi|^2)^{-s}$, then by \eqref{eq:newnative} every continuous $f\in L^2$ with
$\|f\|_{H^s}<\infty$ lies in $\Hh$, with $\|f\|_\Hh^2\le\|f\|_{H^s}^2/c$. Such a kernel is characteristic, as
$\widehat{k_0}>0$ \citep{sriperumbudur2010hilbert}. The approximant is a mollified indicator. For a set $E$ we use the
translation bound $\mathrm{Leb}(E\,\triangle\,(E+h))\le\Pi|h|$ for all $h\in\R^d$, which holds with $\Pi$ the perimeter of $E$
\citep[Chapter~5]{evans2015measure}, and, in $\R$, with $\Pi=2N$ for a union of $N$ disjoint intervals.

\begin{lemma}[Norm of a mollified indicator]
\label{lem:newsobnorm}
Let $s>1/2$, let $E\subset\R^d$ be bounded and measurable with $\mathrm{Leb}(E\,\triangle\,(E+h))\le\Pi|h|$ for all $h$, let
$b\in C_c^\infty$ be a probability density supported in the closed unit ball, $b_w=w^{-d}b(\cdot/w)$, and
$f_w=\mathbf 1_E*b_w$. Let $m$ be the smallest integer with $4m>2s-1$, $C_b=\|\Delta^mb\|_{L^1}$,
$j_d(y)$ the mean of $\cos(yv_1)$ for $v$ uniform in the unit ball, and $c_d=1-\sup_{y\ge1}j_d(y)$. Then $c_d>0$,
$0\le f_w\le1$, $f_w\in C_c^\infty$, and for $0<w\le1$
\[
\|f_w\|_{H^s}^2\ \le\ 2^s\,\mathrm{Leb}(E)+C_{d,s,b}\,\Pi\,w^{1-2s},
\qquad
C_{d,s,b}=\frac{2^{3s}}{2c_d}\Big[\frac1{1-2^{1-2s}}+\frac{C_b^2}{1-2^{-(4m+1-2s)}}\Big].
\]
\end{lemma}

\begin{proof}
The indicator has a Fourier tail of order $1/t$ beyond frequency $t$, because translating $E$ by $h$ changes it on a set of
measure $\Pi|h|$, and mollification at width $w$ cuts the frequencies beyond $1/w$. Write $g=\mathbf 1_E$, so that
$\widehat f_w(\xi)=\widehat g(\xi)\widehat b(w\xi)$. As $b$ is a probability density, $|\widehat b|\le1$, and
$(\Delta^mb)^\wedge(\xi)=(-|\xi|^2)^m\widehat b(\xi)$ gives $|\widehat b(\xi)|\le\min(1,C_b|\xi|^{-2m})$.

We first bound the Fourier tail of $g$. For $h\in\R^d$, Plancherel's identity gives
\[
(2\pi)^{-d}\int|\widehat g(\xi)|^2|e^{ih\cdot\xi}-1|^2\,d\xi=\|g(\cdot+h)-g\|_{L^2}^2=\mathrm{Leb}(E\,\triangle\,(E-h))\le\Pi|h| ,
\]
since $g(\cdot+h)-g$ takes values in $\{-1,0,1\}$. To isolate the high frequencies we average over $h$ uniform in the
ball of radius $1/t$. The mean of $|e^{ih\cdot\xi}-1|^2=2-2\cos(h\cdot\xi)$ is $2\{1-j_d(|\xi|/t)\}$ by rotation
invariance, and the mean of $|h|$ is at most $1/t$. The function $j_d$ is continuous, smaller than $1$ on $(0,\infty)$,
since $\cos(yv_1)<1$ off a null set, and tends to $0$ at infinity by the Riemann--Lebesgue lemma, so $c_d>0$; in $\R$,
$j_1(y)=\sin y/y$ and $c_1=1-\sin1$. As $1-j_d\ge0$ everywhere and $1-j_d(|\xi|/t)\ge c_d$ for $|\xi|\ge t$, the
averaged identity gives
\[
\Upsilon(t)\coloneqq(2\pi)^{-d}\int_{|\xi|\ge t}|\widehat g(\xi)|^2\,d\xi\ \le\ \frac{\Pi}{2c_d\,t},\qquad t>0 .
\]
We now split the frequencies into the unit ball and dyadic shells. On $\{|\xi|<1\}$, $(1+|\xi|^2)^s\le2^s$ and
$|\widehat b|\le1$, which contributes at most $2^s(2\pi)^{-d}\int|\widehat g|^2=2^s\mathrm{Leb}(E)$, the first term of
the bound. On the shell $S_j=\{2^j\le|\xi|<2^{j+1}\}$, $j\ge0$, we have $(1+|\xi|^2)^s\le(2|\xi|^2)^s\le2^{3s}4^{sj}$,
$|\widehat b(w\xi)|^2\le\min\{1,C_b^2(w2^j)^{-4m}\}$ and $(2\pi)^{-d}\int_{S_j}|\widehat g|^2\le\Upsilon(2^j)\le\Pi2^{-j}/(2c_d)$,
so the shells contribute at most $\{2^{3s}\Pi/(2c_d)\}\sum_{j\ge0}2^{j(2s-1)}\min\{1,C_b^2w^{-4m}2^{-4mj}\}$. The sum grows geometrically up to the frequency $1/w$ and decays beyond it. Let $j_0\ge0$ be the integer with $2^{j_0}\le1/w<2^{j_0+1}$. As $2s-1>0$,
$\sum_{j\le j_0}2^{j(2s-1)}\le2^{j_0(2s-1)}/(1-2^{1-2s})\le w^{1-2s}/(1-2^{1-2s})$, and as $2s-1-4m<0$,
\[
\sum_{j>j_0}C_b^2w^{-4m}2^{j(2s-1-4m)}\le\frac{C_b^2w^{-4m}2^{(j_0+1)(2s-1-4m)}}{1-2^{2s-1-4m}}\le\frac{C_b^2w^{1-2s}}{1-2^{-(4m+1-2s)}} .
\]
Adding the parts gives the bound, the two geometric series giving the two terms of $C_{d,s,b}$. Finally $0\le f_w\le1$ because $0\le g\le1$ and $b_w$ is a probability density, and
$f_w$ is smooth with compact support because $b$ is and $E$ is bounded.
\end{proof}

The laws enter through an open set $G$ that carries the sign of $p-q$, the size of its boundary near the contact, and
the order at which $p-q$ vanishes there. We use the open balls $B_\rho$ of radius $\rho$ about a fixed centre.

\begin{theorem}[Margin for Sobolev kernels]
\label{thm:newsobolev}
Let $k(x,y)=k_0(x-y)$ on $\R^d$ with $\widehat{k_0}(\xi)\ge c(1+|\xi|^2)^{-s}$, $s>d/2$, and let $P,Q$ have Lebesgue
densities $p,q$ with $0<\alpha^\star<1$. Let $G\subseteq\R^d$ be open with $p\ge q$ almost everywhere on $G$ and $p\le q$
almost everywhere off $G$, and let $\rho\ge1$, $r>0$, $\ell>0$ and $w_0\in(0,1]$ be such that
\begin{enumerate}[label=(G\arabic*),leftmargin=*]
\item $\mathrm{Leb}(E_\rho\,\triangle\,(E_\rho+h))\le\Pi_\rho|h|$ for all $h$, where $E_\rho=G\cap B_\rho$.
\item $\mathrm{Leb}(\{x\in B_\rho:\dist(x,\partial G)\le t\})\le V_\rho t$ for $0<t\le w_0$.
\item $|p(x)-q(x)|\le\ell\,\dist(x,\partial G)^r$ for almost every $x\in B_\rho$ with $\dist(x,\partial G)\le w_0$.
\end{enumerate}
Let $T(\rho)=|P-Q|(\R^d\setminus B_{\rho-1})$, and for $0<w\le w_0$ let $f_w=\mathbf 1_{E_\rho}*b_w$ as in
Lemma~\ref{lem:newsobnorm}. Then $f_w\in\Hh$, $0\le f_w\le1$, $\mathrm{err}(f_w)\le\ell V_\rho w^{r+1}+T(\rho)$, and for every
$\epsilon\in[0,\alpha^\star]$
\[
D_{\alpha^\star-\epsilon}(P,Q)\ \ge\ \bar D_{\alpha^\star-\epsilon}\ \ge\ \frac{c^{1/2}\,\{\epsilon-\ell V_\rho w^{r+1}-T(\rho)\}_+}
{\{2^s\mathrm{Leb}(B_\rho)+C_{d,s,b}\Pi_\rho w^{1-2s}\}^{1/2}} .
\]
\end{theorem}

\begin{proof}
The function $f_w$ is the approximate witness of Proposition~\ref{prop:newapproxgen}(ii), so we need its norm and its
error. Lemma~\ref{lem:newsobnorm} and \eqref{eq:newnative} give $f_w\in\Hh$, $0\le f_w\le1$ and
$\|f_w\|_\Hh^2\le c^{-1}\{2^s\mathrm{Leb}(B_\rho)+C_{d,s,b}\Pi_\rho w^{1-2s}\}$, as $E_\rho\subseteq B_\rho$. For the error,
since $\mathbf 1_G=\mathbf 1_A$ almost everywhere
for $|P-Q|$, $\mathrm{err}(f_w)=\int|\mathbf 1_G-f_w|\,d|P-Q|$. Off $B_{\rho-1}$ we use $|\mathbf 1_G-f_w|\le1$, which contributes at
most $T(\rho)$. Inside $B_{\rho-1}$ the mollified indicator differs from $\mathbf 1_G$ only near $\partial G$. Let
$x\in B_{\rho-1}$. The density $b_w$ is supported in the closed ball of radius $w\le1$ about $0$, and the
closed ball $\bar B(x,w)$ lies in $B_\rho$, so $f_w(x)=\int\mathbf 1_G(x-y)b_w(y)\,dy$. If $\dist(x,\partial G)>w$, the connected
set $\bar B(x,w)$ misses $\partial G$, so it lies in the open set $G$ or in the open set $\R^d\setminus\bar G$, and
$f_w(x)=\mathbf 1_G(x)$. Hence $|\mathbf 1_G-f_w|\le\mathbf 1\{\dist(\cdot,\partial G)\le w\}$ on $B_{\rho-1}$, and (G2) and (G3) give
$\int_{B_{\rho-1}}|\mathbf 1_G-f_w|\,|p-q|\le\ell w^r\,\mathrm{Leb}(\{x\in B_\rho:\dist(x,\partial G)\le w\})\le\ell V_\rho w^{r+1}$.
Proposition~\ref{prop:newapproxgen}(ii) finishes the proof.
\end{proof}

\begin{corollary}[Exponent of the Sobolev margin]
\label{cor:newsobolev}
In the setting of Theorem~\ref{thm:newsobolev} put $\gamma=1+(2s-1)/\{2(r+1)\}$. Suppose either that $T(\rho)=0$, that
is, $|P-Q|$ vanishes off $B_{\rho-1}$, or that $d=1$, $G=(x_0,\infty)$, the balls are centered at $x_0$, (G3) holds for
every $\rho$ with the same $\ell$ and $w_0$, and $T(\rho)\le C_Te^{-\rho/a}$ for some $C_T,a>0$ and all $\rho\ge1$. Then there
are $c_1>0$ and $\epsilon_1>0$ with $\bar D_{\alpha^\star-\epsilon}\ge c_1\epsilon^\gamma$ for $\epsilon\le\epsilon_1$, and
Assumption~\ref{ass:margin} holds with this $\gamma$. Hence, with $\widehat\alpha$ computed as in
Theorem~\ref{thm:loratemain} on a grid of $J$ levels fixed in advance with mesh $\Delta_{\mathcal G}$ below
$\alpha^\star$, $\widehat\alpha\ge\alpha^\star-\Delta_{\mathcal G}-O_\Pp((n\wedge m)^{-1/(2\gamma)})$
for fixed $J$, with $1/(2\gamma)=(r+1)/\{2(r+1)+2s-1\}$, equal to $2/(2s+3)$ for $r=1$.
\end{corollary}

\begin{proof}
The width $w$ is chosen so that the error of $f_w$ is half of $\epsilon$, which balances error and norm in
Theorem~\ref{thm:newsobolev}. If $T(\rho)=0$, write $V$ and $\Pi$ for $V_\rho$ and $\Pi_\rho$, put
$w(\epsilon)=\{\epsilon/(2\ell V)\}^{1/(r+1)}$, so that $\ell Vw^{r+1}=\epsilon/2$, and let $w_1\in(0,w_0]$ be such that
$C_{d,s,b}\Pi w^{1-2s}\ge2^s\mathrm{Leb}(B_\rho)$ for $w\le w_1$, so that the second term of the denominator dominates. For
$\epsilon\le\epsilon_1=2\ell Vw_1^{r+1}$, Theorem~\ref{thm:newsobolev} gives
$\bar D_{\alpha^\star-\epsilon}\ge c^{1/2}(\epsilon/2)(2C_{d,s,b}\Pi)^{-1/2}w(\epsilon)^{s-1/2}=c_1\epsilon^\gamma$, the exponent being $1+(s-\tfrac12)/(r+1)=\gamma$. In the second case,
$E_\rho=(x_0,x_0+\rho)$, so $\Pi_\rho=2$, and $\{x:\dist(x,\partial G)\le t\}=[x_0-t,x_0+t]$, so $V_\rho=2$. Take
$\rho(\epsilon)=\max\{2,1+a\log(4C_T/\epsilon)\}$, so that $T(\rho(\epsilon))\le\epsilon/4$, and $w(\epsilon)=\{\epsilon/(8\ell)\}^{1/(r+1)}$, so
that $2\ell w^{r+1}=\epsilon/4$. The denominator in Theorem~\ref{thm:newsobolev} is
$\{2^{s+1}\rho(\epsilon)+2C_{d,s,b}w(\epsilon)^{1-2s}\}^{1/2}$, in which the second term, a power of $1/\epsilon$, dominates the
first, of order $\log(1/\epsilon)$, for small $\epsilon$. This gives $c_1$ and $\epsilon_1$ as before. In both cases, for
$\epsilon\in[\epsilon_1,\alpha^\star]$ the path is nonincreasing, so $D_{\alpha^\star-\epsilon}\ge \bar D_{\alpha^\star-\epsilon_1}\ge
c_1(\epsilon_1/\alpha^\star)^\gamma\epsilon^\gamma$, and Assumption~\ref{ass:margin} holds with $c_0=c_1(\epsilon_1/\alpha^\star)^\gamma$, since
$D\ge \bar D$. The rate is Theorem~\ref{thm:lorate}, whose part (i) alone, with parts (ii) and (iii) for the size of
$t_{n,m}$, gives the bound with probability at least $1-\eta$.
\end{proof}

A transversal crossing of a compact $C^2$ hypersurface $\partial G$, with $p-q$ continuously differentiable near it and
$|P-Q|$ compactly supported, satisfies (G1)--(G3) with $r=1$. Indeed $\Pi_\rho$ is at most the sum of the areas of $\partial G$
and $\partial B_\rho$, by the submodularity of the perimeter, $V_\rho$ is finite by the tube formula, and $\ell$ can be
taken as a Lipschitz constant of $p-q$ near $\partial G$.

\subsection{The Gaussian kernel}

For $k(x,y)=\exp\{-(x-y)^2/(2\sigma^2)\}$ on $\R$, $\widehat{k_0}(\xi)=(2\pi)^{1/2}\sigma e^{-\sigma^2\xi^2/2}$, so by
\eqref{eq:newnative} and Plancherel's identity every continuous $f\in L^2$ whose Fourier transform vanishes outside
$[-\xi_0,\xi_0]$ lies in $\Hh$, with
\begin{equation}
\|f\|_\Hh^2\le\frac{e^{\sigma^2\xi_0^2/2}}{(2\pi)^{1/2}\sigma}\,\|f\|_{L^2}^2 ,
\label{eq:newbandlim}
\end{equation}
the elementary half of the description of this RKHS by \citet{steinwart2006explicit}. Let $b$ be an even probability
density with $\widehat b=0$ outside $[-1,1]$ and $m_2=\int t^2b(t)\,dt<\infty$, let $\bar b(t)=\int_t^\infty b$ for $t\ge0$, so that
$\bar b\le1/2$, and $b_{\xi_0}(t)=\xi_0 b(\xi_0 t)$. The density $b(t)=4\pi\cos^2(t/2)/(\pi^2-t^2)^2$ qualifies: it is $h^2/\|h\|_{L^2}^2$
for $h(t)=\cos(t/2)/(\pi^2-t^2)$, whose Fourier transform is $\cos(\pi\xi)$ on $[-1/2,1/2]$ and $0$ elsewhere, so that
$\widehat b$, a multiple of the autocorrelation of $\widehat h$, vanishes outside $[-1,1]$. Moreover $\|h\|_{L^2}^2=1/(4\pi)$ and
$m_2=\int|\widehat h'|^2/\int|\widehat h|^2=\pi^2$. For $t\ge2^{1/2}\pi$, $t^2-\pi^2\ge t^2/2$, so $b(t)\le16\pi t^{-4}$ and
$\bar b(t)\le16\pi/(3t^3)$.

The laws cross once, at $x_0$: $p\ge q$ almost everywhere on $(x_0,\infty)$ and $p\le q$ almost everywhere on
$(-\infty,x_0)$, so that $A=(x_0,\infty)$ up to a null set of $|P-Q|$, and $|p-q|(x)\le\ell|x-x_0|$ for almost every
$x$ with $|x-x_0|\le r_0$, for some $\ell>0$ and $r_0\in(0,\infty]$; we read $\bar b(\infty)$ as $0$. We write
$T_+(\rho)=|P-Q|((x_0+\rho,\infty))$.

\begin{lemma}[A band-limited step]
\label{lem:newgaussstep}
For $\xi_0>0$ and $\rho>0$, $f=\mathbf 1_{(x_0,x_0+\rho)}*b_{\xi_0}$ lies in $\Hh$, $0\le f\le1$,
$\|f\|_\Hh^2\le\rho\,e^{\sigma^2\xi_0^2/2}/\{(2\pi)^{1/2}\sigma\}$, and
\[
\mathrm{err}(f)\ \le\ \frac{\ell m_2}{2\xi_0^2}+2T_+(\rho/2)+2\alpha^\star\{\bar b(\xi_0 r_0)+\bar b(\xi_0\rho/2)\} .
\]
\end{lemma}

\begin{proof}
Let $g=\mathbf 1_{(x_0,x_0+\rho)}$. As $b_{\xi_0}$ is a probability density, $f$ is continuous, in $L^2$, with
values in $[0,1]$, and $\widehat f=\widehat g\,\widehat b(\cdot/\xi_0)$ vanishes outside $[-\xi_0,\xi_0]$. Since $|\widehat b|\le1$,
$\|f\|_{L^2}\le\|g\|_{L^2}=\rho^{1/2}$, and \eqref{eq:newbandlim} gives the norm bound.

For the error we compare $f$ with $g$, and $g$ with $\mathbf 1_A$: $|\mathbf 1_A-f|\le|\mathbf 1_A-g|+|g-f|$, and $\mathbf 1_A-g=\mathbf 1_{[x_0+\rho,\infty)}$ almost everywhere, which
contributes $T_+(\rho)\le T_+(\rho/2)$. Next, $|g-f|(x)\le\int|g(x)-g(x-y)|\,b_{\xi_0}(y)\,dy$. For $x_0<x<x_0+\rho$, $g(x-y)\neq g(x)$
exactly when $y\ge x-x_0$ or $y\le x-x_0-\rho$, a set of $b_{\xi_0}$-mass $\bar b(\xi_0(x-x_0))+\bar b(\xi_0(x_0+\rho-x))$. For
$x<x_0$ only $y<-|x-x_0|$ can change $g$, and for $x>x_0+\rho$ only $y>x-x_0-\rho$. Hence, for almost every $x$, $|g-f|(x)\le\bar b(\xi_0|x-x_0|)+\bar b(\xi_0|x-x_0-\rho|)$, the leakage of the mollifier across the two ends of the interval. Integrated against $|P-Q|$, it is at most
\begin{align*}
\ell\int_{|t|\le r_0}|t|\,\bar b(\xi_0|t|)\,dt+\bar b(\xi_0 r_0)\,|P-Q|(\R)
&\le\frac{2\ell}{\xi_0^2}\int_0^\infty v\,\bar b(v)\,dv+2\alpha^\star\bar b(\xi_0 r_0)\\
&=\frac{\ell m_2}{2\xi_0^2}+2\alpha^\star\bar b(\xi_0 r_0),
\end{align*}
because $\bar b$ is nonincreasing and $\int_0^\infty v\,\bar b(v)\,dv=\int_0^\infty b(y)\,y^2/2\,dy=m_2/4$. The second term is at most
$1/2$ on $(x_0+\rho/2,\infty)$, a set of $|P-Q|$-mass $T_+(\rho/2)$, and at most $\bar b(\xi_0\rho/2)$ elsewhere, where
$|x-x_0-\rho|\ge\rho/2$. Adding the contributions gives the bound.
\end{proof}

\begin{theorem}[Margin for the Gaussian kernel]
\label{thm:newgauss}
In this setting let $\theta\in(0,1)$, $\epsilon\in(0,\alpha^\star]$ and $\xi_\epsilon=\{\ell m_2/(2\theta\epsilon)\}^{1/2}$, and let
$\rho>0$ satisfy $2T_+(\rho/2)+2\alpha^\star\{\bar b(\xi_\epsilon r_0)+\bar b(\xi_\epsilon\rho/2)\}\le(1-\theta)\epsilon/2$. Then
\[
D_{\alpha^\star-\epsilon}(P,Q)\ \ge\ \bar D_{\alpha^\star-\epsilon}\ \ge\ \frac{(1-\theta)\epsilon}2\Big\{\frac{(2\pi)^{1/2}\sigma}{\rho}\Big\}^{1/2}
\exp\Big(-\frac{\sigma^2\ell m_2}{8\theta\epsilon}\Big) .
\]
If $T_+(t)\le C_Te^{-t/a}$, then $\rho_\epsilon=\max\{2,2a\log(8C_T/((1-\theta)\epsilon))\}$ satisfies the condition for all small
$\epsilon$ when $b$ is the density above, and $\log\{1/\bar D_{\alpha^\star-\epsilon}\}\le\sigma^2\ell m_2/(8\theta\epsilon)+\log(1/\epsilon)+O(\log\log(1/\epsilon))$.
\end{theorem}

\begin{proof}
The bandwidth $\xi_\epsilon$ is chosen so that the smoothing error near $x_0$ is the fraction $\theta$ of $\epsilon$,
and the norm of the band-limited step then grows like $\exp\{\sigma^2\xi_\epsilon^2/4\}$. By
Lemma~\ref{lem:newgaussstep} with $\xi_0=\xi_\epsilon$, $\ell m_2/(2\xi_\epsilon^2)=\theta\epsilon$, so
$\mathrm{err}(f)\le\theta\epsilon+(1-\theta)\epsilon/2$ and $\epsilon-\mathrm{err}(f)\ge(1-\theta)\epsilon/2$, while
$\|f\|_\Hh\le\{\rho/((2\pi)^{1/2}\sigma)\}^{1/2}\exp\{\sigma^2\ell m_2/(8\theta\epsilon)\}$, and Proposition~\ref{prop:newapproxgen}(ii) gives
the bound. With the density above, $\bar b(\xi_\epsilon t)\le16\pi/\{3(\xi_\epsilon t)^3\}$ once $\xi_\epsilon t\ge2^{1/2}\pi$, which is
of order $\epsilon^{3/2}$ for $t=r_0$ and for $t=\rho_\epsilon/2\ge1$, so the two terms in $\bar b$ are at most $(1-\theta)\epsilon/4$
for small $\epsilon$, and $2T_+(\rho_\epsilon/2)\le2C_T\exp\{-\log(8C_T/((1-\theta)\epsilon))\}=(1-\theta)\epsilon/4$.
\end{proof}

\begin{proof}[Proof of Proposition~\ref{prop:newgauss}]
The hypotheses of Proposition~\ref{prop:newsobolev} give the setting of Theorem~\ref{thm:newgauss}, with exponential tails,
for every $\ell>\ell_0$ and a small enough $r_0$: $(p-q)(x)/(x-x_0)$ is continuous and nonnegative near $x_0$, away from $x_0$,
and tends to $\ell_0$ at $x_0$. Given $c>\pi^2\sigma^2\ell_0/8$, choose $\theta<1$ and $\ell>\ell_0$ with $\pi^2\sigma^2\ell/(8\theta)<c$. Then the
bound of Theorem~\ref{thm:newgauss}, with $m_2=\pi^2$, exceeds $\exp(-c/\epsilon)$ for small $\epsilon$. The upper bound is
Proposition~\ref{prop:newmoments}(b) below with $2j$ vanishing moments, which gives the order $\epsilon^{1+j}$, as
$M_{4j}=(4j-1)!!\,\sigma^{-4j}<\infty$ for every $j$.
\end{proof}

\begin{proof}[Proof of Proposition~\ref{prop:newrates}(iii)]
Given $c>\pi^2\sigma^2\ell_0/4$, take $c'\in(\pi^2\sigma^2\ell_0/8,c/2)$. By the proof of Proposition~\ref{prop:newgauss},
the bound of Theorem~\ref{thm:newgauss}, with $m_2=\pi^2$, $\rho=\rho_\epsilon$ and suitable $\theta<1$ and
$\ell>\ell_0$, exceeds $\exp(-c'/\epsilon)$ for $0<\epsilon\le\epsilon_0$, for some $\epsilon_0\in(0,\alpha^\star]$. The
proof of Theorem~\ref{thm:newgauss} obtains this bound as a lower bound on $(\epsilon-\mathrm{err}(f))/\|f\|_\Hh$ for a
function $f\in\mathcal F$, so it is also a lower bound on $\omega_\Hh(\epsilon)$. By
Proposition~\ref{prop:newapproxgen}(iii), $\omega_\Hh$ is nondecreasing and at most
$\bar D_{\alpha^\star-\epsilon}=(1-\alpha^\star+\epsilon)D_{\alpha^\star-\epsilon}$, so part (ii) applies to $\omega=\omega_\Hh$,
and $c>2c'$.
\end{proof}

\subsection{Explicit trimmings and sharpness}

Upper bounds on the path come from explicit trimmings: nonnegative measures $P'\le P$ and $Q'\le Q$ of equal mass
$1-\alpha$ give $\bar D_{\alpha}\le\|\int\phi\,d(P'-Q')\|_\Hh$. If $\nu=P'-Q'$ has total variation $|\nu|$, support in $[x_0-W,x_0+W]$
and $\int(x-x_0)^j\,d\nu=0$ for $0\le j<N$, where $N\ge1$ and $j=0$ holds because the masses agree, then
$|e^{it}-\sum_{j<N}(it)^j/j!|\le|t|^N/N!$ gives
\begin{equation}
|\widehat\nu(\xi)|\le|\nu|\min\{1,(|\xi|W)^N/N!\} .
\label{eq:newmomentdecay}
\end{equation}

\begin{proposition}[Sobolev order below $3/2$]
\label{prop:newupperone}
Let $d=1$, $k(x,y)=k_0(x-y)$ with $\widehat{k_0}(\xi)\le C|\xi|^{-2s}$ for some $s\in(1/2,3/2)$, and suppose that
$p-q\ge\ell'(x-x_0)^r$ almost everywhere on $(x_0,x_0+r_0)$ and $q-p\ge\ell'(x_0-x)^r$ almost everywhere on $(x_0-r_0,x_0)$,
for some $\ell',r,r_0>0$. Then for $0<\epsilon\le\ell'r_0^{r+1}/(r+1)$, with $W_\epsilon=\{(r+1)\epsilon/\ell'\}^{1/(r+1)}$,
\[
\bar D_{\alpha^\star-\epsilon}\ \le\ 2\epsilon\,W_\epsilon^{s-1/2}\Big\{\frac C{2\pi}\Big(\frac2{3-2s}+\frac2{2s-1}\Big)\Big\}^{1/2} .
\]
\end{proposition}

\begin{proof}
We keep the common part together with the excess of $P$ on a short interval to the right of $x_0$ and that of $Q$ on
one to the left, so that the two trimmed parts differ by a measure of mass zero carried near $x_0$, whose Fourier
transform vanishes at the origin. The map $w\mapsto\int_{x_0}^{x_0+w}(p-q)$ is continuous, nondecreasing and at least $\ell'w^{r+1}/(r+1)$ on $[0,r_0]$, so some
$w_P\le W_\epsilon$ gives it the value $\epsilon$, and likewise some $w_Q\le W_\epsilon$ on the left. As $p-q=(p-q)_+$ almost everywhere
on $(x_0,x_0+w_P)$, the measure $P'=P\wedge Q+(p-q)\mathbf 1_{(x_0,x_0+w_P)}\,dx$ satisfies $P'\le P\wedge Q+(P-P\wedge Q)=P$ and has mass
$1-\alpha^\star+\epsilon$. The measure $Q'$ is built in the same way, and $\nu=P'-Q'=(p-q)\mathbf 1_{(x_0-w_Q,x_0+w_P)}\,dx$ has $|\nu|=2\epsilon$, mass
$0$ and support in $[x_0-W_\epsilon,x_0+W_\epsilon]$. By \eqref{eq:newbochner} and \eqref{eq:newmomentdecay} with $N=1$,
$\|\int\phi\,d\nu\|_\Hh^2\le(2\pi)^{-1}C(2\epsilon)^2\int\min(1,\xi^2W_\epsilon^2)|\xi|^{-2s}\,d\xi$, and the integral is
$W_\epsilon^{2s-1}\{2/(3-2s)+2/(2s-1)\}$.
\end{proof}

For larger $s$, and for the Gaussian kernel, the difference of the trimmings must have more vanishing moments. They
are created by removing from the common part a small polynomial density near $x_0$, which needs the core to have
positive density there.

\begin{proposition}[Trimmings with vanishing moments]
\label{prop:newmoments}
Let $d=1$, let $p,q$ be continuous on $(x_0-r_0,x_0+r_0)$, let $e=p-q$ be differentiable at $x_0$ with
$e'(x_0)=\ell_0>0$, and let $p(x_0)=q(x_0)>0$. For every integer $N\ge1$ there are $T_N\ge1$ and $w_N>0$ such that for
$0<w\le w_N$ there are nonnegative measures $P'\le P$ and $Q'\le Q$ of equal mass $1-\alpha^\star+\epsilon(w)$, with
$\epsilon(w)\ge\ell_0w^2/8$, whose difference $\nu$ satisfies $|\nu|\le2\ell_0w^2$, $\supp\nu\subseteq[x_0-T_Nw,x_0+T_Nw]$ and
$\int(x-x_0)^j\,d\nu=0$ for $0\le j<N$. Consequently, with $w=(8\epsilon/\ell_0)^{1/2}$ and $0<\epsilon\le\ell_0w_N^2/8$:
\begin{enumerate}[label=(\alph*),leftmargin=*]
\item if $\widehat{k_0}(\xi)\le C(1+\xi^2)^{-s}$ with $s>1/2$ and $N>s-1/2$, then
$\bar D_{\alpha^\star-\epsilon}\le2\ell_0w^2(T_Nw)^{s-1/2}\{CI_{N,s}/(2\pi)\}^{1/2}$, which is at most a constant times
$\epsilon^{(2s+3)/4}$, where $I_{N,s}=\int\min(1,v^{2N}/N!^2)|v|^{-2s}\,dv<\infty$.
\item if $M_{2N}=(2\pi)^{-1}\int\xi^{2N}\widehat{k_0}(\xi)\,d\xi<\infty$, then
$\bar D_{\alpha^\star-\epsilon}\le2\ell_0w^2(T_Nw)^NM_{2N}^{1/2}/N!$, which is at most a constant times $\epsilon^{1+N/2}$. For the
Gaussian kernel $M_{2N}=(2N-1)!!\,\sigma^{-2N}$.
\end{enumerate}
\end{proposition}

\begin{proof}
We add the excess $e$ on $(x_0-w,x_0+w)$ to the common part and cancel its first $N$ moments with a polynomial density
removed from the common part on a fixed multiple of that interval. For $|t|<1$ let $F_w(t)=e(x_0+wt)/(\ell_0w)$, and
$F_w(t)=0$ otherwise. Differentiability at $x_0$ gives $\sup_t|F_w(t)-t\mathbf 1\{|t|<1\}|\to0$ as $w\to0$.

Let $T\ge1$ and let $\mathcal G_T$ map $M\in\R^N$ to the polynomial $G$ of degree less than $N$ with
$\int_{-T}^Tt^jG(t)\,dt=M_j$ for $0\le j<N$, which exists and is unique because the Gram matrix of the monomials on
$[-T,T]$ is positive definite. The map $\mathcal G_T$ is linear. Put $M_j(w)=-\int t^jF_w(t)\,dt$, so that $M(w)\to M(0)$,
with $M_0(0)=0$ and $|M_j(0)|=|\int_{-1}^1t^{j+1}dt|\le2/3$. For $G_0=\mathcal G_T[M(0)]$, the polynomial $v\mapsto TG_0(Tv)$ has
moments $T^{-j}M_j(0)$ on $[-1,1]$, so it is $\mathcal G_1$ applied to a vector of Euclidean norm at most
$(2/3)N^{1/2}/T$, as $M_0(0)=0$. Hence $\int_{-T}^T|G_0|=\int_{-1}^1T|G_0(Tv)|\,dv\le c_N/T$ with
$c_N=(2/3)N^{1/2}\|\mathcal G_1\|$, the operator norm from $\R^N$ to $L^1[-1,1]$. Fix $T=T_N=\max(1,8c_N)$, so that
$\int|G_0|\le1/8$, and let $B_N=\sup_{[-T,T]}|G_0|+1$.

The polynomials $G_w=\mathcal G_T[M(w)]$ converge to $G_0$ uniformly on $[-T,T]$. Choose $w_N>0$ so small that for $w\le w_N$ the following hold. First, $\int|G_w|\le1/4$
and $\sup|G_w|\le B_N$. Next, $Tw\le r_0$, $\min(p,q)\ge p(x_0)/2$ on $[x_0-Tw,x_0+Tw]$ and $\ell_0wB_N\le p(x_0)/2$. Finally,
$e>0$ on $(x_0,x_0+w)$, $e<0$ on $(x_0-w,x_0)$, and both $\int_0^we(x_0+t)\,dt$ and $-\int_{-w}^0e(x_0+t)\,dt$ lie between
$3\ell_0w^2/8$ and $5\ell_0w^2/8$, which is possible because both equal $\ell_0w^2\{1/2+o(1)\}$.

Let $a_P=e\,\mathbf 1_{(x_0,x_0+w)}$, $a_Q=-e\,\mathbf 1_{(x_0-w,x_0)}$, $g(x)=\ell_0wG_w((x-x_0)/w)\mathbf 1\{|x-x_0|\le Tw\}$, and
$P'=P\wedge Q+(a_P-g_-)\,dx$, $Q'=P\wedge Q+(a_Q-g_+)\,dx$. As $|g|\le\ell_0wB_N\le\min(p,q)$ where $g\neq0$, both measures are nonnegative,
and $P'\le P\wedge Q+(p-q)_+\,dx=P$, $Q'\le Q$. Their difference is $\nu=(e\,\mathbf 1_{(x_0-w,x_0+w)}+g)\,dx$, and for $j<N$
\begin{align*}
\int(x-x_0)^j\,d\nu&=\ell_0w^{j+2}\Big\{\int t^jF_w(t)\,dt+\int_{-T}^Tt^jG_w(t)\,dt\Big\}\\
&=\ell_0w^{j+2}\{-M_j(w)+M_j(w)\}=0 .
\end{align*}
The case $j=0$ says that the masses agree, and they equal $1-\alpha^\star+\epsilon(w)$ with $\epsilon(w)=\int a_P-\int g_-$. As
$\int|g|=\ell_0w^2\int|G_w|\le\ell_0w^2/4$, $\epsilon(w)\ge3\ell_0w^2/8-\ell_0w^2/4=\ell_0w^2/8$ and
$|\nu|\le5\ell_0w^2/4+\ell_0w^2/4\le2\ell_0w^2$. The support of $\nu$ lies in $[x_0-Tw,x_0+Tw]$ because $T\ge1$.

For the bounds, $w=(8\epsilon/\ell_0)^{1/2}\le w_N$ gives $\epsilon\le\epsilon(w)$, and $\bar D$ is nonincreasing, so
$\bar D_{\alpha^\star-\epsilon}\le \bar D_{\alpha^\star-\epsilon(w)}\le\|\int\phi\,d\nu\|_\Hh$. By \eqref{eq:newbochner} and \eqref{eq:newmomentdecay}
with $W=Tw$, in case (a) $\|\int\phi\,d\nu\|_\Hh^2\le(2\pi)^{-1}C|\nu|^2\int\min\{1,(|\xi|Tw)^{2N}/N!^2\}|\xi|^{-2s}\,d\xi
=(2\pi)^{-1}C|\nu|^2(Tw)^{2s-1}I_{N,s}$, where $I_{N,s}$ is finite because $2N-2s>-1$ near $0$ and $2s>1$ at infinity. In case
(b), $\|\int\phi\,d\nu\|_\Hh^2\le|\nu|^2(Tw)^{2N}M_{2N}/N!^2$. With $|\nu|\le2\ell_0w^2$ and $w^2=8\epsilon/\ell_0$ these are the
bounds. For the Gaussian kernel, $M_{2N}=(-1)^Nk_0^{(2N)}(0)=(2N-1)!!\,\sigma^{-2N}$.
\end{proof}

\begin{proof}[Proof of Proposition~\ref{prop:newsobolev}]
For the lower bound, $(p-q)(x)/(x-x_0)$ is continuous near $x_0$ with limit $\ell_0>0$, so for some $w_0\le1$ and $\ell<\infty$,
$|p-q|(x)\le\ell|x-x_0|$ for $|x-x_0|\le w_0$, and Corollary~\ref{cor:newsobolev}, second case, applies with $r=1$, since
$\widehat{k_0}\ge c(1+\xi^2)^{-s}$ with $s>1/2=d/2$. For the upper bound, Proposition~\ref{prop:newmoments}(a) with any
$N>s-1/2$ applies, and $D_{\alpha^\star-\epsilon}\le \bar D_{\alpha^\star-\epsilon}/(1-\alpha^\star)$. For $s<3/2$ and $\alpha^\star<1$,
Proposition~\ref{prop:newupperone} with $r=1$ already gives the upper bound, without $p(x_0)>0$.
\end{proof}

\subsection{The converse}

For a version of the analytic baseline run with the larger threshold $\tau^{\mathrm S}$ of \eqref{eq:tauA} alone, an
upper bound on the path gives a lower bound on the error. For $\tau^{\mathrm P}$, the threshold the baseline uses, no
such converse is known.

\begin{proposition}[The margin governs the rate]
\label{prop:newconverse}
Let (A1) and (A2) of Assumption~\ref{ass:regularity} hold, let $\alpha^\star<1$ and $0<\delta<\delta'<1$, let
$\widehat\alpha_{\mathrm{an}}^{\mathrm S}$ be \eqref{eq:alphaest} computed with the threshold $\tau^{\mathrm S}(\cdot,\delta)$, and put
$g_{n,m}=\Delta_\phi\{(1/n+1/m)/2\}^{1/2}\{\log^{1/2}(1/\delta)-\log^{1/2}(1/\delta')\}$. Then, with probability at
least $1-\delta'$,
\[
\alpha^\star-\widehat\alpha_{\mathrm{an}}^{\mathrm S}\ \ge\ \sup\big\{\epsilon\in[0,\alpha^\star]:\ D_{\alpha^\star-\epsilon}(P,Q)\le g_{n,m}\big\} ,
\]
a set that contains $0$, since the path vanishes at $\alpha^\star$.
\end{proposition}

\begin{proof}[Proof of Proposition~\ref{prop:newconverse}]
On the event where the statistic exceeds the path by at most $\tau^{\mathrm S}(\cdot,\delta')$, every level at which
the path lies below the gap between the thresholds at $\delta$ and at $\delta'$ is accepted. Let $E^+$ be the event of Theorem~\ref{thm:twosided} at level $\delta'$, of probability at least $1-\delta'$. On $E^+$, for every
$\epsilon\in[0,\alpha^\star]$, $\widehat D_{\alpha^\star-\epsilon}\le D_{\alpha^\star-\epsilon}+\tau^{\mathrm S}_{n,m}(\alpha^\star-\epsilon,\delta')$. With
$\beta=(1-\alpha^\star+\epsilon)^{-1}\ge1$,
\begin{align*}
&\tau^{\mathrm S}_{n,m}(\alpha^\star-\epsilon,\delta)-\tau^{\mathrm S}_{n,m}(\alpha^\star-\epsilon,\delta')\\
&\qquad=\beta\Delta_\phi\{(1/n+1/m)/2\}^{1/2}\{\log^{1/2}(1/\delta)-\log^{1/2}(1/\delta')\}\ge g_{n,m} .
\end{align*}
So if $D_{\alpha^\star-\epsilon}\le g_{n,m}$, then $\widehat D_{\alpha^\star-\epsilon}\le\tau^{\mathrm S}_{n,m}(\alpha^\star-\epsilon,\delta)$, the level
$\alpha^\star-\epsilon$ is accepted, and $\widehat\alpha_{\mathrm{an}}^{\mathrm S}\le\alpha^\star-\epsilon$ by the definition \eqref{eq:alphaest}. The event does
not depend on $\epsilon$, so this holds for all such $\epsilon$ at once.
\end{proof}

Under the hypotheses of Proposition~\ref{prop:newsobolev}, let $\epsilon_0>0$ be such that
$D_{\alpha^\star-\epsilon}\le c_2\epsilon^{(2s+3)/4}$ for $\epsilon\le\epsilon_0$; then on $E^+$,
$\alpha^\star-\widehat\alpha_{\mathrm{an}}^{\mathrm S}\ge\min\{\alpha^\star,\epsilon_0,(g_{n,m}/c_2)^{4/(2s+3)}\}$, of order $(n\wedge m)^{-2/(2s+3)}$ since
$1/n+1/m\ge1/(n\wedge m)$. Under the hypotheses of Proposition~\ref{prop:newgauss}, the bound
$D_{\alpha^\star-\epsilon}\le C_j\epsilon^j$ gives $\alpha^\star-\widehat\alpha_{\mathrm{an}}^{\mathrm S}\ge c(n\wedge m)^{-1/(2j)}$ on $E^+$ for every
$j$ and $n\wedge m$ large, so the error is not $O((n\wedge m)^{-a})$ for any $a>0$.

\subsection{The excess of the population ramp}
\label{sec:excessproofs}

This subsection proves Proposition~\ref{prop:excess}. Its arguments use $\alpha^\star<1$ only in part (ii) of the
proposition: Lemmas~\ref{lem:excessbathtub} and~\ref{lem:excessid} and parts (i) and (iii) hold for any pair of
probability measures with $\alpha<\TV(P,Q)\le1$, which Section~\ref{sec:laplaceproof} uses for empirical laws. Fix
$\alpha<\alpha^\star$, with $\bar D$ and $\bar{\mathcal C}_\alpha$ as in
Section~\ref{sec:generalmargin}; $\bar D_\alpha>0$ by \eqref{eq:zeroeq}. By Lemma~\ref{lem:attainment} there are parts
$P'$ of $P$ and $Q'$ of $Q$ of mass $1-\alpha$ such that
$d=\mu_{P'}-\mu_{Q'}$ is the element of minimal norm of $\bar{\mathcal C}_\alpha(P)-\bar{\mathcal C}_\alpha(Q)$, with $\|d\|_\Hh=\bar D_\alpha$, and
$f^\star=d/\bar D_\alpha$ is the witness of Lemma~\ref{lem:popramp}, with the same cut points $t^\star$ and $u^\star$.

\begin{lemma}[Bathtub for the closest parts]
\label{lem:excessbathtub}
Every measure $P''\le P$ satisfies $\int f^\star\,dP''\ge\int f^\star\,dP'+t^\star\{P''(\mathcal X)-(1-\alpha)\}$, and
every $Q''\le Q$ satisfies $\int f^\star\,dQ''\le\int f^\star\,dQ'+u^\star\{Q''(\mathcal X)-(1-\alpha)\}$. Moreover
$P'=P$ on $\{f^\star<t^\star\}$, $P'(\{f^\star>t^\star\})=0$, $Q'=Q$ on $\{f^\star>u^\star\}$ and
$Q'(\{f^\star<u^\star\})=0$.
\end{lemma}

\begin{proof}
As $d$ has minimal norm in a convex set, $\langle d,c-d\rangle_\Hh\ge0$ for every $c$ in it. With
$c=\mu_{P''}-\mu_{Q'}$ for a part $P''$ of $P$ of mass $1-\alpha$ and the reproducing property, $P'$ minimizes
$\int f^\star\,dP''$ over these parts; likewise $Q'$ maximizes $\int f^\star\,dQ''$ over the parts of $Q$ of mass
$1-\alpha$. Suppose $(P-P')(\{f^\star<t^\star\})>0$. As $t^\star$ is a $(1-\alpha)$-quantile,
$P(\{f^\star<t^\star\})\le1-\alpha$, so $P'(\{f^\star\ge t^\star\})=(1-\alpha)-P'(\{f^\star<t^\star\})\ge
(P-P')(\{f^\star<t^\star\})>0$. Moving a small mass from $P'$ on $\{f^\star\ge t^\star\}$ to $P-P'$ on
$\{f^\star<t^\star\}$ gives another part of mass $1-\alpha$ with a smaller integral of $f^\star$, a contradiction; so
$P'=P$ on $\{f^\star<t^\star\}$. If $P'(\{f^\star>t^\star\})>0$, then, as $P(\{f^\star\le t^\star\})\ge1-\alpha$, the
measure $P-P'$ has at least that mass on $\{f^\star=t^\star\}$, and moving mass there from $\{f^\star>t^\star\}$ lowers
the integral; so $P'(\{f^\star>t^\star\})=0$. The claims on $Q'$ follow in the same way from
$Q(\{f^\star>u^\star\})\le1-\alpha\le Q(\{f^\star\ge u^\star\})$.

For the inequality, let $P''\le P$ have any mass and $\lambda=P''-P'$. On $\{f^\star<t^\star\}$, $\lambda=P''-P\le0$
and $f^\star<t^\star$; on $\{f^\star>t^\star\}$, $\lambda=P''\ge0$ and $f^\star>t^\star$. So
$\int_Af^\star\,d\lambda\ge t^\star\lambda(A)$ for each of the sets $A=\{f^\star<t^\star\}$,
$\{f^\star=t^\star\}$ and $\{f^\star>t^\star\}$, and adding gives
$\int f^\star\,d\lambda\ge t^\star\lambda(\mathcal X)=t^\star\{P''(\mathcal X)-(1-\alpha)\}$. The bound for
$Q''$ is the same argument with the inequalities reversed.
\end{proof}

\begin{lemma}[Excess identity]
\label{lem:excessid}
\[
\E_P\rp^\star-\E_Q\rp^\star-\alpha
=\frac{\bar D_\alpha+\int(u^\star-f^\star)_+\,dP+\int(f^\star-t^\star)_+\,dQ}{t^\star-u^\star} .
\]
\end{lemma}

\begin{proof}
By Lemma~\ref{lem:excessbathtub}, $P-P'$ has mass $\alpha$ and is carried by $\{f^\star\ge t^\star\}$, where
$\rp^\star=1$, and $Q-Q'$ is carried by $\{f^\star\le u^\star\}$, where $\rp^\star=0$. Hence the left side
is $\int\rp^\star\,d(P'-Q')$. With the clipped witness $\tilde f=(f^\star\vee u^\star)\wedge t^\star$,
$\rp^\star=(\tilde f-u^\star)/(t^\star-u^\star)$, and as $P'$ and $Q'$ have the same mass, the left side is
$\int\tilde f\,d(P'-Q')/(t^\star-u^\star)$. On $\{f^\star\le t^\star\}$, which carries $P'$,
$\tilde f=f^\star+(u^\star-f^\star)_+$, and on $\{f^\star\ge u^\star\}$, which carries $Q'$,
$\tilde f=f^\star-(f^\star-t^\star)_+$. So $\int\tilde f\,d(P'-Q')=\langle f^\star,d\rangle_\Hh+\int(u^\star-f^\star)_+\,dP'
+\int(f^\star-t^\star)_+\,dQ'$, the first term is $\bar D_\alpha$, and $P'$ and $Q'$ may be replaced by $P$ and $Q$ in the
other two, because $P'=P$ on $\{f^\star<u^\star\}\subseteq\{f^\star<t^\star\}$ and $Q'=Q$ on
$\{f^\star>t^\star\}\subseteq\{f^\star>u^\star\}$.
\end{proof}

The two integrals are overshoots of the witness beyond its cut points, each against the law that is not trimmed there.

\begin{proof}[Proof of Proposition~\ref{prop:excess}]
The first inequality of (i) is Lemma~\ref{lem:excessid}, the integrals being nonnegative. For the second, let
$\alpha'\in[0,1)$. As $f^\star$ has unit norm, the Cauchy--Schwarz inequality gives
$\bar D_{\alpha'}\ge\int f^\star\,dP''-\int f^\star\,dQ''$ for the closest parts $P''$ of $P$ and $Q''$ of $Q$ of mass
$1-\alpha'$, and by Lemma~\ref{lem:excessbathtub} the right side is at least
$\int f^\star\,d(P'-Q')-(t^\star-u^\star)(\alpha'-\alpha)=\bar D_\alpha-(t^\star-u^\star)(\alpha'-\alpha)$.

For (ii), let $\alpha=\alpha^\star-\epsilon$ with $\epsilon\le\epsilon_0/2$ and take $\alpha'=\alpha^\star-2\epsilon$ in
(i): $t^\star-u^\star\le\{\bar D_{\alpha^\star-2\epsilon}-\bar D_{\alpha^\star-\epsilon}\}/\epsilon\le
D_{\alpha^\star-2\epsilon}/\epsilon\le c_2(2\epsilon)^\gamma/\epsilon$, while
$\bar D_{\alpha^\star-\epsilon}\ge(1-\alpha^\star)D_{\alpha^\star-\epsilon}\ge(1-\alpha^\star)c_1\epsilon^\gamma$. The first
inequality of (i) gives the bound. Proposition~\ref{prop:newsobolev} gives the two-sided bound with
$\gamma=(2s+3)/4$.

For (iii), write $d(x)=\int k(x,y)\,d\nu(y)$ with $\nu=P'-Q'$, a finite signed measure with $\nu(\R)=0$. Since
$(1-\sigma^2\partial_x^2)e^{-|x|/\sigma}=2\sigma\delta_0$ in the sense of distributions, $d-\sigma^2d''=2\sigma\nu$. The function
$d$ is continuous and tends to $0$ at $\pm\infty$, and it is not zero, as $\|d\|_\Hh=\bar D_\alpha>0$; by Fubini's theorem
$\int d(x)\,dx=2\sigma\nu(\R)=0$, so $d$ takes positive values and attains its maximum $M>0$ at some $z_0$. Suppose that
$M>\bar D_\alpha t^\star$. The open set $O=\{f^\star>t^\star\}$ contains $z_0$, and on it $P'=0$ and $Q'=Q$ by
Lemma~\ref{lem:excessbathtub}, as $t^\star>u^\star$, so $\nu=-Q\le0$ on $O$. On an open interval $I\ni z_0$ inside $O$
on which $d>M/2$, the distribution $\sigma^2d''=d-2\sigma\nu$ is at least $M/2$, so $\tilde d(x)=d(x)-M(x-z_0)^2/(4\sigma^2)$ is
convex on $I$. But $\frac12\{\tilde d(z_0+h)+\tilde d(z_0-h)\}\le d(z_0)-Mh^2/(4\sigma^2)<\tilde d(z_0)$ for small $h>0$, a contradiction.
Hence $f^\star\le t^\star$ on $\R$, and the same argument at the minimum of $d$, where $\nu=P\ge0$ on
$\{f^\star<u^\star\}$, gives $f^\star\ge u^\star$. Both overshoots in Lemma~\ref{lem:excessid} vanish.
\end{proof}

In (iii) the witness is therefore flat on the set it trims from each law, $P(\{f^\star=t^\star\})\ge\alpha$ and
$Q(\{f^\star=u^\star\})\ge\alpha$. The argument rests on $k$ being the Green function of a second-order operator with a
maximum principle.

\subsection{Proof of Theorem~\ref{thm:laplacerate}}
\label{sec:laplaceproof}

Theorem~\ref{thm:laplacerate} is the following theorem with its constants absorbed in the $O$: as $\eta<1$,
$L_{n,m}\ge\log(n\vee m)$, so $(\log(n\vee m))^{1/4}L_{n,m}^{1/2}\le L_{n,m}^{3/4}$ and
$(\log(n\vee m))^{1/2}L_{n,m}\le L_{n,m}^{3/2}$.

\begin{theorem}[Theorem~\ref{thm:laplacerate} with its constants]
\label{thm:laplaceconst}
Let $P,Q$ satisfy the hypotheses of Proposition~\ref{prop:newsobolev}, let $k(x,y)=e^{-|x-y|/\sigma}$ with $\sigma>0$
fixed in advance, and let $n\wedge m\ge10$. Compute $\widehat\alpha$ at level $\delta$ as in
Proposition~\ref{prop:loconsistency}, with the batches and fractions of Section~\ref{sec:betting},
$\lambda_{\max}=\tfrac12$, initial variance $v_0^2\le4$ in \eqref{eq:fractions} and the running variance floored at
$v^2_{\min}\in(0,4]$, an exact minimizer of \eqref{eq:qp} at each level, and a grid of $J$ levels in $(0,1)$ fixed
in advance whose mesh below $\alpha^\star$ is $\Delta_{\mathcal G}$, but with cut points that are genuine empirical
quantiles: any $(1-\alpha)$-quantile of the witness values at the $X$'s and any $\alpha$-quantile of those at the
$Y$'s of the fitting batches, computed with or without the own terms. Let $\eta\in(0,1)$, $L_\delta=\log(1/\delta)$
and $L_{n,m}=\log\{(n\vee m)/\eta\}$.
\begin{enumerate}[label=(\roman*),leftmargin=*]
\item If \eqref{eq:condR} holds for some $r_\ast\in(0,1]$ and $c_\ast>0$, then with probability at least $1-\delta-\eta$,
\[
\alpha^\star-\Delta_{\mathcal G}-C\,\frac{(\log(n\vee m))^{1/4}L_{n,m}^{1/2}
+v_{\min}^{-2}\{L_\delta^{1/2}+(\log(2J/\eta))^{1/2}\}}{(n\wedge m)^{1/2}}
\ \le\ \widehat\alpha\ \le\ \alpha^\star .
\]
\item If instead $(P+Q)(\{x:|x-x_0|>t\})\le C'e^{-t/C'}$ for all $t>0$, then with probability at least
$1-\delta-\eta$,
\[
\alpha^\star-\Delta_{\mathcal G}-C\bigg[\bigg\{\frac{(\log(n\vee m))^{1/2}L_{n,m}}{n\wedge m}\bigg\}^{4/9}
+\frac{L_\delta^{1/2}+(\log(2J/\eta))^{1/2}}{v_{\min}^{2}(n\wedge m)^{1/2}}\bigg]
\le\widehat\alpha\le\alpha^\star .
\]
\end{enumerate}
Here $C$ depends only on $\sigma$ and on the pair through $\alpha^\star$, the constants of
Proposition~\ref{prop:newsobolev} with $s=1$, the tail constant $C'$ (that of $P+Q$ in (ii)), and, in (i), $c_\ast$, $r_\ast$ and
the supremum of $p+q$ on $[x_0-r_\ast,x_0+r_\ast]$, which is finite once $r_\ast$ is small enough.
\end{theorem}

The constant $C$ degenerates when $c_\ast$ is small, as for narrow contaminations on a wide core, and it involves the
constants $c_1$, $c_2$, $\epsilon_0$ of Proposition~\ref{prop:newsobolev}, which come from limiting arguments
(Corollary~\ref{cor:newsobolev} and Proposition~\ref{prop:newmoments}) and are not explicit; whether $C$ is uniform over
a natural class is open, and we return to it after Lemma~\ref{lem:laplocal}. The theorem needs exact minimizers of
\eqref{eq:qp}, as self-clipping (Lemma~\ref{lem:lapclip}) is a first-order property whose survival under an inexact
solve is open. In $\R^d$ the margin that Theorem~\ref{thm:newsobolev} gives a transversal crossing guarantees at best
$(n\wedge m)^{-2/(d+3)}$, as a Sobolev kernel needs $s>d/2$, and the minimax rate is slower than parametric for
$L_1$-type functionals \citep{lepski1999estimation,cai2011testing,han2020estimation}. The simulation of
Section~\ref{sec:excessmc} agrees with the rate, but its cut points are interpolated between order statistics, which
the theorem does not cover, and its grid approaches the known $\alpha^\star$ geometrically.

The proof rests on a property special to the Laplace kernel on the line: the fitted witness never leaves the band
between its cut points, so the fitted ramp maximizes the ratio of empirical excess to norm over the functions of oscillation at most
one (Lemmas~\ref{lem:lapratio} and~\ref{lem:lapclip}). Comparing this ratio with the population path at the levels
$\alpha^\star-\epsilon$ and $\alpha^\star-2\epsilon$ shows that the fitted ramp keeps a fixed fraction of $\epsilon$ as
excess, up to two fluctuation terms (Lemma~\ref{lem:lapcompare}), which we bound through the norm and the error of the
function, localized near $x_0$ by \eqref{eq:condR} (Lemmas~\ref{lem:lapintervals} to~\ref{lem:lapdev}); a fixed-point
argument closes the bound (Proposition~\ref{prop:lapfitted}), and the bets are those of Theorem~\ref{thm:lorate}. Rescaling the data by $1/\sigma$ leaves the witness values, cut points and ramps unchanged and
preserves the hypotheses with rescaled constants, so we take $\sigma=1$. The reproducing kernel Hilbert space of
$e^{-|x-y|}$ is the Sobolev space $H^1(\R)$ with $\|f\|_\Hh^2=\tfrac12\int(f^2+f'^2)$, since $e^{-|x|}$ has Fourier
transform $2/(1+\omega^2)$ \citep[Theorem~10.12]{wendland2005scattered}. Every $f\in\Hh$ is absolutely continuous,
tends to $0$ at $\pm\infty$, and satisfies
\begin{equation}
|f(x)|\le\|f\|_\Hh,\qquad\int_a^b|f'|\le\sqrt{2(b-a)}\,\|f\|_\Hh .
\label{eq:laphone}
\end{equation}
Write $\operatorname{osc}(f)=\sup f-\inf f$; for $f\in\Hh$ with $\operatorname{osc}(f)\le1$, $|f|\le1$, as $f$ tends to
$0$ at infinity. For bounded $f$ let
$L(f)=\int f\,d(P-Q)$ and $\mathrm{err}(f)=\alpha^\star-L(f)$, the error of Section~\ref{sec:linearmargin} when $f$ takes
values in $[0,1]$. If $\operatorname{osc}(f)\le1$, then $0\le\mathrm{err}(f)\le2$, and with $M_f=\sup f$, $m_f=\inf f$,
\begin{equation}
\mathrm{err}(f)=\alpha^\star\{1-(M_f-m_f)\}+\int(M_f-f)(p-q)_+\,dx+\int(f-m_f)(q-p)_+\,dx ,
\label{eq:laperr}
\end{equation}
all three terms being nonnegative: the error of a function of oscillation at most one is its loss of range plus its
distances from its maximum where $p>q$ and from its minimum where $q>p$. With $\bar D$ as in
Section~\ref{sec:generalmargin}, $\bar D_\alpha$ is the smallest $\|\mu_{P'}-\mu_{Q'}\|_\Hh$ over parts $P'\le P$,
$Q'\le Q$ of mass $1-\alpha$, and we put
$V(\epsilon)=\bar D_{\alpha^\star-\epsilon}$. For other laws $\tilde P,\tilde Q$ we write $\bar D_\alpha(\tilde P,\tilde Q)=(1-\alpha)D_\alpha(\tilde P,\tilde Q)$ for
their rescaled path and $\tilde L(f)=\int f\,d(\tilde P-\tilde Q)$.

By Proposition~\ref{prop:newsobolev} with $s=1$ there are $0<c_1\le c_2$ and $\epsilon_0\in(0,\alpha^\star]$ with
$c_1\epsilon^{5/4}\le D_{\alpha^\star-\epsilon}\le c_2\epsilon^{5/4}$ for $0<\epsilon\le\epsilon_0$. Put
\[
\rho_0=\frac{(1-\alpha^\star)c_1}{2^{5/4}c_2},\quad c_V=(1-\alpha^\star)c_1\Big(\frac{\epsilon_0}{\alpha^\star}\Big)^{5/4},
\quad\kappa_0=\min\Big\{\rho_0,\frac{(1-\alpha^\star)c_1(\epsilon_0/2)^{5/4}}{2\alpha^\star}\Big\},
\]
so that $\rho_0,\kappa_0\in(0,1)$ and $c_V>0$. Here $\rho_0$ is a lower bound on $V(\epsilon)/V(2\epsilon)$
for $\epsilon<\epsilon_0/2$, by the two-sided bound and $(1-\alpha^\star)D_{\alpha^\star-\epsilon}\le V(\epsilon)\le
D_{\alpha^\star-\epsilon}$; $c_V$ extends the lower bound to all levels, $V(\epsilon)\ge c_V\epsilon^{5/4}$ for
$\epsilon\in(0,\alpha^\star]$, as $V(\epsilon)\ge(1-\alpha^\star)D_{\alpha^\star-\epsilon}$ and $V$ is nondecreasing; and
$\kappa_0$ is the fraction of $\epsilon$ that the comparison below guarantees as excess. In (i) put $\bar p=\sup_{[x_0-r_\ast,x_0+r_\ast]}(p+q)$, which is finite once $r_\ast$ is small, as $p,q$ are continuous near
$x_0$, and shrinking $r_\ast$ keeps \eqref{eq:condR} with the same $c_\ast$; off this
interval $p+q\le|p-q|/(c_\ast r_\ast)$, so $P+Q$ inherits the exponential tails of $|P-Q|$.

\begin{lemma}[Ratio inequality]
\label{lem:lapratio}
For any probability measures $\tilde P,\tilde Q$, any $\alpha\in[0,1)$ and any $\varphi\in\Hh\setminus\{0\}$ with
$\operatorname{osc}(\varphi)\le1$, $\bar D_\alpha(\tilde P,\tilde Q)\ge\{\tilde L(\varphi)-\alpha\}/\|\varphi\|_\Hh$. In particular
$V(\epsilon')\ge\{\epsilon'-\mathrm{err}(\varphi)\}/\|\varphi\|_\Hh$ for $\epsilon'\in[0,\alpha^\star]$.
\end{lemma}

\begin{proof}
For parts $P''\le\tilde P$, $Q''\le\tilde Q$ of mass $1-\alpha$, $\int\varphi\,dP''\ge\int\varphi\,d\tilde P-\alpha\sup\varphi$ and
$\int\varphi\,dQ''\le\int\varphi\,d\tilde Q-\alpha\inf\varphi$, so $\|\varphi\|_\Hh\|\mu_{P''}-\mu_{Q''}\|_\Hh\ge
\int\varphi\,d(P''-Q'')\ge\tilde L(\varphi)-\alpha$; take the infimum over the parts.
\end{proof}

At a level $\alpha$ and a batch $b$, let $\tilde P,\tilde Q$ be the empirical laws of the fitting batches, of sizes
$n',m'$, with the pooled points pairwise distinct (almost surely, as $P$ and $Q$ have densities), let $\widehat f$ be the witness direction, $\bar{\widehat D}=\bar
D_\alpha(\tilde P,\tilde Q)$, $\widehat t=\widehat f(X)_{(\lceil n'(1-\alpha)\rceil)}$ and $\widehat u=\widehat f(Y)_{(\lfloor m'\alpha\rfloor+1)}$,
the smallest $(1-\alpha)$-quantile of the values at the $X$'s and the largest $\alpha$-quantile of those at the $Y$'s.
By Lemma~\ref{lem:lapclip}, $\widehat t$ is also the $(n'_{1-\alpha}+1)$-th smallest value at the $X$'s,
$n'_{1-\alpha}=\lfloor n'(1-\alpha)\rfloor$, the cut of Proposition~\ref{prop:bangbang}, and likewise for $\widehat u$.

\begin{lemma}[The fitted witness clips itself]
\label{lem:lapclip}
Let $\alpha\in(0,1)$. Then $\bar{\widehat D}>0$, $\widehat u\le\widehat f\le\widehat t$ on $\R$, $\widehat s=\widehat t-\widehat u
=\operatorname{osc}(\widehat f)>0$, the values of $\widehat f$ at the $X$'s of rank at least $\lceil n'(1-\alpha)\rceil$
equal $\widehat t$, and those at the $Y$'s of rank at most $\lfloor m'\alpha\rfloor+1$ equal $\widehat u$. Hence every
genuine $(1-\alpha)$-quantile of the values at the $X$'s is $\widehat t$, every genuine $\alpha$-quantile of those at the
$Y$'s is $\widehat u$, and the plain ramp is $\widehat\rp=(\widehat f-\widehat u)/\widehat s$. Moreover
$\widehat\varphi=\widehat f/\widehat s$ has $\operatorname{osc}(\widehat\varphi)=1$, $\widehat N=\|\widehat\varphi\|_\Hh=
1/\widehat s\ge\tfrac12$ and
\[
\tilde L(\widehat\varphi)-\alpha=\bar{\widehat D}\,\widehat N ,
\]
so by Lemma~\ref{lem:lapratio} $\widehat\varphi$ maximizes $\{\tilde L(\varphi)-\alpha\}/\|\varphi\|_\Hh$ over
$\operatorname{osc}(\varphi)\le1$. With the own terms removed, the values at the $X$'s of rank at least
$\lceil n'(1-\alpha)\rceil$ lie in $[\widehat t-1/(n'\bar{\widehat D}),\widehat t]$ and those at the $Y$'s of rank at most
$\lfloor m'\alpha\rfloor+1$ in $[\widehat u,\widehat u+1/(m'\bar{\widehat D})]$.
\end{lemma}

\begin{proof}
The empirical laws are themselves a pair to which the results on the population ramp apply. Since
$\TV(\tilde P,\tilde Q)=1>\alpha$, $\bar{\widehat D}>0$ by \eqref{eq:zeroeq}. As noted at the start of
Section~\ref{sec:excessproofs}, Lemmas~\ref{lem:excessbathtub} and~\ref{lem:excessid} and
Proposition~\ref{prop:excess}(iii) use only Assumption~\ref{ass:regularity}, Lemmas~\ref{lem:attainment}
and~\ref{lem:popramp}, and $\alpha<\TV(P,Q)$. They therefore hold for $\tilde P,\tilde Q$ and any
$\alpha<1=\TV(\tilde P,\tilde Q)$,
and the distributional argument of (iii) applies to the atomic measure $\nu=P''-Q''$ unchanged, as $\nu\le0$ on
$\{\widehat f>\widehat t\}$. Applied to $\tilde P,\tilde Q$ they give the two facts from which everything else is read
off: $\widehat u\le\widehat f\le\widehat t$ on $\R$, and $\tilde P\widehat\rp-\tilde Q\widehat\rp-\alpha=\bar{\widehat D}/\widehat s$. Both bounds are values of $\widehat f$ at sample points, so
$\operatorname{osc}(\widehat f)=\widehat s$, and $\widehat s>0$, as $\int\widehat f\,dx=0$ forces
$\max\widehat f>0>\min\widehat f$. The order statistics of rank at least $\lceil n'(1-\alpha)\rceil$ are at least
$\widehat t$ and at most $\max\widehat f\le\widehat t$, so they equal $\widehat t$, and likewise those of the $Y$'s of
rank at most $\lfloor m'\alpha\rfloor+1$ equal $\widehat u$. A genuine $(1-\alpha)$-quantile lies between the order
statistics of ranks $\lceil n'(1-\alpha)\rceil$ and $\lfloor n'(1-\alpha)\rfloor+1$, and a genuine $\alpha$-quantile
between those of ranks $\lceil m'\alpha\rceil\ge1$ and $\lfloor m'\alpha\rfloor+1$, so they are $\widehat t$ and
$\widehat u$. As $\widehat\rp$ and $\widehat\varphi$ differ by a constant, $\tilde L(\widehat\varphi)=\tilde
P\widehat\rp-\tilde Q\widehat\rp$, which gives the identity, and $\widehat N\ge\sup|\widehat\varphi|\ge\tfrac12$ by
\eqref{eq:laphone}, as $\operatorname{osc}(\widehat\varphi)=1$. Finally, in the normalization of \eqref{eq:witness} the
own term at $X_i$ is $w_i/\widehat D_\alpha=w_i(1-\alpha)/\bar{\widehat D}\in[0,1/(n'\bar{\widehat D})]$, and at $Y_j$ it
is at most $1/(m'\bar{\widehat D})$; lowering each value by at most $c$ lowers each order statistic by at most $c$.
\end{proof}

At the population level, let $f^\star$, $t^\star$, $u^\star$ and $\rp^\star$ be as in
Proposition~\ref{prop:excess} at $\alpha=\alpha^\star-\epsilon\in(0,\alpha^\star)$, $s^\star=t^\star-u^\star$ and
$\varphi^\star=f^\star/s^\star$. By Proposition~\ref{prop:excess}(iii), $\operatorname{osc}(\varphi^\star)\le1$,
$\|\varphi^\star\|_\Hh=1/s^\star$ and
\begin{equation}
L(\varphi^\star)-\alpha=\E_P\rp^\star-\E_Q\rp^\star-\alpha=V(\epsilon)/s^\star\le\epsilon ,
\label{eq:lappop}
\end{equation}
the last inequality because $L(\varphi^\star)\le\alpha^\star$. So $s^\star\ge V(\epsilon)/\epsilon$,
$\|\varphi^\star\|_\Hh\le c_V^{-1}\epsilon^{-1/4}$, $\mathrm{err}(\varphi^\star)\le\epsilon$, and $s^\star\le\operatorname{osc}(f^\star)\le2$.

\begin{lemma}[Comparison]
\label{lem:lapcompare}
Let $\Delta(\varphi)=\int\varphi\,d\{(\tilde P-P)-(\tilde Q-Q)\}$, $\widehat\Delta=\Delta(\widehat\varphi)$,
$\Delta^\star=\Delta(\varphi^\star)$, $\widehat T=\mathrm{err}(\widehat\varphi)$, $y=V(\epsilon)+s^\star\Delta^\star$ and
$e=\E_P\widehat\rp-\E_Q\widehat\rp-\alpha=\epsilon-\widehat T$. Then:
\begin{enumerate}[label=(\roman*),leftmargin=*]
\item $e+\widehat\Delta=\bar{\widehat D}\widehat N$, $\bar{\widehat D}\ge y$, and $\widehat Ny+\widehat T\le\epsilon+\widehat\Delta$;
\item if $\kappa_0\epsilon-|\Delta^\star|-\widehat\Delta>0$, then $e\ge\kappa_0\epsilon-|\Delta^\star|-\widehat\Delta$;
\item if $|\Delta^\star|\le\kappa_0\epsilon/4$, then $y\ge V(\epsilon)/2\ge c_V\epsilon^{5/4}/2$.
\end{enumerate}
\end{lemma}

\begin{proof}
Part (i) collects the basic relations. The identity is Lemma~\ref{lem:lapclip}, as
$\tilde L(\widehat\varphi)-L(\widehat\varphi)=\widehat\Delta$. By Lemma~\ref{lem:lapratio} for $\tilde P,\tilde Q$ and
$\varphi^\star$, and \eqref{eq:lappop}, $\bar{\widehat D}\ge s^\star\{L(\varphi^\star)-\alpha+\Delta^\star\}=y$, and
multiplying by $\widehat N>0$ gives the last inequality, since $\bar{\widehat D}\widehat N=e+\widehat\Delta=\epsilon-\widehat
T+\widehat\Delta$.

Parts (ii) and (iii) compare the path at two levels. A fitted ramp with a small excess $e$ has error
$\widehat T=\epsilon-e$ close to $\epsilon$, so at the level $\alpha^\star-2\epsilon$ it still has an excess of about
$\epsilon$. Lemma~\ref{lem:lapratio} at that level then forces its norm $\widehat N$ up, and by (i) a large $\widehat N$
forces $e$ up. The population enters through Proposition~\ref{prop:excess}(i) at $\alpha'=\alpha^\star-2\epsilon$, by
which the population gap is a supporting slope of the path,
\[
s^\star\epsilon\ \le\ V(2\epsilon)-V(\epsilon)\qquad(2\epsilon\le\alpha^\star).
\]
Suppose first $\epsilon<\epsilon_0/2$, so that $V(\epsilon)/V(2\epsilon)\ge\rho_0\ge\kappa_0$.
If $|\Delta^\star|\ge\epsilon V(\epsilon)/V(2\epsilon)$, then $\kappa_0\epsilon-|\Delta^\star|-\widehat\Delta\le-\widehat\Delta<e$
by (i), as $\bar{\widehat D}\widehat N>0$, and (ii) holds; so assume $|\Delta^\star|<\epsilon V(\epsilon)/V(2\epsilon)$. Then
$s^\star|\Delta^\star|<\{V(2\epsilon)-V(\epsilon)\}V(\epsilon)/V(2\epsilon)$, hence $0<y<V(2\epsilon)$ and
$r=y/V(2\epsilon)\in(0,1)$. If $e+\epsilon\le0$, then $\widehat\Delta>\epsilon$ by (i), as $\bar{\widehat D}\widehat N>0$,
contradicting the hypothesis of (ii). Otherwise $2\epsilon-\widehat T=e+\epsilon>0$, Lemma~\ref{lem:lapratio} at the
level $\alpha^\star-2\epsilon$ gives $\widehat N\ge(e+\epsilon)/V(2\epsilon)$, and with (i)
\[
e+\widehat\Delta=\bar{\widehat D}\widehat N\ \ge\ y\widehat N\ \ge\ r(e+\epsilon),
\qquad\text{that is,}\qquad(1-r)e\ \ge\ r\epsilon-\widehat\Delta .
\]
As $s^\star\epsilon\le V(2\epsilon)$, $r\epsilon\ge\epsilon V(\epsilon)/V(2\epsilon)-|\Delta^\star|$, so
$(1-r)e\ge\kappa_0\epsilon-|\Delta^\star|-\widehat\Delta>0$, whence $e>0$ and $e\ge(1-r)e$. For (iii),
$s^\star|\Delta^\star|\le\kappa_0V(2\epsilon)/4\le V(\epsilon)/4$, as $s^\star\le V(2\epsilon)/\epsilon$.

If $\epsilon\ge\epsilon_0/2$, the path is bounded below by a constant: $V(\epsilon)\ge(1-\alpha^\star)c_1(\epsilon_0/2)^{5/4}\ge
2\kappa_0\alpha^\star\ge2\kappa_0\epsilon$, and $s^\star\le2$. For (iii), $s^\star|\Delta^\star|\le\kappa_0\epsilon/2\le
V(\epsilon)/4$. For (ii), if $y\ge0$, (i) and $\widehat N\ge\tfrac12$ give $e+\widehat\Delta\ge y/2\ge
V(\epsilon)/2-|\Delta^\star|$, which is enough; if $y<0$, then $|\Delta^\star|>V(\epsilon)/2\ge\kappa_0\epsilon$, while
$e+\widehat\Delta=\bar{\widehat D}\widehat N>0$, so $e>-\widehat\Delta>\kappa_0\epsilon-|\Delta^\star|-\widehat\Delta$.
\end{proof}

We bound $\Delta^\star$ and $\widehat\Delta$ next, the second for a function fitted on the same data. We do it for
all functions of oscillation at most one at once, in terms of their norm and error, on one event that does not depend
on the level.

\begin{lemma}[Intervals]
\label{lem:lapintervals}
Let $\tilde P$ be the empirical law of $n$ independent draws from $P$, which has a continuous distribution function,
let $\eta'\in(0,1)$, $x=\log\{2(n+1)^2/\eta'\}$ and $c_x=2x+4$. With probability at least $1-\eta'$, every interval
$I\subseteq\R$ satisfies $|\tilde P(I)-P(I)|\le\{2P(I)x/n\}^{1/2}+c_x/n$.
\end{lemma}

\begin{proof}
After the probability integral transform, a grid of at most $(n+1)^2$ intervals brackets every interval to within
$2/n$, so Bernstein's inequality and a union bound over the grid suffice. Almost surely no observation lies in an interval of
constancy of the distribution function $F$ or at one of its endpoints, so $X_i\in I$ if and only if $F(X_i)\in J$ for the interval $J$ with endpoints $F(\inf I),F(\sup I)$ of the
same types, and $P(I)=|J|$; it suffices to treat uniform variables. For each of the at most $(n+1)^2$ intervals $G$ with
endpoints in $\{0,1/n,\dots,1\}$, Bernstein's inequality \citep[Section~2.8]{boucheron2013concentration} for the binomial
count, with deviation $\tau=(2n|G|x)^{1/2}+2x/3$, which satisfies $\tau^2\ge2x(n|G|+\tau/3)$, bounds
$|\tilde P(G)-|G||$ by $(2|G|x/n)^{1/2}+2x/(3n)$ outside an event of probability $2e^{-x}$; a union bound covers all of
them. A general $J$ lies between grid intervals $G^-\subseteq J\subseteq G^+$ with $|G^+|-|J|\le2/n$ and $|J|-|G^-|\le2/n$,
which costs at most $(2\sqrt x+2)/n$; and $2\sqrt x+2x/3+2\le c_x$ as $x>2$.
\end{proof}

\begin{lemma}[Localized deviation]
\label{lem:lapbv}
On the event of Lemma~\ref{lem:lapintervals}, for every bounded interval $I$ and every $h:\R\to[0,1]$ continuous on
the closure of $I$ with total variation $\mathrm V_I(h)$ there,
\[
\Big|\int_Ih\,d(\tilde P-P)\Big|\le\Big[2\{\mathrm V_I(h)+1\}\Big(\int_Ih\,dP\Big)\frac xn\Big]^{1/2}+\{\mathrm V_I(h)+1\}\frac{c_x}n .
\]
\end{lemma}

\begin{proof}
A function of bounded variation is an average of indicators of its upper level sets, each a union of few intervals,
to which Lemma~\ref{lem:lapintervals} applies. By Tonelli's theorem, $\int_Ih\,d(\tilde P-P)=\int_0^1(\tilde P-P)(A_s)\,ds$ and $\int_Ih\,dP=\int_0^1P(A_s)\,ds$, with
$A_s=\{z\in I:h(z)>s\}$. Each component $C$ of the relatively open set $A_s$ other than the rightmost one has
$\sup C\in I\setminus A_s$ and $h(\sup C)=s$, so $A_s$ has at most $N(s)+1$ components, $N(s)$ being the number of
points of $\bar I$ where $h=s$. By Banach's indicatrix theorem \citep{banach1925lignes}, $N$ is measurable and
$\int N(s)\,ds=\mathrm V_I(h)$. Where $N(s)<\infty$, the event and the Cauchy--Schwarz inequality over the components give
$|(\tilde P-P)(A_s)|\le[2\{N(s)+1\}P(A_s)x/n]^{1/2}+\{N(s)+1\}c_x/n$; integrate over $s$ and use the Cauchy--Schwarz
inequality once more.
\end{proof}

\begin{lemma}[Deviation of functions of $\Hh$]
\label{lem:lapdev}
Let $\tilde P,\tilde Q$ be empirical laws of $n$ draws from $P$ and $m$ draws from $Q$ on whose events of
Lemma~\ref{lem:lapintervals} the same $x\ge\log8$ holds, let $K\ge1$, $\pi_K=(P+Q)(\{|z-x_0|>K\})$,
$s_{n,m}=(x/n)^{1/2}+(x/m)^{1/2}$ and $r_{n,m}=1/n+1/m$. Let $\varphi\in\Hh$ with $\operatorname{osc}(\varphi)\le1$,
$N=\max(\|\varphi\|_\Hh,\tfrac12)$ and $T=\mathrm{err}(\varphi)$.
\begin{enumerate}[label=(\alph*),leftmargin=*]
\item $|\Delta(\varphi)|\le a'+\varpi'N^{1/2}+bN$, with $a'=5s_{n,m}+4c_xr_{n,m}+2\pi_K$, $\varpi'=2K^{1/4}s_{n,m}$
and $b=6K^{1/2}c_xr_{n,m}$.
\item Under \eqref{eq:condR}, $|\Delta(\varphi)|\le a+\varpi N^{1/2}T^{3/10}+bN$, with $a=A_0s_{n,m}+7c_xr_{n,m}+2\pi_K$,
$\varpi=A_1K^{1/4}s_{n,m}$, $A_0=4+2(2\bar p)^{1/2}+4(c_\ast r_\ast)^{-1/2}$ and
$A_1=2\{(2\sqrt2\,\bar p)^{1/2}+2^{1/5}(2\sqrt2/(c_\ast r_\ast))^{1/2}\}$.
\end{enumerate}
\end{lemma}

\begin{proof}
We bound $\int\varphi\,d(\tilde P-P)$; the bound for $\tilde Q$ is the same, and the two add. On the window
$W=[x_0-K,x_0+K]$ Lemma~\ref{lem:lapbv} applies, with the variation of $\varphi$ controlled by its norm, and off $W$ only
the tail mass enters. Let $M_\varphi=\sup\varphi$ and $m_\varphi=\inf\varphi$. By \eqref{eq:laphone} the variation of $\varphi$ on
an interval of length $\lambda$ is at most $(2\lambda)^{1/2}N$, and off $W$, as $|\varphi|\le1$,
$|\int_{W^c}\varphi\,d(\tilde P-P)|\le\tilde P(W^c)+P(W^c)\le2\pi^P_K+2(\pi^P_Kx/n)^{1/2}+2c_x/n$, with
$\pi^P_K=P(W^c)$, by the event of Lemma~\ref{lem:lapintervals} for the two half-lines. For (a), write $\varphi=m_\varphi+h$ on $W$ and apply
Lemma~\ref{lem:lapbv} with $\mathrm V_W(h)\le2K^{1/2}N$ and $\int_Wh\,dP\le1$, and the event to $m_\varphi(\tilde P-P)(W)$.

For (b), the error $T$ controls the mass that $\varphi$ carries away from its extreme values where $p\neq q$, and
\eqref{eq:condR} turns this into mass under $P$ away from $x_0$. Note that $T>0$: if
$\operatorname{osc}(\varphi)<1$ by \eqref{eq:laperr}; otherwise $T=0$ would force $\varphi=M_\varphi$ almost everywhere on
$\{p>q\}$ and $\varphi=m_\varphi$ on $\{q>p\}$, which by \eqref{eq:condR} contain punctured neighborhoods of $x_0$ on the
right and on the left, and continuity at $x_0$ would give $M_\varphi=m_\varphi$. Write $\varphi=M_\varphi-h$ on $R_+=(x_0,\infty)$ and
$\varphi=m_\varphi+h$ on $R_-=(-\infty,x_0]$, with $h\in[0,1]$. As $(\tilde P-P)(\R)=0$,
\[
\int\varphi\,d(\tilde P-P)=(M_\varphi-m_\varphi)(\tilde P-P)(R_+)-\int_{R_+}h\,d(\tilde P-P)+\int_{R_-}h\,d(\tilde P-P),
\]
and by \eqref{eq:laperr}, $\int_{R_+}h(p-q)\le T$ and $\int_{R_-}h(q-p)\le T$. Put $w=\min(r_\ast,T^{2/5})$ and split $R_+$
into $I_1=(x_0,x_0+w]$, $I_2=(x_0+w,x_0+K]$ and $(x_0+K,\infty)$. On the short interval $I_1$,
$\mathrm V_{I_1}(h)\le(2w)^{1/2}N$ and $\int_{I_1}h\,dP\le\bar pw$. On $I_2$ the densities differ, \eqref{eq:condR} gives
$p\le(p-q)/(c_\ast w)$, so $\int_{I_2}h\,dP\le T/(c_\ast w)$, while $\mathrm V_{I_2}(h)\le(2K)^{1/2}N$.
Lemma~\ref{lem:lapbv} bounds the two pieces by
\begin{multline*}
(2\sqrt2\,\bar p)^{1/2}w^{3/4}N^{1/2}\Big(\frac xn\Big)^{1/2}+(2\bar p)^{1/2}\Big(\frac xn\Big)^{1/2}
+(\sqrt2N+\sqrt2K^{1/2}N+2)\frac{c_x}n\\
+\Big\{\Big(\frac{2\sqrt2}{c_\ast}\Big)^{1/2}K^{1/4}N^{1/2}+\Big(\frac2{c_\ast}\Big)^{1/2}\Big\}\Big(\frac Tw\Big)^{1/2}\Big(\frac xn\Big)^{1/2} .
\end{multline*}
The width $w$ balances the first and the last term: it gives $w^{3/4}\le T^{3/10}$ and
$(T/w)^{1/2}\le2^{1/5}r_\ast^{-1/2}T^{3/10}$, since if $w=T^{2/5}$ the latter is $T^{3/10}$, and if $w=r_\ast<T^{2/5}$ it is
$T^{3/10}T^{1/5}r_\ast^{-1/2}$ with $T\le2$; also $(T/w)^{1/2}\le(2/r_\ast)^{1/2}$. The half-line $R_-$ is split in the same
way, with $q-p$ for $p-q$. Adding the bounds for the two half-lines, for the term $(M_\varphi-m_\varphi)(\tilde
P-P)(R_+)$, which the event of Lemma~\ref{lem:lapintervals} controls, and off $W$, and using
$2\sqrt2(1+K^{1/2})\le6K^{1/2}$, we get
\[
\Big|\int\varphi\,d(\tilde P-P)\Big|\le A_0\Big(\frac xn\Big)^{1/2}+\frac{7c_x}n+2\pi^P_K
+A_1K^{1/4}N^{1/2}T^{3/10}\Big(\frac xn\Big)^{1/2}+\frac{6K^{1/2}Nc_x}n ,
\]
where $A_0$ collects the constant terms, $2(2\bar p)^{1/2}+4(c_\ast r_\ast)^{-1/2}$ from the two half-lines and at most
$4$ from $(\tilde P-P)(R_+)$ and the tails. For $\tilde Q$, use $q\le(p-q)/(c_\ast w)$ on $I_2$ and $Q(I_1)\le\bar pw$.
\end{proof}

\begin{lemma}[Other cut points]
\label{lem:lapown}
Let $\widehat u\le\tilde u<\tilde t\le\widehat t$, $a_o=(\tilde u-\widehat u)/\widehat s$ and $b_o=(\widehat t-\tilde t)/\widehat
s$, so that $a_o+b_o<1$. The ramp $\tilde\rp$ of $\widehat f$ between $\tilde u$ and $\tilde t$ satisfies
$\E_P\tilde\rp-\E_Q\tilde\rp\ge\E_P\widehat\rp-\E_Q\widehat\rp-\max(a_o,b_o)\,\mathrm{err}(\widehat\rp)/(1-a_o-b_o)$. By
Lemma~\ref{lem:lapclip}, own-term cut points at genuine quantiles have $a_o\le1/(m'\bar{\widehat D}\widehat s)$ and
$b_o\le1/(n'\bar{\widehat D}\widehat s)$.
\end{lemma}

\begin{proof}
Moving the cut points inward rescales the ramp, and the error of the ramp controls what the rescaling costs. As $\widehat f=\widehat u+\widehat s\widehat\rp$, $\tilde\rp=\chi(\widehat\rp)$ with
$\chi(s)=\min[1,\max\{0,(s-a_o)/(1-a_o-b_o)\}]$, and checking the three pieces $[0,a_o]$, $[a_o,1-b_o]$, $[1-b_o,1]$
gives $-a_o(1-s)\le(1-a_o-b_o)\{\chi(s)-s\}\le b_os$ on $[0,1]$. Integrate against $(p-q)_+$ and $(q-p)_+$ and use
$\mathrm{err}(\rp)=\int(1-\rp)(p-q)_++\int\rp(q-p)_+$ for $[0,1]$-valued $\rp$.
\end{proof}

\begin{proposition}[The fitted ramp keeps a linear excess]
\label{prop:lapfitted}
On the events of Lemma~\ref{lem:lapintervals} for $\tilde P$ and $\tilde Q$ with a common $x\ge\log8$, and with the
constants of Lemma~\ref{lem:lapdev} for the sizes $n',m'$, put, in case (i) of Theorem~\ref{thm:laplacerate},
\[
\epsilon_{\mathrm{fit}}=\max\Big\{\frac{24a}{\kappa_0},\frac{108\varpi}{\kappa_0c_V^{1/2}},\Big(\frac{300b}{\kappa_0c_V}\Big)^{4/5},
\Big(\frac2{c_V}\Big)^{20/33}\varpi^{40/33},\Big(\frac{1920}{\kappa_0c_V^2(n'\wedge m')}\Big)^{2/3}\Big\},
\]
and in case (ii)
\[
\epsilon_{\mathrm{fit}}=\max\{24a'/\kappa_0,(60\varpi'\kappa_0^{-1}c_V^{-1/2})^{8/9},(156b\kappa_0^{-1}c_V^{-1})^{4/5},
(576\kappa_0^{-1}c_V^{-2}(n'\wedge m')^{-1})^{2/3}\}.
\]
At every level $\alpha=\alpha^\star-\epsilon\in(0,\alpha^\star)$ with
$\epsilon\ge\epsilon_{\mathrm{fit}}$, the ramp fitted on $\tilde P,\tilde Q$ with genuine-quantile cut points, plain or
own-term, has cut points that do not cross and $\E_P\rp-\E_Q\rp-\alpha\ge\kappa_0\epsilon/2$.
\end{proposition}

\begin{proof}
In case (i) we bound the fluctuation of the population ramp, then the norm and the error of the fitted ramp by a
fixed-point argument, and then its fluctuation; the constants $24$, $108$ and $300$ in $\epsilon_{\mathrm{fit}}$ are
chosen so that $a$, $\varpi$ and $b$ each contribute at most $\kappa_0\epsilon/12$ at the end; the fourth term gives $\varpi^5y^{-5/2}\le\epsilon$, and the fifth, with $1920=48\cdot40$, bounds the shifts $a_o,b_o$ of the own-term cut points by $\kappa_0/40$. As
$\|\varphi^\star\|_\Hh\le c_V^{-1}\epsilon^{-1/4}$ and $\mathrm{err}(\varphi^\star)\le\epsilon$, Lemma~\ref{lem:lapdev}(b) and
the first three terms of $\epsilon_{\mathrm{fit}}$ give
\[
|\Delta^\star|\le a+c_V^{-1/2}\varpi\epsilon^{7/40}+c_V^{-1}b\epsilon^{-1/4}\le\kappa_0\epsilon\Big(\frac1{24}+\frac1{108}+\frac1{300}\Big)<\frac{\kappa_0\epsilon}4 .
\]
By Lemma~\ref{lem:lapcompare}(iii), $y\ge c_V\epsilon^{5/4}/2$, so $b\le y/4$. The fitted ramp now appears on both sides:
Lemma~\ref{lem:lapcompare}(i) and Lemma~\ref{lem:lapdev}(b) for $\widehat\varphi$ give
$\widehat Ny+\widehat T\le\epsilon+a+\varpi\widehat N^{1/2}\widehat T^{3/10}+b\widehat N$. The weighted arithmetic--geometric mean inequality $u^{1/2}v^{3/10}w^{1/5}\le u/2+3v/10+w/5$ with $u=\widehat Ny$,
$v=\widehat T$ and $w=\varpi^5y^{-5/2}$ separates the unknowns on the right, and with $b\widehat N\le\widehat Ny/4$ gives
\[
\frac{\widehat Ny}4+\frac{7\widehat T}{10}\le\epsilon+a+\frac{\varpi^5y^{-5/2}}5\le3\epsilon ,
\]
the fourth term of $\epsilon_{\mathrm{fit}}$ giving $\varpi^5y^{-5/2}\le\epsilon$. Hence $\widehat N\le24c_V^{-1}\epsilon^{-1/4}$
and $\widehat T\le5\epsilon$: up to constants, the fitted ramp has the norm and the error of the population one.
Lemma~\ref{lem:lapdev}(b) once more gives $\widehat\Delta\le a+8c_V^{-1/2}\varpi+24c_V^{-1}b\epsilon^{-1/4}$, so that
$|\Delta^\star|+\widehat\Delta\le2a+9c_V^{-1/2}\varpi+25c_V^{-1}b\epsilon^{-1/4}\le\kappa_0\epsilon/4$, and by Lemma~\ref{lem:lapcompare}(ii) the plain ramp has excess at least $3\kappa_0\epsilon/4$. For own-term cut points,
$\bar{\widehat D}\widehat s\ge y/\widehat N\ge c_V^2\epsilon^{3/2}/48$, so by Lemma~\ref{lem:lapown} and the fifth term
$\max(a_o,b_o)\le\kappa_0/40$, and the loss is at most $(\kappa_0/40)(20/19)\,5\epsilon<\kappa_0\epsilon/4$.

Case (ii) is the same argument with part (a) of Lemma~\ref{lem:lapdev}, which has no factor $T^{3/10}$. It gives
$|\Delta^\star|\le a'+c_V^{-1/2}\varpi'\epsilon^{-1/8}+c_V^{-1}b\epsilon^{-1/4}\le\kappa_0\epsilon/4$, so
$y\ge c_V\epsilon^{5/4}/2\ge4b$. With $\varpi'\widehat N^{1/2}\le\widehat Ny/4+\varpi'^2/y$ and $\varpi'^2/y\le\epsilon$
(second term), $\widehat Ny/2+\widehat T\le3\epsilon$, so $\widehat N\le12c_V^{-1}\epsilon^{-1/4}$, $\widehat T\le3\epsilon$ and
$|\Delta^\star|+\widehat\Delta\le2a'+5c_V^{-1/2}\varpi'\epsilon^{-1/8}+13c_V^{-1}b\epsilon^{-1/4}\le\kappa_0\epsilon/4$; the own
terms have $\max(a_o,b_o)\le24/\{c_V^2(n'\wedge m')\epsilon^{3/2}\}\le\kappa_0/24$ and cost less than $\kappa_0\epsilon/4$.
\end{proof}

\begin{proof}[Proof of Theorem~\ref{thm:laplaceconst}]
The structure is that of the proof of Theorem~\ref{thm:lorate}, with Proposition~\ref{prop:lapfitted} in place of the
margin: one event makes every fitted ramp good, and a union over the levels handles the bets.

We first fix the fitting event. Take
\[
x=\log\{16(n\vee m+1)^2/\eta\},\qquad K=\max\{1,C'\log(n\vee m)\},
\]
so that the window of Lemma~\ref{lem:lapdev} leaves a tail mass of order $1/(n\vee m)$. Let $\mathcal E_{\mathrm{fit}}$
be the intersection of the events of Lemma~\ref{lem:lapintervals}, each with $\eta'=\eta/8$, for the $X$'s and the
$Y$'s of batch $0$ and of batches $0$ and $1$. These events weaken as $x$ grows, so all four hold with this $x$, and
$\Pp(\mathcal E_{\mathrm{fit}}^c)\le\eta/2$. The constants of Lemma~\ref{lem:lapdev} and the last term of
$\epsilon_{\mathrm{fit}}$ decrease in the sample sizes, so the $\epsilon_{\mathrm{fit}}$ of batch $0$ serves for both
fits.

Next we choose the distance $\epsilon$ from $\alpha^\star$. Let
\[
\widetilde T_{\mathrm{bet}}(J')=\Big(\frac2N\Big)^{1/2}\Big\{\Big(1+\frac{16}{v^2_{\min}}\Big)L_\delta^{1/2}
+\frac2{v_{\min}}\big(\log(2J'/\eta)\big)^{1/2}\Big\},
\]
the $T_{\mathrm{bet}}(\eta,J')$ of \eqref{eq:loconst} divided by $2\Delta_\phi$. With $J_\epsilon$ and $a_\epsilon$ as in
Theorem~\ref{thm:lorate}, fix $\epsilon\in[\epsilon_{\mathrm{fit}},\alpha^\star)$ with
$\kappa_0\epsilon\ge2\widetilde T_{\mathrm{bet}}(J_\epsilon)$, and put $\varrho_\ast=\kappa_0\epsilon/2$. Write each grid
level $\alpha_j\le\alpha^\star-\epsilon$ as $\alpha_j=\alpha^\star-\epsilon\xi_j$, so that $\xi_j\ge1$ and
$\epsilon\xi_j\ge\epsilon_{\mathrm{fit}}$. On $\mathcal E_{\mathrm{fit}}$, Proposition~\ref{prop:lapfitted} at the level
$\alpha_j$ shows that both batches have a ramp with
\[
\E_P\rp_b-\E_Q\rp_b-\alpha_j\ \ge\ \varrho_\ast\xi_j .
\]

The bets then win at every such level. As $1>\kappa_0\epsilon\ge10(2L_\delta/N)^{1/2}$, we have $N\ge200L_\delta$, and
the first bound of \eqref{eq:betbounds}, with $\mathcal E=\mathcal E_{\mathrm{fit}}$ and $\varrho=\varrho_\ast\xi_j$,
gives
\[
\Pp\big(\mathcal E_{\mathrm{fit}}\cap\{H_{0,\alpha_j}\text{ is not rejected}\}\big)
\le\exp\Big\{-\frac{v^2_{\min}}{4L_\delta}\big(\varrho_\ast\xi_jS-C_0\big)_+^2\Big\},
\]
with $S=(NL_\delta/2)^{1/2}$ and $C_0=L_\delta(1+16/v^2_{\min})$. Multiplied by $S$, the condition
$\varrho_\ast\ge\widetilde T_{\mathrm{bet}}(J_\epsilon)$ reads
$a_\ast=\varrho_\ast S-C_0\ge2v_{\min}^{-1}\{L_\delta\log(2J_\epsilon/\eta)\}^{1/2}$. Since
$\varrho_\ast\xi_jS-C_0\ge a_\ast\xi_j$, each term is at most $e^{-A\xi_j^2}$ with
$A=v^2_{\min}a_\ast^2/(4L_\delta)\ge\log(2J_\epsilon/\eta)>\log2$. The summation over the levels in the proof of
Theorem~\ref{thm:lorate}, with $\gamma=1$, bounds the sum by $\eta/2$, and \eqref{eq:downclosed} gives
$\Pp(\widehat\alpha<a_\epsilon)\le\eta$. Put $\epsilon_{n,m}=\max\{\epsilon_{\mathrm{fit}},2\widetilde
T_{\mathrm{bet}}(J)/\kappa_0\}$, which is either at least $\alpha^\star$ or satisfies the requirements on $\epsilon$.
With Corollary~\ref{cor:lower}, with probability at least $1-\delta-\eta$,
\[
\alpha^\star-\Delta_{\mathcal G}-\min\{\epsilon_{n,m},\alpha^\star\}\ \le\ \widehat\alpha\ \le\ \alpha^\star .
\]

It remains to read the orders of the terms of $\epsilon_{\mathrm{fit}}$ and of $\widetilde T_{\mathrm{bet}}$. As in the
proof of Theorem~\ref{thm:lorate}(iii), $n'\wedge m'\ge(n\wedge m)/5$ and $N\ge(n\wedge m)/2$ for $n\wedge m\ge10$, so
that
\[
s_{n',m'}\le2\Big(\frac{5x}{n\wedge m}\Big)^{1/2},\qquad r_{n',m'}\le\frac{10}{n\wedge m},\qquad
\pi_K\le\frac{C''}{n\vee m},
\]
with $C''$ depending on $C'$ and, in case (i), on $c_\ast r_\ast$, and
$K^{1/4}x^{1/2}\le C'''(\log(n\vee m))^{1/4}L_{n,m}^{1/2}$ with $C'''\ge1$ depending only on $C'$. If
$K^{1/2}x>n\wedge m$, the bounds of the theorem exceed $1>\alpha^\star$ once $C\ge C'''$, and there is nothing to prove.
Otherwise $u=K^{1/4}\{x/(n\wedge m)\}^{1/2}\le1$, and $u^{8/5}=\{K^{1/2}x/(n\wedge m)\}^{4/5}$, $u^{40/33}$,
$(n\wedge m)^{-2/3}$ and $x/(n\wedge m)$, the other quantities through which $a$, $\varpi$, $b$ and the last term enter
$\epsilon_{\mathrm{fit}}$ in case (i), are all at most $u$. Hence
\[
\epsilon_{\mathrm{fit}}\le Cu\ \text{ in case (i)},\qquad
\epsilon_{\mathrm{fit}}\le C\Big\{\frac{K^{1/2}x}{n\wedge m}\Big\}^{4/9}\ \text{ in case (ii)},
\]
and, as $v^2_{\min}\le4$,
\[
\widetilde T_{\mathrm{bet}}(J)\le\frac{2\{20L_\delta^{1/2}+4(\log(2J/\eta))^{1/2}\}}{v^2_{\min}(n\wedge m)^{1/2}}.
\]
\end{proof}

Near every pair covered by part (i) no valid lower bound does better than $(n\wedge m)^{-1/2}$. The pairs of Proposition~\ref{prop:lower2pt} are separated
and do not satisfy these hypotheses, so we perturb $P$ away from $x_0$ instead.

\begin{lemma}[A local two-point bound]
\label{lem:laplocal}
Let $(P,Q)$ satisfy the hypotheses of Theorem~\ref{thm:laplacerate}(i), let $r\in(0,r_\ast]$ be such that $p,q$ are continuous
and positive on $[x_0-2r,x_0+2r]$, put $A_1=(x_0+r,x_0+2r)$, $B_1=(x_0-2r,x_0-r)$, $\pi=\min\{P(A_1),P(B_1)\}>0$ and, for
$h\in[0,1]$,
\[
P_h=P+h\pi\Big\{\frac{P(\cdot\cap A_1)}{P(A_1)}-\frac{P(\cdot\cap B_1)}{P(B_1)}\Big\}.
\]
Then $P_h$ is a probability measure, $\TV(P_h,Q)=\alpha^\star+h\pi$, and $(P_h,Q)$ satisfies the hypotheses of
Theorem~\ref{thm:laplacerate}(i), with $c_\ast/2$, $2\bar p$ and a tail constant depending only on $C'$ and $r$. Every
$\tilde\alpha$, a measurable function of the samples and of an independent randomization, with
$\Pp_{P,Q}(\tilde\alpha>\alpha^\star)\le\delta$ satisfies
$\Pp_{P_h,Q}\{\TV(P_h,Q)-\tilde\alpha\ge h\pi\}\ge1-\delta-h(n\pi)^{1/2}$. The same holds for the second sample, with $Q$
moved from $A_1$ to $B_1$, $m$ in place of $n$ and $\pi'=\min\{Q(A_1),Q(B_1)\}$ in place of $\pi$.
\end{lemma}

\begin{proof}
Moving mass $h\pi$ from $B_1$ to $A_1$, away from $x_0$, raises the total variation by $h\pi$ without touching the
crossing, and costs little in $\chi^2$-divergence. The density of $P_h$ is $p\{1+h\pi/P(A_1)\}$ on $A_1$, $p\{1-h\pi/P(B_1)\}\ge0$ on $B_1$ and $p$ elsewhere; it jumps
at $x_0\pm r$ and $x_0\pm2r$, which is harmless, as the hypotheses ask continuity near $x_0$ only. By
\eqref{eq:condR}, $p>q$ almost everywhere on $A_1$ and $q>p$ on $B_1$, so $p_h-q$ has the sign of $p-q$ everywhere and
$\TV(P_h,Q)=\int_{\{p>q\}}(p_h-q)=\alpha^\star+h\pi$. Near $x_0$ nothing changes, so the hypotheses of
Proposition~\ref{prop:newsobolev} hold; on $A_1$, $|p_h-q|\ge|p-q|$ and $p_h+q\le2(p+q)$, and on $B_1$, $|p_h-q|\ge|p-q|$ and
$p_h\le p$, which gives \eqref{eq:condR} with $c_\ast/2$. Further, $\chi^2(P_h\|P)=(h\pi)^2\{1/P(A_1)+1/P(B_1)\}\le2h^2\pi$, so,
by $\mathrm{KL}\le\chi^2$ for each of the $n$ observations and Pinsker's inequality, the laws of the data and the randomization under $(P,Q)$ and
$(P_h,Q)$ are at total variation at most $\{n\chi^2(P_h\|P)/2\}^{1/2}\le h(n\pi)^{1/2}$, and on $\{\tilde\alpha\le\alpha^\star\}$ the
error at $(P_h,Q)$ is at least $h\pi$. For the second sample, $Q_h=Q+h\pi'\{Q(\cdot\cap B_1)/Q(B_1)-Q(\cdot\cap A_1)/Q(A_1)\}$
with $\pi'=\min\{Q(A_1),Q(B_1)\}$ gives $\TV(P,Q_h)=\alpha^\star+h\pi'$ in the same way.
\end{proof}

With $h=(1-\delta)/\{2(n\pi)^{1/2}\}$, which is at most $1$ once $n\pi\ge1/4$, a valid lower bound falls short at $(P_h,Q)$
by at least $\tfrac12(1-\delta)(\pi/n)^{1/2}$ with probability at least $(1-\delta)/2$. The constants of
Proposition~\ref{prop:newsobolev} for $(P_h,Q)$ exist; whether they stay bounded uniformly in $h$, as a minimax
statement over a class containing these pairs would need, is open.

\subsection{The two-point lower bound and the minimax rate}
\label{sec:optproofs}

This subsection proves Proposition~\ref{prop:lower2pt} and Corollary~\ref{cor:minimax} of the main text. The
perturbed pairs of Proposition~\ref{prop:lower2pt} satisfy the margin condition with a constant that does not depend on
$a$ or $h$.

\begin{lemma}[The two-point pairs satisfy the margin]
\label{lem:twopointmargin}
Let (A1) and (A2) of Assumption~\ref{ass:regularity} hold, let $R$, $P_0$ and $Q_0$ be probability measures
concentrated on pairwise disjoint measurable sets $S_R$, $S_P$ and $S_Q$, define $P_a$, $Q_a$, $P_{a,h}$ and $Q_{a,h}$
as in Proposition~\ref{prop:lower2pt}, and suppose some $f\in\Hh$ takes the value $0$ on $S_R$, $1$ on $S_P$ and $-1$
on $S_Q$. Then for every $a\in(0,1)$ and $h\in[0,1-a)$, each of the pairs $(P_{a,h},Q_a)$ and $(P_a,Q_{a,h})$, with
$\alpha^\star=a+h$, satisfies $D_{\alpha^\star-\epsilon}\ge\epsilon/\|f\|_\Hh$ for all $\epsilon\in[0,\alpha^\star]$,
that is, Assumption~\ref{ass:margin} with $\gamma=1$ and $c_0=1/\|f\|_\Hh$.
\end{lemma}

\begin{proof}
Consider $(P_{a,h},Q_a)$ and fix $\alpha\in[0,\alpha^\star]$. Under $P_{a,h}$, $f(X)\in\{0,1\}$ almost surely with
$P_{a,h}(f(X)=1)=a+h$. Since $1-\alpha\ge1-a-h$, an $\alpha$-trimming $P'=gP_{a,h}$ puts mass at most
$\beta(1-a-h)\le1$ on $\{f=0\}$, so $\E_{P'}f\ge1-\beta(1-a-h)=(a+h-\alpha)/(1-\alpha)$. Under $Q_a$,
$f(Y)\in\{0,-1\}$ almost surely with $Q_a(f(Y)=-1)=a$, and an $\alpha$-trimming $Q'$ puts mass at most
$\min\{1,\beta(1-a)\}$ on $\{f=0\}$, so $\E_{Q'}f\le-(1-\beta(1-a))_+=-(a-\alpha)_+/(1-\alpha)\le0$. By the reproducing
property and the Cauchy--Schwarz inequality, for all such $P'$ and $Q'$,
\[
\MMD(P',Q')\ge\frac{\E_{P'}f-\E_{Q'}f}{\|f\|_\Hh}\ge\frac{a+h-\alpha}{(1-\alpha)\|f\|_\Hh}\ge\frac{\alpha^\star-\alpha}{\|f\|_\Hh} .
\]
Taking the infimum over $P'$ and $Q'$ and putting $\alpha=\alpha^\star-\epsilon$ gives the claim. The pair
$(P_a,Q_{a,h})$ is the same computation with $-f$ in place of $f$ and the roles of the two laws exchanged.
\end{proof}

\begin{proof}[Proof of Proposition~\ref{prop:lower2pt}]
The two pairs are close in $\chi^2$, so a bound valid at the first must often fall short at the second, by Le Cam's
two-point method. The three measures are mutually singular. Hence $P_{a,h}\wedge Q_a$ has mass $\min(1-a-h,1-a)=1-a-h$ on $S_R$ and
none elsewhere, so $\TV(P_{a,h},Q_a)=a+h$, and likewise $\TV(P_a,Q_{a,h})=a+h$ and $\TV(P_a,Q_a)=a$. The density of
$P_{a,h}$ with respect to $P_a$ is $(1-a-h)/(1-a)$ on $S_R$ and $(a+h)/a$ on $S_P$, so
\[
\chi^2(P_{a,h}\,\|\,P_a)=(1-a)\Big(\frac h{1-a}\Big)^2+a\Big(\frac ha\Big)^2=\frac{h^2}{a(1-a)} .
\]
Let $\mathbb Q_0$ and $\mathbb Q_1$ be the laws of the data and the randomization under $(P_a,Q_a)$ and under
$(P_{a,h},Q_a)$. They differ only in the law of the $n$ observations of the first sample, and the Kullback--Leibler
divergence adds over independent coordinates, so with $\mathrm{KL}\le\log(1+\chi^2)\le\chi^2$,
$\mathrm{KL}(\mathbb Q_1\|\mathbb Q_0)\le nh^2/\{a(1-a)\}$. By Pinsker's inequality
$\TV(\mathbb Q_0,\mathbb Q_1)\le\{\mathrm{KL}(\mathbb Q_1\|\mathbb Q_0)/2\}^{1/2}\le h\{n/(2a(1-a))\}^{1/2}$. Therefore
\[
\Pp_{P_{a,h},Q_a}(\tilde\alpha\le a)\ge\Pp_{P_a,Q_a}(\tilde\alpha\le a)-\TV(\mathbb Q_0,\mathbb Q_1)
\ge1-\delta-h\sqrt{n/\{2a(1-a)\}},
\]
and on $\{\tilde\alpha\le a\}$ the error at $(P_{a,h},Q_a)$, whose total variation is $a+h$, is at least $h$. The same
argument with the second sample perturbed gives the bound with $m$.
\end{proof}

\begin{proof}[Proof of Corollary~\ref{cor:minimax}]
Write $\bar n=n\wedge m$. For (i), let $R$, $P_0$ and $Q_0$ be the point masses at $x_2$, $x_1$ and $x_3$, put $a=a_0$ and
$h=c_{\mathrm{lo}}\bar n^{-1/2}$, so that $a_0+h\le a_1<1$ by the condition on $\bar n$. The pair $(P_{a_0},Q_{a_0})$ belongs to
$\mathcal M$ by Example~\ref{ex:separated}, since $c_0\le1/\|f\|_\Hh\le2/\|f\|_\Hh$, and the two perturbed pairs belong
to $\mathcal M$ by Lemma~\ref{lem:twopointmargin}. If $\bar n=n$, Proposition~\ref{prop:lower2pt} at the pair
$(P_{a_0,h},Q_{a_0})$ gives an error of at least $h$ with probability at least
$1-\delta-h\{n/(2a_0(1-a_0))\}^{1/2}=(1-\delta)/2$; if $\bar n=m$, the pair $(P_{a_0},Q_{a_0,h})$ gives the same with $m$.

For (ii), validity is Corollary~\ref{cor:lower}. On the grid $\{j\bar n^{-1/2}\}$ the smallest gap is $\Delta_{\min}=\bar n^{-1/2}$ and
the mesh below $\alpha^\star$ is at most $\bar n^{-1/2}$. Apply Theorem~\ref{thm:lorate}(i) with $\epsilon=C_\epsilon
\bar n^{-1/2}$: then $J_\epsilon\le2(1+C_\epsilon)$ does not grow with $\bar n$, and by Theorem~\ref{thm:lorate}(iii),
$r_1\le12/\bar n$ and the number of pairs is at least $\bar n/2$, so $\beta_\star\max\{T_{\mathrm{fit}}(\eta),
T_{\mathrm{bet}}(\eta,J_\epsilon)\}\le C''(C_\epsilon)\bar n^{-1/2}$ with $\beta_\star\le(1-a_1)^{-1}$ and $C''(C_\epsilon)$
growing like $\{\log(1+C_\epsilon)\}^{1/2}$. Choose $C_\epsilon$ with $c_0C_\epsilon\ge C''(C_\epsilon)$ and let
$\bar n\ge10$, so that Theorem~\ref{thm:lorate}(iii) applies; every step of the proof holds for these sizes, so $n_1=10$
will do. If $\epsilon>\alpha^\star$, then $\widehat\alpha\ge0>\alpha^\star
-\epsilon$ and there is nothing to prove. Otherwise $\epsilon\in(0,\alpha^\star]$, Theorem~\ref{thm:lorate}(i) gives
$\Pp(\widehat\alpha<a_\epsilon)\le\eta$, and $a_\epsilon\ge\alpha^\star-\epsilon-\bar n^{-1/2}$. In both
cases $\Pp(\alpha^\star-\widehat\alpha>(C_\epsilon+1)\bar n^{-1/2})\le\eta$. Take $C=C_\epsilon+1$.
\end{proof}
\space
\section{Proofs for Section~\ref{sec:trimming} and Proposition~\ref{prop:bangbang}}
\label{app:trimming}

This section proves Proposition~\ref{prop:bangbang} and the results of Section~\ref{sec:trimming}, and with them the
uniform bound \eqref{eq:hausdorff} (Corollary~\ref{cor:uniform}). The witnesses, cuts, kept sets, trimming densities,
the constants $c_f$, $C_f$, $c'_f$, $r_Q$ and the modulus $\Omega$ are those of Section~\ref{sec:trimming}.

\subsection{Structure of the trimming}

\begin{proof}[Proof of Proposition~\ref{prop:bangbang}]
The minimizing difference of two polytopes is unique, and the gradient in the weights is the witness, so each
optimal weight vector solves the bathtub problem of Lemma~\ref{lem:trimmedmean}. The objective is continuous and convex
on a compact convex set, so $\mathcal T_\alpha$ is nonempty, compact and convex.

(i) With $u=\sum_iw_i\phi(X_i)$ and $z=\sum_jv_j\phi(Y_j)$ the objective is $\|u-z\|_\Hh^2$, and $u-z$ ranges over the
convex compact set $\mathcal C_\alpha(P_n)-\mathcal C_\alpha(Q_m)$ (Proposition~\ref{prop:georep}), whose element of
minimal norm is unique and has norm $\widehat D_\alpha$.

(ii) The objective $F$ is convex and differentiable, so $(w^\star,v^\star)\in\mathcal T_\alpha$ if and only if $w^\star$ minimizes $w\mapsto\langle\nabla_wF(w^\star,v^\star),w\rangle$ over the capped simplex, and likewise for $v^\star$. By the reproducing property and part (i), at every minimizer
\[
(\nabla_wF)_i=2(K_{XX}w^\star-K_{XY}v^\star)_i=2\langle\phi(X_i),u^\star-z^\star\rangle_\Hh=2\,\widehat d(X_i),
\]
the last equality because $u^\star-z^\star=\widehat d$ is the same at every minimizer, and Lemma~\ref{lem:trimmedmean}, in its
minimum form, describes the minimizers of this linear function. The statement for $v^\star$ follows from $(\nabla_vF)_j=-2\,\widehat d(Y_j)$ and the maximum form
of the same lemma.

(iii) $\widehat D_\alpha=0$ forces $\widehat d=0$ by (i), that is, $u^\star=z^\star$. Conversely, every feasible pair
with $u=z$ attains the value $0$, which is the minimum.
\end{proof}

The population trimming has the same threshold form \eqref{eq:popdensity}.

\begin{lemma}[Threshold form of the population trimming]
\label{lem:popthreshold}
Let Assumption~\ref{ass:regularity} hold, let $\alpha\in[0,1)$, and suppose that
$\dist(\mathcal C_\alpha(P),\mathcal C_\alpha(Q))>0$. Let $d^\star$ be the element of minimal norm of
$\mathcal C_\alpha(P)-\mathcal C_\alpha(Q)$, write $d^\star=a^\star-b^\star$ with $a^\star=\E_P[g^\star\phi]$ for some
$g^\star\in\mathcal G_\alpha(P)$ and $b^\star\in\mathcal C_\alpha(Q)$, and let $t$ be any $(1-\alpha)$-quantile of
$d^\star(X)$ for $X\sim P$. Then $g^\star=\beta$ $P$-almost surely on $\{d^\star<t\}$ and $g^\star=0$ $P$-almost surely
on $\{d^\star>t\}$.
\end{lemma}

\begin{proof}
We show that $g^\star$ minimizes a linear functional of the density, and then that every minimizer of that
functional has the threshold form. The set $\mathcal C_\alpha(P)-\mathcal C_\alpha(Q)$ is convex and weakly compact, as the image of the weakly compact
product of the two regions (Lemma~\ref{lem:attainment}) under the weakly continuous map $(a,b)\mapsto a-b$, hence
norm-closed; so $d^\star$ exists, and it has a decomposition of the stated form by \eqref{eq:region}. For
$a\in\mathcal C_\alpha(P)$, $a-b^\star$ lies in the same set, so $\langle a-a^\star,d^\star\rangle_\Hh\ge0$, and by the
reproducing property $g^\star$ minimizes $g\mapsto\E_P[gU]$ over $\mathcal G_\alpha(P)$, where $U=d^\star(X)$.

The threshold density is such a minimizer, and every minimizer agrees with it off the cut. Since $\Pp(U<t)\le1-\alpha\le\Pp(U\le t)$, there is
$c\in[0,\beta]$ such that $h=\beta\mathbf 1\{U<t\}+c\mathbf 1\{U=t\}$ lies in $\mathcal G_\alpha(P)$, as in the proof of
Lemma~\ref{lem:cvar}. For every $g\in\mathcal G_\alpha(P)$, $(g-h)(U-t)\ge0$ pointwise, because $h=\beta\ge g$ where $U<t$
and $h=0\le g$ where $U>t$, and $\E[(g-h)(U-t)]=\E[gU]-\E[hU]$ since $\E g=\E h=1$. So $h$ is a minimizer, and $(g^\star-h)(U-t)\ge0$ has
expectation at most $0$, hence vanishes almost surely: $g^\star=\beta$ where $U<t$ and $g^\star=0$ where $U>t$.
\end{proof}

\subsection{The parametric rate}

Throughout this subsection $\alpha\in(0,1)$, $\mathcal A=\mathcal C_\alpha(P)$, $\mathcal B=\mathcal C_\alpha(Q)$,
$\mathcal A_n,\mathcal B_m$ are the empirical regions, $\mathrm{sep}(f)=\mathrm{sep}(f;\mathcal A,\mathcal B)$ and
$\mathrm{sep}_n(f)=\mathrm{sep}(f;\mathcal A_n,\mathcal B_m)$ as in \eqref{eq:sep}, and $\xi(f)=\mathrm{sep}_n(f)-
\mathrm{sep}(f)$ is the fluctuation of the separation. For a probability measure $R$ and a bounded measurable $f$ write
\begin{equation}
\begin{aligned}
L_R(f)&=\inf_{g\in\mathcal G_\alpha(R)}\E_R[gf]=\sup_{t\in\R}\big\{t-\beta\,\E_R(t-f)_+\big\},\\
U_R(f)&=\sup_{g\in\mathcal G_\alpha(R)}\E_R[gf]=\inf_{u\in\R}\big\{u+\beta\,\E_R(f-u)_+\big\},
\end{aligned}
\label{eq:twotail}
\end{equation}
the equalities being Lemma~\ref{lem:cvar} applied to $-f$ and to $f$. For $f\in\Hh$ these are
$L_R(f)=-h_{\mathcal C_\alpha(R)}(-f)$ and $U_R(f)=h_{\mathcal C_\alpha(R)}(f)$, with the support functions of
Section~\ref{sec:regions}. The function maximized in $L_R$ is concave with
right and left derivatives $1-\beta R(f\le t)$ and $1-\beta R(f<t)$, so its maximizers are the $t$ with
$R(f<t)\le1-\alpha\le R(f\le t)$, the $(1-\alpha)$-quantiles of $f$ under $R$; likewise the minimizers in $U_R$ are the
$\alpha$-quantiles. By Proposition~\ref{prop:georep}, $\mathrm{sep}(f)=L_P(f)-U_Q(f)$ and
$\mathrm{sep}_n(f)=L_{P_n}(f)-U_{Q_m}(f)$.

\begin{lemma}[A quadratic margin on the unit sphere]
\label{lem:sepmargin}
Let $K\subset\Hh$ be nonempty, convex and closed, write $\mathrm{sep}_K(f)=\inf_{z\in K}\langle f,z\rangle_\Hh$, let
$d^\star$ be its element of minimal norm, suppose $D=\|d^\star\|_\Hh>0$, and put $f^\star=d^\star/D$. Then
$\mathrm{sep}_K(f)\le\langle f,d^\star\rangle_\Hh$ for every $f\in\Hh$, $\mathrm{sep}_K(f^\star)=D=\max_{f\in B_\Hh}
\mathrm{sep}_K(f)$, and
\[
\mathrm{sep}_K(f^\star)-\mathrm{sep}_K(u)\ \ge\ \tfrac D2\,\|u-f^\star\|_\Hh^2
\qquad\text{for every }u\in\Hh\text{ with }\|u\|_\Hh=1 .
\]
For $d'\in\Hh$ with $D'=\|d'\|_\Hh>0$ and $f'=d'/D'$,
$\|d'-d^\star\|_\Hh^2=(D'-D)^2+D'D\,\|f'-f^\star\|_\Hh^2$.
\end{lemma}

\begin{proof}
Since $d^\star\in K$, $\mathrm{sep}_K(f)\le\langle f,d^\star\rangle_\Hh$. For $z\in K$ and $t\in(0,1]$,
$\|d^\star\|_\Hh^2\le\|d^\star+t(z-d^\star)\|_\Hh^2$ by convexity, and letting $t\downarrow0$ gives $\langle d^\star,z-d^\star\rangle_\Hh\ge0$, that is, $\langle f^\star,z\rangle_\Hh\ge D$ on $K$. So
$\mathrm{sep}_K(f^\star)\ge D=\langle f^\star,d^\star\rangle_\Hh\ge\mathrm{sep}_K(f^\star)$, and for $f\in B_\Hh$,
$\mathrm{sep}_K(f)\le\langle f,d^\star\rangle_\Hh\le D$ by Cauchy--Schwarz. For a unit vector $u$,
$\langle u,f^\star\rangle_\Hh=1-\tfrac12\|u-f^\star\|_\Hh^2$, so $\mathrm{sep}_K(u)\le D\langle u,f^\star\rangle_\Hh=
D-\tfrac D2\|u-f^\star\|_\Hh^2$. Finally $\|D'f'-Df^\star\|_\Hh^2=D'^2+D^2-2D'D\langle f',f^\star\rangle_\Hh$, and
substituting $\langle f',f^\star\rangle_\Hh=1-\tfrac12\|f'-f^\star\|_\Hh^2$ gives the identity.
\end{proof}

\begin{lemma}[A basic inequality at two cuts]
\label{lem:basicineq}
Let (A1) and (A2) of Assumption~\ref{ass:regularity} hold, let $\alpha\in(0,1)$, and suppose $D=D_\alpha(P,Q)>0$ and
$\widehat D=\widehat D_\alpha>0$. Let $f^\star=d^\star/D$ and $\widehat f=\widehat d/\widehat D$, let $t^\star$ be a
$(1-\alpha)$-quantile of $f^\star(X)$ for $X\sim P$ and $u^\star$ an $\alpha$-quantile of $f^\star(Y)$ for $Y\sim Q$,
and let $\widehat t$ be the $(n_{1-\alpha}+1)$-th smallest of the values $\widehat f(X_i)$ and $\widehat u$ the
$(m_{1-\alpha}+1)$-th largest of the values $\widehat f(Y_j)$. Then, for every realization of the samples,
\[
\begin{aligned}
\tfrac D2\|\widehat f-f^\star\|_\Hh^2\ \le\ \xi(\widehat f)-\xi(f^\star)
\ \le\ \beta\Big[&(P-P_n)\big\{(\widehat t-\widehat f)_+-(t^\star-f^\star)_+\big\}\\
&+(Q-Q_m)\big\{(\widehat f-\widehat u)_+-(f^\star-u^\star)_+\big\}\Big].
\end{aligned}
\]
\end{lemma}

\begin{proof}
The population separation loses at $\widehat f$ at most the difference of the fluctuations, and the
quadratic margin turns that loss into a squared distance; the right-hand side then comes from evaluating each trimmed
mean at a cut that is optimal for the other measure. The set $\mathcal A-\mathcal B$ is nonempty, convex and closed (proof of
Lemma~\ref{lem:popthreshold}), with element of minimal norm $d^\star$, and $\mathrm{sep}=\mathrm{sep}_{\mathcal
A-\mathcal B}$; the same holds for the polytope $\mathcal A_n-\mathcal B_m$ and $\widehat d$
(Proposition~\ref{prop:bangbang}(i)). By Lemma~\ref{lem:sepmargin}, $\widehat f$ maximizes $\mathrm{sep}_n$ over
$B_\Hh\ni f^\star$, so $\mathrm{sep}(f^\star)-\mathrm{sep}(\widehat f)=[\mathrm{sep}_n(f^\star)-\mathrm{sep}_n(\widehat
f)]+\xi(\widehat f)-\xi(f^\star)\le\xi(\widehat f)-\xi(f^\star)$, and Lemma~\ref{lem:sepmargin} at $u=\widehat f$ bounds
the left-hand side below by $\tfrac D2\|\widehat f-f^\star\|^2_\Hh$.

At $\widehat f$ we bound the fluctuation from above, using the empirical cuts. Since $P_n(\widehat f<\widehat t)\le n_{1-\alpha}/n\le1-\alpha<(n_{1-\alpha}+1)/n\le
P_n(\widehat f\le\widehat t)$, $\widehat t$ is a $(1-\alpha)$-quantile of $\widehat f$ under $P_n$. By the
characterization after \eqref{eq:twotail}, $L_{P_n}(\widehat f)=\widehat t-\beta P_n(\widehat t-\widehat f)_+$, while
$L_P(\widehat f)\ge\widehat t-\beta P(\widehat t-\widehat f)_+$, so $L_{P_n}(\widehat f)-L_P(\widehat f)\le\beta(P-P_n)
(\widehat t-\widehat f)_+$. Likewise $\widehat u$, the $k$-th smallest of the $\widehat f(Y_j)$ with $k=m-m_{1-\alpha}$, is an $\alpha$-quantile of
$\widehat f$ under $Q_m$, as $\alpha\le k/m<\alpha+1/m$ gives $Q_m(\widehat f<\widehat u)<\alpha\le Q_m(\widehat f\le\widehat u)$. So
$U_{Q_m}(\widehat f)=\widehat u+\beta Q_m(\widehat f-\widehat u)_+$ and $U_Q(\widehat f)\le\widehat u+\beta Q(\widehat
f-\widehat u)_+$. Adding, since $\xi=(L_{P_n}-L_P)+(U_Q-U_{Q_m})$,
$\xi(\widehat f)\le\beta[(P-P_n)(\widehat t-\widehat f)_++(Q-Q_m)(\widehat f-\widehat u)_+]$.

At $f^\star$ we bound it from below in the same way, using the population cuts. Now $t^\star$ maximizes the population objective, so $L_P(f^\star)=t^\star-\beta
P(t^\star-f^\star)_+$, while $L_{P_n}(f^\star)\ge t^\star-\beta P_n(t^\star-f^\star)_+$; and $U_Q(f^\star)=u^\star+\beta
Q(f^\star-u^\star)_+$, while $U_{Q_m}(f^\star)\le u^\star+\beta Q_m(f^\star-u^\star)_+$. Hence
$\xi(f^\star)\ge\beta[(P-P_n)(t^\star-f^\star)_++(Q-Q_m)(f^\star-u^\star)_+]$, and subtracting gives the claim.
\end{proof}

The next lemma uses, as in Section~\ref{app:analytic}, $n_\star=nm/(n+m)$ and $e_k=\E|\varepsilon_1+\dots+
\varepsilon_k|/k\le k^{-1/2}$ for independent Rademacher signs $\varepsilon_i$.

\begin{lemma}[A local modulus for the ramps]
\label{lem:localmod}
Let (A1) and (A2) of Assumption~\ref{ass:regularity} hold, let $f^\star\in\Hh$ and $t^\star,u^\star\in\R$ be fixed, and
let $r,\lambda_1,\lambda_2\ge0$. Let $\mathcal H(r)$ be the set of pairs of functions $h_1=(t-f)_+-(t^\star-f^\star)_+$
and $h_2=(f-u)_+-(f^\star-u^\star)_+$ with $f\in\Hh$, $\|f-f^\star\|_\Hh\le r$, $|t-t^\star|\le\kappa r+\lambda_1$ and
$|u-u^\star|\le\kappa r+\lambda_2$, and put $\Gamma(r)=\sup_{(h_1,h_2)\in\mathcal H(r)}[(P-P_n)h_1+(Q-Q_m)h_2]$. Then $\Gamma(r)$
is measurable, nonnegative and nondecreasing in $r$,
\[
\E \Gamma(r)\le2\big[\kappa r/\sqrt{n_\star}+(\kappa r+\lambda_1)e_n+(\kappa r+\lambda_2)e_m\big],
\]
and for every $\chi>0$, with probability at least $1-e^{-\chi}$,
\[
\Gamma(r)\le\kappa r\,L(\chi)+M(\chi),
\qquad
\begin{aligned}
L(\chi)&=2/\sqrt{n_\star}+2(e_n+e_m)+2\sqrt{2\chi/n_\star},\\
M(\chi)&=2(\lambda_1e_n+\lambda_2e_m)+\sqrt{2\chi\big(\lambda_1^2/n+\lambda_2^2/m\big)} .
\end{aligned}
\]
\end{lemma}

\begin{proof}
Write $f=f^\star+v$, $t=t^\star+a$, $u=u^\star+b$, and for each $x$ let $\varphi^1_x(z)=(t^\star-f^\star(x)+z)_+-(t^\star-
f^\star(x))_+$ and $\varphi^2_x(z)=(f^\star(x)-u^\star+z)_+-(f^\star(x)-u^\star)_+$. Both are $1$-Lipschitz and vanish
at $0$, and $h_1(x)=\varphi^1_x(a-v(x))$, $h_2(y)=\varphi^2_y(v(y)-b)$, so $\Gamma(r)$ is the supremum $\Gamma$ of
Lemma~\ref{lem:hinge} with these maps, radius $r$ and $\rho_i=\kappa r+\lambda_i$. That lemma gives the measurability and
the bound on $\E \Gamma(r)$. The choice $(v,a,b)=0$ gives $\Gamma(r)\ge0$, and the classes increase with $r$. Since
$|\varphi^i_x(z)|\le|z|$, every pair has $|h_1|\le2\kappa r+\lambda_1$ and $|h_2|\le2\kappa r+\lambda_2$, so the tail
bound of the lemma holds with $\Delta_i=2(2\kappa r+\lambda_i)$. With probability at least $1-e^{-\chi}$, by
Minkowski's inequality in $\R^2$,
\[
\begin{aligned}
\Gamma(r)&\le\E \Gamma(r)+\sqrt{2\chi\big[(2\kappa r+\lambda_1)^2/n+(2\kappa r+\lambda_2)^2/m\big]}\\
&\le\E \Gamma(r)+\sqrt{2\chi}\,\big[2\kappa r/\sqrt{n_\star}+(\lambda_1^2/n+\lambda_2^2/m)^{1/2}\big],
\end{aligned}
\]
and adding the bound on $\E \Gamma(r)$ and collecting the terms in $\kappa r$ gives the claim.
\end{proof}

\begin{proof}[Proof of Theorem~\ref{cor:trimrate}]
The basic inequality of Lemma~\ref{lem:basicineq} bounds $\tfrac D2\widehat r^{\,2}$ by a supremum whose mean grows
linearly in $\widehat r$, which forces $\widehat r$ to be of order $(n\wedge m)^{-1/2}$; dyadic shells and Markov's
inequality turn this into a rate, as in the proof of \citet[Theorem~3.2.5]{vandervaart1996weak}. We work with the
direction $f^\star=d^\star/D$, so cuts and radii on the scale of $d^\star$ are divided by $D$. Precisely, we show
that for every $\eta\in(0,1)$ there are a constant $C_\eta$, depending on $\eta$ alone, and a size $n_\eta$, depending on
$\eta$, $(P,Q,k,\alpha)$ and $r_Q$, such that for $n\wedge m\ge n_\eta$ each of the four bounds holds with $C_\eta$ in
place of $O_\Pp$ with probability at least $1-\eta$.

We write $\widehat D=\widehat D_\alpha$, $f^\star=d^\star/D$, $\widehat f=\widehat d/\widehat D$ when $\widehat D>0$,
$\widehat r=\|\widehat f-f^\star\|_\Hh$, $t^\star=t_\alpha/D$ and $u^\star=u_\alpha/D$, which are a
$(1-\alpha)$-quantile of $f^\star(X)$ and an $\alpha$-quantile of $f^\star(Y)$, and let $\widehat t$ and $\widehat u$ be
as in Lemma~\ref{lem:basicineq}, so that $\widehat t=\widehat t_n/\widehat D$. Write $\underline c=\min(c_f,c'_f)$ and
$\underline n=n\wedge m$, fix $\eta\in(0,1)$ and put $\eta'=\eta/2$,
\[
\lambda_X=\frac1{c_f}\Big(\varepsilon_n+\frac1n\Big),\qquad
\lambda_Y=\frac1{c'_f}\Big(\varepsilon_m+\frac1m\Big),\qquad
\varepsilon_n=\Big\{\frac{\log(10/\eta')}{2n}\Big\}^{1/2},
\]
with $\varepsilon_m$ defined likewise, and $\tau^{\mathrm S}=\tau^{\mathrm S}_{n,m}(\alpha,\eta'/5)$ as in
\eqref{eq:tauA}. Let $\Gamma$ be the process of Lemma~\ref{lem:localmod} for $f^\star,t^\star,u^\star$,
$\lambda_1=\lambda_X/D$ and $\lambda_2=\lambda_Y/D$. The proof uses four events, each of probability at least
$1-\eta'/5$: the events $E^+$ and $E^-$ of Theorem~\ref{thm:twosided} at level $\eta'/5$, and, by the
Dvoretzky--Kiefer--Wolfowitz inequality with Massart's constant \citep{massart1990dkw},
$E_X=\{\sup_t|F_n(t)-F(t)|\le\varepsilon_n\}$ for the distribution function $F$ of $d^\star(X)$ and the empirical one
$F_n$ of the $d^\star(X_i)$, and the analogous $E_Y$ for the $d^\star(Y_j)$, with $G$, $G_m$ and $\varepsilon_m$. Let
$E=E^-\cap E_X\cap E_Y$, so $\Pp(E^c)\le3\eta/10$. For $n\wedge m$ large enough, $\tau^{\mathrm S}<D$ and $\max(\lambda_X,\lambda_Y)\le r_Q$.

On $E^-$ the problem is nondegenerate: $\widehat D\ge D-\tau^{\mathrm S}>0$. So $\widehat f$ is a unit vector,
$\widehat r\le2$, and Proposition~\ref{prop:bangbang}(ii) applies, with $\widehat A_n=\{\widehat f<\widehat t\}$; also
$A^\star=\{f^\star<t^\star\}$.

The lower density bounds keep the empirical cuts of $f^\star$ near the population ones, within $\lambda_X$ or
$\lambda_Y$, the radius of the Dvoretzky--Kiefer--Wolfowitz event plus one rank, divided by $c_f$ or $c'_f$. By \eqref{eq:qdens}, $F$ is continuous at $t_\alpha$, so $F(t_\alpha)=1-\alpha$, and
integrating the lower density bound gives $F(t_\alpha+\varepsilon)\ge1-\alpha+c_f\varepsilon$ and
$F(t_\alpha-\varepsilon)\le1-\alpha-c_f\varepsilon$ for $0\le\varepsilon\le r_Q$. Let $d^\star_{(n_{1-\alpha}+1)}$ be the
$(n_{1-\alpha}+1)$-th smallest of the $d^\star(X_i)$, so that $F_n(t)\ge(n_{1-\alpha}+1)/n$ exactly when $t\ge
d^\star_{(n_{1-\alpha}+1)}$. On $E_X$, since $c_f\lambda_X=\varepsilon_n+1/n$ and $\lambda_X\le r_Q$,
$F_n(t_\alpha+\lambda_X)\ge1-\alpha+1/n\ge(n_{1-\alpha}+1)/n$ and $F_n(t_\alpha-\lambda_X)\le1-\alpha-1/n<(n_{1-\alpha}+1)/n$,
using $n(1-\alpha)-1<n_{1-\alpha}\le n(1-\alpha)$; so $|d^\star_{(n_{1-\alpha}+1)}-t_\alpha|\le\lambda_X$. On the second sample
let $d^\star_{[k]}$ be the $k$-th smallest of the $d^\star(Y_j)$, $k=m-m_{1-\alpha}$, which is the $(m_{1-\alpha}+1)$-th largest,
with $\alpha\le k/m<\alpha+1/m$. In the same way \eqref{eq:qdensY} gives, on $E_Y$, $G_m(u_\alpha+\lambda_Y)\ge
\alpha+1/m>k/m$ and $G_m(u_\alpha-\lambda_Y)\le\alpha-1/m<k/m$, so $|d^\star_{[k]}-u_\alpha|\le\lambda_Y$. Dividing by
$D$, the corresponding order statistics of the $f^\star(X_i)$ and $f^\star(Y_j)$ are within $\lambda_1$ of $t^\star$ and
within $\lambda_2$ of $u^\star$. Order statistics are $1$-Lipschitz in the sup-norm of the values, and
$|\widehat f(x)-f^\star(x)|\le\kappa\widehat r$ by Cauchy--Schwarz, so
$|\widehat t-t^\star|\le\kappa\widehat r+\lambda_1$ and $|\widehat u-u^\star|\le\kappa\widehat r+\lambda_2$. The pair of
functions in Lemma~\ref{lem:basicineq} therefore belongs to $\mathcal H(\widehat r)$, and the lemma gives
\begin{equation}
\tfrac D2\,\widehat r^{\,2}\le\beta\,\Gamma(\widehat r) .
\label{eq:basicZ}
\end{equation}

A peeling turns \eqref{eq:basicZ} into a rate. We peel the values of $\underline n^{1/2}\widehat r$ into dyadic shells. For
$j\in\mathbb Z$ let $S_j=\{2^{j-1}<\underline n^{1/2}\widehat r\le2^j\}$. On $S_j\cap E$, \eqref{eq:basicZ} and the monotonicity of
$\Gamma$ give $\Gamma(2^j\underline n^{-1/2})\ge D4^j/(8\beta\underline n)$, so $S_j\cap E$ lies in the measurable
event $\{\Gamma(2^j\underline n^{-1/2})\ge D4^j/(8\beta\underline n)\}$. The bound on $\E \Gamma$ in Lemma~\ref{lem:localmod}, with $n_\star^{-1/2}\le(2/\underline n)^{1/2}$ and $e_n,e_m\le\underline n^{-1/2}$, gives
$\underline n\,\E \Gamma(2^j\underline n^{-1/2})\le a2^j+b$ with $a=2(2+\sqrt2)\kappa$, the linear term, and
$b=2\underline n^{1/2}(\lambda_1+\lambda_2)$, the cost
of the cuts, so by Markov's inequality, since
$\Gamma\ge0$, that event has probability at most $8\beta(a2^j+b)/(D4^j)$. Write $\Xi_{\eta'}=\sqrt{\log(10/\eta')/2}+1$.
Since $\underline n\le n$, $\underline n^{1/2}\lambda_X\le\Xi_{\eta'}/c_f$, and likewise for $\lambda_Y$, so
$b\le4\Xi_{\eta'}/(\underline cD)$. Let $j_0$ be the smallest integer with $2^{j_0}\ge\Lambda V_\alpha/D$, for a
number $\Lambda\ge1$ chosen below. The shells $S_j$ with $j>j_0$ cover $\{\underline n^{1/2}\widehat r>2^{j_0}\}$, so
$\{\underline n^{1/2}\widehat r>2^{j_0}\}\cap E$ lies in the union $\mathcal B_{j_0}$ of the events above over $j>j_0$, and summing the geometric
series,
\[
\Pp(\mathcal B_{j_0})\le\frac{8\beta a}{D2^{j_0}}+\frac{8\beta b}{3D4^{j_0}}
\le\frac{8\beta a}{\Lambda V_\alpha}+\frac{32\beta\Xi_{\eta'}}{3\underline c\Lambda^2V_\alpha^2}
\le\frac{16(2+\sqrt2)}\Lambda+\frac{32\Xi_{\eta'}}{3\Lambda^2},
\]
using $V_\alpha\ge\beta\kappa$ and $V_\alpha^2\ge\beta/\underline c$. Choose $\Lambda$, depending only on $\eta$, so
that the right-hand side is at most $\eta/2$. Since $2^{j_0}<2\Lambda V_\alpha/D$, the event
$\{\widehat r>2\Lambda V_\alpha/(D\underline n^{1/2})\}$ lies in $E^c\cup\mathcal B_{j_0}$, of probability at most $3\eta/10+\eta/2<
\eta$ for $n\wedge m$ large, which is the rate of the direction.

The bounds on the witness and the kept set follow from that on the direction. For the witness, Lemma~\ref{lem:sepmargin} gives $\|\widehat d-d^\star\|_\Hh\le|\widehat
D-D|+\sqrt{\widehat DD}\,\widehat r\le\tau^{\mathrm S}+\sqrt2D\widehat r$ on $E^+\cap E^-$ when $\tau^{\mathrm S}<D$, and \eqref{eq:tauA} gives
$\tau^{\mathrm S}_{n,m}(\alpha,\eta'/5)\le\beta\kappa\big(2\sqrt2+4+2\sqrt{\log(5/\eta')}\big)\underline n^{-1/2}$, a multiple of $V_\alpha\underline n^{-1/2}$ depending only on $\eta$. For the kept set, on $E$, if $\widehat f(x)<\widehat t$ and $f^\star(x)\ge t^\star$, then
$f^\star(x)\le\widehat f(x)+\kappa\widehat r<\widehat t+\kappa\widehat r\le t^\star+2\kappa\widehat r+\lambda_1$; if
$f^\star(x)<t^\star$ and $\widehat f(x)\ge\widehat t$, then $f^\star(x)\ge\widehat t-\kappa\widehat r\ge
t^\star-2\kappa\widehat r-\lambda_1$. So $\widehat A_n\triangle A^\star$ lies in the band
$\{|f^\star-t^\star|\le2\kappa\widehat r+\lambda_1\}=\{|d^\star-t_\alpha|\le2\kappa D\widehat r+\lambda_X\}$, and the same computation puts $\{\widehat d=\widehat t_n\}=
\{\widehat f=\widehat t\}$ in it. If its half-width $\ell=2\kappa D\widehat r+\lambda_X$ is at most $r_Q$, the upper density bound of
\eqref{eq:qdens} gives the band mass at most $2C_f\ell$. Finally, $\widehat g_n$ and $g^\star$ take values in
$[0,\beta]$ and agree outside $(\widehat A_n\triangle A^\star)\cup\{\widehat d=\widehat t_n\}\cup\{d^\star=t_\alpha\}$:
both equal $\beta$ on $\widehat A_n\cap A^\star$ and $0$ where $\widehat d>\widehat t_n$ and $d^\star>t_\alpha$. All three
sets lie in the band, which gives the bound on $\int|\widehat g_n-g^\star|\,dP$. On $E$ the half-width is at most $r_Q$ with probability tending to one, and
$\lambda_X\le\Xi_{\eta'}/(c_f\underline n^{1/2})$, while the rate of the direction bounds $2\kappa D\widehat r$ by
$4\Lambda\kappa V_\alpha\underline n^{-1/2}$
outside probability $\eta$.
\end{proof}

\subsection{A sufficient condition for \eqref{eq:qdens}}

The density conditions of Section~\ref{sec:trimming} hold, for instance, under the following condition on the witness.

\begin{remark}[A sufficient condition for \eqref{eq:qdens}]
\label{rem:qdens}
Let $\mathcal X=\R^d$ with the Gaussian kernel, so that $d^\star$ is infinitely differentiable and vanishes at
infinity. If $t_\alpha\neq0$, the level set $\{d^\star=t_\alpha\}$ is nonempty, the gradient of $d^\star$ does not
vanish on it, and $P$ has a Lebesgue density $p$ that is continuous near the level set and positive on it, then
\eqref{eq:qdens} holds, by the coarea formula \citep{evans2015measure}. The same conditions at $u_\alpha$ under $Q$
give \eqref{eq:qdensY} (proof below). For the Laplace kernel on $\R$, by contrast, \eqref{eq:qdens} fails for every
pair and every level $0<\alpha<\alpha^\star$, as the law of $d^\star(X)$ has an atom of mass at least $\alpha$ at the
cut (Proposition~\ref{prop:excess}(iii) and the remark after its proof); whether it can hold for the Laplace kernel in
$\R^d$ is open.
\end{remark}

\begin{proof}[Proof of Remark~\ref{rem:qdens}]
For the Gaussian kernel every function in $\Hh$ is infinitely differentiable \citep[Corollary~4.36]{steinwart2008svm}
and vanishes at infinity \citep[Theorem~4.21 and Lemma~4.23]{steinwart2008svm}, so for $0<\rho<|t_\alpha|/2$ the band
$B_\rho=\{|d^\star-t_\alpha|\le\rho\}$ is compact, and for $\rho$ small enough $\nabla d^\star$ does not vanish and $p$ is
continuous and positive on it, by continuity and compactness. On $B_\rho$ the coarea formula \citep{evans2015measure}
gives $d^\star(X)$ the density $h(t)=\int_{\{d^\star=t\}}p/|\nabla d^\star|\,d\mathcal H^{d-1}$ for $|t-t_\alpha|\le\rho$,
which is continuous, as the flow of $\nabla d^\star/|\nabla d^\star|^2$ carries the level sets smoothly onto one another,
and positive. So \eqref{eq:qdens} holds for $r_Q<\rho$ with $c_f$ and $C_f$ the extremes of $h$ on
$[t_\alpha-r_Q,t_\alpha+r_Q]$, and the same argument for $Q$, $Y$ and $u_\alpha$ gives \eqref{eq:qdensY}.
\end{proof}

\subsection{The trimming at the estimated level}

The regions touch in a single point at $\alpha^\star$, and concentrate in Hausdorff distance $d_H$. Recall the modulus
of contact $\Omega(\epsilon)=\sup\{\|a-\mu_{R^\star}\|_\Hh:a\in\mathcal C_{\alpha^\star}(P),\
\dist(a,\mathcal C_{\alpha^\star}(Q))\le\epsilon\}$.

\begin{lemma}[The contact point]
\label{lem:contact}
Let Assumption~\ref{ass:regularity} hold. Then $\mathcal C_\alpha(P)$ is norm-compact for every $\alpha\in[0,1)$. If
$\alpha^\star<1$, then $R^\star=(P\wedge Q)/(1-\alpha^\star)$ is the only common element of $R_{\alpha^\star}(P)$ and
$R_{\alpha^\star}(Q)$, $\mathcal C_{\alpha^\star}(P)\cap\mathcal C_{\alpha^\star}(Q)=\{\mu_{R^\star}\}$, and
$\Omega(\epsilon)\to0$ as $\epsilon\downarrow0$.
\end{lemma}

\begin{proof}
Norm-compactness is part of Lemma~\ref{lem:attainment}.

The common trimming is unique by the maximality of $P\wedge Q$, and the common embedding by (A3). If $R\in R_{\alpha^\star}(P)\cap R_{\alpha^\star}(Q)$, then $(1-\alpha^\star)R\le P\wedge Q$
(Section~\ref{app:regions}), and both have mass $1-\alpha^\star$, so $R=R^\star$; conversely $R^\star$ lies in both. If $a$ lies in both regions, then
$a=\E_P[g\phi]=\E_Q[h\phi]$ with $g\in\mathcal G_{\alpha^\star}(P)$ and $h\in\mathcal G_{\alpha^\star}(Q)$, the
probability measures $gP$ and $hQ$ have the same embedding, and they are equal by (A3). So $a$ is the embedding of a
common trimming, which is $R^\star$.

That the modulus tends to zero is a compactness argument. If $\Omega(\epsilon)\not\to0$, there are $\epsilon_j\to0$, $c>0$ and
$a_j\in\mathcal C_{\alpha^\star}(P)$ with $\dist(a_j,\mathcal C_{\alpha^\star}(Q))\le\epsilon_j$ and
$\|a_j-\mu_{R^\star}\|_\Hh\ge c$. By compactness a subsequence converges in norm to some
$\bar a\in\mathcal C_{\alpha^\star}(P)$, and $\dist(\bar a,\mathcal C_{\alpha^\star}(Q))=0$, so $\bar a$ lies in the
closed set $\mathcal C_{\alpha^\star}(Q)$ as well. Then $\bar a=\mu_{R^\star}$, which contradicts
$\|\bar a-\mu_{R^\star}\|_\Hh\ge c$.
\end{proof}

Let $\mathcal G=\{x\mapsto(f(x)-t)_+:\ f\in B_\Hh,\ t\in[-\kappa,\kappa]\}$, a class of hinge functions, not to be
confused with the sets $\mathcal G_\alpha(P)$ of trimming densities.

\begin{corollary}[Concentration of the regions, simultaneously in $\alpha$]
\label{cor:uniform}
Under Assumption~\ref{ass:regularity}, for every $n\ge1$ and $\delta\in(0,1)$ there is an event of probability at least
$1-\delta$, namely $\{\sup_{g\in\mathcal G}|(P_n-P)g|\le4\kappa/\sqrt n+\kappa\sqrt{2\log(2/\delta)/n}\}$, which does
not depend on $\alpha$ and on which, for every $\alpha\in[0,1)$,
\begin{equation}
d_H\big(\mathcal C_\alpha(P_n),\mathcal C_\alpha(P)\big)\le r_n(\alpha,\delta)
:=\frac{\kappa}{1-\alpha}\bigg[\frac4{\sqrt n}+\sqrt{\frac{2\log(2/\delta)}n}\,\bigg].
\label{eq:radius}
\end{equation}
\end{corollary}

\begin{proof}
The Hausdorff distance between two convex sets is the largest difference of their support functions, each support
function is an infimum over cuts of means of hinge functions, and so one supremum over the hinge class controls every
level. Write $\mathcal A=\mathcal C_\alpha(P)$ and $\mathcal A_n=\mathcal C_\alpha(P_n)$. For $a\in\mathcal A_n$, \eqref{eq:distmax}
for $\mathcal A$ and $\{a\}$, with $-\langle f,a\rangle_\Hh\le h_{\mathcal A_n}(-f)$, gives
\[
\dist(a,\mathcal A)=\max_{f\in B_\Hh}\{-h_{\mathcal A}(-f)-\langle f,a\rangle_\Hh\}
\le\sup_{f\in B_\Hh}\{h_{\mathcal A_n}(-f)-h_{\mathcal A}(-f)\},
\]
and in the same way
$\dist(a,\mathcal A_n)\le\sup_{f\in B_\Hh}\{h_{\mathcal A}(-f)-h_{\mathcal A_n}(-f)\}$ for $a\in\mathcal A$. As in
the deterministic reduction in the proof of Theorem~\ref{thm:twosided}, Lemmas~\ref{lem:cvar} and~\ref{lem:tcompact} bound
$h_{\mathcal A_n}(-f)-h_{\mathcal A}(-f)$ by $\beta(P_n-P)(-f-t)_+$ at the population minimizer $t\in[-\kappa,\kappa]$, and
$h_{\mathcal A}(-f)-h_{\mathcal A_n}(-f)$ by $\beta(P-P_n)(-f-t_n)_+$ at the empirical one. Since $B_\Hh$ and
$[-\kappa,\kappa]$ are symmetric, these functions lie in $\mathcal G$, so $d_H(\mathcal A_n,\mathcal A)\le\beta
\sup_{g\in\mathcal G}|(P_n-P)g|$, in which only $\beta$ depends on $\alpha$.
Lemma~\ref{lem:hinge} for one sample, with $\varphi=(\cdot)_+$, $r=1$, $\rho_1=\kappa$ and $\Delta_1=2\kappa$ (the functions in
$\mathcal G$ take values in $[0,2\kappa]$), bounds the means of $\sup_g(P_n-P)g$ and $\sup_g(P-P_n)g$ by $4\kappa/\sqrt n$,
and at $e^{-\chi}=\delta/2$ each exceeds its mean by more than $\kappa\sqrt{2\log(2/\delta)/n}$ with probability at most
$\delta/2$.
\end{proof}

\begin{proof}[Proof of Proposition~\ref{prop:alphahat}]
The empirical trimming is nearly in both population regions at $\alpha^\star$, and $\Omega$ says such a
point is close to $\mu_{R^\star}$. That $\Omega(\epsilon)\to0$ is Lemma~\ref{lem:contact}, and one event serves all
levels. Let $E$ be the intersection of the events of
Corollary~\ref{cor:uniform} for the two samples at level $\delta'$, of probability at least $1-2\delta'$, with
$r_n=r_n(\alpha^\star,\delta')$ and $r_m=r_m(\alpha^\star,\delta')$. Let $\hat a=\sum_iw_i\phi(X_i)\in\mathcal C_{\tilde\alpha}(P_n)$ and $\hat b=\sum_jv_j\phi(Y_j)\in\mathcal C_{\tilde\alpha}(Q_m)$,
so that $\|\hat a-\hat b\|_\Hh=\widehat D_{\tilde\alpha}$ by Proposition~\ref{prop:bangbang}(i) and
$\MMD(P_n^w,R^\star)=\|\hat a-\mu_{R^\star}\|_\Hh$.

We pass from the sample regions to the population ones at $\alpha^\star$. On $E$, $\dist(\hat a,\mathcal C_{\tilde\alpha}(P))
\le r_n(\tilde\alpha,\delta')\le r_n$ for every $\tilde\alpha\le\alpha^\star$ simultaneously, and likewise for $\hat b$.
The regions increase with the level and are weakly compact, so there are $a\in\mathcal C_{\alpha^\star}(P)$ with
$\|\hat a-a\|_\Hh\le r_n$ and $b\in\mathcal C_{\alpha^\star}(Q)$ with $\|\hat b-b\|_\Hh\le r_m$. Then $\dist(a,\mathcal C_{\alpha^\star}(Q))\le\|a-b\|_\Hh\le r_n+\widehat D_{\tilde\alpha}+r_m$, by the
triangle inequality through $\hat a$ and $\hat b$, and by the definition of $\Omega$,
$\|a-\mu_{R^\star}\|_\Hh\le\Omega(\widehat D_{\tilde\alpha}+r_n+r_m)$, and the triangle inequality gives the bound.
The second sample is the same argument with $P$ and $Q$ exchanged.
\end{proof}

\begin{proof}[Proof of Corollary~\ref{cor:alphalo}]
Consistency of $\widehat\alpha$ makes the trimmed distance at $\widehat\alpha$ small, as the path falls to zero at most
linearly, and Proposition~\ref{prop:alphahat} turns a small trimmed distance into closeness to $R^\star$. Fix
$\epsilon>0$ and $\delta'\in(0,1)$, and let $E$ be the event of the proof of Proposition~\ref{prop:alphahat} at
level $\delta'/3$, of probability at least $1-2\delta'/3$, with its radii $r_n,r_m$. On $E$, for every level
$\alpha\le\alpha^\star$ the empirical regions are within $r_n$ and $r_m$ of the population ones in Hausdorff distance,
by Corollary~\ref{cor:uniform} and the monotonicity of $r_n(\alpha,\delta')$ in $\alpha$, and the distance between two
sets moves by at most the sum of these, so $\widehat D_\alpha\le D_\alpha+r_n+r_m$. On $E\cap\{\widehat\alpha\le\alpha^\star\}$,
Lemma~\ref{lem:pathconvex}(ii) at $\alpha^\star-\widehat\alpha\in[0,\alpha^\star]$ then gives
\[
\widehat D_{\widehat\alpha}\le\frac{D_0(P,Q)}{\alpha^\star(1-\alpha^\star)}\,
\big(\alpha^\star-\widehat\alpha\big)+r_n+r_m ,
\]
and Proposition~\ref{prop:alphahat} at $\tilde\alpha=\widehat\alpha$, a level at most $\alpha^\star$,
gives $\MMD(P_n^w,R^\star)\le r_n+\Omega(\widehat D_{\widehat\alpha}+r_n+r_m)$ for every
$(w,v)\in\mathcal T_{\widehat\alpha}$, and the same bound for $\MMD(Q_m^v,R^\star)$. Since
$\Omega(t)\to0$ as $t\downarrow0$ and $\Omega$ is nondecreasing, there is $\varepsilon_1>0$ with $\varepsilon_1+\Omega(3\varepsilon_1)\le\epsilon$. By
(A3), $D_0(P,Q)=\MMD(P,Q)>0$ as $\alpha^\star>0$. For $n,m$ large enough, $r_n+r_m\le\varepsilon_1$, and by
Proposition~\ref{prop:loconsistency} the event
$A=\{|\widehat\alpha-\alpha^\star|\le\varepsilon_1\,\alpha^\star(1-\alpha^\star)/D_0(P,Q)\}$ has probability at least
$1-\delta'/3$. On $E\cap A\cap\{\widehat\alpha\le\alpha^\star\}$ both distances are at most
$r_n+\Omega(\varepsilon_1+2(r_n+r_m))\le\varepsilon_1+\Omega(3\varepsilon_1)\le\epsilon$, and by Corollary~\ref{cor:lower} and a union bound this event has
probability at least $1-\delta-\delta'$.
\end{proof}
\space
\section{Experimental details}
\label{app:experiments}

This section gives the details of Sections~\ref{sec:analytic}, \ref{sec:rates} and~\ref{sec:numerical}, with
$\delta=0.05$ throughout.

\subsection{Synthetic designs}

Table~\ref{tab:hardern80} and Figure~\ref{fig:recovery} of the main text use the designs of
Table~\ref{tab:harderdesigns}. Each follows \eqref{eq:contam} with a mixture weight $a$ in place of $\alpha$, so that
$\TV(P,Q)=a\,\TV(P_0,Q_0)$ whatever the common component $R$; in the overlap and shift designs $R=N(0,I)$ and
$a=\tfrac14$, and mutually singular $P_0$ and $Q_0$ give $\alpha^\star=a$. The module
\texttt{explore/harder\_sims/designs.py} of the repository of Section~\ref{sec:computationdetails} defines every design,
its core, weight and contaminations, and \texttt{explore/harder\_sims/DESIGN.md} derives each $\alpha^\star$.

\begin{table}[t]
\centering
\scriptsize
\setlength{\tabcolsep}{4pt}
\begin{tabular}{llcccc}
\toprule
design & $P_0$ vs $Q_0$ & $d$ & $n$ & $\alpha^\star$ & $\MMD$ \\
\midrule
RHC-shaped & RHC core, boxes at $\pm$3 SD & $33$ & $100$--$800$ & $0.2$ & $0.201$ \\
overlap, $d=2$ & $N(\pm\mu e_1,I)$ & $2$ & $100$--$800$ & $0.1$ & $0.091$ \\
overlap, $d=2$ & $N(\pm\mu e_1,I)$ & $2$ & $100$--$1600$ & $0.2$ & $0.181$ \\
overlap, $d=10$ & $N(\pm\mu e_1,I)$ & $10$ & $100$--$800$ & $0.1$ & $0.042$ \\
overlap, $d=10$ & $N(\pm\mu e_1,I)$ & $10$ & $100$--$800$ & $0.2$ & $0.103$ \\
dense shift, $d=20$ & $N(\pm\mu\mathbf 1/\sqrt d,I)$ & $20$ & $100$--$800$ & $0.2$ & $0.077$ \\
dense shift, $d=50$ & $N(\pm\mu\mathbf 1/\sqrt d,I)$ & $50$ & $100$--$800$ & $0.2$ & $0.048$ \\
dense shift, $d=100$ & $N(\pm\mu\mathbf 1/\sqrt d,I)$ & $100$ & $100$--$800$ & $0.2$ & $0.038$ \\
sparse shift, $d=50$ & $N(\pm\mu(e_1+e_2)/\sqrt2,I)$ & $50$ & $100$--$800$ & $0.2$ & $0.049$ \\
correlation flip, $d=50$ & $\operatorname{corr}(x_1,x_2)=\pm\rho$ & $50$ & $100$--$800$ & $0.2$ & $\approx0$ \\
small contamination, $d=2$ & $N(\pm2.5e_1,0.25I)$, $a=0.05$ & $2$ & $100$--$1600$ & $0.05$ & $0.050$ \\
mixed types, $m=5n$ & integer and binary coordinate & $5$ & $100$--$400$ & $0.2$ & $0.135$ \\
$t_3$ tails, $d=5$ & scale $0.5$ vs $3$ of $x_1$ & $5$ & $100$--$800$ & $0.2$ & $0.076$ \\
curves, wide bump & $\pm$ bump of width $0.2$ & $50$ & $100$--$800$ & $0.2$ & $0.118$ \\
curves, narrow bump & $\pm$ bump of width $0.05$ & $50$ & $100$--$800$ & $0.2$ & $0.043$ \\
oscillating, $\nu=1$ & $U[0,1]$ vs density $1+\varsigma(\nu x)$ & $1$ & $100$--$800$ & $0.5$ & $0.492$ \\
oscillating, $\nu=4$ & $U[0,1]$ vs density $1+\varsigma(\nu x)$ & $1$ & $100$--$800$ & $0.5$ & $0.069$ \\
oscillating, $\nu=16$ & $U[0,1]$ vs density $1+\varsigma(\nu x)$ & $1$ & $100$--$800$ & $0.5$ & $0.019$ \\
oscillating, $\nu=64$ & $U[0,1]$ vs density $1+\varsigma(\nu x)$ & $1$ & $100$--$800$ & $0.5$ & $\approx0$ \\
\bottomrule
\end{tabular}
\space
\caption{The synthetic designs of Section~\ref{sec:numharder}: $P_0$ against $Q_0$ ($P$ against $Q$ for the oscillating
designs); sizes $n=m$ at which every method that certifies a level was run, doubling ($m=5n$ for mixed types);
$\alpha^\star$, exact (small contamination: $0.05\,(2\Phi(5)-1)$); population MMD at the analytic test's bandwidth, by
Monte Carlo ($\approx0$: estimate of $\MMD^2$ not positive).}
\label{tab:harderdesigns}
\end{table}

All methods see the same $\harderReps$ replicates per design and size. The permutation test and MMDAgg, which uses
$\abAggL$ Gaussian bandwidths, were run up to $n=m=\harderNmax$, and to $1{,}600$ on the narrow bump. The tests of
$H_{0,0}$ that certify a level were continued: every design and method below power $0.8$ at $n=m=\harderNmax$ ($n=400$
for mixed types) was run at $n=m=1600$, $3200$, $6400$ and $\harderNcap$ (from $n=800$, with $m=5n$, for mixed types),
with $\harderExtReps$ new replicates per size, until its power was at least $0.8$ at two consecutive sizes, in
$\harderExtCells$ design and size cells, and the four nulls with $P=Q$ below in $\harderNullExtCells$ cells
($\harderNullExtReplicates$ replicates). The methods of Section~\ref{sec:baselines} see the replicates of the main size,
with the same data, folds and bandwidths. Gradient boosting's drop on the correlation flip at $n=\harderNcap$ is
scikit-learn's default early stopping above $10{,}000$ training points; without it its power is $\harderFlipEsPowOff$
instead of $\harderFlipEsPowOn$ ($\harderFlipEsReps$ replicates).

The analytic test and the permutation test take their Gaussian bandwidth from one pilot draw per design, $1000+1000$
points ($400+2000$ for mixed types), as the median pairwise distance of the pooled pilot. Every analytic number of the
main text and of this section uses $\tau=\min\{\tau^{\mathrm P},\tau^{\mathrm S}\}$, which equals $\tau^{\mathrm P}$ at
every size and level of these runs, so the analytic test is that of Theorem~\ref{thm:plugin};
Section~\ref{sec:kernelfallback} says where another threshold is used. The Laplace ramp cuts its witness
at genuine empirical quantiles (Algorithm~\ref{alg:split}), and the Gaussian ramp, the untrimmed witness and the check
of Section~\ref{sec:excessmc} between order statistics, within one rank of genuine quantiles \eqref{eq:cutrank}. The
classifiers, histogram gradient boosting with default settings and logistic regression with unit $L^2$ penalty, label a
point as from $P$ when its fitted probability exceeds $n_{\mathrm{tr}}/(n_{\mathrm{tr}}+m_{\mathrm{tr}})$, with
$n_{\mathrm{tr}}$ and $m_{\mathrm{tr}}$ the numbers of $X$'s and $Y$'s they are trained on. The simulations test the levels $j\alpha^\star/8$, $j=0,\dots,10$, which use the known
$\alpha^\star$ and serve only to measure recovery; the real data use the grid $0,0.05,\dots,0.95$.

The null boundary of Section~\ref{sec:analytic} is that of a standard Gaussian core in $d=2$ with contaminations
uniform on two disjoint boxes and $\alpha^\star=0.2$. With $\calibReps$ replicates per level at $n=m=300$, no replicate
rejected $H_{0,\alpha}$ at $\alpha\ge\alpha^\star$ (one-sided $95\%$ Clopper--Pearson upper limit $\calibCPUpper$), and at
$\alpha=\alpha^\star$ the threshold $\tau^{\mathrm P}$ is $\stCalibRatioTauP$ times the empirical $1-\delta$ quantile of
$\widehat D_\alpha$.

In the designs, the Laplace ramp rejected the true null $H_{0,\alpha^\star}$ in at most a fraction $\mnBdMax$ of the
replicates of a design (binomial $p=\mnBdMaxP$), its rate under $P=Q$ lies between $\mnNullLoPct$ and $\mnNullHiPct\%$ in every cell, and its
bound exceeded $\alpha^\star$ in at most a fraction $\mnAloGtMax$ of the replicates of any design. The analytic test
rejected in none of its $\mnAnBdCells$ boundary cells and never under $P=Q$. The other methods were checked in the
same way. At the boundary, in the $\harderBoundaryCells$ cells of the eighteen designs other than the RHC-shaped one,
the Gaussian ramp rejected at most $\hvBdWorstRamp$ times in a cell, gradient boosting $\hvBdWorstGbm$ and logistic
regression $\hvBdWorstLogit$. On the RHC-shaped design the three rejected $\dNineRampBdMax$, $\dNineGbmBdMax$ and
$\dNineLogitBdMax$ times. All these counts are below the $\harderFlagK$ rejections at which a binomial test against
$\delta$ at level $0.01$ flags a cell.
Four nulls with $P=Q$, on which every bet has mean exactly $\alpha=0$, use the law $P$ of the dense-shift ($d=100$),
mixed-type, $t_3$ and narrow-bump designs; up to $n=800$ the Gaussian ramp rejected in $\hvNullRamp\%$ of the
replicates, gradient boosting in $\hvNullGbm\%$ and logistic regression in $\hvNullLogit\%$, and up to $n=\harderNcap$
at most $\harderNullExtMaxRamp$, $\harderNullExtMaxGbm$ and $\harderNullExtMaxLogit$ times in a cell of
$\harderExtReps$. The lower bound of every method exceeded $\TV(P,Q)$ in at most a fraction $\harderAloGtMax$ of the
replicates of a cell on the eighteen designs, $\dNineAloGtMax$ on the RHC-shaped one and $\harderAloGtMaxNull$ on the
nulls.

\subsection{The kernel and the cut-point rule}
\label{sec:kernelfallback}

The Gaussian kernel and the Laplace kernel $e^{-\|x-y\|/\sigma}$ are compared at the same bandwidth rule. The
Gaussian ramp cuts its witness between order statistics, and the Laplace ramp at genuine quantiles; on the same Laplace
fits the two cut rules give the same recovery in $\mnOrderLinearTies$ designs, slightly less with genuine quantiles in
$\mnOrderLinearLosses$, and the same real-data bounds, so the comparison is one of kernels. On all $\mnRecReps$
replicates of each design at the main size, at two paired standard errors, the Laplace ramp recovers more than the
Gaussian ramp in $\mnLapGaussWins$ designs, the $t_3$ tails and $\mnLapGaussWinsLine$ oscillating densities, as much in
$\mnLapGaussTies$, and less in $\mnLapGaussLosses$, the RHC-shaped design, overlap designs, a dense shift and the narrow
bump, by at most $\mnLapGaussLossMax$. Its median recovery is $\mnRecRampMedian$ against $\mnRecGaussMedian$, and
$\mnRecRampMedianNoNfour$ against $\mnRecGaussMedianNoNfour$ without the oscillating density with $\nu=4$, on which it
is $\mnRecNfourLap$ against $\mnRecNfourGauss$. On the first $\fbSyReps$ replicates of each design, the analytic
estimator certifies less with the Laplace kernel than with the Gaussian kernel on $\dvKerAnLapBelow$ designs and more
on $\dvKerAnLapAbove$; this comparison uses a threshold that is valid and nowhere larger than $\tau^{\mathrm P}$, the
minimum of $\tau^{\mathrm P}$, $\tau^{\mathrm S}$ and $\tau^{\mathrm P}$ with the tails of its sums inverted
numerically, with and without coupled kernel constants.

On the real data, every bound of Table~\ref{tab:realdata} uses the Laplace kernel. With the Gaussian kernel the
minima over the eight splits are $\mnRdChildGaussMin$, $\mnRdAgeGaussMin$ and $\mnRdBmiGaussMin$ on the child, adult age
and body-mass contrasts, the same as with the Laplace kernel on the two age contrasts and lower on body mass. On RHC
they are $\mnRdRhcFullGaussMin$ on all $\rrRhcNcovFull$ covariates, a run that cuts the Gaussian witness at genuine
quantiles, as for the Laplace kernel, against $\rrRhcFullMin$ with the Laplace kernel,
and $\mnRdRhcGaussMin$ on the $\rrRhcNcovPaper$ covariates stored as numbers, where the Laplace kernel also gives
$\rrRhcPaperMin$. Reporting the larger of the bounds of the two kernels would be valid only at level $2\delta$, by the
union bound.

With the own term removed, the cut points can cross at a level where the fitted witness still separates the samples,
and the batch is then skipped. Two other rules compare with this one as follows: plain empirical quantile cuts
whenever the own-term cuts cross, the fallback, and plain cuts always. On the first $\fbSyReps$ replicates of each
design, at two paired standard errors, the fallback wins on $\fbSyFallbackWins$ design, ties on $\fbSyFallbackTies$ and
loses on $\fbSyFallbackLosses$ against the own-term rule, and plain cuts always win on $\fbSyPlainWins$, tie on
$\fbSyPlainTies$ and lose on $\fbSyPlainLosses$. On the real data, with RHC on its $\rrRhcNcovPaper$ covariates stored as
numbers, the fallback with the Gaussian kernel bets where every split stops with no bet under the own-term rule, but
moves no minimum (largest wealth $e^{\fbRdRhcFallbackLogWMax}$
on RHC against $1/\delta=e^{\fbLogInvDelta}$), and with the Laplace kernel it lowers the medians on RHC and the adult age
contrast, from $\kRdRhcLaplaceOwnMed$ to $\kRdRhcLaplaceFallbackMed$ and from $\kRdAgeLaplaceOwnMed$ to
$\kRdAgeLaplaceFallbackMed$: where the own cuts cross in one batch only, the own rule bets on the other batch alone,
while the fallback also bets on the crossed batch and loses wealth there. No rule and no kernel was flagged at
$\alpha=\alpha^\star$ (exact binomial $p\ge0.01$).

\subsection{Recovery and further baselines}
\label{sec:baselines}

\begin{figure}[t]
\centering
\includegraphics[width=0.7\textwidth]{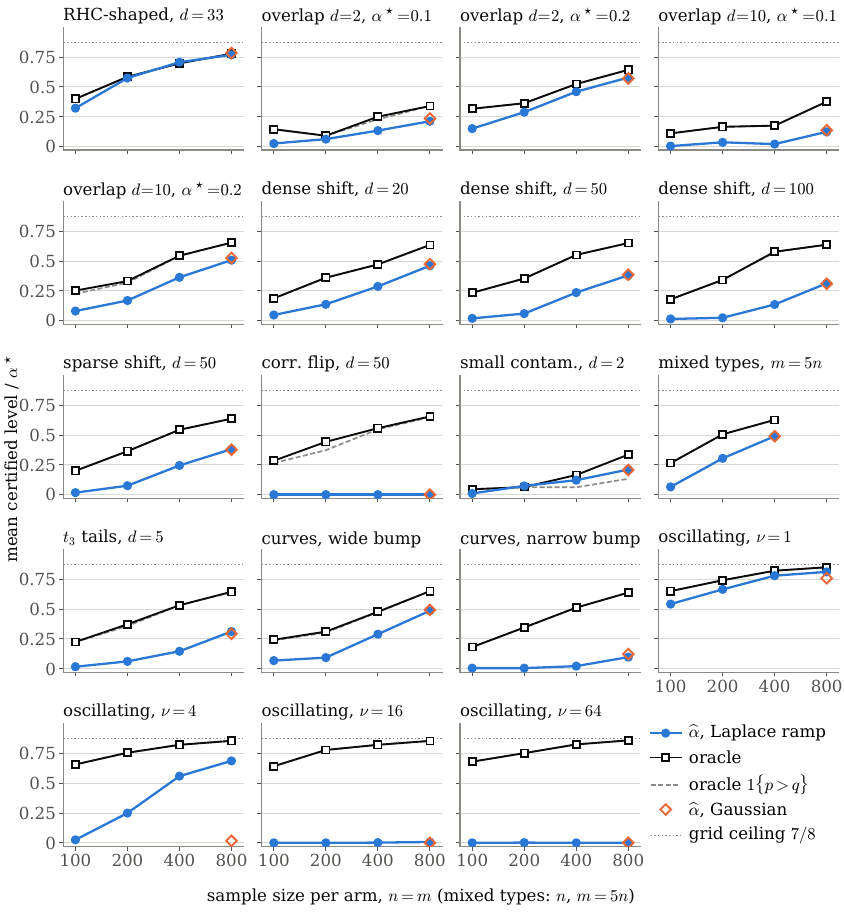}
\caption{Mean certified level over $\alpha^\star$ against $n=m$ (mixed types: $m=5n$): Laplace ramp, oracle reference,
betting on $\mathbf 1\{p>q\}$, and Gaussian ramp at the main size; $\rcReps$ replicates per size below $n=800$ (mixed
types: $400$), $\rcRepsMain$ at it; dotted: largest grid level below $\alpha^\star$.}
\label{fig:recoverycurves}
\end{figure}

Figure~\ref{fig:recoverycurves} follows the recovery of Figure~\ref{fig:recovery} against $n$, with $\rcReps$
replicates per size below $n=800$. Its oracle reference runs the same test, with the same batches, fractions
and grid, on a fixed function known only to an oracle, and reports the better of two choices: $\psi=\mathbf 1\{p>q\}$,
whose drift $\E_P\psi-\E_Q\psi=\alpha^\star$ is the largest of any $[0,1]$-valued function, and, at each level, the
level set $\{p/q>t\}$ whose drift has the largest ratio to its standard deviation, the quantity this betting rule
rewards to first order ($t$ from $\rcLevelsetMC$ draws of each law). Both tests are valid, but neither is a proven ceiling: the
rule is not log-optimal and $\psi$ pays for its noise, so on the small contamination at $n=800$, where $\psi$ halves the
common component, betting on it recovers $\rcSmallPsi$, the ramp $\rcSmallRamp$ and the best level set $\rcSmallLevelset$. At
$n=800$ the ramp reaches $0.8$ of the reference in $\rcEightyEightHundred$ designs (\rcEightyDesignsEightHundred), and the
tenth percentile of $\widehat\alpha/\alpha^\star$, median over the designs, is $\rcRampTenthMedEightHundred$, against
$\rcOracleTenthMedEightHundred$ for the reference. The grid still uses the known $\alpha^\star$.

Gradient boosting and logistic regression refitted on the batches of the ramp, which
removes the handicap of Table~\ref{tab:hardern80}, default or tuned by threefold cross-validation, bet their labels or
their probabilities corrected for the sample sizes; the best of them has more power against $H_{0,0}$ than the Gaussian
ramp in $\abBestMatchedAboveRampPow$ designs and less in $\abRampAboveBestMatchedPow$, and than the Laplace ramp in
$\mnPowBestClfAboveLap$ and $\mnPowLapAboveBestClf$, all at $n=m=\harderNmax$. With default settings the better of the
two classifiers recovers more than the Laplace ramp in $\mnBestMatchedAboveRamp$ designs and less in
$\mnRampAboveBestMatched$. Tuned logistic labels recover more than the Laplace ramp in
$\wtllosses{\mnWtlRecLogitTuned}$ designs and less in $\wtlwins{\mnWtlRecLogitTuned}$ at two paired standard errors,
and tuned gradient boosting labels recover more than it, at two decimals, in $\dvGbmTunedAboveLapN$ designs
(\dvGbmTunedAboveLapDesigns).
Probabilities lose to labels (median recovery $\abRecMedLogitMatchedProb$ against $\abRecMedLogitMatchedLabels$ for
logistic regression), as calibrated probabilities shrink toward $1/2$ on the common part. The normal of a $\nu$-support
vector machine on the Gram matrices of the ramp ($\nu$ by cross-validation, median $\abSvmNuMed$), bet as a ramp with
the cut points of the tested level, certifies less than the ramp except on $\abSvmAboveRamp$ designs. The untrimmed
witness of Section~\ref{sec:numharder} is the Gaussian witness fitted at $\alpha=0$ on the fitting batches and used at
every level, which needs no quadratic program; Section~\ref{sec:numharder} gives its recoveries, and cut at the tested
level it is equal to the Laplace ramp at two decimals in the remaining $\dvZeroCutEqualLap$ designs. A
fixed-sample empirical Bernstein bound on the ramp fitted on one half certified less than betting on the other half in
$\abSingleBetAboveEb$ of the $\abNdesigns$ designs and more in none. On the line, the Kolmogorov--Smirnov statistic
minus the two Dvoretzky--Kiefer--Wolfowitz radii, a lower bound on $\TV(P,Q)$, exceeded the better of the analytic
estimator and the Gaussian ramp in $\dvKsAboveKernel$ of $\dvKsCells$ cells of the oscillating designs and of a shift and
a change of scale of a $t_3$ law. Of further certificates, the trimmed-mean statistic of the witness
with a Dvoretzky--Kiefer--Wolfowitz bound certified nothing except on the oscillating density with $\nu=1$, and nearest
neighbors bet like the ramp certified less than it except on the oscillating densities; the trimmed-Wasserstein test of \citet{alvarez2012similarity}, asymptotic and pointwise, is not
compared. MMDAgg \citep{schrab2023aggregated} uses $\abAggL$ Gaussian bandwidths. None of these tests was flagged, at the boundary in any of the
$\abNdesigns$ designs or under $P=Q$ on the four nulls; the largest rates were $\abBoundaryMax$ ($\abBoundaryMaxK$ of
$\abReps$, logistic labels, RHC-shaped design) at the boundary and $\abNullMax$ (binomial $p=\abNullMaxP$, MMDAgg,
mixed-type null) under $P=Q$.

\subsection{Numerical checks on the line}
\label{sec:newnumerics}
\label{sec:excessmc}

For the shift $N(\pm1,1)$ and the mixtures $0.8\,N(0,1)+0.2\,N(\pm\mu,1)$, $\mu\in\{1.5,2,3\}$, which cross at $x_0=0$,
the rescaled path $\bar D_{\alpha^\star-\epsilon}$ of Section~\ref{sec:rates}, bracketed to certified accuracy between the separation
\eqref{eq:sep} in a computed direction and the norm of an explicit pair of trimmings, with the kernels $e^{-|x-y|}$ and
$e^{-(x-y)^2/2}$, lay at all $\mgRows$ levels above the lower bound of Proposition~\ref{prop:newapproxgen} and below the
explicit trimming of Proposition~\ref{prop:newupperone}. With the Laplace kernel its local log-log slope for the shift
reaches $\mgLapShiftSlope$ at $\epsilon=\mgLapShiftEps$, approaching $5/4$; with the Gaussian kernel and $\mu=3$ it falls
from $\mgGaussPathHi$ to $\mgGaussPathLo$ between $\epsilon=\mgGaussEpsHi$ and $\mgGaussEpsLo$, and a fit of
$\log\bar D=A+B\log\epsilon-c'/\epsilon$ gives $c'/\ell_0$ between $\mgFitPathLo$ and $\mgFitPathHi$ over $\mu$, with
$\ell_0=0.4\mu\varphi(\mu)$, below the constant $\pi^2/8\approx\mgPiEight$ of Theorem~\ref{thm:newgauss}.

The split-sample bound was computed on the shift and the mixture with $\mu=1.5$, which satisfy \eqref{eq:condR}, with
both kernels, $n=m$, the batches of Section~\ref{sec:betting}, own-term cut points interpolated between order
statistics, which Theorem~\ref{thm:laplacerate}, stated for genuine quantiles, does not cover, levels approaching the
known $\alpha^\star$ geometrically (from $\exGridS$ and $\exGridM$ below it to $\exGridMinS$ and $\exGridMinM$) and
$\exRepMin$ to $\exRepMax$ replicates per cell. With the Laplace kernel the mean error of $\widehat\alpha$ falls like $n^{-\exLapSlopeS}$
and $n^{-\exLapSlopeM}$ up to $n=\exNmax$ (least-squares slopes on $\log n$), as fast as that of the
Kolmogorov--Smirnov lower bound, $n^{-\exKSSlopeS}$ and $n^{-\exKSSlopeM}$, and is smaller by factors $\exKSRatioS$ and
$\exKSRatioM$ at $n=\exNmax$. The analytic estimator with the Laplace kernel and the threshold
$\min\{\tau^{\mathrm P},\tau^{\mathrm S}\}$ falls like $n^{-\exAnSlopeS}$ and $n^{-\exAnSlopeM}$; we prove no rate for
it, and with $\tau^{\mathrm S}$ alone its error is at least of order $n^{-2/5}$ (Proposition~\ref{prop:newconverse}).
With the Gaussian kernel the split-sample bound falls like $n^{-\exGaussSlopeS}$ and $n^{-\exGaussSlopeM}$ up to
$n=\exNmaxGauss$, against $n^{-\exLapSlopeCommonS}$ and $n^{-\exLapSlopeCommonM}$ for the Laplace kernel over the same
range.

\subsection{Real data, bandwidths and the cohort}
\label{sec:rdbw}

For RHC \citep{connors1996rhc}, the first contrast of Table~\ref{tab:realdata}, we use $\rrRhcNcovFull$ baseline
covariates, named as in \texttt{rhc.csv}. The $\rrRhcNcovPaper$ stored as numbers are the measurements \texttt{age},
\texttt{edu}, \texttt{surv2md1}, \texttt{das2d3pc}, \texttt{aps1}, \texttt{scoma1}, \texttt{meanbp1}, \texttt{wblc1},
\texttt{hrt1}, \texttt{resp1}, \texttt{temp1}, \texttt{pafi1}, \texttt{alb1}, \texttt{hema1}, \texttt{bili1},
\texttt{crea1}, \texttt{sod1}, \texttt{pot1}, \texttt{paco21}, \texttt{ph1} and \texttt{wtkilo1} and the history
indicators \texttt{cardiohx}, \texttt{chfhx}, \texttt{dementhx}, \texttt{psychhx}, \texttt{chrpulhx}, \texttt{renalhx},
\texttt{liverhx}, \texttt{gibledhx}, \texttt{malighx}, \texttt{immunhx}, \texttt{transhx} and \texttt{amihx}; the
$\rrRhcNcovAdded$ binary ones stored as text and coded $0/1$ are \texttt{sex} (female $1$), the do-not-resuscitate
status \texttt{dnr1} and the admission-diagnosis indicators \texttt{resp}, \texttt{card}, \texttt{neuro},
\texttt{gastr}, \texttt{renal}, \texttt{meta}, \texttt{hema}, \texttt{seps}, \texttt{trauma} and \texttt{ortho}
(yes $1$),
all standardized by the moments of the data each test fits on (batch $0$ of each split, the pilot for the analytic
test, the pooled sample for the cohort of Section~\ref{sec:appcohort}, which bets on nothing). We exclude the treatment, the dated or
outcome-derived variables (\texttt{death}, \texttt{dth30}, \texttt{t3d30} and the dates), \texttt{adld3p} and
\texttt{urin1}, missing for more than half of the patients, and the categorical variables \texttt{cat1}, \texttt{cat2},
\texttt{ca}, \texttt{ninsclas}, \texttt{race} and \texttt{income}, whose one-hot indicators would drive the median
bandwidth. The weight \texttt{wtkilo1} keeps the code $0$ that the data use for a missing weight; the zeros are a group of patients that the trimming may keep or discard like any other. The lower bounds hold for any longer vector, since total variation can only grow
with it, but the balance of Section~\ref{sec:appcohort} concerns the covariates used. Of these $\rrRhcNcovAdded$ covariates $\rrAddedSmdAbove$ have an unweighted standardized difference above $0.1$, up to
$\rrAddedSmdMax$, and two are rare ($\rrOrthoN$ and $\rrTraumaN$ patients coded $1$). On the $\rrRhcNcovPaper$ covariates
the averaged e-value certifies $\rrEvRhcAvgFinal$, the analytic baseline again no positive level, and the Laplace cohort at
$\alpha=\rrCohPaperAlpha$ had largest standardized difference $\rrCohPaperSmdMax$ on them but $\rrCohPaperOmitSmdMax$
on the others, and mortality differences $\rrCohPaperRD$ for the trimming and $\estKernelRD$ to $\estAtmRD$ for the
other weightings of Table~\ref{tab:cohort}.

The propensity estimate of Section~\ref{sec:appcohort} uses a ridge logistic fit, refitted on each bootstrap
resample; fivefold cross-fitting gives $\rrPropFullPlugCf$, and on the $\rrRhcNcovPaper$ covariates it is
$\rrPropPaperPlug$. Seeing the arms only through the fitted score, a
misspecified model tends to understate $\alpha^\star$, while fitting in sample overstates it. The classifiers of
Section~\ref{sec:data} are those of Section~\ref{sec:baselines}; their labels certify $\rrClfFullLogitHardMin$
(logistic) and $\rrClfFullGbmHardMin$ (boosting) with default settings and $\rrClfFullLogitTunedHardMin$ and
$\rrClfFullGbmTunedHardMin$ tuned, and on the $\rrRhcNcovPaper$ covariates $\rrClfPaperLogitHardMin$,
$\rrClfPaperGbmHardMin$, $\rrClfPaperLogitTunedHardMin$ and $\rrClfPaperGbmTunedHardMin$. Their probabilities, corrected
for the sample sizes as in Section~\ref{sec:baselines}, certify $\rrClfFullLogitSoftMin$ and $\rrClfFullGbmSoftMin$
with default settings and $\rrClfFullLogitTunedSoftMin$ and $\rrClfFullGbmTunedSoftMin$ tuned, and on the
$\rrRhcNcovPaper$ covariates $\rrClfPaperLogitSoftMin$, $\rrClfPaperGbmSoftMin$, $\rrClfPaperLogitTunedSoftMin$ and
$\rrClfPaperGbmTunedSoftMin$.

For NHANES \citep{nhanes2012,john2019mims} we pool the cycles 2011--2012 and 2013--2014, reconstruct the
clock time of each hourly record, drop invalid and quality-flagged hours, and require at least $20$ hours of wear on at
least four days; the profiles are used without standardization and without survey weights.
The public files carry no household identifier. With whole masked variance units (SDMVPSU within SDMVSTRA,
$\rrClChildUnits$ units) assigned at random to the batches, which makes the fitting batches independent of the batches
bet on if the units are, the minima are $\rrClChildMin$, $\rrClAgeMin$ and $\rrClBmiMin$ for the child, adult age and
body-mass contrasts, against $\mnRdChildMin$, $\mnRdAgeMin$ and $\mnRdBmiMin$ with participants assigned to batches.

In Table~\ref{tab:realdata} the pilot of each group is a $30\%$ random subsample (seed fixed in advance), the
scales are $0.1$, $0.15$, $0.2$, $0.3$, $0.5$, $0.7$ and $1$ times the pilot median, the scale whose analytic bound on the pilot is largest is chosen, ties going to the scale closest to $0.5$, and the chosen
scales, under the threshold $\tau^{\mathrm P}$, are $\mnPvRhcScale$ for RHC, $\mnPvChildScale$ for the child contrast,
$\mnPvAgeScale$ for the adult age contrast and $\mnPvBmiScale$ for body mass. In the analytic columns of
Table~\ref{tab:realdata} nothing the test uses depends on the held-out $70\%$, so the theorems cover it.

For the cohort of Section~\ref{sec:appcohort}, with the Gaussian kernel at the pooled median
distance $\lcPoolSigma$, the program at $\alpha=\lcCohAlpha$ is solved by
accelerated projected gradient followed by $\lcPoolSmoSteps$ steps of sequential minimal optimization, to a duality
gap of $\lcPoolGap$; the smallest eigenvalue of the Gram matrix is $\lcPoolLamMin$, so the minimizer is unique, and
the gap bounds the distance of the computed weights to it by $\lcPoolLtwoBound$ and the error of the witness by
$\lcPoolWitnessBound$, which places every patient on the same side of the cut as at the exact minimizer. The
$\lcPoolFracX$ and $\lcPoolFracY$ fractionally weighted patients carry fractions $\lcPoolFracMassX$ and
$\lcPoolFracMassY$ of the trimmed samples, and their witness values agree to within $\lcPoolFracSpreadX$. The summaries of
Table~\ref{tab:cohort} for the trimming row are those of this minimizer. Standardized by batch $0$ of one split
instead, the minimizer retains $\lcCohKeptX$ and $\lcCohKeptY$ patients, and over the eight splits of
Table~\ref{tab:realdata} between $\lcSensKeptXlo$ and $\lcSensKeptXhi$ catheterized and between $\lcSensKeptYlo$ and
$\lcSensKeptYhi$ other patients, with the largest standardized difference between $\lcSensSmdMaxLo$ and
$\lcSensSmdMaxHi$. Table~\ref{tab:cohort} compares the trimmed cohort with five weightings that differ in target as much as in method:
matching and overlap weights target overlap populations like the common core, energy and kernel balancing the pooled one.

\begin{table}[!htb]
\centering
\footnotesize
\setlength{\tabcolsep}{1.9pt}
\begin{tabular}{llcccccc}
\toprule
RHC & target & \multicolumn{2}{c}{effective fraction} & max $|$SMD$|$ & mean $|$SMD$|$ & overlap & risk diff. \\
 & population & cath. & others & & & & (95\% int.) \\
\midrule
all patients & -- & 1.00 & 1.00 & 0.501 & 0.160 & 0.58 & 0.074 (0.047, 0.098) \\
trimming, $\alpha=\lcCohAlpha$ & common core & 0.66 & 0.65 & 0.032 & 0.008 & 0.81 & 0.040 (0.011, 0.072) \\
\addlinespace
matching weights & overlap (ATM) & 0.84 & 0.62 & 0.012 & 0.004 & 0.93 & 0.055 (0.028, 0.085) \\
overlap weights & overlap (ATO) & 0.83 & 0.68 & 0.000 & 0.000 & 0.93 & 0.053 (0.024, 0.081) \\
\addlinespace
energy balancing & pooled (ATE) & 0.51 & 0.77 & 0.018 & 0.004 & 0.90 & 0.047 (0.013, 0.080) \\
kernel balancing & pooled (ATE) & 0.25 & 0.47 & 0.017 & 0.005 & 0.86 & 0.056 (0.011, 0.102) \\
\bottomrule
\end{tabular}
\space
\caption{The RHC arms under six weightings: trimming with the Gaussian kernel at the level of the Laplace lower bound; matching
\citep{ligreene2013matching} and overlap weights \citep{li2018balancing} from a logistic propensity model; energy
\citep{huling2024energy} and kernel balancing \citep{hazlett2020kernel,kallus2020optimal} of each arm to the pooled
sample. Effective fraction: Kish effective sample size over arm size; absolute standardized mean differences (SMD) over
the $\rrRhcNcovFull$ covariates; overlap: overlapping coefficient of the propensity distributions; risk difference:
$30$-day mortality, catheterized minus others, $95\%$ percentile interval from $\lcBoot$ bootstrap resamples within each
arm, weights fixed.}
\label{tab:cohort}
\end{table}

\subsection{Computation, code and data}
\label{sec:computationdetails}

Throughout Section~\ref{sec:numerical} we solve \eqref{eq:qp} as in Algorithm~\ref{alg:split} and evaluate
\eqref{eq:alphaest} on its grid, with the threshold of Section~\ref{app:experiments}. The iterates are feasible, so
their value bounds $\widehat D_\alpha^{\,2}$ from above, and the Frank--Wolfe gap bounds it from below; a level is
rejected only when the lower bound exceeds $\tau(\alpha)^2$. Every level rejected in Section~\ref{sec:data} is at a converged solve, and an
unconverged solve changes only the direction of the split-sample estimator, not its level. Where
the data do not determine the minimizer, we return the minimizer $\bar z$ of the strictly convex
$G(w,v)=\|w-\tfrac1n\mathbf 1\|_2^2+\|v-\tfrac1m\mathbf 1\|_2^2$ over $\mathcal T_\alpha$, the limit as
$\lambda\downarrow0$ of the minimizers of $F+\lambda G$, with $F$ the objective of \eqref{eq:qp} (every cluster point
lies in $\mathcal T_\alpha$ and has $G\le G(\bar z)$). On one core, a replicate of the synthetic study at
$n=\harderNmax$ took $\harderRunAnalyticEight$ seconds for the analytic test and $\harderRunSplitEight$ for the betting
test, and one of all the tests of $H_{0,0}$ at $n=m=\harderNcap$ about $\harderRunExtCap$, $\harderRunExtKernelCap$ of
them for kernel sums computed in blocks.

The RHC data are \url{https://hbiostat.org/data/repo/rhc.csv}; the NHANES data are the files PAXHR, PAXHD, DEMO and BMX
of the cycles G and H (2011--2014) of the National Center for Health Statistics. The repository
\url{https://github.com/HristoInouzhe/trimmed-mmd} holds the package \texttt{trimmedmmd} and one script for each table
and figure, and \texttt{scripts/fetch\_data.py} downloads the data. The Laplace ramp and the analytic test of
Table~\ref{tab:hardern80} and Figure~\ref{fig:recovery} come from \texttt{scripts/aos\_main\_synth.py},
\texttt{aos\_main\_alpha0.py} and \texttt{aos\_main\_analytic.py}, the classifiers on the batches of the ramp, the
permutation test and MMDAgg from \texttt{aos\_baselines\_run.py}, the Gaussian ramp and the classifiers trained once
from the study in \texttt{explore/harder\_sims}, and the tables and the figure from \texttt{aos\_main\_tables.py},
\texttt{aos\_baselines\_tables.py} and \texttt{aos\_main\_figs.sh}; Table~\ref{tab:realdata} comes from
\texttt{aos\_main\_realdata.py} and \texttt{exp\_valid\_bandwidth.py}, Table~\ref{tab:cohort} from
\texttt{aos\_main\_cohort.py}, Figure~\ref{fig:excess} from \texttt{aos\_fig\_excess.py},
Figure~\ref{fig:recoverycurves} from \texttt{aos\_ref\_recovery\_run.py} and \texttt{aos\_ref\_recovery\_fig.py},
Section~\ref{sec:kernelfallback} from \texttt{aos\_fallback\_synth.py}, \texttt{aos\_fallback\_realdata.py} and
\texttt{aos\_fallback\_tables.py}, Section~\ref{sec:excessmc} from \texttt{aos\_margin\_macros.py} and \texttt{aos\_excess\_table.py}, and the boundary
calibration of Section~\ref{app:experiments} from \texttt{aos\_supp\_threshold.py}. The simulated data are drawn from
fixed seeds. The runs also read the per-replicate records of the study in \texttt{explore/harder\_sims/out}, which the
repository holds: \texttt{scripts/aos\_main\_synth.py} takes its bandwidths from them, and
\texttt{scripts/aos\_main\_tables.py} the Gaussian ramp and the classifiers trained once. So the repository, not the
seeds alone, reproduces the runs.
\space

\FloatBarrier
\makeatletter
\let\toclevel@section\unified@toclevel@section
\let\toclevel@subsection\unified@toclevel@subsection
\let\toclevel@subsubsection\unified@toclevel@subsubsection
\makeatother

\begin{acks}[Acknowledgments]
The author thanks Eustasio del Barrio and Carlos Matr\'an for their comments after a talk on this work in Valladolid,
which helped sharpen Sections~\ref{sec:intro} and~\ref{sec:rates}. AI, primarily Claude Opus 5.5 and GPT 5.6 SOL, was used in developing the theory, software and numerical
experiments. The author formulated the research questions, assessed the results and directed their practical
implementation. The numbered results are formalized in Lean~4; the repository
\url{https://github.com/HristoInouzhe/trimmed-mmd-lean} maps each to its formal statement. The author takes full responsibility for the paper.
\end{acks}

\begin{funding}
The author was supported by the Spanish MCyT grants PID2023-148081NB-I00 and PID2024-162240NB-I00.
\end{funding}

\begin{supplement}
\stitle{Code and data}
\sdescription{The package \texttt{trimmedmmd} and the scripts behind every table and figure, at
\url{https://github.com/HristoInouzhe/trimmed-mmd}.}
\end{supplement}
\space

\space

\end{document}